\documentclass[a4paper, 11pt, twoside, onecolumn, openright, final]{memoir}

\isopage
\setbinding{0cm}
\setlrmargins{*}{*}{1.0}
\checkandfixthelayout
\chapterstyle{verville}
\captionnamefont{\small\bfseries}
\captiontitlefont{\small}
\maxtocdepth{subsection}
\maxsecnumdepth{subsection}

\usepackage[T1]{fontenc}
\usepackage[ngerman,english]{babel}
\usepackage{csquotes}
\usepackage{graphicx}
\usepackage{placeins}
\usepackage{afterpage}
\usepackage{amsmath}
\usepackage{amssymb}
\usepackage{mathtools}
\usepackage{bbold}
\usepackage[backend=biber, style=numeric-comp, sorting=none, maxbibnames=8, maxcitenames=8, giveninits=true]{biblatex}
\usepackage{hyperref}
\usepackage[capitalise]{cleveref} 

\MakeOuterQuote{"}
\allowdisplaybreaks[1]
\counterwithout{footnote}{chapter}
\abstractintoc

\hypersetup{colorlinks=false,citecolor=magenta,urlcolor=blue,linkcolor=blue,pdftitle={(3+1)D dilute Glasma initial conditions in simulations of heavy-ion collisions},pdfauthor={Kayran Schmidt}}

\DeclareFieldFormat[misc]{title}{\mkbibquote{#1}}

\DeclareMathAlphabet\mathbfcal{OMS}{cmsy}{b}{n}

\newcommand{\bperp}{\mathbf b}
\newcommand{\xperp}{\mathbf x}
\newcommand{\yperp}{\mathbf y}
\newcommand{\zperp}{\mathbf z}
\newcommand{\uperp}{\mathbf u}
\newcommand{\vperp}{\mathbf v}
\newcommand{\wperp}{\mathbf w}
\newcommand{\pperp}{\mathbf p}
\newcommand{\qperp}{\mathbf q}
\newcommand{\kperp}{\mathbf k}
\newcommand{\lperp}{\mathbf l}
\newcommand{\kappaperp}{\boldsymbol \kappa}
\newcommand{\zeroperp}{\mathbf 0}
\newcommand{\deltaperp}{\delta^{(2)}}
\newcommand{\Sp}{\mathtt{S}_\perp}
\newcommand{\Sop}{\mathtt{S}^\between_\perp}
\newcommand{\Zp}{\mathtt{Z}_\perp}
\newcommand{\Zop}{\mathtt{Z}^\between_\perp}

\newcommand{\xpm}{x^\pm}
\newcommand{\xmp}{x^\mp}
\newcommand{\ypm}{y^\pm}
\newcommand{\ymp}{y^\mp}
\newcommand{\zpm}{z^\pm}

\newcommand{\upm}{u^\pm}
\newcommand{\vpm}{v^\pm}

\newcommand{\pmp}{p^\mp}
\newcommand{\ppm}{p^\pm}
\newcommand{\qmp}{q^\mp}
\newcommand{\qpm}{q^\pm}
\newcommand{\kmp}{k^\mp}
\newcommand{\kpm}{k^\pm}
\newcommand{\lmp}{l^\mp}

\newcommand{\bi}{\mathbf i}
\newcommand{\bj}{\mathbf j}
\newcommand{\bk}{\mathbf k}
\newcommand{\bl}{\mathbf l}

\newcommand{\Y}{\mathtt{Y}}

\newcommand{\A}{\mathrm{A}}
\newcommand{\B}{\mathrm{B}}

\newcommand{\Gret}{G_\mathrm{ret}}
\newcommand{\tGret}{\tilde{G}_\mathrm{ret}}

\newcommand{\xB}{\mathbb{x}}
\newcommand{\xG}{\xB G}
\newcommand{\xGeff}{\xB G^\mathrm{eff}}

\newcommand{\tBeff}{\tilde{B}^\mathrm{eff}}

\newcommand{\Ep}{E_\perp}
\newcommand{\ep}{\varepsilon_\perp}
\newcommand{\elrf}{\epsilon_\mathrm{LRF}}

\newcommand{\ee}{\mathrm{e}} 
\newcommand{\ii}{\mathtt{i}} 
\newcommand{\dd}{\mathrm{d}}
\newcommand{\nn}{\nonumber}

\DeclareMathOperator\artanh{artanh}
\DeclareMathOperator\arcosh{arcosh}

\DeclareMathOperator\tr{tr}

\newcommand{\WSxi}{\text{WS-}\zeta}
\newcommand{\Gxi}{\text{Gauss-}\xi}

\renewenvironment{abstract}
  {\phantomsection\addcontentsline{toc}{chapter}{Abstract}
  \begin{center}
   \textbf{Abstract}
   \end{center}}
  {\cleardoublepage}
  
\newenvironment{kurzfassung}
  {\phantomsection\addcontentsline{toc}{chapter}{Kurzfassung}
  \begin{otherlanguage}{ngerman}
  \begin{center}
   \textbf{Kurzfassung}
   \end{center}}
  {\end{otherlanguage}\cleardoublepage}

\newenvironment{acknowledgments}
  {\phantomsection\addcontentsline{toc}{chapter}{Acknowledgments}
  \begin{center}
   \textbf{Acknowledgments}
   \end{center}}
  {\cleardoublepage}

\begin{document}

\frontmatter
\thispagestyle{empty}

    \begin{center}
	\vspace*{0.5cm}%
	{\LARGE DISSERTATION\\[1.0cm]}
	{\LARGE\textbf{(3+1)D dilute Glasma initial conditions in simulations of heavy-ion collisions}\\[1.0cm]}
	\end{center}
	
	\begin{center}    
		{\normalsize ausgef\"uhrt zum Zwecke der Erlangung des akademischen Grades eines\\
        Doktors der Technischen Wissenschaften unter der Leitung von}\\[0.8cm]
	\end{center}
	\begin{center}    
		{\normalsize Privatdoz.\ Dipl.-Ing.\ Dr.techn.\ Andreas Ipp\\
			Institut f\"ur Theoretische Physik,\\
			Technische Universit\"at Wien, \"Osterreich \\[0.8cm]}
	\end{center}
 \begin{center}    
		{\normalsize mitbetreut durch}\\[0.8cm]
	\end{center}
 \begin{center}    
		{\normalsize Dipl.-Ing.\ Dr.techn.\ David I. M\"uller, BSc\\
			Institut f\"ur Theoretische Physik,\\
			Technische Universit\"at Wien, \"Osterreich \\[0.8cm]}
\end{center}
	\begin{center}    
		{\normalsize eingereicht an der Technischen Universit{\"a}t Wien,\\
			Fakult{\"a}t f{\"u}r Physik} \\[0.8cm]     
	\end{center} 
	\begin{center}    
		{\normalsize von\\[0.1cm]
			\textbf{Dipl.-Ing.\ Kayran Schmidt, BSc BSc}}\\[0.1cm]
   Matrikelnummer 01604789\\[2.5cm]
	\end{center} 
	\begin{center} 
	\noindent\begin{tabular}{llr}
		{Wien, am \makebox[2.5cm]{\hrulefill}}&\hspace*{4cm} & \makebox[4cm]{\hrulefill}  \\
	\end{tabular}\\
    \null\vfill\null
	\leftskip -1.7cm
     \begin{tabular}{p{\textwidth/4+1cm}@{}p{\textwidth/4+1cm}@{}p{\textwidth/4+1cm}@{}p{\textwidth/4+1cm}@{}}
		\makebox[\textwidth/4]{\hrulefill} & \makebox[\textwidth/4]{\hrulefill} & \makebox[\textwidth/4]{\hrulefill}  & \makebox[\textwidth/4]{\hrulefill}\\
		Dr.\ Andreas Ipp&  Dr.\ David I. M\"uller&  Dr.\ Tuomas Lappi & Dr.\ Sangyong Jeon\\
		(Hauptbetreuer)  &  (Zweitbetreuer) & (Gutachter) & (Gutachter)  \\
	\end{tabular}
	\end{center}

 \cleardoublepage

\begin{abstract}
\noindent
The theoretical description of the initial stages of relativistic heavy-ion collisions is based on effective theories that capture the relevant degrees of freedom of the underlying fundamental theory of Quantum Chromodynamics (QCD).
The colliding nuclei in the initial state are modeled by the Color Glass Condensate, where the valence color charges of the nuclei are described as recoilless, classical color currents that source strong, classical color fields.
The interaction of these fields gives rise to the dynamical Glasma stage, which evolves according to the non-Abelian classical Yang-Mills equations.
The established procedure for extracting the properties of the Glasma relies on resource-intensive numerical simulations and the simplification of the boost-invariant limit.
This reduces the system to (2+1)D, where the structure along rapidity is not included.

In this thesis, an approximation for the full (3+1)D dynamics of the Glasma is presented, which breaks boost-invariance on the level of the nuclear fields and leads to rapidity dependence in the final results.
For this treatment, the Yang-Mills equations are linearized in covariant gauge, where lower-order, nonlinear contributions are neglected and the dynamics are captured by the (3+1)D dilute Glasma.
The analytic solutions of the (3+1)D dilute Glasma are derived in both position and momentum space formulations, providing a comprehensive understanding of the involved (3+1)D dynamics.

In position space, the field strength tensor results from the integration of free-streaming gluons that are produced in $2\rightarrow1$ scattering processes where the initial nuclear fields overlap.
An efficient numerical implementation is developed to calculate the energy-momentum tensor of the (3+1)D dilute Glasma on an event-by-event basis.
This enables state-of-the-art simulation frameworks for heavy-ion collisions to utilize a rapidity-dependent description of the Glasma stage.

In momentum space, the event-averaged gluon number distribution for the (3+1)D dilute Glasma is derived in Coulomb gauge.
The dynamics correspond to free streaming gluons at asymptotically late times.
The resulting expression generalizes the established $k_T$-factorization formula used in perturbative QCD calculations with genuine rapidity dependence from the longitudinal structure of the nuclei.

A generalized, three-dimensional McLerran-Venugopalan nuclear model is developed for nuclei with realistic envelopes and intrinsic longitudinal correlations.
The physical viability of the resulting color charge correlator is carefully examined in terms of the envelope and correlation scales.
Numerical results are presented for the novel spacetime and momentum rapidity structure of the energy-momentum tensor, the gluon number distribution, and the transverse energy of the (3+1)D dilute Glasma.
In position space, the extended longitudinal collision geometry and finite longitudinal correlation length break boost-invariance.
In momentum space, the results each follow universal parametrizations and are fixed by the values of two scaling parameters, each of which combines a transverse momentum scale with the longitudinal correlation length scale.
Furthermore, the numerical results exhibit limiting fragmentation where the rapidity profiles approach a limiting distribution at large rapidities.
This feature is also derived locally in position space for the analytic expressions of the field strength tensor and, in momentum space, for the transverse energy of the (3+1)D dilute Glasma.

\end{abstract}

\begin{kurzfassung}
\noindent
Die theoretische Beschreibung der frühen Stadien relativistischer Schwerionenkollisionen basiert auf effektiven Theorien, die die relevanten Freiheitsgrade der zugrunde liegenden fundamentalen Theorie der Quantenchromodynamik (QCD) erfassen.
Die kollidierenden Atomkerne im Anfangszustand werden durch das „Color Glass Condensate“ modelliert.
Dabei bilden die Valenz-Farbladungen der Atomkerne klassische Ströme, die ohne Rückstoß als Quellen von starken Farbfeldern agieren.
Die Wechselwirkung dieser Felder führt zur Dynamik im Stadium des Glasma, das sich als Lösung der nichtabelschen klassischen Yang-Mills-Gleichungen ergibt.
Das etablierte Verfahren zur Bestimmung der Eigenschaften des Glasma beruht auf ressourcenintensiven numerischen Simulationen und der Vereinfachung des boostinvarianten Grenzfalls.
Dies führt zu einer Reduktion des Systems auf (2+1)D, wobei die Struktur entlang der Rapidität nicht berücksichtigt wird.

In dieser Dissertation wird eine Approximation für die vollständige (3+1)D Dynamik des Glasma vorgestellt.
Sie führt zur Brechung der Boostinvarianz auf Ebene der Kernfelder und stellt die Rapiditätsabhängigkeit der Endergebnisse wieder her.
Dafür werden die Yang-Mills-Gleichungen in kovarianter Eichung linearisiert, sodass nichtlineare Beiträge niedrigerer Ordnung vernachlässigt werden.
Die Dynamik dieses Systems wird durch das (3+1)D dünne Glasma erfasst.
Die analytischen Lösungen des (3+1)D dünnen Glasma werden sowohl im Orts- als auch im Impulsraum hergeleitet und ermöglichen ein umfassendes Verständnis der (3+1)D Dynamik.

Im Ortsraum werden frei strömende Gluonen bei $2\rightarrow1$ Stoßprozessen im Gebiet, wo die Kernfelder überlappen, erzeugt und zum Feldstärketensor aufintegriert.
Der Energie-Impuls-Tensor des (3+1)D dünnen Glasma wird dann für einzelne Kollisionen mit einem effizienten numerischen Verfahren ausgerechnet.
Dies ermöglicht es, dass modernste Simulationsprogramme für die Berechnung von Schwerionenkollisionen die Rapiditätsabhängigkeit im Stadium des Glasma berücksichtigen können.

Im Impulsraum wird die Verteilung der Gluonenzahlen des (3+1)D dünnen Glasma in Coulomb-Eichung als Mittel über viele Kollisionen hergeleitet.
Die Dynamik entspricht frei strömenden Gluonen zu asymptotisch späten Zeiten.
Im Vergleich zu etablierten Ergebnissen im $k_T$-faktorisierten Ansatz der perturbativen QCD ergibt sich eine Generalisierung der Rapiditätsabhängigkeit, die direkt auf die longitudinale Struktur der Kerne zurückgeführt werden kann.

Als Beschreibung der Atomkerne wird das McLerran-Venugopalan-Modell um realistische dreidimensionale Kernprofile und intrinsische longitudinale Korrelationen erweitert.
In diesem Modell sind die klar definierten und separierten Skalen des Profils und der Korrelationslänge entscheidend, um die physikalische Anwendbarkeit zu gewährleisten.
Es werden numerische Ergebnisse für die Rapiditätsprofile des Energie-Impuls-Tensors, der Verteilung der Gluonenzahlen und der transversalen Energie der Gluonen des (3+1)D-dünnen Glasma präsentiert.
Die Boostinvarianz im Ortsraum wird durch die ausgedehnte longitudinale Geometrie der Kollision und die endliche longitudinale Korrelationslänge gebrochen.
Im Impulsraum folgen die Kurven universellen Parametrisierungen und skalieren jeweils mit einem Parameter, der eine transversale Impulsskala und die longitudinale Korrelationslänge kombiniert.
Außerdem zeigen die numerischen Ergebnisse „limiting fragmentation“.
Dabei folgen die Kurven bei hohen Rapiditäten einer limitierenden Verteilung.
Diese Eigenschaft des (3+1)D dünnen Glasma wird lokal im Ortsraum für den Feldstärketensor und im Impulsraum für die transversale Energie analytisch hergeleitet.

\end{kurzfassung}
\begin{acknowledgments}
\noindent
I thank my supervisor, Andreas Ipp, for his mentorship and support since my undergraduate studies.
I am equally indebted to my co-supervisor, David I.\ M\"uller, for countless discussions and late-night sessions.
Without him, my PhD would not have been possible.
I owe special thanks to my office mate and dear friend, Markus Leuthner.
Working together with him was truly fulfilling and there was always a plan for how to tackle the next problem.
I further thank my collaborators S\"oren Schlichting and Pragya Singh for many insightful discussions and our joint publications.
\newline

\noindent
I am grateful for the welcoming atmosphere at our institute and for the many lunches with my colleagues.
I will not forget the light‑hearted conversations with Ali, Florian, Florian, Jonas, Liane, Ludwig, Paul, and Thomas.
Certainly, I thank my friend Florian Lindenbauer for various technical and inspiring discussions.
I had the privilege of meeting many wonderful people during my travels and sharing personal memories with fellow PhD students and postdocs, including Adam, Carlisle, Carlos, Dana, Jarrko, Rachel, and Patricia.
\newline

\noindent
I want to express my deepest gratitude to my parents for their unconditional support.
\newline

\noindent
My research was funded by the Austrian Science Fund FWF No.\ P34764\@.
I was a fellow of the Doktoratskolleg Particles and Interactions (DK-PI, FWF doctoral program No.\ W1252) and I am grateful for the summer schools and retreats I could attend.
I acknowledge funding for travel expenses by the TU International Office.
\newline

\noindent
My research would not have been possible without the services provided by the ASC infrastructure.
I thank the ASC staff for managing the cluster and cleaning up the aftermath I left behind.
I acknowledge the generous hardware gifts received from the ASC, which allowed me to set up our working group with powerful servers.

\end{acknowledgments}

\tableofcontents
\cleardoublepage
\listoffigures
\cleardoublepage
\listoftables
\cleardoublepage

\mainmatter

\chapter{Introduction}\label{sec:introduction}

The theoretical treatment of relativistic heavy-ion collisions (HICs) is anchored on the fundamental theory of quantum chromodynamics (QCD)~\cite{Brambilla:2014jmp,Gross:2022hyw,Arslandok:2023utm} that describes the strong interaction.
As part of the Standard Model of particle physics, the elementary particles that carry the color charge of the strong interaction are the quarks and gluons (together partons).
The gluons are the force-carrying bosons that mediate the interactions between the partons.
The mathematical formulation of QCD incorporates the gauge symmetry of the non-Abelian gauge group $SU(3)$, which results in the property of asymptotic freedom.
Here, the strong coupling constant $\alpha_s$ decreases when the energy scale of the momentum exchange between the interacting particles is increased, and QCD becomes weakly coupled in the high-energy limit.
In the opposite limit, the vacuum ground state is characterized by confinement, where quarks are strongly bound into hadrons.
In HIC experiments, which were performed, for example, at the Relativistic Heavy Ion Collider (RHIC)~\cite{PHENIX:2004vcz,BRAHMS:2004adc,STAR:2005gfr} and the Large Hadron Collider (LHC)~\cite{Roland:2014jsa,ALICE:2022wpn,CMS:2024krd}, nuclear matter is studied under extreme conditions and at very high energies.
This allows to break apart the hadrons into their constituent quarks and gluons.

Over the course of the last decades, experimental and theoretical research led to an overarching understanding of the matter produced during HICs\@.
Different effective theories for the strong interaction are applied to distinct evolutionary stages of the collision~\cite{Heinz:2013wva,Pasechnik:2016wkt,Busza:2018rrf,Bzdak:2019pkr,Adolfsson:2020dhm,Elfner:2022iae,Du:2024wjm,Busza:2025gpg}.
A major milestone on this journey was the announcement of the discovery of the quark-gluon plasma (QGP)~\cite{Heinz:2000bk,Heinz:2002gs,Gyulassy:2004zy}, an exotic state of matter where quarks and gluons propagate freely on length scales comparable to the size of the system.
Still, the QGP is strongly coupled and behaves according to relativistic viscous hydrodynamics~\cite{Bass:1998vz,Shuryak:2004cy,Muller:2025qof,Wang:2025lct} with a very low shear viscosity to entropy density ratio comparable to an almost perfect fluid.
The hydrodynamic evolution of the QGP leaves imprints on experimentally measurable particle distributions.
The collective behavior of the system, dominated by transverse (to the beam) and longitudinal expansion, predicts particular flow patterns that are identified via the coefficients of the Fourier decomposition of the azimuthal particle distribution~\cite{Heinz:2013th,Dusling:2015gta,Schlichting:2016sqo,Schenke:2017bog,Altinoluk:2020wpf}.
For example, off-central collisions lead to the typical ``almond'' shape of the fireball in the transverse plane and elliptic flow develops as a response to the azimuthal anisotropy in the initial state.

The hydrodynamic modeling of the QGP introduces model parameters that characterize the behavior of the system~\cite{Kolb:2003dz,Gale:2013da,Jeon:2015dfa,Schenke:2021mxx,Rocha:2023ilf,Heinz:2024jwu,Weller:2017tsr}.
The aforementioned shear viscosity is part of a set of transport coefficients that arise from higher orders in the gradient expansion scheme.
Different implementations of various schemes are used throughout the literature.
Among them are MUSIC~\cite{Gale:2012rq,Schenke:2010nt,Schenke:2010rr,Schenke:2011bn,McDonald:2016vlt,Schenke:2019pmk,Schenke:2020mbo}, VISHNU~\cite{Shen:2014vra}, EPOS~\cite{Werner:2010aa}, Trajectum \cite{Nijs:2020roc}, and EKRT~\cite{Niemi:2015qia,Kuha:2024kmq,Hirvonen:2024zne}.
Their success lies in their ability to predict the experimentally obtained particle distributions for a wide range of HIC setups by calibrating the transport coefficients once to a single setup~\cite{Molnar:2014zha,Bernhard:2015hxa,Bernhard:2016tnd,Bernhard:2019bmu,Denicol:2015nhu,Denicol:2014vaa,Denicol:2012cn,Ollitrault:1992bk}.

However, the theoretical description of HICs with relativistic hydrodynamics relies on accurate initial conditions to provide the distribution of energy density.
On the one hand, energy deposition can be derived from the positions of the participant nucleons in the transverse plane.
This approach is realized, for example, in the Monte Carlo Glauber~\cite{Miller:2007ri,dEnterria:2020dwq,Alver:2008aq,Shen:2020jwv} or the T\raisebox{-.5ex}{R}ENTo~\cite{Moreland:2014oya} frameworks.
The strength of these models lies in their accurate geometric treatment of the colliding nuclei while taking event-by-event statistics into account.

On the other hand, the Color Glass Condensate (CGC) effective theory~\cite{Iancu:2003xm,Iancu:2012xa,Gelis:2010nm,Gelis:2012ri,Fukushima:2011ca,Fukushima:2016xgg,Gelis:2021zmx,Garcia-Montero:2025hys} introduces a dynamical stage in the initial condition of HICs.
The valence partons of the relativistic nuclei give rise to highly saturated, classical gluon fields, which evolve according to the classical Yang-Mills (YM) equations into a state called Glasma~\cite{Lappi:2006fp,Dumitru:2008wn,Chen:2013ksa}.
The successful treatment of these early-stage dynamics led to initial conditions facilitated by frameworks such as the IP-Glasma~\cite{Bartels:2002cj,Kowalski:2003hm,Schenke:2012wb,Schenke:2012hg,Schenke:2013dpa,Schenke:2014tga,Schenke:2016ksl,Mantysaari:2025tcg}, MC-KLN~\cite{Kharzeev:2000ph,Kharzeev:2002ei,Kharzeev:2001yq,Drescher:2006ca}, and McDIPPER~\cite{Garcia-Montero:2023gex,Garcia-Montero:2025bpn}.
The gluonic fields of the Glasma stage are then used to construct the energy-momentum tensor~\cite{Schenke:2019ruo,Schenke:2020mbo,Mantysaari:2017cni} that is evolved further in the hydrodynamic QGP stage.
The energy-momentum tensor includes information about the flow dynamics and is used to initialize the stress tensor of the viscous contribution, in addition to the energy density.

The changeover is commonly realized at times $0.1\div1.0$~fm/$c$ into the collisions.
At this point, the dynamics of Glasma remain anisotropic and far from equilibrium, posing conceptual challenges to the applicability of hydrodynamics under the assumption of local thermodynamic equilibrium.
Recent theoretical advancements propose an intermediate stage between the Glasma and the QGP, based on QCD effective kinetic theory (EKT)~\cite{Baier:2000sb,Arnold:2002zm,Arnold:2000dr,Arnold:2003zc,Berges:2013fga,Kurkela:2015qoa,Keegan:2016cpi,Kurkela:2018wud,Gale:2021emg,Boguslavski:2024ezg,Boguslavski:2024jwr,Barata:2025agq,Greif:2017bnr,Kurkela:2018vqr,Schlichting:2019abc,Berges:2020fwq,Ambrus:2021fej,Du:2025bhb}, where a quasi-particle picture is evolved according to the Boltzmann equation.
The resulting universal behavior follows limiting attractors that drive the system toward local equilibrium.

As an initial stage to the later evolution of the fireball, the Glasma only contributes indirectly to experimentally measurable observables.
Direct experimental probes of the Glasma, based on the modification of initial hard scatterings in the presence of the medium, are actively being researched.
The ``quenching'' of back-to-back jets and the diffusion of heavy quarks in the early Glasma enable promising theoretical predictions for the final particle distribution~\cite{Sun:2019fud,Ipp:2020nfu,Ipp:2020mjc,Carrington:2022bnv,Barata:2024xwy,Avramescu:2024poa,Avramescu:2024xts,Parisi:2025slf}.

The relevant degrees of freedom in the Glasma are captured by a classical YM field.
This Glasma field is produced by the interaction of the highly occupied gluon fields in the incoming nuclei, as predicted by the CGC.
The gluons in the nuclei only carry a small fraction $\xB$ of the longitudinal momentum of the nucleons.
The large-$\xB$ valence partons are treated as the sources of the gluon fields.
In the small\nobreakdash-$\xB$ regime at high energies, the gluon dynamics are dominated by gluon splitting and the occupation number of the gluon fields $\sim 1/\alpha_s \gg 1$ and becomes very large.
The non-perturbative nature of these fields is captured by the CGC in a coherent sum of the small-$\xB$ gluon cascade.
This gives rise to a semi-hard, transverse momentum scale $Q_s(\xB)$ where the gluon distribution saturates and gluon recombination balances the splittings.
The saturation scale $Q_s(\xB)$ grows with decreasing $\xB$.
The nonlinear dynamics in the small-$\xB$ regime are formalized via the BK/JIMWLK~\cite{Balitsky:1995ub,Kovchegov:1999yj,Rummukainen:2003ns,Iancu:2002aq,Lam:2001ax,Jalilian-Marian:1996mkd,Jalilian-Marian:1997qno,Jalilian-Marian:1997jhx,Jalilian-Marian:1998tzv,Weigert:2000gi,Iancu:2000hn,Ferreiro:2001qy,Iancu:2001ad,Weigert:2005us,Blaizot:2002np,Iancu:2013uva} evolution equations which predict the dynamical saturation scale of the gluon density.

The Glasma is extensively studied in the literature.
Robust techniques to evolve the Glasma on discretized spacetime lattices
~\cite{Krasnitz:1999wc,Krasnitz:1998ns,Krasnitz:2000gz,Krasnitz:2001qu,Krasnitz:2002ng,Krasnitz:2002mn,Lappi:2003bi,Lappi:2004sf,Lappi:2009xa,Lappi:2011ju,Epelbaum:2013ekf,Dumitru:2014nka,Schenke:2012wb,Schenke:2012hg,Schenke:2013dpa,Schenke:2014tga,Schenke:2015aqa,Dumitru:2001ux,Blaizot:2008yb,McLerran:2016snu,Lappi:2006fp,Dumitru:2008wn,Lappi:2017skr,Lappi:2006hq,Lappi:2007ku} are established as the standard for the non-perturbative solutions of the system.
A competing scheme based on the expansion of the YM equations for small proper times allows limited analytic access~\cite{Chen:2013ksa,Fries:2006pv,Fujii:2008km,Chen:2015wia,Guerrero-Rodriguez:2021ask,Carrington:2020ssh,Carrington:2021qvi,Carrington:2025xws}.
These references consider a simplification of the Glasma where the longitudinal dynamics are neglected in favor of a boost-invariant description, which is found to be phenomenologically relevant for the mid-rapidity region of the final particle distributions.
However, the longitudinal dynamics of the Glasma are crucial for describing the rapidity structure of the experimental results~\cite{PHOBOS:2006mfc,CMS:2015xmx,ATLAS:2017rij,ALICE:2023tvh,STAR:2025vmb}.
For example, long-range rapidity correlations, known as the ``ridge phenomenon'' can only originate in the initial stage of a HIC, where the correlated regions are still causally connected.
Decorrelations and fluctuations along the longitudinal direction are also linked to the longitudinal dynamics of the initial stage~\cite{Dusling:2013oia,Bozek:2015bna,Bozek:2015tca,Pang:2014pxa,Pang:2015zrq,Monnai:2015sca,Behera:2020mol,Sakai:2021rug,Schenke:2022mjv,Dusling:2009ni,Gelis:2008sz}.

Several mechanisms for breaking boost invariance in the Glasma are discussed in the literature.
The small-$\xB$ dynamics given by the JIMWLK evolution are included in the IP-Glasma framework~\cite{Mantysaari:2025tcg,Schenke:2016ksl,McDonald:2017eml,McDonald:2018wql,McDonald:2020oyf,McDonald:2020xrz,McDonald:2023qwc}.
Sub-eikonal corrections to the nuclear currents are considered in~\cite{Altinoluk:2014oxa,Altinoluk:2015gia,Altinoluk:2015xuy,Agostini:2019avp,Agostini:2022ctk}.
The importance of the finite longitudinal extent of the nuclei is discussed by~\cite{Lam:2000nz,Fukushima:2007ki,Ozonder:2012vw,Ozonder:2013moa,Shen:2017bsr,Shen:2022oyg}.
The time evolution of the Glasma on three-dimensional lattices with dynamical sources, including the longitudinal dynamics, is developed in~\cite{Gelfand:2016yho,Ipp:2017lho,Ipp:2018hai,Ipp:2020igo,Muller:2019bwd,Schlichting:2020wrv,Singh:2021hct,Matsuda:2023gle,Matsuda:2024moa,Matsuda:2024mmr}.
While these lattice calculations contain the full non-perturbative evolution of the Glasma, resolving the longitudinal dynamics with the necessary detail is computationally very expensive.

An alternative scheme is based on the linearization of the YM equations, which enables the semi-analytic calculation of the (3+1)D dynamics of the Glasma.
The earliest reports for solutions of linearized YM equations in the context of the boost-invariant description of HICs can be found in~\cite{Kovner:1995ts,Kovner:1995ja}.
The linearization approach is reconciled with perturbative QCD techniques in~\cite{Kovchegov:1997ke,Guo:1998pe}.
The dynamics along the rapidity direction are first studied using parametrizations for the longitudinal momentum dependence in~\cite{Szczurek:2003fu,Gyulassy:1997vt,Kovchegov:1998bi,Kovchegov:2001sc,Dumitru:2008wn} and are further studied in the context of proton-nucleus collisions in~\cite{Dumitru:2001ux,Blaizot:2004wu,McLerran:2016snu}.
Higher orders in the linearization, together with non-perturbative lattice results, are compared to the lowest order results in~\cite{Blaizot:2010kh,Avsar:2012hj,Chirilli:2015tea,Li:2021zmf,Li:2021yiv,Schlichting:2019bvy}.
The lowest order is found to accurately describe single-inclusive gluon production and total deposited energy in the regime of large transverse momenta of the produced gluons $k_T > Q_s$~\cite{Gribov:1983ivg,Kharzeev:2001gp,Gelis:2008rw,Gelis:2008ad}.

In this thesis, the linearized YM equations are applied to realistic three-dimensional nuclear models that include structure along the longitudinal direction and are shaped with finite-size envelopes.
The longitudinal structure of the nuclei is essential for the rapidity dependence of the observables.
The resulting (3+1)D dilute Glasma~\cite{Ipp:2021lwz,Singh:2021hct,Ipp:2022lid,Ipp:2024ykh,Ipp:2025sbt,Ipp:2025sbc,Ipp:2025cdh,Leuthner:2025vsd} is studied in great detail.
Compared to non-perturbative lattice calculations, the semi-analytic (3+1)D dilute Glasma is computationally favorable and can be easily integrated into event-by-event simulation frameworks of HICs.

This thesis contains the following chapters.
In \cref{sec:description-initial-state}, the description of the initial state of HICs is introduced.
This includes the geometry of the collision and the commonly used coordinates.
The treatment of the incoming nuclei using the CGC is reviewed and the boost-invariant Glasma is briefly discussed.
The McLerran-Venugopalan (MV) nuclear model is introduced and generalized to nuclei with finite three-dimensional envelopes and intrinsic correlations.
In particular, the physical viability of the resulting color charge correlators is inspected.
In \cref{sec:dilute-glasma-approx}, the dilute approximation is derived as a linearization of the classical YM equations.
This leads to the solutions of the (3+1)D dilute Glasma in position space, allowing for the interpretation of gluon production within this framework.
The solutions of the (3+1)D dilute Glasma in momentum space are derived in \cref{ch:momentum-space-picture}.
The result is used to calculate the gluon number distribution, and in \cref{sec:TMDs-QCD}, it is reframed in terms of effective TMDs and the squared Lipatov vertex.
An illustrative limit, where the nuclear envelopes are assumed to be the largest scales in the system, is discussed and reconciled with the $k_T$-factorization formula for gluon production.
In \cref{ch:limiting-fragmentation}, the limiting fragmentation behavior of the position and momentum space results is derived.
Concrete realizations of the generalized MV nuclear model are introduced in \cref{ch:nuclear-models}.
These nuclear models are then used for the numerical evaluation of the energy-momentum tensor in position space and the gluon number distribution and transverse energy in momentum space.
The results are discussed in \cref{ch:numerical-results}, where the focus is on the rapidity structure of the inspected observables.
Finally, \cref{ch:conclusion} provides the summary and preliminary results obtained from coupling the (3+1)D dilute Glasma to the hydrodynamic stage using MUSIC.

\chapter{Description of the initial state}\label{sec:description-initial-state}

\begin{figure}
    \centering
    \includegraphics[width=0.7\linewidth]{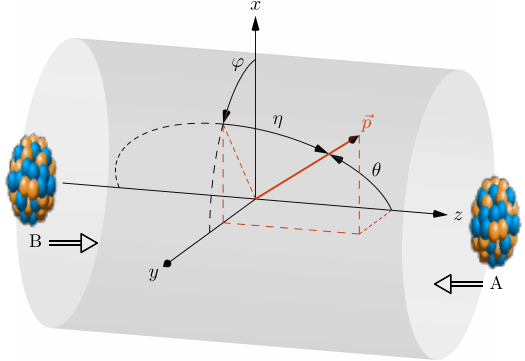}
    \caption[Coordinate system for collider geometry.]{\label{fig:col-geom}
    The cylindrical geometry of HICs is described by a coordinate system where the $z$-axis is aligned with the beam axis.
    The $x$ and $y$-axes span the plane transverse to the beam with the azimuthal angle $\varphi \in [0, 2\pi)$.
    The polar angle $\theta \in [0,\pi]$ is measured from the $z$-axis and is closely related to the pseudo-rapidity $\eta=\artanh(p^z/|\vec p|) \in (-\infty, \infty)$.
    The nucleus with label $\A$ moves in the negative $z$ direction, while the nucleus with label $\B$ moves in the positive $z$ direction.
    }
\end{figure}

The typical geometry of a HIC experiment with two colliding beams is dictated by the acceleration of the beams along a certain (longitudinal) axis and the cylindrical arrangement of detectors around the collision point.
We place our coordinate system such that the $z$-axis is aligned with the beam axis and the nucleus with label $\A$ moves in the negative $z$ direction, while nucleus $\B$ moves in the positive $z$ direction.
The coordinate origin is placed at the collision point, which corresponds to the center of mass for identical collision partners.
This setup is illustrated in \cref{fig:col-geom}.
The $x$ and $y$-axes span the plane transverse to the beam and define the azimuthal angle $\varphi$ that measures the angle from the $x$-axis in a full circle.
In contrast, the polar angle $\theta$ measures the angle from the $z$-axis and only takes the values $\theta \in [0,\pi]$.

In this thesis, we use letters from the Greek alphabet for the spacetime indices of four-vectors.
These indices take the values $\mu \in \{t, x, y, z\}$ for the Cartesian coordinate system, or the corresponding labels for any given coordinate system.
For example, the coordinate vector $x^\mu$ has the Cartesian components%
\footnote{%
We use the natural unit system where the vacuum speed of light $c=1$ and the reduced Planck constant $\hbar =1$. Further conventions are listed in \cref{appx:conventions}.}
$(t,x,y,z)$.
In the main text, the spacetime index of four-vectors is usually suppressed.
Indices denoting only the spatial part of a vector are written with letters from the Latin alphabet, i.e., $x^i$ with $i \in \{x,y,z\}$ has the Cartesian components $(x,y,z)$.
We write spatial-only vectors using the arrow, $\vec x =x^i$.
Additionally, we use transverse-only vectors set in bold face and with bold-face Latin indices as $\xperp = x^\bi$ with $\bi \in \{x,y\}$.

We use the mostly-minus metric convention for the Minkowski metric,
\begin{align}
    g_{\mu\nu} = \mathrm{diag}(+1,-1,-1,-1)_{\mu\nu},
\end{align}
which defines the inner product of vectors.
Using Einstein's sum convention, we write
\begin{align}
    x^\mu x_\mu = x^\mu x^\nu g_{\mu\nu}, \qquad x^i x^i = \vec x \cdot \vec x = \vec x^2, \qquad x^\bi x^\bi = \xperp \cdot \xperp = \xperp^2,
\end{align}
for the inner products of vectors.
Squaring a spatial or transverse vector is defined via this inner product and the modulus is given by
\begin{align}
    |\vec x| = \sqrt{\vec x^{\,2}}, \qquad |\xperp| = \sqrt{\xperp^2}.
\end{align}

The momentum vector $p$ with the components $(p^t,p^x,p^y,p^z)$, and the Lorentz scalar
\begin{align}
   p_\mu p^\mu = (p^t)^2 - {\vec p\,}^2 = m^2,
\end{align}
where $m$ is the rest mass, are used to describe the kinematics.
In experiments, it is useful to parametrize the momentum of the detected particles via the pseudo-rapidity
\begin{align}
    \eta = \artanh(p^z/|\vec p\,|)
\end{align}
because it only requires the measurement of the particles' momentum, rather than the identification of their species and mass.
As such, $\eta$ is a geometric quantity that is closely related to the polar angle $\theta$ (see \cref{fig:col-geom}).
Pseudo-rapidity is positive where $\theta \in [0,\pi/2)$, negative where $\theta \in (\pi/2,\pi]$ and takes the extremal values
\begin{align}
    \eta = 0 \text{\ \ for\ \ } \theta=\frac{\pi}{2}, \qquad \eta \rightarrow \infty \text{\ \ for\ \ } \theta = 0, \qquad \eta\rightarrow-\infty \text{\ \ for\ \ } \theta=\pi.
\end{align}
A different notion of rapidity is provided by the momentum rapidity
\begin{align}
    \Y = \artanh(p^z/p^t), \label{eq:Y-def}
\end{align}
which is used to parametrize a particle's energy and momentum component $p^z$ as
\begin{align}
    p^t = \sqrt{m^2 + \pperp^2} \cosh(\Y), \qquad p^z = \sqrt{m^2 + \pperp^2}\sinh(\Y).\label{eq:Y-with-mass}
\end{align}
For ultra-relativistic particles, $p^t \approx |\vec p\,|$ and pseudo-rapidity and momentum rapidity coincide.
Furthermore, for gluons on shell, $m=0$ and Eqs.~\eqref{eq:Y-with-mass} reduce to
\begin{align}
    |\vec p\,| = p^t = |\pperp|\cosh(\Y), \qquad p^z = |\pperp|\sinh(\Y),
\end{align}
which becomes particularly useful when parametrizing the $p^z$ component in phase space integrals with the volume element
\begin{align}
    \dd^3\vec p = \dd^2\pperp\,\dd\Y\,|\vec p\,|.
\end{align}

The energy of a HIC is typically given in terms of $\sqrt{s}$, the square root of the Mandelstam variable 
\begin{align}
    s = (p^\mu_\A + p^\mu_\B)(p_\mu^\A+p_\mu^\B),
\end{align}
which characterizes the center of mass energy in the system.
For two nuclei traveling in opposite directions along the $z$-axis, their momenta in the lab frame read $p^\mu_{\A/\B}=(m\gamma,0,0,\mp p^z)$.
The mass of each nucleus is $m$ and $\gamma = \left(1-(v^z)^2\right)^{-1/2}$ is the standard Lorentz factor.
Then, $\sqrt{s}=2\gamma m$, which is often translated to the energy per nucleon-nucleon pair $\sqrt{s_\mathrm{NN}}=2\gamma m_n$, with the mass of a nucleon $m_n \approx 1$ GeV.

\begin{figure}
    \centering
    \includegraphics{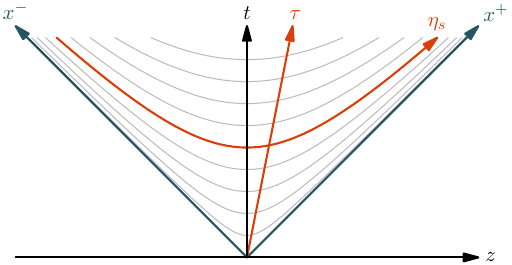}
    \caption[Light cone and Milne coordinates.]{\label{fig:coordinates}
    The $t$-$z$ plane is additionally parametrized by light cone coordinates $\xpm=(t\pm z)/\sqrt{2}$ (blue) and the curvilinear Milne coordinates (orange) with proper time $\tau^2=t^2-z^2$ and spacetime rapidity $\eta_s=\artanh(z/t)$. Grey curves are lines of constant $\tau$.}
\end{figure}

For particles moving close to the speed of light, the light cone (LC) coordinates
\begin{align}
    \xpm = \frac{1}{\sqrt{2}}(t\pm z)
\end{align}
in the $t$-$z$ plane prove to be convenient.
The metric for this frame is
\begin{align}
    g^\mathrm{LC}_{\mu\nu} = \begin{pmatrix}
        0 & 1 & 0 & 0\\
        1 & 0 & 0 & 0\\
        0 & 0 & -1 & 0\\
        0 & 0 & 0 & -1
    \end{pmatrix}_{\mu\nu},
\end{align}
and contains an off-diagonal block for the $+$ and $-$ components, and the Minkowski "minus" signature for $x$ and $y$.
It leads to the identification of co- and contravariant light cone components when exchanging $+\leftrightarrow-$, i.e., $\xpm = x_\mp$.
Consider an ultra-relativistic particle moving in the positive $z$ direction.
Its world line is given by $x^-=0$ and the component $x^+ =\sqrt{2}\,t$ plays the role of light cone time.
Its four-momentum has just one component $p^+ = \sqrt{2}\,p^t$.
The description of a particle moving in the negative $z$ direction is analogous with the roles of the $+$ and $-$ components swapped.

Another frame for the $t$-$z$ plane that is widely used to describe the flowing particles produced after a HIC is the so-called Milne frame.
It defines the proper time $\tau$ and spacetime rapidity%
\footnote{%
To distinguish spacetime rapidity from pseudo-rapidity, we use the subscript "$s$" and stick to common convention for the symbols used for pseudo-rapidity ($\eta$) and momentum rapidity ($\Y$).}
$\eta_s$ as
\begin{align}
    \tau &= \sqrt{t^2-z^2} = \sqrt{2x^+x^-}, \\
    \eta_s &= \artanh(z/t) = \frac{1}{2}\ln(x^+/x^-), \label{eq:eta-s-def}
\end{align}
with the inverse relations
\begin{align}
    t&=\tau\cosh(\eta_s), \qquad z=\tau\sinh(\eta_s), \\
    \xpm &= \frac{1}{\sqrt{2}}\tau\ee^{\pm\eta_s}.
\end{align}
In \cref{fig:coordinates}, the standard Minkowski (black), light cone (blue) and Milne coordinates (orange) are compared.
The Milne frame is curvilinear and the transformation of contravariant vector components%
\footnote{%
We label the spacetime index of the spacetime rapidity components of vectors and tensors with the symbol $\eta$, where we drop the subscript "$s$" to improve readability.
This should not cause confusion with pseudo-rapidity, which is purely used to parametrize momentum components.}
is given by
\begin{align}
    \renewcommand*{\arraystretch}{1.5}\begin{pmatrix}
        p^\tau \\
        p^\eta
    \end{pmatrix} &= \renewcommand*{\arraystretch}{1.5}\begin{pmatrix}
        \frac{\partial\tau}{\partial t} & \frac{\partial\tau}{\partial z} \\
        \frac{\partial\eta_s}{\partial t} & \frac{\partial\eta_s}{\partial z}
    \end{pmatrix}\renewcommand*{\arraystretch}{1.5}\begin{pmatrix}
        p^t \\
        p^z
    \end{pmatrix} = \renewcommand*{\arraystretch}{1.5}\begin{pmatrix}
        \cosh(\eta_s)p^t-\sinh(\eta_s)p^z \\
        -\frac{\sinh(\eta_s)}{\tau} p^t + \frac{\cosh(\eta_s)}{\tau}p^z
    \end{pmatrix}. \label{eq:Milne-vec-transf}
\end{align}
From this transformation matrix, the Jacobian for the volume element results in
\begin{align}
    \dd p^t\,\dd p^z = \dd p^\tau\,\dd p^\eta\,\tau .
\end{align}
The metric in the Milne frame is
\begin{align}
    g^\mathrm{Mi}_{\mu\nu} = \mathrm{diag}(+1,-1,-1,-\tau^2)_{\mu\nu}, \label{eq:Milne-metric}
\end{align}
and is now time-dependent.
However, the metric is singular at $\tau=0$, which corresponds to the light cone where either $x^+=0$ or $x^-=0$.
This set of Milne coordinates, therefore, can only be used inside the future light cone of the coordinate origin, which puts special emphasis on the choice of the origin.

The metric in \cref{eq:Milne-metric} describes the geometry of a system expanding along the $z$ direction.
This symmetry is perfectly adapted to the special four-velocity field
\begin{align}
    \begin{pmatrix}
        u^t \\
        u^z
    \end{pmatrix} = \gamma \begin{pmatrix}
        1 \\
        z/t
    \end{pmatrix},
\end{align}
with no flow in the transverse plane, i.e., $\uperp = 0$.
This flow pattern is known as Bjorken flow~\cite{Bjorken:1982qr} and is commonly used to approximate the dynamics of the fireball after the collision.
In the Milne frame, the components of the four-velocity transform according to
\begin{align}
    \renewcommand*{\arraystretch}{1.5}\begin{pmatrix}
        u^\tau \\
        u^\eta
    \end{pmatrix} &= \renewcommand*{\arraystretch}{1.5}\begin{pmatrix}
        \cosh(\eta_s) & -\sinh(\eta_s) \\
        -\frac{\sinh(\eta_s)}{\tau} & \frac{\cosh(\eta_s)}{\tau}
    \end{pmatrix}\renewcommand*{\arraystretch}{1.5}\begin{pmatrix}
        \gamma \\
        \gamma \frac{z}{t}
    \end{pmatrix} = \renewcommand*{\arraystretch}{1.5}\begin{pmatrix}
        1 \\
        0
    \end{pmatrix},
\end{align}
which follows from a straightforward calculation using \cref{eq:eta-s-def}.
This is the four-velocity of a system at rest.
We can, therefore, interpret the Milne frame as the local rest frame of a Bjorken-flowing system.
That is, an observer using the Minkowski coordinates and who is boosted to the local rest frame (where he reports $u'^{t'} = 1$ and $\vec u' =0$ with primed coordinates after the boost) reports the same components as an observer at rest in the lab frame but who is using Milne coordinates.

Furthermore, we may compare the Milne frame of an observer at rest and the Milne frame of an observer boosted along the $z$-axis with the boost angle $\xi$, the latter of which we label as the primed system.
Then,
\begin{align}
\begin{pmatrix}
    t' \\ z'
\end{pmatrix} = \begin{pmatrix}
    \cosh(\xi) & -\sinh(\xi) \\
    -\sinh(\xi) & \cosh(\xi)
\end{pmatrix}\begin{pmatrix}
    t \\ z
\end{pmatrix} = \begin{pmatrix}
    t \cosh(\xi) - z \sinh(\xi) \\
    -t \sinh(\xi) + z \cosh(\xi)
\end{pmatrix},
\end{align}
from which we obtain
\begin{align}
    \tau' &= \sqrt{t'^2-z'^2} = \sqrt{t^2 -z^2} = \tau, \\
    \eta_s' &= \artanh(z'/t') = \artanh(\tanh(\eta_s-\xi)) = \eta_s - \xi.
\end{align}
The proper times $\tau'$ and $\tau$ reported by the two observers are identical.
The spacetime rapidity $\eta_s'$ in the boosted frame is $\eta_s$ shifted by $\xi$.
These properties under boosts allow for an elegant description of boost-invariant systems, where the (3+1)D dynamics effectively reduce to (2+1)D, with two spatial dimensions for the transverse plane and one temporal dimension.
If a system is boost invariant, it does not depend on $\eta_s$.

The spacetime rapidity $\eta_s$ is closely related to the momentum rapidity $\Y$.
Consider a free-streaming particle with velocity $v$ along the $z$-axis.
We place the coordinate origin such that at time $t=0$ the particle's position is $z=0$.
The $z$ coordinate of this particle at a time $t$ is then given by $z=v\,t$.
From \cref{eq:eta-s-def,eq:Y-def}, it follows that
\begin{align}
    \eta_s = \artanh(z/t) = \artanh(v) = \artanh(p^z/p^t) = \Y,
\end{align}
i.e., the particle with momentum rapidity $\Y$ is located at spacetime rapidity $\eta_s =\Y$.
The momentum of the particle in the Milne frame reads
\begin{align}
    \renewcommand*{\arraystretch}{1.5}\begin{pmatrix}
        p^\tau \\
        p^\eta \\
        \pperp
    \end{pmatrix} = \renewcommand*{\arraystretch}{1.5}\begin{pmatrix}
        \sqrt{m^2 + \pperp^2}\cosh(\Y-\eta_s) \\
        \frac{1}{\tau}\sqrt{m^2 + \pperp^2}\sinh(\Y-\eta_s) \\
        \pperp
    \end{pmatrix} = \renewcommand*{\arraystretch}{1.5}\begin{pmatrix}
        \sqrt{m^2 + \pperp^2} \\
        0 \\
        \pperp
    \end{pmatrix},
\end{align}
from which we see that a free-streaming particle has zero momentum in the $p^\eta$ component.
We take note of this important property.
In the special case where $\pperp=0$, the Milne frame describes the local rest frame of the moving particle.
However, in the general case, nonzero transverse momentum breaks this identification.

\section{Classical Yang-Mills theory}

We briefly discuss Yang-Mills (YM) field theory to the extent relevant for this thesis.
The details can be found in many textbooks, such as~\cite{Srednicki:2007qs,Peskin:1995ev} and the historic significance is reviewed in~\cite{Gross:2022hyw}.

The fundamental symmetry of QCD is the local gauge group $SU(N_c)$.
The $N_c=3$ colors are interpreted as the charges of the strong interaction, which is mediated by $N_c^2-1$ force-carrying bosons ("gluons").
The group transformation $U(x) \in SU(N_c)$ depends locally on the spacetime coordinate $x$.
It is special ("S"), meaning it has determinant 1 and unitary ("U"), i.e., $U U^\dagger = \mathbb{1}$.
A basis for the $\mathfrak{su}(N_c)$ algebra is given by the generators $t^a$.
The color index $a = 1,\ldots, N_c^2-1$, and each generator is an $N_c\times N_c$-dimensional matrix in the fundamental representation of the group.
From the properties of $U$, it follows that the generators are hermitian and traceless,
\begin{align}
    (t^a)^\dagger = t^a, \qquad \tr t^a = 0. \label{eq:generators-herm-trace}
\end{align}
We adopt the common normalization of the generators given as
\begin{align}
    \tr t^a t^b = \frac{1}{2} \delta^{ab}. \label{eq:generators-norm}
\end{align}
The Lie bracket of the generators is
\begin{align}
    [t^a,t^b] = \ii f^{abc} t^c, \label{eq:generators-comm}
\end{align}
where repeated indices are understood to be summed over.
The non-commutativity of YM theory is encoded in the real, totally anti-symmetric structure constants $f^{abc}$.
Thus, YM theory is non-Abelian, in contrast to the commuting algebra of Abelian electrodynamics.

The YM Lagrangian for the field content reads
\begin{align}
    \mathcal{L}_\mathrm{YM} = - \frac{1}{2} \tr F^{\mu\nu}(x) F_{\mu\nu}(x),
\end{align}
where the anti-symmetric field strength tensor $F_{\mu\nu}$ takes values in the $\mathfrak{su}(N_c)$ algebra and the trace acts in color space.
This Lagrangian is invariant under transformations in the $SU(N_c)$ group.
To this end, the field strength tensor transforms as
\begin{align}
    F_{\mu\nu}(x) \rightarrow U(x)F_{\mu\nu}(x)U^\dagger(x). \label{eq:field-strength-gauge-trafo}
\end{align}
The degrees of freedom of $F_{\mu\nu}$ are captured by the $\mathfrak{su}(N_c)$-valued gauge field $A$,
\begin{align}
    F_{\mu\nu}(x) = \partial_\mu A_\nu(x) - \partial_\nu A_\mu(x) - \ii g [A_\mu(x), A_\nu(x)]. \label{eq:Fmunu-def-general}
\end{align}
Here, $g$ is the YM coupling, which, in the case of QCD, decreases for larger energy scales of the interaction and leads to so-called asymptotic freedom.
It is related to the strong coupling constant $\alpha_s = g^2/(4\pi)$.
The gauge field $A$ transforms as
\begin{align}
    A_\mu(x) \rightarrow U(x) \left( A_\mu(x) - \frac{1}{\ii g} \partial_\mu \right) U^\dagger(x),
\end{align}
which can be rewritten using the generator of the transformation $\omega(x) \in \mathfrak{su}(N_c)$ for $U(x) = \exp(\ii g \omega(x))$.
To leading order in the infinitesimal $\omega$,
\begin{align}
    A_\mu(x) \rightarrow A_\mu(x) + \partial_\mu \omega(x) + \ii g [ \omega(x), A_\mu(x)].
\end{align}
Expressing the field strength and gauge field in the basis of the generators $t^a$ allows us to read off the color components,
\begin{align}
    A_\mu(x) &= t^a A^a_\mu(x), \qquad F_{\mu\nu} = t^a F^a_{\mu\nu}(x), \\
    F^a_{\mu\nu}(x) &= \partial_\mu A^a_\nu(x) - \partial_\nu A^a_\mu(x) + g f^{abc} A^b_\mu(x) A^c_\nu(x), \\
    \mathcal{L}_\mathrm{YM} &= -\frac{1}{4} F^a_{\mu\nu}(x) F^{\mu\nu,a}(x), \label{eq:CYM-L}
\end{align}
where we used \cref{eq:generators-comm} to simplify the commutator of generators for the field strength and accumulated a factor $1/2$ from the normalization in \cref{eq:generators-norm} in the Lagrangian.

The dynamics of the field content is governed by the YM equations of motion (in short: YM equations),
\begin{align}
    \left[ D_\mu , F^{\mu\nu}(x) \right] = 0.
\end{align}
Here, we introduced the gauge covariant derivative given as
\begin{align}
    D_\mu = \mathbb{1}\partial_\mu - \ii g A_\mu(x),
\end{align}
which can be used to express the field strength tensor via
\begin{align}
    F_{\mu\nu}(x) = \frac{\ii}{g} \left[ D_\mu, D_\nu \right].
\end{align}
In general, the commutator of gauge fields in the YM equations leads to a nonlinear system of differential equations.

The fermionic part of the QCD Lagrangian, which is not included in \cref{eq:CYM-L}, couples gauge fields to fermionic degrees of freedom.
Instead of introducing fermions, we add the matter current $J^{\mu,a}$ coupled to the gauge field to the YM Lagrangian,
\begin{align}
    \mathcal{L}_\mathrm{CYM} = -\frac{1}{4} F^a_{\mu\nu}(x) F^{\mu\nu,a}(x) - A^a_\mu(x) J^{\mu,a}(x).
\end{align}
This allows for a convenient description of the dynamics relevant for HICs in terms of the classical YM equations
\begin{align}
    \left[ D_\mu , F^{\mu\nu}(x) \right] = \partial_\mu F^{\mu\nu}(x) - \ii g \left[ A_\mu(x), F^{\mu\nu}(x) \right] = J^\nu(x). \label{eq:CYM}
\end{align}
Note that the addition of the matter current breaks the gauge invariance of the Lagrangian, because $J$ transforms the same way as the field strength in \cref{eq:field-strength-gauge-trafo} under gauge transformations.
Still, the classical equations of motion are gauge-covariant.
The charges described by $J$ are conserved according to the non-Abelian charge conservation equation,
\begin{align}
    \left[ D_\mu, J^\mu(x) \right] = \partial_\mu J^\mu(x) - \ii g \left[ A_\mu(x), J^\mu(x) \right] = 0. \label{eq:current-conservation}
\end{align}

\section{The Color Glass Condensate}\label{sec:cgc}

The Color Glass Condensate (CGC) is an effective theory for the description of ultra-relativistic "cold" QCD matter (see~\cite{Iancu:2003xm,Iancu:2012xa,Gelis:2010nm,Gelis:2012ri,Fukushima:2011ca,Fukushima:2016xgg,Gelis:2021zmx,Garcia-Montero:2025hys} for more details), such as the incoming nuclei in HICs.\
It is considered to describe the relevant degrees of freedom in the initial condition for the "hot" QCD matter produced during the collision.

In the CGC, the nucleus is conceptually separated into hard and soft partons.
The hard partons are treated as sources of classical color charge for the soft partons.
In a reference frame, such as the lab frame, where the nucleus is boosted to relativistic speeds, the hard partons appear to carry almost all of the longitudinal momentum of the nucleus.
Their structure is effectively frozen in time due to time dilation.
This fixes the classical color current associated with the hard partons to be static.

We denote the current four-vector of a nucleus with label $\A$ moving with the speed of light along the negative $z$ direction as $\mathcal{J}_\A$.
In the following, calligraphic letters are used for fields associated with a single nucleus.
Using light cone coordinates, the components of $\mathcal{J}_\A$ are
\begin{align}
    &\mathcal{J}^-_\A(x^+,\xperp) = \rho_\A(x^+,\xperp), \\
    &\mathcal{J}^+_\A = 0, \qquad \mathcal{J}^\bi_\A = 0. \label{eq:single-nucl-J-zero-comp}
\end{align}
Here, the color charge distribution $\rho_\A$ describes the configuration of the hard partons and takes values in the $\mathfrak{su}(N_c)$ algebra.
Only the component $\mathcal{J}^-_\A$ is nonzero in light cone coordinates.
The current and color charge density do not depend on the light cone time $x^-$.
Their dependence on the remaining longitudinal coordinate $x^+$ will be sharply localized around the light cone, where $x^+ \sim 0$, due to Lorentz contraction.
In fact, the idealized scenario where the charges are moving with the speed of light would only permit Dirac-delta-like support along $x^+$.
Enforcing this feature leads to boost-invariant dynamics, which will be discussed in \cref{sec:boost-inv-glasma}.
The main content in this thesis, however, revolves around a generalized description where the support along $x^+$ and, therefore, longitudinal structure, is incorporated in the description of the nuclei.
Note that in this case, the velocities of the charges are considered to be close to the speed of light, where the components in \cref{eq:single-nucl-J-zero-comp} are subleading.

The soft partons are identified with a highly occupied gluon field treated classically as the gauge field $\mathcal{A}_\A$ and associated field strength $\mathcal{F}_\A$ given in \cref{eq:Fmunu-def-general}.
We adopt covariant gauge,
\begin{align}
    \partial_\mu \mathcal{A}^\mu_\A(x) = 0,
\end{align}
and are left to solve the YM equations~\eqref{eq:CYM} for $\mathcal{A}_\A$,
\begin{align}
   \partial_\mu \mathcal{F}_\A^{\mu\nu} - \ii g \left[ \mathcal{A}_\mu^\A , \mathcal{F}_\A^{\mu\nu} \right] & \nn\\
   &\hspace{-34.3pt}= \partial_\mu \partial^\mu \mathcal{A}^\nu_\A - 2\ii g \left[ \mathcal{A}^\mu_\A , \partial_\mu \mathcal{A}^\nu_\A \right] + \ii g \left[ \mathcal{A}_\mu^\A , \partial^\nu \mathcal{A}^\mu_\A \right] - g^2 \left[ \mathcal{A}_\mu^\A , \left[ \mathcal{A}^\mu_\A , \mathcal{A}^\nu_\A \right] \right] \nn\\
   &\hspace{-34.3pt}= \mathcal{J}^\nu_\A, \label{eq:nucl-A-CYM}
\end{align}
with the restriction that the current is preserved according to \cref{eq:current-conservation},
\begin{align}
    \partial_\mu \mathcal{J}^\mu_\A(x) - \ii g \left[ \mathcal{A}^\A_\mu(x), \mathcal{J}^\mu_\A(x) \right] = -\ii g \left[  \mathcal{A}^+_\A(x), \mathcal{J}^-_\A(x) \right] = 0 . \label{eq:nucl-A-current-cons}
\end{align}
The conservation of the current reduces to a color rotation.
A suitable ansatz for the gauge field reads
\begin{align}
    &\mathcal{A}^-_\A(x) = \mathcal{A}^-_\A(x^+,\xperp), \\
    &\mathcal{A}^+_\A = 0, \qquad \mathcal{A}^\bi_\A = 0,
\end{align}
which trivially solves \cref{eq:nucl-A-current-cons} and mirrors the dependence of the color charge distribution on the spacetime coordinates for the only nonzero component $\mathcal{A}^-$.
Inserting this ansatz into \cref{eq:nucl-A-CYM}, we see that all the nonlinear contributions are eliminated.
For the first commutator,
\begin{align}
   \left[ \mathcal{A}^\mu_\A(x^+,\xperp) , \partial_\mu \mathcal{A}^\nu_\A(x^+,\xperp) \right] = \left[ \mathcal{A}^-_\A(x^+,\xperp) , \partial_- \mathcal{A}^\nu_\A(x^+,\xperp) \right] = 0
\end{align}
because $\partial_- \mathcal{A}_\A = 0$.
The other two commutators vanish because $\mathcal{A}^\mu_\A \mathcal{A}_\mu^\A = 0$.
As a result, \cref{eq:nucl-A-CYM} only contains one nontrivial equation where the value of the index $\nu=-$,
\begin{align}
    \partial_\mu\partial^\mu \mathcal{A}^-_\A(x^+,\xperp) = - \partial^\bi\partial^\bi \mathcal{A}^-_\A(x^+,\xperp) = \rho_\A(x^+,\xperp). \label{eq:nucl-A-poisson}
\end{align}

The structure of \cref{eq:nucl-A-poisson} is that of a Poisson equation in the transverse plane.
It is straightforward to write down the solution in Fourier space,%
\footnote{%
The conventions for the Fourier transformation are defined in \cref{appx:conventions}.}
\begin{align}
    \tilde{\mathcal{A}}^-_\A(x^+,\kperp) = \frac{1}{\kperp^2} \tilde{\rho}_\A(x^+,\kperp). \label{eq:nucl-A-poisson-FT}
\end{align}
However, this solution is potentially infrared divergent when $\kperp\rightarrow 0$.
The established procedure to cure this divergence while still allowing for the most general color charge distributions is to modify the charge density by a suitable factor,
\begin{align}
    \tilde{\rho}_\A(x^+,\kperp) \rightarrow \frac{\kperp^2}{\kperp^2 + m^2}\tilde{\rho}_\A(x^+,\kperp),
\end{align}
where we added the infrared (IR) regulator $m$ in the denominator, such that for $\kperp \gg m$ the solution is unmodified.
This factor also puts the zero mode to zero and ensures color neutrality.
It can be interpreted as a screening mass that is part of the model assumptions used to describe the color charge distribution.
We then define the solution of the gauge field in position space as
\begin{align}
    \mathcal{A}^-_\A(x^+,\xperp) = \intop \frac{\dd^2\kperp}{(2\pi)^2}\frac{\ee^{+\ii \kperp\cdot\xperp}}{\kperp^2+m^2}\, \tilde{\rho}_\A(x^+,\kperp).
\end{align}
We will take care always to include the explicit regulator $m$ whenever we express the solution for $\mathcal{A}^-_\A$ in terms of the color charge distribution.

This is the single-nucleus solution that arises within the CGC effective description of charges moving with highly relativistic speeds.
The structure of this solution is remarkably simple.
There is only one nonzero light cone component for the gauge field $\mathcal{A}^-_\A$.
The gauge field is independent of the light cone time $x^-$ and appears frozen, just as the current.
The longitudinal structure along $x^+$ is directly given by the charge distribution $\rho_\A$.
The transverse structure is linked to $\rho_\A$ via a transverse Poisson equation for every value of $x^+$.
Given a realization of the color charge distribution $\rho_\A$ and the gauge choice of covariant gauge, the colored field dynamics of a relativistic nucleus are described by the so-called Weizs\"acker-Williams fields for each of the $N_c^2-1$ color components, in full analogy to classical Abelian electrodynamics.

From the gauge field, we can compute the light cone and Minkowski components of the field strength tensor,
\begin{align}
    \mathcal{F}^{-\bi}_\A(x^+,\xperp) &= - \partial^\bi \mathcal{A}^-_\A(x^+,\xperp), \\
    \mathcal{F}^{t\bi}_\A(x^+,\xperp) &= -\mathcal{F}^{z\bi}_\A(x^+,\xperp) = \frac{1}{\sqrt{2}} \mathcal{F}^{-\bi}_\A(x^+,\xperp).
\end{align}
There are only two independent, nonzero tensor components, $\mathcal{F}^{t\bi}_\A$ with the transverse index $\bi \in \{x,y\}$, which are related to the light cone components via a factor of $\sqrt{2}$.
This leads to a particular structure for the chromo-electric and chromo-magnetic fields, whose components read
\begin{align}
    \mathcal{E}^t_\A &= \mathcal{E}^z_\A = 0, \qquad \mathcal{E}^\bi_\A(x^+,\xperp) = \frac{1}{\sqrt{2}} \mathcal{F}^{-\bi}_\A(x^+,\xperp), \label{eq:nucl-A-E-field} \\
    \mathcal{B}^t_\A &= \mathcal{B}^z_\A = 0, \qquad \mathcal{B}^\bi_\A(x^+,\xperp) = \frac{1}{\sqrt{2}} \varepsilon^{\bi\bj}\mathcal{F}^{-\bj}_\A(x^+,\xperp). \label{eq:nucl-A-B-field}
\end{align}
Here, we introduced the transverse Levi-Civita symbol $\varepsilon^{\bi\bj}$, which is totally anti-symmetric in its transverse indices, i.e., $\varepsilon^{xy} = - \varepsilon^{yx} = 1$.
Note that the chromo-electric and chromo-magnetic fields are equal in magnitude and orthogonal to each other.
Additionally, they are aligned with the transverse plane and, therefore, both are also orthogonal to the beam axis.

\subsection{The collision of two nuclei}

\begin{figure}
    \centering
    \includegraphics{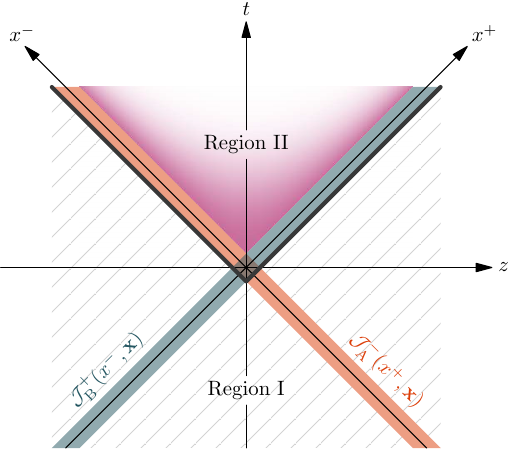}
    \caption[Setup of two colliding nuclei in the $t$-$z$ plane.]{\label{fig:two-ncl-collision-regions}
    The collision of two nuclei, as described by the CGC, can be divided into two distinct regions along the thick black lines.
    In Region I, the covariant gauge, single-nucleus solutions are localized to the support given by the $\xpm$-dependence of the charge distributions.
    All interactions occur in Region II, where one set of nonlinear YM equations has to be solved for both nuclei.
    Figure adapted from~\cite{Ipp:2024ykh}.}
\end{figure}

With the results from the previous section, we can assemble the initial conditions for the collision of two nuclei in a HIC.
In addition to nucleus $\A$, we also consider nucleus $\B$, which moves along the beam axis in the positive $z$ direction.
This setup is illustrated in \cref{fig:two-ncl-collision-regions}.
In the asymptotic past of the collision, each of the nuclei can be described by the CGC without any influence from the other nucleus.
We may write the only nonzero components of the currents in combined notation as
\begin{align}
    \mathcal{J}^\mp_{\A/\B}(\xpm,\xperp) = \rho_{\A/\B}(\xpm,\xperp), \label{eq:single-nucleus-currents}
\end{align}
where the color charge distributions $\rho_{\A/\B}$ are understood to have compact support along their longitudinal directions $\xpm$.
The currents give rise to the gauge fields and field strengths, with the covariant gauge solutions
\begin{align}
    \mathcal{A}^\mp_{\A/\B}(\xpm,\xperp) &= \intop \frac{\dd^2\kperp}{(2\pi)^2}\frac{\ee^{+\ii \kperp\cdot\xperp}}{\kperp^2+m^2}\, \tilde{\rho}_{\A/\B}(\xpm,\kperp), \label{eq:single-nucleus-gauge-fields} \\
    \mathcal{F}^{\mp\bi}_{\A/\B}(\xpm,\xperp) &= - \partial^\bi \mathcal{A}^\mp_{\A/\B}(\xpm,\xperp). \label{eq:single-nucleus-field-strengths}
\end{align}
Note the symmetry of the $+$ and $-$ light cone components between nucleus $\A$ and $\B$, which amounts to exchanging $+\leftrightarrow-$ for all components.

Due to causality, any interaction between nucleus $\A$ and $\B$ can only arise as soon as the current of one nucleus enters the past light cone of the other.
In covariant gauge, the interactions between the gauge fields and currents are even localized to the interaction region where the currents overlap in the $t$-$z$ plane.
This is because the longitudinal support along $\xpm$ of the gauge fields is directly inherited from the color charge distributions.
Therefore, we can split the $t$-$z$ plane into two regions marked in \cref{fig:two-ncl-collision-regions}.
In Region I, the single-nucleus solutions obtained from the CGC are valid for both nuclei.
In the forward light cone of the first interaction point, annotated as Region II, we expect new dynamics to emerge as a result of the nonlinear structure of the YM equations.
There, it is necessary to treat the currents and gauge fields of both nuclei within the same YM equations and it is no longer possible to separate the fields into independent solutions.

We do, however, restrict the dynamics of the currents to simplify the YM equations.
Each current is assumed to be recoilless, meaning that its four-vector structure remains unchanged for all times.
As a result, the non-abelian conservation of the currents can only contain a color rotation induced by the gauge field of the other nucleus.
This is motivated by the so-called eikonal approximation, which entails that colored charges moving with the speed of light are not deflected when passing through a gauge field.

\section{The boost-invariant Glasma}\label{sec:boost-inv-glasma}

The collision of two nuclei in the setup discussed in the previous section was treated extensively in the boost-invariant approximation.
The earliest studies~\cite{Kovner:1995ts,Kovner:1995ja,Kovchegov:1997ke,Gribov:1983ivg,Gyulassy:1997vt,Guo:1998pe,Kovchegov:1998bi,Kovchegov:2001sc,Kharzeev:2001gp} focused on a perturbative treatment and linearization of the YM equations.
More details about the initial field configuration were worked out in~\cite{Blaizot:2008yb,Dumitru:2001ux,Blaizot:2008yb,McLerran:2016snu,Lappi:2006fp,Dumitru:2008wn,Chen:2013ksa,Lappi:2017skr,Lappi:2006hq,Lappi:2007ku} and subsequently used to describe gluon production from the initial state~\cite{Gelis:2008rw,Gelis:2008ad,Blaizot:2004wu,Blaizot:2010kh}.
Incorporating more aspects, such as impact parameter dependence and energy dependence, is still an active field of research~\cite{Kowalski:2003hm,Schenke:2012wb,Schenke:2012hg,Schenke:2013dpa,Schenke:2014tga,Mantysaari:2025tcg}.

The goal in this section is to compute the gauge field $A_\mathrm{(II)}$ that develops in the future light cone of the collision, marked as Region II in \cref{fig:two-ncl-collision-regions}.
The gauge field $A_\mathrm{(II)}$ is not a simple superposition of the single-nucleus fields $\mathcal{A}_{\A/\B}$ because the YM equations that govern the interaction of these fields are, in general, nonlinear.
It is these nonlinear dynamics and properties that describe the Glasma.
Note that we use calligraphic letters with the labels $\A/\B$ for the solutions of the single-nucleus fields in Region I.
The boost-invariant limit allows deriving the initial conditions along the boundary between Regions I and II, which are used to calculate the evolution of $A_\mathrm{(II)}$ in an idealized setup.

The starting point is to reduce the longitudinal support of the color charge densities to a Dirac-delta distribution,
\begin{align}
    \mathcal{J}^\mp_{\A/\B}(\xpm,\xperp) = \rho_{\A/\B}(\xpm,\xperp) = \delta(\xpm) \rho^\perp_{\A/\B}(\xperp). \label{eq:boost-inv-rho}
\end{align}
Here, we introduced the transverse color charge distributions $\rho^\perp_{\A/\B}$.
\Cref{eq:boost-inv-rho} is invariant under boosts along the longitudinal directions $\xpm$ because of the scaling property of the delta function.
We assume the currents to be recoilless, which prevents any change and mixing of the vector structure of $\mathcal{J}_\A$ and $\mathcal{J}_\B$.
We may write the total current for all of spacetime as a sum of the single-nucleus currents,
\begin{align}
    J^\mu(x) = \mathcal{J}^\mu_\A(x^+,\xperp) + \mathcal{J}^\mu_\B(x^-,\xperp).
\end{align}
Additionally, the conservation of the total current given by \cref{eq:current-conservation} holds for $\mathcal{J}_\A$ and $\mathcal{J}_\B$ individually,
\begin{align}
    \left[ D_\mu, \mathcal{J}^\mu_{\A/\B}(\xpm,\xperp) \right] = \partial_\mp \mathcal{J}^\mp_{\A/\B}(\xpm,\xperp) - \ii g \left[ A^\pm_\mathrm{(II)}(x), \mathcal{J}^\mp_{\A/\B}(\xpm,\xperp) \right], \label{eq:boost-inv-color-rot}
\end{align}
where the first term is zero and the currents only undergo color rotation due to the gauge field $A_\mathrm{(II)}$ in Region II.
The special choice of Fock-Schwinger gauge for Region II, given by the gauge fixing condition
\begin{align}
    x^+ A^-_\mathrm{(II)}(x) + x^- A^+_\mathrm{(II)}(x) = 0, \label{eq:boost-inv-FS-gauge}
\end{align}
sets the $\pm$ components of the gauge field $A_\mathrm{(II)}$ to zero.
Hence, the color rotation in \cref{eq:boost-inv-color-rot} can be "gauged away" and the current $J$ is truly static.

The solutions of the single-nucleus fields $\mathcal{A}_{\A/\B}$ in \cref{eq:single-nucleus-gauge-fields} are given in covariant gauge.
Therefore, it is necessary to perform the appropriate gauge transformation of these fields, valid in Region I, to smoothly connect them to the Fock-Schwinger gauge in Region II.
This leads to the light cone gauge condition
\begin{align}
    \mathcal{A}^\mp_{\A/\B}(\xpm,\xperp) \rightarrow 0, \label{eq:light-cone-gauge-cond}
\end{align}
that has to be solved by the gauge transformation
\begin{align}
    \mathcal{A}^\mu_{\A/\B}(\xpm,\xperp) \rightarrow U_{\A/\B}(\xpm,\xperp)\left( \mathcal{A}^\mu_{\A/\B}(\xpm,\xperp) - \frac{1}{\ii g} \partial^\mu \right) U^\dagger_{\A/\B}(\xpm,\xperp).
\end{align}
Here, we enforce the same spacetime dependencies after the gauge transformation, which preserves the vanishing fields $A^\pm_{\A/\B} = 0$,
\begin{align}
    A^\pm_{\A/\B}(\xpm,\xperp) \rightarrow - \frac{1}{\ii g} U_{\A/\B}(\xpm,\xperp) \partial^\pm U^\dagger_{\A/\B}(\xpm,\xperp) = 0,
\end{align}
because the derivatives acting on $U^\dagger_{\A/\B}$ yield 0.
The solution to \cref{eq:light-cone-gauge-cond} is given by the solution to the differential equation
\begin{align}
    \partial^\mp U^\dagger_{\A/\B}(\xpm,\xperp) = \ii g \mathcal{A}^\mp_{\A/\B}(\xpm,\xperp) U^\dagger_{\A/\B}(\xpm,\xperp),
\end{align}
which defines the lightlike Wilson line
\begin{align}
    U^\dagger_{\A/\B}(\xpm,\xperp) = \mathtt{P}\, \ee^{\ii g \int_{-\infty}^{\xpm} \dd z^\pm \mathcal{A}^\mp_{\A/\B}(z^\pm,\xperp)}. \label{eq:light-like-wilson-line}
\end{align}
This exponential is defined via its series expansion, where the path ordering operator $\mathtt{P}$ ensures the correct order of the non-commuting fields $\mathcal{A}_{\A/\B}$.

\begin{figure}[p]
    \centering
    \includegraphics{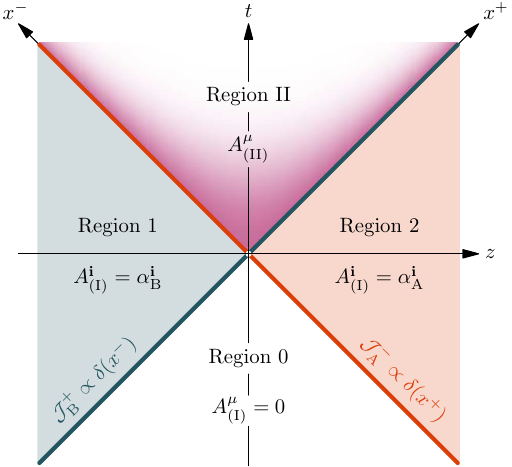}
    \caption[Setup of the boost-invariant Glasma in the $t$-$z$ plane.]{\label{fig:boost-invariant-regions}
    The gauge field $A_\mathrm{(II)}$ in Region II develops as a result of the collision of two boost-invariant currents $\mathcal{J}_\A$ and $\mathcal{J}_\B$.
    Region I from \cref{fig:two-ncl-collision-regions} is split up into Regions 0, 1, and 2 with distinct solutions for $A_\mathrm{(I)}$ in light cone gauge.
    Each current fills the spacetime in its wake with the pure gauge fields $\alpha^\bi_{\A/\B}$ in the transverse direction.
    Figure adapted from~\cite{Leuthner:2025vsd}.}
\end{figure}

The transformation to light cone gauge for the single-nucleus fields yields a new field configuration for Region I in \cref{fig:two-ncl-collision-regions}.
The previously nonzero light cone components are set to zero and, instead, the previously vanishing transverse components acquire nonzero contributions as
\begin{align}
    \mathcal{A}^\bi_{\A/\B}(\xpm,\xperp) \rightarrow -\frac{1}{\ii g} U_{\A/\B}(\xpm,\xperp) \partial^\bi U^\dagger_{\A/\B}(\xpm,\xperp). \label{eq:boost-inv-single-nucl-transv-A}
\end{align}
Due to the boost-invariant structure of the color charge distributions, the Wilson line in \cref{eq:light-like-wilson-line} has a particularly simple dependence on the longitudinal coordinate,
\begin{align}
    U^\dagger_{\A/\B}(\xpm,\xperp) = \theta(\xpm)U^\dagger_{\A/\B}(\xperp) + \theta(-\xpm)\mathbb{1},
\end{align}
with the asymptotic Wilson lines $U^\dagger_{\A/\B}(\xperp)$ that only depend on the transverse coordinate and are given as
\begin{align}
    U^\dagger_{\A/\B}(\xperp) = U^\dagger_{\A/\B}(\xpm\rightarrow\infty,\xperp) = \mathtt{P}\, \ee^{\ii g \int_{-\infty}^{\infty} \dd z^\pm \mathcal{A}^\mp_{\A/\B}(z^\pm,\xperp)}.
\end{align}
We may write the nonzero components of the gauge field for Region I as
\begin{align}
    A^\bi_\mathrm{(I)}(x) = \theta(x^+)\theta(-x^-) \alpha^\bi_\A(\xperp) + \theta(x^-)\theta(-x^+)\alpha^\bi_\B(\xperp),
\end{align}
where we used the asymptotic Wilson lines to express the single-nucleus fields in \cref{eq:boost-inv-single-nucl-transv-A} via the auxiliary fields
\begin{align}
    \alpha^\bi_{\A/\B}(\xperp) = -\frac{1}{\ii g} U_{\A/\B}(\xperp) \partial^\bi U^\dagger_{\A/\B}(\xperp).
\end{align}
The Heaviside step functions make it explicit that now the $t$-$z$ plane is split up into four distinct regions, illustrated in \cref{fig:boost-invariant-regions}.
The regions are separated by the tracks of the nuclei, which form the infinitesimally thin boundaries.
Region 0 is causally disconnected from any of the currents and fields and, therefore, contains no field.
The fields sourced by the boost-invariant currents $\mathcal{J}_{\A/\B}$ are transverse only and form a pure gauge configuration (i.e., the field strength is zero) in the wakes of the nuclei that smoothly connects to the field $A_\mathrm{(II)}$ in Region II.
However, the field $A_\mathrm{(I)}$ is discontinuous when crossing the nuclear tracks, and the contribution to the field strength in Region I is localized to the same Dirac-delta support as the currents.
Region II is similar to the setup in \cref{fig:two-ncl-collision-regions}, with the conceptual change that the currents now reside on the boundary, rather than inside the future light cone of the collision region.
Additionally, the collision region of the two nuclei is reduced to a single point in the $t$-$z$ plane.

\begin{figure}[p]
    \centering
    \includegraphics[width=0.7\textwidth]{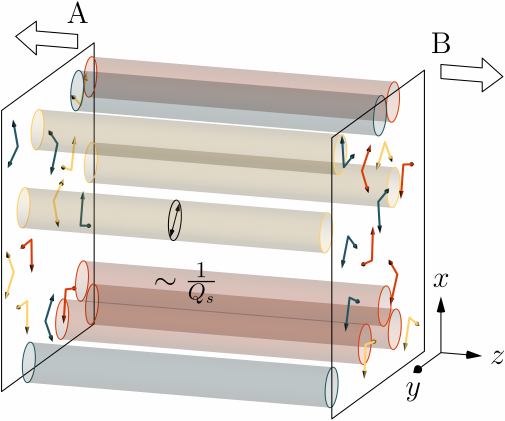}
    \caption[Flux-tubes in the boost-invariant Glasma.]{\label{fig:flux-tubes}
    Illustration of the field configuration shortly after the collision of nucleus $\A$ and $\B$.
    Chromo-electric and chromo-magnetic fields point in orthogonal directions in the transverse plane and are distributed according to the color charge density of each nucleus.
    Between the infinitesimally thin sheets of CGC for each nucleus, longitudinal "flux-tubes" are formed, with a typical transverse size given by the inverse saturation scale $1/Q_s$.}
\end{figure}

The gauge field in Region II is conveniently described in the Milne frame.
Using Milne coordinates, boost-invariance can be implemented by dropping all dependence on spacetime rapidity $\eta_s$.
Then, 
\begin{align}
    A^\tau_\mathrm{(II)}(x) &= \frac{1}{\tau} \left(x^+ A^-_\mathrm{(II)}(x) + x^-A^+_\mathrm{(II)}(x) \right) = 0, \label{eq:AII-tau} \\
    A^\eta_\mathrm{(II)}(x) &= \alpha^\eta(\tau,\xperp), \label{eq:AII-eta} \\
    A^\bi_\mathrm{(II)}(x) &= \alpha^\bi(\tau,\xperp), \label{eq:AII-i}
\end{align}
where the first equation reflects the Fock-Schwinger gauge from \cref{eq:boost-inv-FS-gauge}.
We introduced the new field $\alpha$ for the nonzero components of $A_\mathrm{(II)}$ that are dynamic, i.e., depend on proper time $\tau$, but not on $\eta_s$.
Inserting the ansatz from \cref{eq:AII-tau,eq:AII-eta,eq:AII-i} into the YM equations, one can derive initial conditions for the dynamic field $\alpha$ by demanding continuity along the boundary of Region II, yielding
\begin{align}
    \alpha^\eta(\tau\rightarrow 0,\xperp) &= \frac{\ii g}{2} \left[ \alpha^\bi_\A(\xperp), \alpha^\bi_\B(\xperp) \right], \label{eq:alphaII-eta} \\
    \alpha^\bi(\tau\rightarrow 0,\xperp) &= \alpha^\bi_\A(\xperp) + \alpha^\bi_\B(\xperp), \label{eq:alphaII-i} \\
    \partial_\tau A^\mu_\mathrm{(II)}(x)\Big|_{\tau\rightarrow 0} &= 0.
\end{align}
Note that the transverse components are a superposition of the single-nucleus transverse fields $\alpha_{\A/\B}$, but the $\eta$-component is only nonzero for non-commuting YM fields.

The initial conditions in \cref{eq:alphaII-eta,eq:alphaII-i} give rise to the following chromo-electric and chromo-magnetic fields 
\begin{align}
    E^z_\mathrm{(II)}(\tau\rightarrow 0,\xperp) &= \ii g \delta^{\bi\bj} \left[ \alpha^\bi_\A(\xperp), \alpha^\bj_\B(\xperp) \right], & \quad E^\bi_\mathrm{(II)}(\tau\rightarrow 0,\xperp) &= 0, \label{eq:boost-inv-EzII} \\
    B^z_\mathrm{(II)}(\tau\rightarrow 0,\xperp) &= \ii g \epsilon^{\bi\bj} \left[ \alpha^\bi_\A(\xperp), \alpha^\bj_\B(\xperp) \right], & \quad B^\bi_\mathrm{(II)}(\tau\rightarrow 0,\xperp) &= 0, \label{eq:boost-inv-BzII}
\end{align}
which are purely longitudinal.
They complement the purely-transverse, single-nucleus fields $\mathcal{E}_{\A/\B}$ and $\mathcal{B}_{\A/\B}$ from \cref{eq:nucl-A-E-field,eq:nucl-A-B-field} (and analogously for nucleus $\B$).
The picture that arises in terms of these fields is illustrated in \cref{fig:flux-tubes}.
The fields $\mathcal{E}_{\A/\B}$ and $\mathcal{B}_{\A/\B}$ (colored arrows) are aligned along the transverse plane and orthogonal to each other at every point.
They are constrained to the infinitesimally thin sheets that contain the charges of the nuclei and follow their color charge distributions.
Between the receding sheets of color charge, the longitudinal $E_\mathrm{(II)}$ and $B_\mathrm{(II)}$ fields emerge in Region II.
These longitudinal fields are focused in coherent domains in the transverse plane, which leads to the "flux-tube"~\cite{Lappi:2006fp,Dumitru:2008wn,Chen:2013ksa,Lappi:2017skr} structure in \cref{fig:flux-tubes}.
The characteristic size of these domains is given by the inverse of the saturation scale $Q_s$.
In the later evolution of the fields in Region II, these flux-tubes expand and decay into particles.

Finally, we summarize the properties of the energy-momentum tensor $T_\mathrm{(II)}$ for Region II in \cref{fig:boost-invariant-regions}.
We will consider event-averaged quantities denoted by $\langle \cdot \rangle_\mathrm{ev}$.
This average conceptually corresponds to the ensemble average of all possible charge distributions $\rho_{\A/\B}$ for the nuclei, which we will discuss in detail in the next section.
In experiment, this is realized by averaging independent collision events.
Evaluating the general expression,
\begin{align}
    \langle T^{\mu\nu}_\mathrm{(II)}(x)\rangle_\mathrm{ev} &= 2 \tr \langle F^{\mu\rho}F_{\rho}^{\hphantom{\rho}\nu} + \frac{1}{4}g^{\mu\nu}F^{\rho\sigma}F_{\rho\sigma} \rangle_\mathrm{ev} \nn\\
    &= \mathrm{diag}(\varepsilon(x),p_\perp(x),p_\perp(x),p_\mathrm{L}(x)),
\end{align}
one finds that $\langle T_\mathrm{(II)}\rangle_\mathrm{ev}$ for the boost-invariant Glasma is diagonal.
One may parametrize the energy-momentum tensor with the energy density $\varepsilon$, the transverse pressure $p_\perp$ and the longitudinal pressure $p_L$.
Due to the cylindrical symmetry of the collision, the transverse pressures along the $x$ and $y$ directions are expected to be equal to each other, although on an event-by-event basis, they will not be identical.
As for any conformal field theory (gluons are massless), the trace is zero, i.e., $\varepsilon = 2p_\perp + p_L$.
On the boundary of Region II, the particular structure of the initial fields in \cref{eq:boost-inv-EzII,eq:boost-inv-BzII} leads to
\begin{align}
    \langle T^{\mu\nu}_\mathrm{(II)}(x)\rangle_\mathrm{ev}\Big|_{\tau\rightarrow 0} = \mathrm{diag}(\varepsilon_0(x),\varepsilon_0(x),\varepsilon_0(x),-\varepsilon_0(x)), \label{eq:boost-inv-T-init}
\end{align}
where the energy-momentum tensor only contains a single independent component given by
\begin{align}
    \varepsilon_0 = \tr \langle E^2_{z\mathrm{(II)}}(\tau\rightarrow 0,\xperp) + B^2_{z\mathrm{(II)}}(\tau\rightarrow 0,\xperp) \rangle_\mathrm{ev}.
\end{align}
Note that the longitudinal pressure identified with the last component of $\langle T_\mathrm{(II)}\rangle_\mathrm{ev}$ in \cref{eq:boost-inv-T-init} is negative, i.e., $p_L = -\varepsilon_0 < 0$ because the energy density $\varepsilon_0 > 0$.
This shows that the initial conditions of the collision lead to a highly anisotropic system along the beam axis.
In the later evolution of the Glasma, $p_L$ approaches zero from below, whereas the energy density $\varepsilon \propto \varepsilon_0 / \tau$ because of the longitudinal expansion of the system.
Full studies of the time evolution of the boost-invariant Glasma are carried out using an expansion for small times in~\cite{Chen:2013ksa,Fries:2006pv,Fujii:2008km,Chen:2015wia,Guerrero-Rodriguez:2021ask,Carrington:2020ssh,Carrington:2021qvi,Carrington:2025xws} or fully non-perturbatively on the lattice in~\cite{Krasnitz:1999wc,Krasnitz:1998ns,Krasnitz:2000gz,Krasnitz:2001qu,Krasnitz:2002mn,Krasnitz:2002ng,Lappi:2003bi,Lappi:2004sf,Lappi:2009xa,Lappi:2011ju,Epelbaum:2013ekf,Dumitru:2014nka,Schenke:2015aqa}.
The isotropization of the boost-invariant Glasma and its behavior along the longitudinal direction is studied in~\cite{Romatschke:2005ag,Romatschke:2006nk,Romatschke:2005pm,Fukushima:2007yk,Fujii:2008dd,Fukushima:2011nq,Epelbaum:2013waa,Bazak:2023kol}, where the role of plasma instabilities in this process is worked out.

\section{The McLerran--Venugopalan model}\label{sec:MV-nucl-model}

The color charge densities $\rho_{\A/\B}$, which describe the distribution of color charge within the nuclei $\A$ and $\B$, are stochastic quantities.
A single experiment would probe one concrete realization, and repeating the experiment would probe another one.
To characterize the mean of these fluctuations, we introduced the event average $\langle \cdot \rangle_\mathrm{ev}$ in the previous section.
Now, we will formalize the fluctuations of the color charge density and its expectation value in the usual treatment within the CGC effective description for theoretical modeling applications~\cite{Iancu:2003xm,Iancu:2012xa,Gelis:2010nm,Gelis:2012ri,Fukushima:2011ca,Fukushima:2016xgg,Gelis:2021zmx,Garcia-Montero:2025hys}.
This yields the CGC expectation value as the equivalent of the event average.

The ensemble of possible color charge configurations is assumed to be distributed according to the CGC weight function $W[\rho_{\A/\B}]$.
It is normalized, 
\begin{align}
    \intop \mathcal{D}\rho_{\A/\B}\, W[\rho_{\A/\B}]= 1, \label{eq:CGC-weight-norm}
\end{align}
where $\mathcal{D}$ denotes functional integration.
Using the weight function, any observables $\mathcal{O}$ will depend on the charge distributions of the nuclei via the CGC expectation value defined as
\begin{align}
    \langle \mathcal{O}[\rho_\A,\rho_\B] \rangle = \intop \mathcal{D}\rho_\A\,\mathcal{D}\rho_\B\, \mathcal{O}[\rho_\A,\rho_\B] W[\rho_\A] W[\rho_\B].
\end{align}

Determining the weight function from first principles requires accurate knowledge of the non-perturbative regime of QCD.
To date, this has not been achieved.
Therefore, we rely on physically motivated models for the weight function.
The most famous and widely used model is named after McLerran and Venugopalan (MV)~\cite{McLerran:1993ka,McLerran:1993ni}, who proposed that the weight function takes the form of a Gaussian functional.
As an immediate result, $W[\rho_{\A/\B}]$ is completely fixed by the one-point and two-point functions%
\footnote{%
In the following, we use the terms "correlator" and "two-point function" interchangeably because no correlator with more than two fields enters the model.}
\begin{align}
    \langle \rho^{\perp,a}_{\A/\B}(\xperp) \rangle &= 0, \label{eq:MV-correlator-transverse-1pt} \\
    \langle \rho^{\perp,a}_{\A/\B}(\xperp) \rho^{\perp,b}_{\A/\B}(\yperp) \rangle &= g^2 \mu^2 \delta^{ab} \deltaperp(\xperp-\yperp),\label{eq:MV-correlator-transverse}
\end{align}
where in the original description the color charge densities $\rho^\perp_{\A/\B}$ have no longitudinal extent, i.e., this model is boost-invariant.
Here, $g$ is the coupling and $\mu$ is the phenomenological MV parameter given in units of energy that characterizes the strength of color charge.
It is assumed to scale $\mu^2 \sim A^{1/3}$, with the atomic number $A$ of the nucleus, and is related to the non-perturbative saturation scale $\mu \propto Q_s/g^2$.
The special choice of a Gaussian functional also leads to the simplification that any observables are given in terms of the correlator in \cref{eq:MV-correlator-transverse}, in analogy to Wick's theorem.
Additionally, many calculations become analytically tractable.

The properties of the color charge distribution in the MV model are encoded in \cref{eq:MV-correlator-transverse-1pt,eq:MV-correlator-transverse}.
The average color charge for each color component $a$ for a given nucleus is required to be zero due to the color neutrality of the nuclei.
This leads to \cref{eq:MV-correlator-transverse-1pt}.
On the other hand, the local fluctuations of the color charge are controlled by \cref{eq:MV-correlator-transverse} where the Dirac-delta function in the transverse separation $(\xperp-\yperp)$ leads to a completely random configuration in the transverse plane.
Two points with nonzero separation are not correlated, such that there is no transverse structure.
This holds for each of the color components individually due to the Kronecker delta in color space.

The validity of the Gaussian weight functional was studied extensively in the literature~\cite{Iancu:2002aq,Lam:2001ax,Jalilian-Marian:1996mkd,Jalilian-Marian:1997qno,Jalilian-Marian:1997jhx,Jalilian-Marian:1998tzv,Weigert:2000gi,Iancu:2000hn,Ferreiro:2001qy,Iancu:2001ad,Weigert:2005us,Blaizot:2002np,Iancu:2013uva}, which led to the development of the JIMWLK renormalization group equation.
It governs the evolution of the CGC weight function with saturation built in.
However, the two-dimensional MV model had to be extended to incorporate the dependence on the momentum rapidity of the gluons in the CGC.
On the one hand, this led to phenomenological parametrizations extended by the momentum rapidity~\cite{Lappi:2004sf,Lappi:2011ju,Schenke:2012wb,Schenke:2012hg,Schenke:2013dpa,Schenke:2014tga,Mantysaari:2025tcg,Gelis:2008sz,Dusling:2009ni,Schenke:2016ksl,McDonald:2017eml,McDonald:2018wql,McDonald:2020oyf,McDonald:2020xrz,McDonald:2023qwc}.
On the other hand, the color charge correlator in the MV model was generalized to finite longitudinal support~\cite{Lam:2000nz,Fukushima:2007ki,Ozonder:2012vw,Ozonder:2013moa,Shen:2017bsr,Shen:2022oyg} and to nuclei with finite three-dimensional envelopes on the lattice~\cite{Gelfand:2016yho,Ipp:2017lho,Ipp:2018hai,Ipp:2020igo,Muller:2019bwd,Schlichting:2020wrv,Singh:2021hct,Matsuda:2023gle,Matsuda:2024moa,Matsuda:2024mmr}.

Finite longitudinal support can be incorporated in the MV model from \cref{eq:MV-correlator-transverse} by allowing the MV parameter $\mu$ to depend on $\xpm$,
\begin{align}
    \langle \rho^a_{\A/\B}(\xpm,\xperp) \rho^b_{\A/\B}(\ypm,\yperp) \rangle &= \delta^{ab} g^2 \mu(\xpm)^2 \delta(\xpm-\ypm)\deltaperp(\xperp-\yperp).\label{eq:MV-correlator-longitudinal}
\end{align}
The longitudinal support of the envelope given by $\mu(\xpm)$ scales $\sim R/\gamma$, where $R$ is the nuclear radius and $\gamma$ the Lorentz factor, and is highly peaked on the light cone for ultra-relativistic nuclei.
This also introduces the dependence on collider energy (via $\gamma$) into the model.
To stay consistent with the two-dimensional MV model in \cref{eq:MV-correlator-transverse}, the profile $\mu(\xpm)$ is normalized to the original MV parameter $\mu$,
\begin{align}
    \intop \dd\xpm \mu(\xpm)^2 = \mu^2.
\end{align}
Then, integrating \cref{eq:MV-correlator-longitudinal} over both longitudinal coordinates and exchanging the order of the integrations and the CGC expectation value restores \cref{eq:MV-correlator-transverse},
\begin{align}
    \langle \intop\dd\xpm\,\rho^a_{\A/\B}(\xpm,\xperp) \intop\dd\ypm\,\rho^b_{\A/\B}(\ypm,\yperp) \rangle & \nn\\
    &\hspace{-23pt}= \intop\dd\xpm\dd\ypm\,\delta^{ab} g^2 \mu(\xpm)^2 \delta(\xpm-\ypm)\deltaperp(\xperp-\yperp) \nn\\
    &\hspace{-23pt}= \delta^{ab} g^2 \mu^2 \deltaperp(\xperp-\yperp) \nn\\
    &\hspace{-23pt}= \langle \rho^{\perp,a}_{\A/\B}(\xperp) \rho^{\perp,b}_{\A/\B}(\yperp) \rangle.
\end{align}
Here, we introduced the transverse projected color charge densities and set them equal to the transverse densities from the two-dimensional MV model,
\begin{align}
    \rho^\perp_{\A/\B}(\xperp) \coloneqq \int\dd\xpm\,\rho_{\A/\B}(\xpm,\xperp).
\end{align}

While the longitudinal parametrization in \cref{eq:MV-correlator-longitudinal} does not capture the energy dependence given by the JIMWLK equation, it does lead to broken boost invariance due to the collider energy dependence and finite Lorentz contraction.
The additional longitudinal structure, however, is completely uncorrelated because of the addition of the longitudinal Dirac-delta function.
In simple terms, this model corresponds to stacking independent two-dimensional MV charge distributions along the longitudinal direction.
It needs to be emphasized that performing the boost-invariant limit $\mu(\xpm)^2\rightarrow \delta(\xpm)\mu^2$ requires special care because of the path ordering in the Wilson lines.
The characterization of the fluctuations in the transverse plane remains unchanged.

For completeness, we take note of the explicit expression for the Gaussian functional that takes the role of the CGC weight function in the generalized MV model,
\begin{align}\label{eq:CGC-weight-fct-gauss}
    W[\rho_{\A/\B}] = \mathcal{N} \exp\left\{ -\frac{1}{2} \intop \dd\xpm\,\dd^2\xperp\, \frac{\rho^a_{\A/\B}(\xpm,\xperp)\rho^a_{\A/\B}(\xpm,\xperp)}{g^2\mu(\xpm)^2} \right\},
\end{align}
where the normalization factor $\mathcal{N}$ ensures that \cref{eq:CGC-weight-norm} is satisfied.
Similar generalizations are discussed, for example, in~\cite{Jalilian-Marian:1996mkd}.

\subsection{Finite size nuclei with generalized correlations}\label{sec:generalized-correl}

The original MV correlator in \cref{eq:MV-correlator-transverse} is a fitting model for the center of a large nucleus, where boundary effects due to the actually finite size of the nucleus are negligible.
Adding the longitudinal profile $\mu(\xpm)$ in \cref{eq:MV-correlator-longitudinal} relaxes the assumption of boost invariance but does not constrain the transverse size of the nucleus.
A different approach is to keep the boost-invariant description and introduce a transverse envelope $T^\perp_{\A/\B}$, referred to as the "thickness function",
\begin{align}\label{eq:MV-correlator-thickness-T}
    \langle \rho^{\perp,a}_{\A/\B}(\xperp) \rho^{\perp,b}_{\A/\B}(\yperp) \rangle &= \delta^{ab} g^2 \mu^2\, T^\perp_{\A/\B}(\xperp) \deltaperp(\xperp-\yperp).
\end{align}
This allows studying the transverse geometry of the Glasma (e.g.,~\cite{Lappi:2003bi,Krasnitz:2002ng,Krasnitz:2002mn,Carrington:2020ssh}) and, in particular, incorporates impact parameter dependence~\cite{Kowalski:2003hm,Schenke:2012wb,Schenke:2012hg,Schenke:2013dpa,Schenke:2014tga,Schenke:2016ksl} and control over centrality of the collision.

In recent efforts~\cite{Gelfand:2016yho,Ipp:2017lho,Ipp:2018hai,Ipp:2020igo,Muller:2019bwd,Schlichting:2020wrv,Singh:2021hct,Matsuda:2023gle,Matsuda:2024moa,Matsuda:2024mmr,Ipp:2024ykh,Ipp:2025sbt,Ipp:2025sbc,Leuthner:2025vsd,Ipp:2025cdh}, a fully three-dimensional envelope $T_{\A/\B}$, which depends on the longitudinal $\xpm$ and transverse $\xperp$, was used to constrain the MV model correlator to realistic nuclear sizes.
Given the approximate spherical symmetry of a large nucleus with radius $R$ in its rest frame, the envelope function for a boosted nucleus contains the intrinsic scales $R$ and $R/(\sqrt{2}\gamma)$ in the transverse and longitudinal (light cone) directions.
The difference to combining phenomenological parametrizations of the MV parameter or JIMWLK with transverse envelopes is that here the energy dependence is reduced to the effects of Lorentz contraction with the $\gamma$ factor.
The general ansatz reads
\begin{align}\label{eq:MV-correlator-T(x)}
    \langle \rho^a_{\A/\B}(\xpm,\xperp) \rho^b_{\A/\B}(\ypm,\yperp) \rangle &= \delta^{ab} g^2 \mu^2\, T_{\A/\B}(\xpm,\xperp) \delta(\xpm-\ypm)\deltaperp(\xperp-\yperp),
\end{align}
where $\mu$ is again a phenomenological constant related to the average strength of the color charge.
It is desirable to normalize $T_{\A/\B}$ such that the original MV correlator in \cref{eq:MV-correlator-transverse} is restored when integrating \cref{eq:MV-correlator-T(x)} over both longitudinal coordinates and evaluating the envelope at the center of the nuclei at $\xperp=\zeroperp$.
The resulting normalization condition reads
\begin{align}\label{eq:envelope-norm}
    \intop \dd\xpm\,T_{\A/\B}(\xpm,\zeroperp) = T^\perp_{\A/\B}(\zeroperp) = 1.
\end{align}
Here, we identified the thickness function from \cref{eq:MV-correlator-thickness-T} as the transverse projected envelope (i.e., longitudinally integrated).
Note that \cref{eq:envelope-norm} imposes a different normalization%
\footnote{%
Common normalizations (e.g.~\cite{Miller:2007ri,dEnterria:2020dwq}) are $\int d^2\xperp\, T^\perp_{\A/\B}(\xperp) = 1$, which allows $T^\perp_{\A/\B}(\xperp)$ to be interpreted as the probability per unit area to find a nucleon inside a tube with cross-section $d^2\xperp$, and $\int d^2\xperp\, T^\perp_{\A/\B}(\xperp) = A$, with the nuclear mass number $A$, which corresponds to a scaling with the volume of the nucleus.}
compared to the commonly used thickness functions in Glauber models~\cite{Miller:2007ri,dEnterria:2020dwq}.
Additionally, the envelope function $T_{\A/\B}$ acquires a dimension of inverse length, whereas the thickness function $T^\perp_{\A/\B}$ is dimensionless.

Still, \cref{eq:MV-correlator-T(x)} describes completely random color charges with no structure inside the nuclei.
We now aim to generalize the Dirac-delta functions in the correlator to a correlation function $\Gamma_{\A/\B}$ with non-singular support.
Physically, this corresponds to three-dimensional, coherent domains of color charge.
In such a model, the lumpy structure of a nucleus, given by the assembly of nucleons, can be captured.
However, we cannot replace the Dirac-delta functions with a correlation function%
\footnote{%
Another complication with the generalized correlation function is gauge invariance.
The Dirac-delta functions ensure that the correlator transforms locally, which, together with the Kronecker delta in color space, leads to a gauge-invariant correlator.
Throughout this thesis, we fix covariant gauge for the initial conditions, where the interpretation of charges is straightforward.
See~\cite{Leuthner:2025vsd} for a detailed discussion.}
$\Gamma_{\A/\B}$,
\begin{align}\label{eq:general-correlator-T(x)}
    \langle \rho^a_{\A/\B}(\xpm,\xperp) \rho^b_{\A/\B}(\ypm,\yperp) \rangle &= \delta^{ab}\, T_{\A/\B}(\xpm,\xperp) \Gamma_{\A/\B}(\xpm-\ypm,\xperp-\yperp),
\end{align}
because the right-hand side of \cref{eq:general-correlator-T(x)} is no longer symmetric in $(\xpm,\xperp)$ and $(\ypm,\yperp)$ under the exchange of the color indices.
For \cref{eq:general-correlator-T(x)} and in the following, we absorb all prefactors compared to \cref{eq:MV-correlator-T(x)} into the definition of $\Gamma_{\A/\B}$.
To retain the symmetry, we can symmetrize the contribution from the envelope $T_{\A/\B}$ in two straightforward ways:

First, we change the arguments of the envelope to average coordinates
\begin{align}\label{eq:MV-correlator-T(x+y)}
    \langle \rho^a_{\A/\B}(\xpm,\xperp) \rho^b_{\A/\B}(\ypm,\yperp) \rangle &= \delta^{ab}g^2\mu^2\, T_{\A/\B}(\tfrac{\xpm+\ypm}{2},\tfrac{\xperp+\yperp}{2})\delta(\xpm-\ypm)\deltaperp(\xperp-\yperp),
\end{align}
which is still equivalent to \cref{eq:MV-correlator-T(x)} because of the delta-like correlations.
We now relax the Dirac-delta functions using the correlation function $\Gamma_{\A/\B}$,
\begin{align}\label{eq:general-correlator-T(x+y)}
    \langle \rho^a_{\A/\B}(\xpm,\xperp) \rho^b_{\A/\B}(\ypm,\yperp) \rangle &= \delta^{ab}\, T_{\A/\B}(\tfrac{\xpm+\ypm}{2},\tfrac{\xperp+\yperp}{2}) \Gamma_{\A/\B}(\xpm-\ypm,\xperp-\yperp).
\end{align}
This way of parametrizing the functional dependence has the advantage that the average coordinates are orthogonal to the difference coordinates in the arguments of the correlation function.
Conceptually, the clean separation of the coordinate dependencies allows the correlation function $\Gamma_{\A/\B}$ to entirely fix the structure of the correlations without contributions from the envelopes.
However, as we demonstrate in \cref{sec:pos-semi-definite-correlators}, not all choices for the envelope and correlation function in this formulation lead to physically viable nuclear models.

For the second option, we start by rewriting the correlator in \cref{eq:MV-correlator-T(x)} as
\begin{align}\label{eq:MV-correlator-sqrtT(x)T(y)}
    \langle \rho^a_{\A/\B}(\xpm,\xperp) \rho^b_{\A/\B}(\ypm,\yperp) \rangle & \nn\\
    &\hspace{-60pt}= \delta^{ab}g^2\mu^2 \sqrt{T_{\A/\B}(\xpm,\xperp)T_{\A/\B}(\ypm,\yperp)}\delta(\xpm-\ypm)\deltaperp(\xperp-\yperp),
\end{align}
and introduce a new shorthand for the square roots%
\footnote{%
We will use the term "nuclear envelopes" also to refer to the square roots $t_{\A/\B}$, without explicitly emphasizing that these are square roots of the envelopes $T_{\A/\B}$.}
of the nuclear envelopes $t_{\A/\B} = \sqrt{T_{\A/\B}}$.
Then we replace the delta functions with the correlation function,
\begin{align}\label{eq:general-correlator-t(x)-t(y)}
    \langle \rho^a_{\A/\B}(\xpm,\xperp) \rho^b_{\A/\B}(\ypm,\yperp) \rangle &= \delta^{ab}\, t_{\A/\B}(\xpm,\xperp)t_{\A/\B}(\ypm,\yperp) \Gamma_{\A/\B}(\xpm-\ypm,\xperp-\yperp).
\end{align}
In contrast to \cref{eq:general-correlator-T(x+y)}, the parametrization is no longer cleanly separated into average and difference coordinates.
Therefore, contributions from the envelopes will mix into the correlations.
The advantage of \cref{eq:general-correlator-t(x)-t(y)} over \cref{eq:general-correlator-T(x+y)} is that \cref{eq:general-correlator-t(x)-t(y)} allows for more freedom in the choice of $t_{\A/\B}$, as will become evident in the next \cref{sec:pos-semi-definite-correlators}.
In the following, we will primarily refer back to \cref{eq:general-correlator-t(x)-t(y)}.
In \cref{ch:nuclear-models}, we discuss different realizations of this correlator and contrast different models for $\Gamma_{\A/\B}$.

We stress that after the Dirac-delta functions have been replaced by $\Gamma_{\A/\B}$, the two color charge correlators in \cref{eq:general-correlator-t(x)-t(y),eq:general-correlator-T(x+y)} are, in general, not equivalent.
A practical example where the envelope $T_{\A/\B}$ can be factorized is the choice of a Gaussian,
\begin{align}\label{eq:general-gauss-T}
T_{\A/\B}(\xpm,\xperp) = \beta\, \ee^{-\frac{(\gamma\xpm)^2}{R^2} -\frac{\xperp^2}{2R^2}},
\end{align}
with a suitable normalization factor $\beta$ and the radius parameter $R$.
It is straightforward to check that \cref{eq:general-gauss-T} can be factorized as
\begin{align}\label{eq:T(x+y)-factorized}
    T_{\A/\B}(\tfrac{\xpm+\ypm}{2},\tfrac{\xperp+\yperp}{2}) &= t_{\A/\B}(\xpm,\xperp)t_{\A/\B}(\ypm,\yperp)\, \ee^{+\frac{\gamma^2(\xpm-\ypm)^2}{4R^2}+\frac{(\xperp-\yperp)^2}{8R^2}}.
\end{align}
The explicit exponential factor on the right-hand side is a function of the difference coordinates and can be absorbed into a redefinition of the correlation function $\Gamma_{\A/\B}$, which results in a color charge correlator of the same structure as \cref{eq:general-correlator-t(x)-t(y)}.

We enforce a particular normalization of the correlation function $\Gamma_{\A/\B}$ to fix the phenomenological prefactors.
In the limit where the longitudinal and transverse correlations reduce to delta-like support, we want to recover \cref{eq:MV-correlator-T(x)}.
When we parametrize the longitudinal correlation scale with $r_l$ and the transverse scale with $r$, the condition reads
\begin{align}\label{eq:Gamma-MV-limit-norm}
    \lim_{\substack{r_l\,\rightarrow\,0 \\ r\,\rightarrow\,0}} \Gamma_{\A/\B}(\xpm,\xperp) = g^2\mu^2\delta(\xpm)\deltaperp(\xperp).
\end{align}

For later reference, we calculate the Fourier transformation of the generalized color charge correlator in \cref{eq:general-correlator-t(x)-t(y)},
\begin{align}
    \langle \tilde{\rho}^a_{\A/\B}(\pmp,\pperp)\tilde{\rho}^b_{\A/\B}(\qmp,\qperp)\rangle & \nn\\
    &\hspace{-63pt}= \intop \dd\xpm\,\dd^2\xperp\,\dd\ypm\,\dd^2\yperp\,\ee^{\ii\xpm\pmp -\ii\xperp\cdot\pperp + \ii\ypm\qmp -\ii\yperp\cdot\qperp}\,\langle\rho^a_{\A/\B}(\xpm,\xperp)\rho^b_{\A/\B}(\ypm,\yperp)\rangle \nn\\
    &\hspace{-63pt}= \intop \dd\xpm\,\dd^2\xperp\,\dd\ypm\,\dd^2\yperp\,\ee^{\ii\xpm\pmp -\ii\xperp\cdot\pperp + \ii\ypm\qmp -\ii\yperp\cdot\qperp} \nn\\
    &\hspace{-63pt}\times \delta^{ab} t_{\A/\B}(\xpm,\xperp) t_{\A/\B}(\ypm,\yperp) \Gamma_{\A/\B}(\xpm-\ypm,\xperp-\yperp).
    \label{eq:FT-general-correlator-tx-ty-step1}
\end{align}
The Fourier transformation of the product of the two envelopes and the correlation function leads to a convolution of the Fourier-transformed functions and we obtain
\begin{align}
    \langle \tilde{\rho}^a_{\A/\B}(\pmp,\pperp)\tilde{\rho}^b_{\A/\B}(\qmp,\qperp)\rangle & \nn\\
    &\hspace{-72.1pt}= \delta^{ab}\intop \frac{\dd\kappa^\mp\,\dd^2\kappaperp}{(2\pi)^3}\, \tilde{t}_{\A/\B}(\pmp-\kappa^\mp,\pperp-\kappaperp)\tilde{t}_{\A/\B}(\qmp+\kappa^\mp,\qperp+\kappaperp)\tilde{\Gamma}_{\A/\B}(\kappa^\mp,\kappaperp) .
    \label{eq:FT-general-correlator-t(x)-t(y)}
\end{align}

\subsection{Positive semi-definiteness in the MV nuclear model}\label{sec:pos-semi-definite-correlators}

In the MV nuclear model, the CGC weight function in \cref{eq:CGC-weight-fct-gauss} is assumed to be Gaussian.
For any valid Gaussian probability functional, the corresponding two-point function must be positive semi-definite, i.e.,
\begin{align}\label{eq:positive_definite}
    2\intop\dd\xpm\,\dd^2\xperp\,\dd\ypm\,\dd^2\yperp\, \psi(\xpm,\xperp) \tr\langle \rho_{A/B}(\xpm, \xperp) \rho_{A/B}(\ypm, \yperp)\rangle \psi(\ypm, \yperp) \geq 0
\end{align}
and bounded for any real-valued, square-integrable function $\psi(\xpm, \xperp)$.
In the following, we analyze this condition for the generalized color charge correlator in \cref{eq:general-correlator-t(x)-t(y)}.
We insert \cref{eq:general-correlator-t(x)-t(y)} into \cref{eq:positive_definite} and get
\begin{align}
    &\intop\dd\xpm\,\dd^2\xperp\,\dd\ypm\,\dd^2\yperp\, \psi(\xpm,\xperp)\psi(\ypm, \yperp) \nn\\
    &\times \delta^{aa}\, t_{\A/\B}(\xpm,\xperp)t_{\A/\B}(\ypm,\yperp) \Gamma_{\A/\B}(\xpm-\ypm,\xperp-\yperp) \nn\\
    &= (N_c^2 -1) \intop \dd\xpm\,\dd^2\xperp\,\dd\ypm\,\dd^2\yperp\, \zeta_{\A/\B}(\xpm,\xperp)\zeta_{\A/\B}(\ypm, \yperp) \Gamma_{\A/\B}(\xpm - \ypm, \xperp - \yperp) \nn\\
    &\geq 0.
\end{align}
In the second equation, we defined the shorthand
\begin{align}
    \zeta_{\A/\B}(\xpm, \xperp) = \psi(\xpm, \xperp) t_{\A/\B}(\xpm, \xperp).
\end{align}
Next, we change to Fourier space for each $\zeta_{\A/\B}$ and $\Gamma_{\A/\B}$,
\begin{align}
    &(N_c^2 -1) \intop \frac{\dd\xpm\,\dd^2\xperp\,\dd\ypm\,\dd^2\yperp\,\dd\pmp\,\dd^2\pperp\,\dd\qmp\,\dd^2\qperp\,\dd\kmp\,\dd^2\kperp}{(2\pi)^9} \nn\\
    &\times \ee^{-\ii\xpm\pmp + \ii\xperp\cdot\pperp - \ii\ypm\qmp + \ii\yperp\cdot\qperp - \ii\kmp(\xpm-\ypm) + \ii\kperp\cdot(\xperp-\yperp)} \nn\\
    &\times \tilde{\zeta}_{\A/\B}(\pmp,\pperp)\tilde{\zeta}_{\A/\B}(\qmp,\qperp)\tilde{\Gamma}_{\A/\B}(\kmp,\kperp) \nn\\
    &= (N_c^2 -1) \intop \frac{\dd\pmp\,\dd^2\pperp\,\dd\qmp\,\dd^2\qperp\,\dd\kmp\,\dd^2\kperp}{(2\pi)^3} \nn\\
    &\times \delta(\pmp+\kmp)\deltaperp(\pperp+\kperp)\delta(\qmp-\kmp)\deltaperp(\qperp-\kperp) \nn\\
    &\times \tilde{\zeta}_{\A/\B}(\pmp,\pperp)\tilde{\zeta}_{\A/\B}(\qmp,\qperp)\tilde{\Gamma}_{\A/\B}(\kmp,\kperp) \nn\\
    &= (N_c^2 -1) \intop \frac{\dd\kmp\,\dd^2\kperp}{(2\pi)^3}\tilde{\zeta}_{\A/\B}(-\kmp,-\kperp)\tilde{\zeta}_{\A/\B}(\kmp,\kperp)\tilde{\Gamma}_{\A/\B}(\kmp,\kperp) \nn\\
    &= (N_c^2 -1) \intop \frac{\dd\kmp\,\dd^2\kperp}{(2\pi)^3} \left| \tilde{\zeta}_{\A/\B}(\kmp,\kperp)\right|^2 \tilde{\Gamma}_{\A/\B}(\kmp,\kperp) \geq 0. \label{eq:fourier-pos-semi-def-cond}
\end{align}
Integrating out $(\xpm,\xperp)$ and $(\ypm,\yperp)$ yields six Dirac-delta functions, which reduce the momentum integrations to a three-dimensional convolution.
In the last line, we used that the Fourier transformation of the real-valued function $\zeta_{\A/\B}$ has the symmetry property
\begin{align}
    \tilde{\zeta}_{\A/\B}(\kmp,\kperp)^* = \tilde{\zeta}_{\A/\B}(-\kmp,-\kperp),
\end{align}
where the star denotes complex conjugation.
The condition in \cref{eq:fourier-pos-semi-def-cond} is trivially satisfied if the kernel of the integration is non-negative for all values of $(\kmp,\kperp)$.
Since the absolute square of $\tilde{\zeta}_{\A/\B}$ is non-negative, we can identify
\begin{align}\label{eq:tildeGamma-gt-0}
    \tilde{\Gamma}_{\A/\B}(\kmp,\kperp) \geq 0
\end{align}
as a sufficient criterion for the color charge correlator in \cref{eq:general-correlator-t(x)-t(y)} to describe a valid Gaussian functional.
While \cref{eq:tildeGamma-gt-0} constitutes a concrete restriction for model-building, it is possible that $\Gamma_{\A/\B}$, which violate \cref{eq:tildeGamma-gt-0}, still lead to positive semi-definite correlators, if the envelopes $t_{\A/\B}$ are tuned in accordance.
The usefulness of \cref{eq:tildeGamma-gt-0} lies in the possibility to choose the envelopes freely, as long as the Fourier-transformed correlation function is non-negative.

Note that the other formulation of the color charge correlator in \cref{eq:general-correlator-T(x+y)}, where the envelope $T_{\A/\B}$ is evaluated with average coordinates, does not lead to \cref{eq:fourier-pos-semi-def-cond}.
The test function $\psi$ cannot be recast into a square because of how the arguments are mixed together after the Fourier transformation.
Still, this does not mean that \cref{eq:general-correlator-T(x+y)} is never positive semi-definite.
On the contrary, the example of Gaussian $T_{\A/\B}$ in \cref{eq:general-gauss-T} leads to a valid correlator if
\begin{align}
    \intop \dd\xpm\,\dd^2\xperp\, \ee^{\ii\xpm\kmp - \ii\xperp\cdot\kperp}\Gamma_{\A/\B}(\xpm,\xperp) \ee^{+\frac{\gamma^2(\xpm)^2}{4R^2}+\frac{(\xperp)^2}{8R^2}} \geq 0,
\end{align}
where the additional exponential factors appear because of the factorization in \cref{eq:T(x+y)-factorized}.
Different choices of $\Gamma_{\A/\B}$ or $T_{\A/\B}$ might violate positive semi-definiteness for the correlator in \cref{eq:general-correlator-T(x+y)}.

\chapter{The (3+1)D dilute Glasma approximation}\label{sec:dilute-glasma-approx}

In light of the generalized three-dimensional MV nuclear model discussed in \cref{sec:generalized-correl}, we seek to find solutions for the nonlinear dynamics of the Glasma.
In particular, the goals are to preserve the longitudinal dynamics (i.e., rapidity dependence) and study the properties of rapidity distributions in terms of the model parameters of the nuclear model.
We choose an approximation of the solution to the YM equations, formulated as a linearization where non-Abelian contributions are largely pushed into higher-order terms.
Given a suitable gauge choice, the contributions of these higher-order terms can be minimized.
This technique was first developed using the boost-invariant MV nuclear model in~\cite{Kovner:1995ts,Kovner:1995ja,Kovchegov:1997ke} and is extensively used in~\cite{Dumitru:2001ux,Blaizot:2004wu,Blaizot:2008yb,Blaizot:2010kh,McLerran:2016snu}, where the MV model is extended with parametric rapidity dependence.
In the following, we aim to derive the results for the three-dimensional MV nuclear model, where the rapidity dependence is a consequence of the longitudinal structure of the nuclei.

The core assumption is to treat the source terms that appear in the YM equations as small compared to the other competing scales in the system.
This leads to an expansion in terms of the color charge distributions $\rho_{\A/\B}$ and tremendously simplifies the YM equations.
The common interpretation of the regime of validity for this approximation is argued in terms of the ratio of the saturation scales of the nuclei $Q_s^{\A/\B}$ and the transverse momentum $\kperp$ of the gluons produced in the Glasma.
When $\kperp \gg Q_s^{\A/\B}$, the incoming nuclei are "dilute" compared to the momentum scale set by the gluons.
This is precisely when the non-Abelian interactions are minimal and the dilute approximation becomes valid.
Recent studies~\cite{Schlichting:2019bvy,Blaizot:2010kh,Avsar:2012hj,Chirilli:2015tea,Li:2021zmf,Li:2021yiv} rigorously examine higher order contributions and compare to non-perturbative results obtained from lattice calculations.
They confirm that the dilute regime reproduces gluon spectra for pA collisions and is also applicable to nucleus-nucleus collisions, albeit with a larger discrepancy when $\kperp \lesssim Q_s^{\A/\B}$.

In this chapter, we study the (3+1)D dilute Glasma~\cite{Ipp:2021lwz,Singh:2021hct,Ipp:2022lid,Ipp:2024ykh,Ipp:2025sbt,Ipp:2025sbc,Ipp:2025cdh,Leuthner:2025vsd} in detail.
First, we establish the identification between the lowest-order, non-trivial terms in the linearization of the YM equations and the Glasma that is produced in a HIC.
We then proceed to solve the full dynamics of the linearized YM equations in position space.
Of special focus is the structure of the resulting field-strength tensor of the Glasma.
We conclude this chapter with a short review of the numerical implementation applied to produce the results presented in \cref{ch:numerical-results}.
In the next \cref{ch:momentum-space-picture}, we change to momentum space and rigorously derive gluon production in the (3+1)D dilute Glasma.
The results recover the prediction of $k_T$ factorization from perturbative techniques.
The boost-invariant description is naturally extended to include longitudinal dependence in the unintegrated gluon distributions of the nuclei, highlighting the importance of longitudinal structure in the nuclear model.

\subsubsection{Linearization of the Yang-Mills equations}

Recall the classical YM equations~\eqref{eq:CYM}
\begin{align}
   \partial_\mu \partial^\mu A^\nu - 2\ii g \left[ A^\mu , \partial_\mu A^\nu \right] + \ii g \left[ A_\mu , \partial^\nu A^\mu \right] - g^2 \left[ A_\mu , \left[ A^\mu , A^\nu \right] \right] = J^\nu, \label{eq:YM-eq-dilute}
\end{align}
where we applied covariant gauge $\partial_\mu A^\mu = 0$ (cf.~\cref{eq:nucl-A-CYM}), and the covariant conservation of the current in \cref{eq:current-conservation}
\begin{align}
    \partial_\mu J^\mu - \ii g \left[ A_\mu , J^\mu \right] = 0, \label{eq:YM-current-dilute}
\end{align}
which describe the nonlinear dynamics of the collision of two relativistic nuclei in the CGC theory.
Our primary subject of interest is the Glasma field that develops in the future light cone of the interaction region (Region II in \cref{fig:two-ncl-collision-regions}) and which is described by the gauge field $A^\mu$ that solves the collision problem.

The starting point of the dilute approximation is the ansatz
\begin{align}
    A^\mu(x) &= \mathcal{A}_\A^\mu(x^+,\xperp) + \mathcal{A}_\B^\mu(x^-,\xperp) + \mathtt{a}^\mu(x), \label{eq:dilute-A-ansatz}\\
    J^\mu(x) &= \mathcal{J}_\A^\mu(x^+,\xperp) + \mathcal{J}_\B^\mu(x^-,\xperp) + \mathtt{j}^\mu(x), \label{eq:dilute-J-ansatz}
\end{align}
where we separated $A$ and $J$ into contributions from the non-interacting, single-nucleus fields $\mathcal{A}_{\A/\B}$ and currents $\mathcal{J}_{\A/\B}$ and isolated all nonlinear interactions in the correction terms $\mathtt{a}$ and $\mathtt{j}$.
Conceptually, the idea is to perform a perturbative expansion in the sources $(\mathcal{J}_\A)^n (\mathcal{J}_\B)^m$, where $n,m \in \mathbb{N}^0$.
The single-nucleus fields already solve the full, nonlinear YM equations before any interactions take place or in the absence of the other nucleus (cf.\ \cref{sec:cgc}).
Therefore, $\mathcal{A}_\A$ contains all contributions to order $(\mathcal{J}_\A)^n (\mathcal{J}_\B)^0$ and reduces to a linear functional of the source term in covariant gauge.
Equivalently, $\mathcal{A}_\B$ contains all contributions to order $(\mathcal{J}_\A)^0 (\mathcal{J}_\B)^m$.
We can interpret $\mathcal{A}_{\A/\B}$ as the background fields whose interactions give rise to the correction terms.
Hence, the perturbative expansions
\begin{align}
    \mathtt{a} = \sum_{n,m} (\mathtt{a})^{n,m}, \qquad 
    \mathtt{j} = \sum_{n,m} (\mathtt{j})^{n,m}
\end{align}
will start at the lowest order with the terms
\begin{align}
    a \coloneq (\mathtt{a})^{1,1}, \quad 
    j \coloneq (\mathtt{j})^{1,1},
\end{align}
that capture the contributions of order $\mathcal{J}_\A \mathcal{J}_\B$.
We may assume the source terms to be small and only continue with $a$ and $j$ to leading order.
This weak field limit for the color charge distributions of the colliding nuclei motivates the name \emph{dilute Glasma}~\cite{Ipp:2021lwz}.

In covariant gauge, the single-nucleus fields $\mathcal{A}_{\A/\B}$ are each localized to the track of their respective nucleus along the light cone.
These fields stay localized on the light cone even after the collision, because we assume the nuclei to be recoilless.
The Glasma field, however, develops in the causally connected future of the interaction region of the collision, which entails the entire region encompassed by the future light cone.
In particular, at large distances to the nuclear tracks, the only contribution to the gauge field $A$ in our ansatz in \cref{eq:dilute-A-ansatz} comes from the correction term $a$.
Therefore, we can identify the field $a$ as the Glasma field to the lowest order in the dilute approximation.

The assumption of recoilless currents $\mathcal{J}_{\A/\B}$ does not contradict the existence of the correction term $j$.
We lock the nonzero vector components of the currents to be the light cone components of the vector $J$ and only allow for corrections to those components.
The effect of the correction term is reduced to rotations in color space only.
These color rotations will happen as long as the currents and color fields of both nuclei overlap, i.e., in the interaction region.
For later times, the corrections are static, but still localized to the nuclear tracks where the currents $\mathcal{J}_{\A/\B}$ are nonzero.
We will come back to this interpretation when we derive the expressions for $j$ in \cref{eq:jplus-dilute-solution,eq:jminus-dilute-solution} in \cref{sec:position-space-picture}.

\begin{figure}
    \centering
    \includegraphics[width=\linewidth]{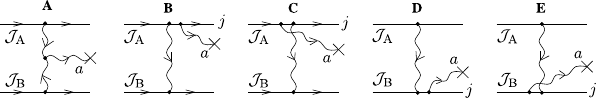}
    \caption[Diagrammatic illustration of the dilute approximation.]{\label{fig:dilute-expansion-diagramm-sketch}
    Diagrams that correspond to the lowest order in the dilute expansion.
    To this order, the contributions are of order $g^3$ in the coupling and lead to the five diagrams \textbf{A} to \textbf{E}.
    Diagram \textbf{A} depicts the dominant $2\rightarrow1$ gluon production mechanism.
    Diagrams \textbf{B} to \textbf{E} are nonlinear corrections to the nuclear currents $\mathcal{J}_{\A/\B}$.
    Figure adapted from~\cite{Kovchegov:1997ke}.}
\end{figure}

A diagrammatic illustration of the perturbative expansion can be found in~\cite{Kovchegov:1997ke} and is drawn in Fig.~\ref{fig:dilute-expansion-diagramm-sketch}.
Therein, the authors reformulate the dilute approximation as a perturbative expansion in the coupling $g$.
They use that the single-nucleus currents and fields are of order $g$.
Then, the lowest non-trivial order for the Glasma field is $g^3$ and gives rise to the five diagrams in Fig.~\ref{fig:dilute-expansion-diagramm-sketch}.
Diagram \textbf{A} describes the $2\rightarrow1$ scattering of one gluon sourced by each of the single-nucleus currents $\mathcal{J}_{A/B}$.
The resulting gluon contributes to the Glasma field $a$ formed in the collision.
Diagrams \textbf{B} to \textbf{E} describe the color rotation of $\mathcal{J}_{A/B}$ and the subsequent radiation of a gluon that is collinear to the track of the radiating current.
Continuing the expansion to higher orders, however, includes loop diagrams that do not contribute to classical calculations.
The authors in~\cite{Kovchegov:1997ke} argue that corrections to the dilute approximation at the next order can only be calculated in a quantum field theory.


\section{Position space picture}\label{sec:position-space-picture}

We continue to carry through with the dilute approximation at leading order for the collision problem.
First, we insert our ansatz from \cref{eq:dilute-A-ansatz,eq:dilute-J-ansatz} into the conservation equation \cref{eq:YM-current-dilute} and drop all terms of higher order than $a$ or $j$
\begin{align}
    \partial_\mu J^\mu - \ii g \left[ A_\mu , J^\mu \right] &= \partial_\mu \mathcal{J}_\A^\mu - \ii g\left[ \mathcal{A}_\mu^\A, \mathcal{J}_\A^\mu \right] \nn\\
    &+ \partial_\mu \mathcal{J}_\B^\mu - \ii g\left[ \mathcal{A}_\mu^\B, \mathcal{J}_\B^\mu \right] \nn\\
    &+ \partial_+ j^+ + \partial_- j^- - \ii g\left[ \mathcal{A}_\A^-, \mathcal{J}_\B^+ \right] - \ii g\left[ \mathcal{A}_\B^+, \mathcal{J}_\A^- \right] \nn\\
     &= 0. \label{eq:YM-current-dilute-leading}
\end{align}
In the first and second lines, we recover the conservation equations for the single-nucleus currents $\mathcal{J}_{\A/\B}$, which are zero by themselves.
There are no terms with $a$ because the gauge field in \cref{eq:YM-current-dilute} enters via the commutator with the current and only produces higher order terms.
The only remaining terms are collected in the third line, where we expanded the contraction of the spacetime indices for the commutator terms and kept only the nonzero vector components of the single-nucleus fields.
We also implicitly set $j^\bi=0$ in accordance with our restriction of recoilless currents so that $j$ is limited to the $j^\pm$ components.
The correction term can only contribute to the nonzero vector components of the currents $\mathcal{J}_{\A/\B}$.

Furthermore, \cref{eq:YM-current-dilute-leading} provides two independent equations for both $j^\pm$ components.
To see this, note that the current $\mathcal{J}_\B^+=\mathcal{J}_\B^+(x^-,\xperp)$ is localized to the track of nucleus $\B$ along the $x^+$ direction.
The particular structure of $\mathcal{J}_\B^+$ along the $x^-$ direction has to be shared by any recoilless corrections from $j^+$.
Then, the only contribution to the change of $j^+$ along $x^+$ can come from the interaction with $\mathcal{A}_\A^-$ from the other nucleus.
The situation is analogous for the $j^-$ component with $\A \leftrightarrow \B$.
We get
\begin{align}
    \partial_+ j^+(x) = \ii g \left[ \mathcal{A}_\A^-(x^+,\xperp) , \mathcal{J}_\B^+(x^-,\xperp) \right], \label{eq:jplus-dilute} \\
    \partial_- j^-(x) = \ii g \left[ \mathcal{A}_\B^+(x^-,\xperp) , \mathcal{J}_\A^-(x^+,\xperp) \right]. \label{eq:jminus-dilute}
\end{align}
This spatial localization can be cleanly interpreted at late times after the collision when the currents are well separated again.
To this end, we can solve for $j^\pm$ by integrating \cref{eq:jplus-dilute,eq:jminus-dilute}
\begin{align}
    j^+(x) = \ii g \intop_{-\infty}^{x^+} \dd y^+\, \left[ \mathcal{A}_\A^-(y^+,\xperp) , \mathcal{J}_\B^+(x^-,\xperp) \right], \label{eq:jplus-dilute-solution} \\
    j^-(x) = \ii g \intop_{-\infty}^{x^-} \dd y^-\, \left[ \mathcal{A}_\B^+(y^-,\xperp) , \mathcal{J}_\A^-(x^+,\xperp) \right]. \label{eq:jminus-dilute-solution}
\end{align}
Here, we integrate over the respective light cone times of the nuclei.
As long as the upper limits of the integrals, where we evaluate the correction terms, are still inside the interaction region, $j^\pm$ accumulate color rotation corrections from the interaction with the gauge field from the other nucleus.
As soon as we evaluate at sufficiently late light cone times for each nucleus, the gauge field from the other nucleus will fall off to zero.
Then, $j^\pm$ do not change with light cone time anymore.

The solutions for $j^\pm$ in \cref{eq:jplus-dilute-solution,eq:jminus-dilute-solution} are expressed exclusively in terms of quantities known prior to the collision.
We can use these expressions for $j^\pm$ in \cref{eq:YM-eq-dilute} to derive solutions for the Glasma field $a$.
We insert our ansatz for $A$ in \cref{eq:dilute-A-ansatz} into \cref{eq:YM-eq-dilute}
\begin{align}
   \partial_\mu \partial^\mu A^\nu - 2\ii g \left[ A^\mu , \partial_\mu A^\nu \right] + \ii g \left[ A_\mu , \partial^\nu A^\mu \right] - g^2 \left[ A_\mu , \left[ A^\mu , A^\nu \right] \right] - J^\nu & \nn\\
   &\hspace{-213.2pt}= \partial_\mu \mathcal{F}_\A^{\mu\nu} - \ii g \left[ \mathcal{A}_\mu^\A , \mathcal{F}_\A^{\mu\nu} \right] - \mathcal{J}_\A^\nu \nn \\
   &\hspace{-213.2pt}+ \partial_\mu \mathcal{F}_\B^{\mu\nu} - \ii g \left[ \mathcal{A}_\mu^\B , \mathcal{F}_\B^{\mu\nu} \right] - \mathcal{J}_\B^\nu \nn \\
   &\hspace{-213.2pt}- \ii g \left[ \mathcal{A}_\mu^\A , \mathcal{F}_\B^{\mu\nu} \right] - \ii g \left[ \mathcal{A}_\mu^\B , \mathcal{F}_\A^{\mu\nu} \right] - \ii g \partial_\mu \left( \left[ \mathcal{A}_\A^\mu , \mathcal{A}_\B^{\nu} \right] +  \left[ \mathcal{A}_\B^\mu , \mathcal{A}_\A^{\nu} \right] \right) \nn \\
   &\hspace{-213.2pt}+ \partial_\mu \left( \partial^\mu a^\nu - \partial^\nu a^\mu \right) - j^\nu \nn\\
   &\hspace{-213.2pt}=0,
\end{align}
where we dropped again orders higher than $a$ and $j$.
Isolating all terms that are only associated with either nucleus leads to the single-nucleus YM equations in the second and third lines (cf.~\cref{eq:nucl-A-CYM}).
Both of them are zero individually.
The mixed terms encode the nonlinear nature of the YM equations at the lowest order in the dilute expansion.
After using the covariant gauge condition $\partial_\mu \mathcal{A}_{\A/\B}^\mu=0$, we are left with
\begin{align}\label{eq:YM-dilute-fmunu}
    \partial_\mu f^{\mu\nu} = j^\nu + \ii g \left[ \mathcal{A}^\A_\mu , \partial^\mu \mathcal{A}_\B^\nu + \mathcal{F}_\B^{\mu\nu} \right] + \ii g \left[ \mathcal{A}^\B_\mu , \partial^\mu \mathcal{A}_\A^\nu + \mathcal{F}_\A^{\mu\nu} \right],
\end{align}
where we introduced the field strength tensor $f^{\mu\nu}$ of the Glasma.
In the dilute approximation, the contribution to $f^{\mu\nu}$ up to the same order as $a$ only consists of the abelian part 
\begin{align}\label{eq:dilute-fmunu-def}
    f^{\mu\nu} = \partial^\mu a^\nu - \partial^\nu a^\mu,
\end{align}
because the dropped commutator term is of higher order.
We may introduce the source terms $S^\nu$ for the right-hand side of \cref{eq:YM-dilute-fmunu}, with the components given as
\begin{align}
    S^+(x) &= \ii g \left[ \mathcal{A}_\A^-(x^+,\xperp), \partial_- \mathcal{A}_\B^+(x^-,\xperp) \right] + \ii g\! \intop_{-\infty}^{x^+}\! \dd y^+ \left[ \mathcal{A}_\A^-(y^+,\xperp) , \mathcal{J}_\B^+(x^-,\xperp) \right],\! \label{eq:dilute-source-S+}\\
    S^-(x) &= \ii g \left[ \mathcal{A}_\B^+(x^-,\xperp), \partial_+ \mathcal{A}_\A^-(x^+,\xperp) \right] + \ii g\! \intop_{-\infty}^{x^-}\! \dd y^- \left[ \mathcal{A}_\B^+(y^-,\xperp) , \mathcal{J}_\A^-(x^+,\xperp) \right],\! \label{eq:dilute-source-S-}\\
    S^\bi(x) &= \ii g \left[ \mathcal{A}_\A^-(x^+,\xperp), \mathcal{F}_\B^{+\bi}(x^-,\xperp) \right] + \ii g \left[ \mathcal{A}_\B^+(x^-,\xperp) , \mathcal{F}_\A^{-\bi}(x^+,\xperp) \right]. \label{eq:dilute-source-Si}
\end{align}
Here, $j$ only contributes to the light cone components and we used the expressions from \cref{eq:jplus-dilute-solution,eq:jminus-dilute-solution}.
The single-nucleus field strengths only contribute to the transverse components.
Because covariant gauge also holds for the Glasma field, $\partial_\mu a^\mu = 0$.
Then, we can rewrite \cref{eq:YM-dilute-fmunu} as the inhomogeneous wave equation 
\begin{align}\label{eq:dilute-a-wave}
    \partial_\mu \partial^\mu a^\nu = S^\nu,
\end{align}
and equivalently
\begin{align}\label{eq:dilute-fmunu-wave}
    \partial_\sigma \partial^\sigma f^{\mu\nu} = \partial^\mu \partial_\sigma \partial^\sigma a^\nu - \partial^\nu \partial_\sigma \partial^\sigma a^\mu = S^{\mu\nu},
\end{align}
using the inhomogeneity
\begin{align}\label{eq:dilute-Smunu-def}
    S^{\mu\nu} = \partial^\mu S^\nu - \partial^\nu S^\mu.
\end{align}

The formal solution to \cref{eq:dilute-a-wave} can be obtained with the method of Green's function
\begin{align}\label{eq:dilute-a(x)-formal-solution}
    a^\mu(x) = \intop \dd^4 y\, \Gret (x-y)S^\mu(y),
\end{align}
where we used the retarded propagator
\begin{align}\label{eq:ret-propagator-x}
    \Gret (x) = -\frac{1}{2\pi} \Theta(t) \delta(x_\mu x^\mu),
\end{align}
that maintains causality.
After a lengthy calculation, which can be found in~\cite{Ipp:2021lwz,Singh:2021hct,Leuthner:2025vsd}, the solution for the Glasma field can be expressed in compact form as\pagebreak
\begin{align}
    a^+(x) &= \frac{\ii g}{2} \intop \frac{\dd^2\pperp\,\dd^2\qperp}{(2\pi)^2} \intop_0^\infty \dd v^+ \intop_0^\infty \dd v^-\, \ee^{-\ii \xperp\cdot(\pperp + \qperp)} \left[ \tilde{\mathcal{A}}_\A^-(x^+ -v^+,\pperp), \tilde{\mathcal{A}}_\B^+(x^- -v^-,\qperp) \right] \nn \\
    &\times v^+ \left((\pperp+\qperp)^2 - 2\qperp^2\right) \frac{J_1(\sqrt{2v^+ v^-}|\pperp+\qperp|)}{\sqrt{2v^+ v^-}|\pperp+\qperp|}, \label{eq:dilute-a+-sol}\\
    a^-(x) &= \frac{\ii g}{2} \intop \frac{\dd^2\pperp\,\dd^2\qperp}{(2\pi)^2} \intop_0^\infty \dd v^+ \intop_0^\infty \dd v^-\, \ee^{-\ii \xperp\cdot(\pperp + \qperp)} \left[ \tilde{\mathcal{A}}_\A^-(x^+ -v^+,\pperp), \tilde{\mathcal{A}}_\B^+(x^- -v^-,\qperp) \right] \nn \\
    &\times v^- \left( 2\pperp^2 - (\pperp+\qperp)^2 \right) \frac{J_1(\sqrt{2v^+ v^-}|\pperp+\qperp|)}{\sqrt{2v^+ v^-}|\pperp+\qperp|}, \label{eq:dilute-a--sol} \\
    a^\bi(x) &= \frac{g}{2} \intop \frac{\dd^2\pperp\,\dd^2\qperp}{(2\pi)^2} \intop_0^\infty \dd v^+ \intop_0^\infty \dd v^-\, \ee^{-\ii \xperp\cdot(\pperp + \qperp)} \left[ \tilde{\mathcal{A}}_\A^-(x^+ -v^+,\pperp), \tilde{\mathcal{A}}_\B^+(x^- -v^-,\qperp) \right] \nn \\
    &\times (p^\bi-q^\bi) J_0(\sqrt{2v^+ v^-}|\pperp+\qperp|) \label{eq:dilute-ai-sol}.
\end{align}
Here, $J_n$ is the $n^\text{th}$ Bessel function of the first kind.
During the calculation, some boundary terms of the $v^\pm$ integrals were dropped.
These terms are nonzero only along the tracks of the nuclei.
By neglecting these terms, we restrict the evaluation point $x$ for $a^\mu(x)$ to be sufficiently far from the nuclear tracks in the $t$-$z$ plane, i.e., well inside the future light cone where the single-nucleus fields $\mathcal{A}_{\A/\B}$ fall off to zero.

The six-dimensional integrations in \cref{eq:dilute-a+-sol,eq:dilute-a--sol,eq:dilute-ai-sol} prove to be impractical to evaluate directly.
Instead, the field strength tensor of the Glasma $f$ can be reduced to only comprise three-dimensional integrals without any oscillating Bessel functions.
The detailed steps of this calculation are discussed in~\cite{Ipp:2024ykh,Leuthner:2025vsd}, where again some boundary terms that are localized to the nuclear tracks are dropped, and the single-nucleus fields are assumed to fall off to zero at infinity in the transverse plane.
The final results for the independent field strength tensor components read
\begin{align}
    f^{+-}(x) &= -\frac{g}{2\pi} \intop \dd \eta'\,\dd^2 \vperp\, V^{\bk\bl}(x-v) \delta^{\bk\bl}, \label{eq:dilute-f+--sol}\\
    f^{+\bi}(x) &= -\frac{g}{2\pi} \intop \dd \eta'\,\dd^2 \vperp\, \frac{v^\bj}{2v^-} V^{\bk\bl}(x-v) ( \delta^{\bi\bj}\delta^{\bk\bl} - \epsilon^{\bi\bj}\epsilon^{\bk\bl}), \label{eq:dilute-f+i-sol}\\
    f^{-\bi}(x) &= -\frac{g}{2\pi} \intop \dd \eta'\,\dd^2 \vperp\, \frac{v^\bj}{2v^+} V^{\bk\bl}(x-v) ( - \delta^{\bi\bj}\delta^{\bk\bl} -\epsilon^{\bi\bj}\epsilon^{\bk\bl}), \label{eq:dilute-f-i-sol}\\
    f^{\bi\bj}(x) &= -\frac{g}{2\pi} \intop \dd \eta'\,\dd^2 \vperp\, V^{\bk\bl}(x-v) \epsilon^{\bi\bj}\epsilon^{\bk\bl}, \label{eq:dilute-fij-sol}
\end{align}
where the integration variables parametrize the lightlike displacement four-vector
\begin{align}
    v^\mu = (v^+,v^-,\vperp)^\mu = (\tfrac{|\vperp|\ee^{\eta'}}{\sqrt{2}},\tfrac{|\vperp|\ee^{-\eta'}}{\sqrt{2}},\vperp)^\mu, \label{eq:dilute-v-param}
\end{align}
and we introduced the shorthand field%
\footnote{%
Compared to~\cite{Ipp:2024ykh,Leuthner:2025vsd}, we define a single tensor $V^{\bi\bj}$ and factor out Kronecker deltas and Levi-Civita symbols using the identity from~\cref{eq:eps-eps-transv-id} in the expressions for $f^{\mu\nu}$.
}
\begin{align}
    V^{\bk\bl}(x) &= -\ii \left[ \mathcal{F}_\A^{-\bk}(x^+,\xperp), \mathcal{F}_\B^{+\bl}(x^-,\xperp) \right], \label{eq:dilute-Vij-integrand}
\end{align}
which is entirely fixed by the single-nucleus field strengths known prior to the collision.

The following picture emerges for the physics encoded in these solutions for the Glasma field.
The value of the field at a given evaluation point $x$ is provided by the integration over a displacement vector $v$.
The evaluation point $x$ needs to be located away from the nuclear tracks inside the future light cone where the single-nucleus fields $\mathcal{A}_{\A/\B}$ fall off to zero and no longer contribute.
The displacement vector $v$ measures the distance between $x$ and the spacetime points at which the sources for the Glasma field are evaluated for the integrands.
As such, the integrands will only be nonzero in the overlap region of the collision where both the single-nucleus field strengths of nucleus $\A$ and $\B$ are nonzero.
This is directly visible from the expression for the integrand $V^{\bk\bl}$ in \cref{eq:dilute-Vij-integrand}.

When comparing $a$ in \cref{eq:dilute-a+-sol,eq:dilute-a--sol,eq:dilute-ai-sol} to $f$ in \cref{eq:dilute-f+--sol,eq:dilute-f+i-sol,eq:dilute-f-i-sol,eq:dilute-fij-sol}, we see that the reduction of the number of integrals also comes with a coordinate transformation for the displacement vector $v$.
The new parametrization is akin to a Milne-like coordinate system with 
\begin{align}
    \tau'=\sqrt{2v^+ v^-}, \qquad \eta'=\frac{1}{2}\ln(v^+/v^-).
\end{align}
By solving the integrals, the retarded propagator forces $v$ to be lightlike and leads to $\tau' = |\vperp|$ for \cref{eq:dilute-v-param}.
Naturally, in this parametrization, the integration over $v$ is limited to the past light cone that is attached to the evaluation point $x$.

\begin{figure}
    \centering
    \includegraphics{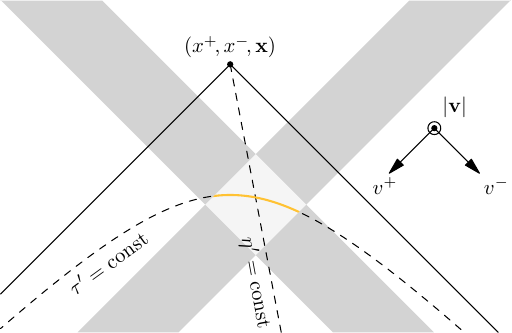}
    \caption[Projection of integration domain for $f$ onto plane of displacement vector.]{\label{fig:fmunu-backwards-milne-integral}
    Projection of the spacetime picture of the collision of two nuclei (dark gray tracks) onto the $v^+$-$v^-$ plane.
    The displacement vector $v$ is parametrized by the Milne-like coordinates $(\tau',\eta')$ on the past light cone attached to the evaluation point $x=(x^+,x^-,\xperp)$ and integrated over along lightlike paths fixed by $\tau'=|\vperp|$.
    Only segments as the yellow-highlighted one contribute to the integrals because the integrands are zero outside of the light grey interaction region.
    Figure adapted from~\cite{Ipp:2024ykh}.
    }
\end{figure}

In \cref{fig:fmunu-backwards-milne-integral}, a projection of this picture onto the $v^+$-$v^-$ plane is shown.
The dark grey bands denote the nuclear tracks with longitudinal extent.
The light grey diamond marks the interaction region where the nuclear tracks, and thus, the single-nucleus fields, overlap.
Now, the displacement vector $v$ connects all points from the interaction region with the evaluation point $x$ along lightlike paths.
The boundaries of the projection of this past light cone spanned by all $v$ are drawn in \cref{fig:fmunu-backwards-milne-integral}, where the dashed lines denote two cuts along $\tau'=\text{const}$ and $\eta'=\text{const}$.
The segment of the $\tau'=\text{const}$ hyperbola that is highlighted in yellow marks the only contributions to the integrands for this value of $|\vperp|=\tau'$.
The complete integration sums over all possible segments of this kind.

In the end, the $2\rightarrow1$ gluon scattering process in \cref{fig:dilute-expansion-diagramm-sketch} discussed at the beginning of this chapter becomes manifest.
The scattering happens only inside the interaction region, where both single-nucleus field strengths are nonzero, and only between gluons that belong to these single-nucleus fields.
Then, the produced gluons travel along lightlike paths to the evaluation point $x$ without any further interactions.
This contribution is the leading order result in the dilute approximation of the Glasma.

The solutions for the Glasma field presented in this section provide genuine longitudinal structure and generalize established results obtained in the boost-invariant limit.
The relevant longitudinal dependence of the integrands can be traced back to the single-nucleus field strengths $\mathcal{F}_{\A/\B}^{\mp\bi}$, which in turn are directly related to the color charge distributions of the nuclei.
Within the (3+1)D dilute Glasma, the nontrivial longitudinal support of the color charge distributions can be readily incorporated and is crucially linked to the final 
longitudinal structure of the Glasma field.

Contrarily, in the boost-invariant limit, the $x^+$ and $x^-$ dependencies of the integrand $V^{\bi\bj}$ in \cref{eq:dilute-Vij-integrand} factorize into delta functions and a remaining part that depends only on the transverse coordinates.
The integrals in \cref{eq:dilute-f+--sol,eq:dilute-f+i-sol,eq:dilute-f-i-sol,eq:dilute-fij-sol} can then be carried out up to one remaining integration over the angle in the transverse plane (see~\cite{Leuthner:2025vsd} for a detailed calculation) and no dependence on longitudinal coordinates remains.

\section{Observables of the Glasma stage}\label{sec:observables-glasma}

While the field strength tensor of the Glasma $f^{\mu\nu}$ is a gauge invariant quantity itself, it is not particularly useful for direct comparison to experimental observables such as particle distributions.
Still, it is the fundamental object required to build \emph{observables} of the Glasma stage that characterize the distribution of energy density.
Note that we call these quantities \emph{observables}, although they are not directly measurable by experiment.
The selective list discussed here consists of \emph{observables} that have a clear interpretation and are important for describing the later stages of the collision.

\subsection{Spacetime Milne coordinates}\label{sec:Milne-frame-shift}

\begin{figure}
    \centering
    \includegraphics{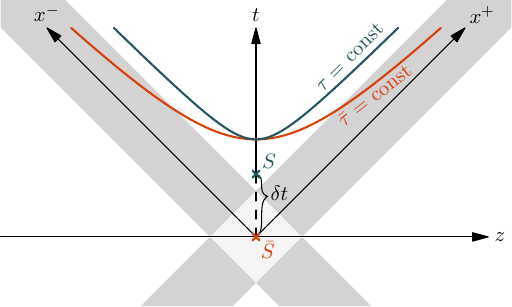}
    \caption[Shift of Milne coordinate frame origin.]{\label{fig:milne-origin-shift}
    The origin of a Milne frame in the $t$-$z$ plane is ambiguous.
    No choice will perfectly capture the symmetries of the proper time evolution for all gluons produced in the light grey overlap region of the dark grey nuclear tracks.
    The orange frame placed at $\bar S$ leads to difficulties interpreting results for large spacetime rapidities.
    The shifted frame with origin $S$ (blue) is used for all observables and prevents evaluation too close to the tracks.
    Figure adapted from~\cite{Ipp:2024ykh}.
    }
\end{figure}

We want to choose a suitable parametrization for the spacetime coordinates in the future light cone of the collision, where we want to evaluate the observables.
Traditionally, the optimal choice is the Milne coordinates $(\tau, \eta_s)$, which are particularly well suited to describe boost-invariant systems (cf.\ \cref{sec:boost-inv-glasma}).
In the formulation of the (3+1)D dilute Glasma, however, the extended longitudinal size of the nuclei leads to ambiguities w.r.t.\ where to place the origin of said Milne frame.
As illustrated in \cref{fig:milne-origin-shift}, the interaction region (light grey) of the dark grey nuclear tracks extends in $t$ and $z$ direction.
There is no possibility to choose the origin such that the $\tau$ coordinate matches the proper time of all gluons that are produced in the interaction region.
A choice motivated by symmetry would be $\bar S$ (orange), placed in the center of the collision region, which corresponds to the spacetime point at which the centers of both nuclei overlap.%
\footnote{%
This definition would also work for asymmetric collision partners, in which case the collision region would not be a perfect square.}
However, in the orange frame, the hyperbolas at constant $\bar \tau$ will cut into the nuclear tracks for sufficiently large spacetime rapidities.
On the one hand, we ought not to evaluate results close to (or even inside) the nuclear tracks, because the formalism discussed in the previous section neglects contributions localized precisely to that region.
On the other hand, the orange $\bar \tau$ hyperbola cuts through equal proper time hypersurfaces of gluons produced at positive times in the $\bar S$ frame in their reverse proper time direction, and, eventually, breaches their light cones.

Therefore, we place the origin of the spacetime Milne coordinates at $S$ (blue in \cref{fig:milne-origin-shift}).
This corresponds to a small shift $\delta t$ in the $t$ direction such that the origin $S$ is placed sufficiently far away from the border of the interaction region.
Note that the components of tensor-valued quantities are also affected by the choice of origin, because the Milne frame is curvilinear.
In the following, the shifted frame $S$ applies to all observables and results in this thesis that are evaluated in terms of spacetime coordinates.
We emphasize, however, that different choices are valid in different formalisms. For example, in the lattice-based time evolution in Milne coordinates presented in~\cite{Matsuda:2023gle}, the authors place the origin at a shift $-\delta t$ such that the entire nuclear tracks, as well as the region of the produced Glasma, are accessible for times after the collision.

\subsection{Energy-momentum tensor and local rest frame energy density}\label{sec:EMT-epsilon(x)}

From the field strength tensor of the Glasma in \cref{eq:dilute-f+--sol,eq:dilute-f+i-sol,eq:dilute-f-i-sol,eq:dilute-fij-sol}, it is straightforward to obtain the energy-momentum tensor $T$ of the Glasma (to leading order in the dilute expansion) as
\begin{align}\label{eq:EMT-def}
    T^{\mu\nu} = 2 \tr \left[ f^{\mu\rho}f_{\rho}^{\hphantom{\rho}\nu} + \frac{1}{4} g^{\mu\nu} f^{\rho\sigma}f_{\rho\sigma} \right],
\end{align}
where the trace acts in color space.
Note that $T^\mu_{\hphantom{\mu}\mu}=0$, i.e., the energy-momentum tensor of the Glasma is traceless.
As discussed in the previous section, the Milne frame tensor components of $T$ are sensitive to the Milne frame origin.
To circumvent this ambiguity, we define the total transverse energy as the sum of the transverse pressures 
\begin{align}
    \Ep(\tau) &= \intop \dd\tau'\,\dd\eta_s\,\dd^2\xperp\,\tau' \delta(\tau'-\tau) \left(T^{xx}(\tau',\eta_s,\xperp) + T^{yy}(\tau',\eta_s,\xperp)\right) , \label{eq:Eperp(x)-def}
\end{align}
where we evaluate $\Ep$ on a hypersurface with fixed proper time $\tau$.
Note the additional Jacobian factor $\tau'$ that enters the volume element in the Milne frame.
The transverse energy differential in spacetime rapidity $\eta_s$ is then given as
\begin{align}\label{eq:dEperp-detas}
    \frac{\dd\Ep(\tau,\eta_s)}{\dd\eta_s} = \tau \intop \dd^2\xperp\,\left(T^{xx}(\tau,\eta_s,\xperp) + T^{yy}(\tau,\eta_s,\xperp)\right).
\end{align}

A different observable for the energy is the local rest frame energy density $\elrf$, which provides the energy density measured by an observer who is at rest in the rest frame that is local to an infinitesimal volume of Glasma.
Formally, $\elrf$ is defined via the Landau condition
\begin{align}\label{eq:Landau-condition}
    T^\mu_{\hphantom{\mu}\nu} u^\nu = \elrf\, u^\mu,
\end{align}
which identifies the specific, positive eigenvalue that corresponds to the only timelike eigenvector $u$ as $\elrf$.
This $u$ then gives the local flow velocity of the medium and, in the Landau frame defined by \cref{eq:Landau-condition}, points in the direction of energy flow.%
\footnote{%
These considerations are usually discussed in the context of relativistic hydrodynamics, see e.g.,~\cite{Rocha:2023ilf,Jeon:2015dfa}.
The existence of a positive $\elrf$ and timelike $u$ is not guaranteed, as $T^\mu_{\hphantom{\mu}\nu}$ is no longer symmetric.
However, any physically consistent system must have an energy-momentum tensor that allows for such solutions of its eigenvalue problem.}
As a Lorentz scalar, $\elrf$ is not affected by the transformation to the Milne frame.
The local rest frame energy density plays an important phenomenological role, as it defines the energy available for particlization~\cite{ALICE:2022wpn,Elfner:2022iae,Cooper:1974mv} after the hydrodynamic evolution in the later stages of the collision~\cite{Kolb:2003dz,Gale:2013da}.
Therefore, it is directly connected to the multiplicities of the experimentally measured particles~\cite{McDonald:2016vlt,McDonald:2017eml,Andronic:2025ylc}.

We can compare $\elrf$ to the energy density accessible via $T^{\tau\tau}$ in the Milne frame.
The $T^{\tau\tau}(\tau,\eta_s,\xperp)$ component provides the local energy density measured by an observer that is boosted by a rapidity parameter identified with the spacetime rapidity $\eta_s$.
Similarly, $\elrf$ is the $T'^{t't'}$ component of the energy-momentum tensor boosted to the local rest frame with velocity $\vec u/u^t$.
In this frame, $T$ is diagonal.
However, considering the (3+1)D dynamics in the general case, these two Lorentz boosts are not identical.%
\footnote{%
In the case of a (1+1)D, Bjorken-expanding system, $T^{\tau\tau}=\elrf$ and $(u^t, u^z)=\gamma\,(1,z/t)$.
Any transverse dynamics would already destroy this identification.}

Yet another observable that strongly influences the bulk dynamics of the later stages is the eccentricity
\begin{align}\label{eq:eccentricity-def}
    \varepsilon_n(\tau,\eta_s) = \frac{\int \dd^2\xperp\, u^\tau(\tau,\eta_s,\xperp)\, \elrf(\tau,\eta_s,\xperp)\, \bar{r}^n\,\ee^{\ii n\bar{\phi}}}{\int \dd^2\yperp\, u^\tau(\tau,\eta_s,\yperp)\, \elrf(\tau,\eta_s,\yperp)\, \bar{r}^n}.
\end{align}
Here, $\bar{r}$ and $\bar{\phi}$ are polar coordinates in the transverse plane shifted by a center of mass $\xperp_0$ and are given as
\begin{align}
    \bar{r}^2 &= (\xperp-\xperp_0)^2, \\
    \bar{\phi} &= \arg\left( \xperp-\xperp_0 \right),
\end{align}
where we define the center of mass in the transverse plane at mid-rapidity
\begin{align}
    \xperp_0 \coloneqq \xperp_0(\tau) = \frac{\int \dd^2\xperp\, u^\tau(\tau,\eta_s,\xperp)\, \elrf(\tau,\eta_s,\xperp)\, \xperp}{\int \dd^2\yperp\, u^\tau(\tau,\eta_s,\yperp)\, \elrf(\tau,\eta_s,\yperp)}\Big|_{\eta_s = 0}.
\end{align}
The order of eccentricity is given by $n\in\mathbb{N}^+$.
Note that $\varepsilon_1$ would be zero if we evaluated $\xperp_0$ for each spacetime rapidity $\eta_s$ instead.
The eccentricity is a geometric observable that characterizes the transverse shape of the Glasma.
Throughout the QGP stage, eccentricity is converted to flow and subsequently measured as angular distributions of particles in the transverse plane.
The response of the QGP to $\varepsilon_2$ in particular leads to characteristic measurements of ``elliptic flow'' of detected particles~\cite{Heinz:2013th,Elfner:2022iae,ALICE:2022wpn}.

\section{Numerical implementation}\label{sec:position-space-numerical-implementation}

This section serves as an overview of the numerical implementation of the integrals in \cref{eq:dilute-f+--sol,eq:dilute-f+i-sol,eq:dilute-f-i-sol,eq:dilute-fij-sol} used to calculate the field strength tensor of the (3+1)D dilute Glasma for the results presented in \cref{ch:numerical-results}.
The details of the implementation can be looked up in~\cite{Ipp:2024ykh,Leuthner:2025vsd}.
Additionally, this section provides numerical details about the Landau matching procedure for solving the Landau condition in \cref{eq:Landau-condition}.

\subsection{Lattice discretization}\label{sec:position-space-lattice}

Given a particular nuclear model, the color charge densities of the colliding nuclei $\A$ and $\B$ are discretized on a three-dimensional lattice.
The three dimensions are spanned by the two transverse coordinates $\xperp$ and the longitudinal light cone coordinates $\xpm$ for $\A/\B$.
Using the color charge densities, the single-nucleus gauge fields $\mathcal{A}_{\A/\B}$ are obtained via fast Fourier transformations by solving the transverse Poisson equation, resulting in \cref{eq:single-nucleus-gauge-fields}.
The discretization in the transverse plane, however, introduces a hard ultraviolet (UV) cutoff for the modes in momentum space that is inversely proportional to the lattice spacing.
To gain parametric control over the UV cutoff scale, \cref{eq:single-nucleus-gauge-fields} is modified by introducing a UV regulator $\Lambda$ via an exponential damping factor as
\begin{align}
    \mathcal{A}^\mp_{\A/\B}(\xpm,\xperp) &= \intop \frac{\dd^2\kperp}{(2\pi)^2}\frac{\ee^{+\ii \kperp\cdot\xperp}}{\kperp^2+m^2}\,\ee^{-\frac{\kperp^2}{2\Lambda^2}} \,\tilde{\rho}_{\A/\B}(\xpm,\kperp). \label{eq:single-nucleus-gauge-fields-UV-regulated}
\end{align}
By choosing $\Lambda$ sufficiently smaller than the lattice UV cutoff, the UV modes are smoothly suppressed and, ideally, already at zero amplitude for values close to the lattice cutoff.
In practice, the lattice spacing dictates the computational resources necessary to perform the calculations.
It is best chosen to maximize the utilization of the hardware executing the instructions.
This provides a hard constraint from which a suitable value for $\Lambda$ can be derived.

The total extent of the lattice has to encompass the entirety of the gauge fields, which themselves depend on the parameters of the nuclear model.
When using smooth envelope profiles for the nuclei, however, the gauge fields only reach a value of zero asymptotically.
To prevent unfeasibly large lattices, the fields are cut off once they reach about 1\% of their maximum value.
This determines the extent of the lattice.
The lattice spacing is then chosen as small as possible, given the hardware restriction, while also ensuring that the features of the discretized fields are properly resolved.
In the transverse direction, the Poisson equation yields long-range fields that extend beyond the extent of the color charge distributions.
In the longitudinal direction, the support of the gauge fields is directly given by the sources.
Therefore, the transverse size of the lattice is typically larger than the longitudinal size.

The lattice sizes used for the results in \cref{ch:numerical-results} are $1024\times1024$ cells spanning $\approx 20$ fm $\times\ 20$ fm in the transverse plane and 256 cells spanning $\approx 14$ fm along the light cone coordinate in the rest frame of the nuclei.%
\footnote{%
When boosting the nuclei to the desired collider energies, the longitudinal lattice spacing is Lorentz contracted by the same $\gamma$ factor as the longitudinal extent.
In the Cartesian coordinates, this lattice is tetragonal.}
A suitable value for $\Lambda$ in this setup is 10~GeV, which is well separated from the lattice cutoff at $\approx 28$~GeV.

The single-nucleus gauge fields enter the integrands for the calculation of the field strength tensor in \cref{eq:dilute-f+--sol,eq:dilute-f+i-sol,eq:dilute-f-i-sol,eq:dilute-fij-sol}.
These integrals have to be solved for each spacetime point at which the field strength is required.
To this end, a four-dimensional evaluation grid in terms of the transverse coordinate $\xperp$ and the Milne grid in $(\tau,\eta_s)$ is set up.
This grid can be chosen completely independently of the lattice for the single-nucleus gauge fields.
The transverse extent is matched to the transverse overlap of the incoming nuclei and depends on the impact parameter.
It is also adjusted for the expansion of the medium for later proper times.
For the results in \cref{ch:numerical-results}, it was determined that transverse lattice sizes between $256\times 256$ and $512\times 512$ yield enough resolution and are complemented by 21 values for $\eta_s$.

From the structure of \cref{eq:dilute-f+--sol,eq:dilute-f+i-sol,eq:dilute-f-i-sol,eq:dilute-fij-sol}, it is clear that the integrations for each evaluation spacetime point of the field strength tensor are fully independent.
Therefore, it is straightforward to parallelize the computation on the four-dimensional evaluation grid.
The custom Monte Carlo integration routine presented in the next section is implemented with capabilities to be executed in parallel on both the CPU and GPU accelerators, where the code has been tediously optimized for the highest performance possible.

\subsection{Custom Monte Carlo integration}

\begin{figure}
    \centering
    \includegraphics{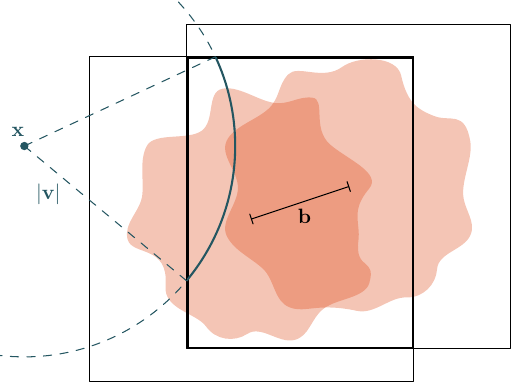}
    \caption[Transverse nuclei overlap and integration domain.]{\label{fig:transverse-MC-sampling}
    Illustration of the transverse overlap (dark orange) of two blobs of nuclear fields (light orange) offset by the impact parameter $\bperp$.
    The thick black rectangle denotes the intersection of the two single-nucleus lattices where the integrand is nonzero.
    Given the evaluation point $\xperp$ and a fixed value of $|\vperp|$, the highlighted arc contributes to the integral.
    Figure adapted from~\cite{Ipp:2024ykh}.}
\end{figure}

The integrals for the field strength tensor in \cref{eq:dilute-f+--sol,eq:dilute-f+i-sol,eq:dilute-f-i-sol,eq:dilute-fij-sol} are approximated using tuned Monte Carlo integration.
Due to the discretization of the single-nucleus fields on finite-sized lattices, only a finite volume of the entire integration domain, which is formally infinite, contributes to the integral.
This allows the derivation of tighter bounds on the sampling intervals for the Monte Carlo process.
The procedure is as follows.
For a given evaluation point $x$, first, the limits for the transverse integration over $\vperp$ are determined.
In polar coordinates $(|\vperp|, \theta_v)$, the limits for the radius $|\vperp|$ are given by a combination of the $\xpm$ coordinate and longitudinal size of the single-nucleus lattices.
The transverse and longitudinal dimensions mix because of the parametrization of the lightlike displacement vector $v$ in \cref{eq:dilute-v-param}.
After a possible value of $|\vperp|$ is sampled, the angular integration is restricted as illustrated in \cref{fig:transverse-MC-sampling}.
Only the highlighted arc on the circle with radius $|\vperp|$, that is centered at the transverse evaluation point $\xperp$, contributes to the integration.
Similarly, the interval for the rapidity-like integration variable $\eta'$ is also limited by the sampled value of $|\vperp|$.
\pagebreak

During the Monte Carlo sampling process, the evaluation of the integrand is not restricted to the exact lattice points on which the single-nucleus fields are discretized.
The values of the integrand at arbitrary points are calculated via linear interpolation from the available lattice points.

The statistics of the Monte Carlo integration are checked using the Jackknife method.
Hereby, the statistical variance of the observables themselves is used as the deciding metric.
Note that any observables calculated from the energy-momentum tensor in \cref{eq:EMT-def} require two statistically independent Monte Carlo integration results for the field strength tensor.
Using a single result would introduce a biased estimator when squaring the individual components of the field strength tensor.

It is important to distinguish between event-by-event fluctuations of the observables and the statistical errors introduced by the Monte Carlo integration process.
While the former are intrinsic to the stochastic nature of the color charge distributions and ultimately governed by the CGC weight functional, the latter can be controlled by the number of samples drawn during the integration process and can be made negligible.

\subsection{Landau matching}

It is straightforward to implement the solution to the Landau condition in \cref{eq:Landau-condition} via standard linear algebra libraries that provide eigenvalue solvers.
In analogy to the Monte Carlo integration, all eigenvalue problems on the evaluation grid for the spacetime point $x$ are independent and can be solved in parallel.

However, the selection of the correct eigenvalues and eigenvectors, which are interpreted as the local rest frame energy density $\elrf$ and flow $u$, requires further treatment.
For every energy-momentum tensor that describes a physical system, the existence of at least one positive eigenvalue with an associated timelike eigenvector is guaranteed.
This follows from the identification that the diagonalization of the energy-momentum tensor corresponds to a boost into the local rest frame of the medium, which has to exist (for a detailed discussion see~\cite{Leuthner:2025vsd}).
Still, due to the numerical nature of the calculation and unavoidable use of inexact machine-representable numbers, the numerical eigenvalue problem does not yield physical results in every case.
It is therefore necessary to check for invalid values and perform cleanup, as discussed below.

The first cleanup step is to discard any imaginary parts of the eigenvalues and eigenvectors.
The magnitudes of these imaginary parts are usually negligible.
They necessarily appear and guarantee that every eigenvalue problem has a solution over the complex domain (cf.~the fundamental theorem of algebra).
Next, it is checked whether there is a physical pair of eigenvalues and eigenvectors.
If no such pair exists for a given evaluation point, it is marked for later treatment.
It has proven to be necessary to also mark any solution with a value of the Lorentz factor $\gamma = u^0$ that exceeds the beam energy, even though the result might be physical.
This situation is encountered primarily when values close to zero are involved, leading to a critical loss of accuracy.

\begin{figure}
    \centering
    \includegraphics[width=\linewidth]{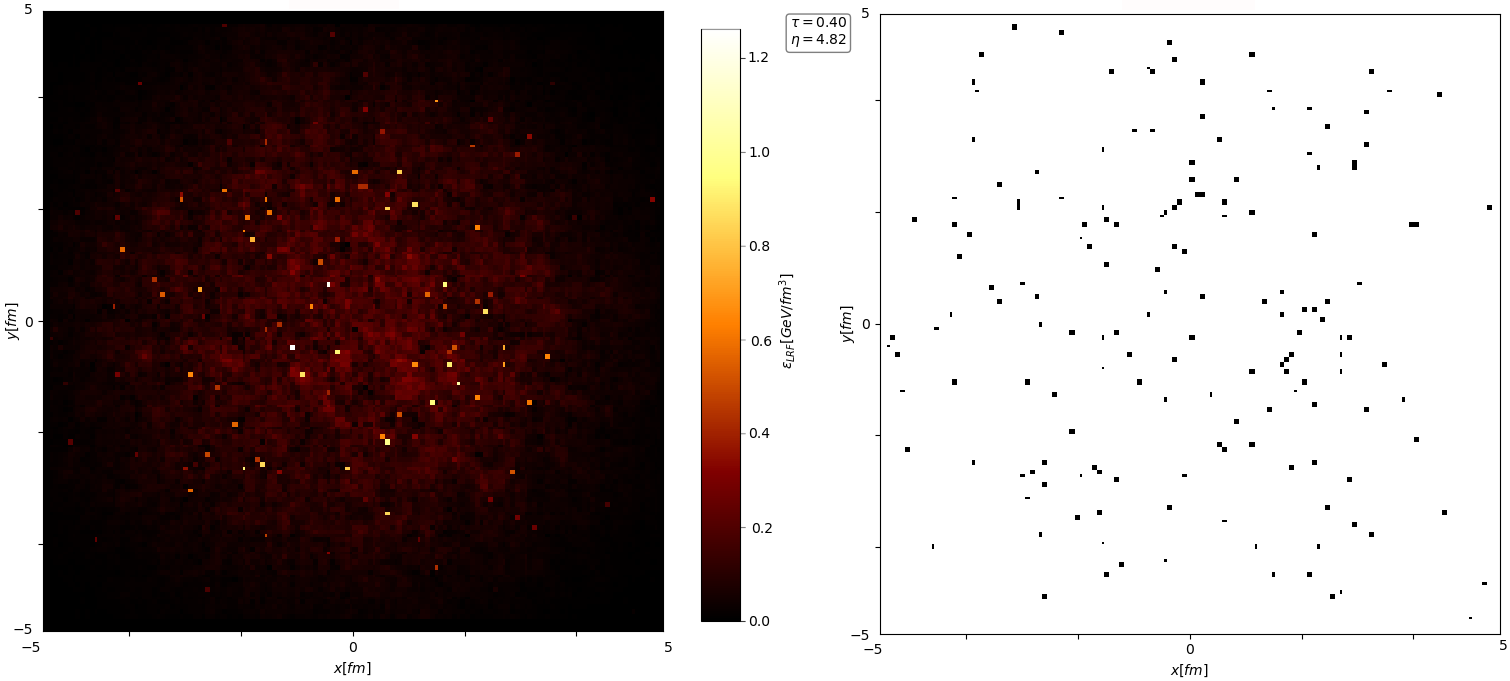}
    \caption[Transverse slice of unphysical Landau matching results.]{\label{fig:landau-discarded-slice}
    Data points in the transverse plane for a single event at $\tau=0.4$ fm/c and $\eta_s = 4.82$, where Landau matching resulted in unphysical solutions for the eigenvalue problem.
    Left: The obtained values for $\elrf$ where clear outliers are visible.
    Right: The mask where the identified unphysical values are marked with black pixels.
    The shown data correspond to the numerical results presented in \cref{ch:numerical-results}.
    }
\end{figure}

In \cref{fig:landau-discarded-slice}, the raw result of the eigenvalues that still include the marked, unphysical results is shown.
In the left panel, the transverse slice of $\elrf$ at a large $\eta_s=4.82$ is plotted.
The right panel represents the matching mask of the evaluation points that were marked for later treatment after checking the solutions of the eigenvalue problem.
The data corresponds to one event of the results presented in \cref{ch:numerical-results}.
Here, the magnitude of the imaginary parts, averaged over the entire transverse plane, is $\approx 10^{-5}$.
This value drops below machine precision close to mid-rapidity, where $|\eta_s| \lesssim 1$.
The total number of masked values constitutes 0.002\% of the total number of evaluation points in the transverse plane.
This number drops to 0 close to mid-rapidity as well.
Therefore, \cref{fig:landau-discarded-slice} demonstrates a worst-case scenario, where the number of masked points is the largest.

The masked values are replaced by a next-neighbor averaging process, where only unmasked neighbors are used.
The local rest frame energy density $\elrf'$ and flow $u'$ for a masked point are calculated via
\begin{align}
    \elrf' u'^\mu = \frac{1}{\mathrm{NN}} \sum^{\mathrm{NN}}_{i} \elrf^i u^\mu_i, \label{eq:landau-nn-avg}
\end{align}
where NN is the number of unmasked next neighbors, and the sum only includes these known-good neighbors.
The averaging strictly includes only cells in the transverse plane, i.e., the averaging of neighboring cells along $\eta_s$ or temporal averaging is not performed.
Instead of averaging the values of $\elrf$ and $u$ independently, combining them in \cref{eq:landau-nn-avg} is equivalent to averaging the four-momenta of the respective cells.
From the result of \cref{eq:landau-nn-avg}, $\elrf'$ is identified as the factor that is required to normalize the four-velocity $u'$.
This automatically ensures that the averaging process returns physical $\elrf'$ and $u'$, and prevents an ambiguous normalization factor that would be introduced by averaging the four-velocities independently.

\begin{figure}
    \centering
    \includegraphics[width=\linewidth]{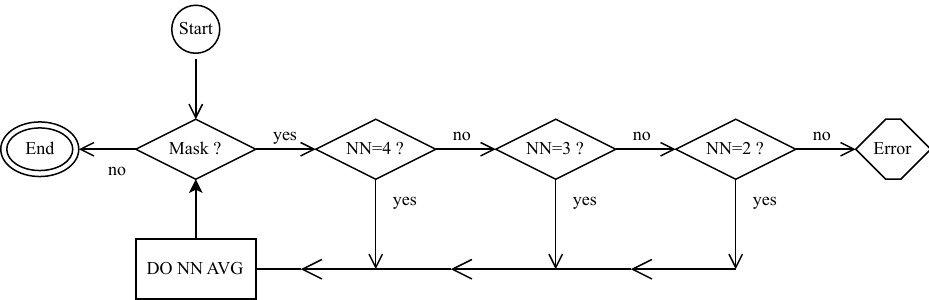}
    \caption[Next-neighbor averaging flowchart for Landau matching.]{\label{fig:landau-matching-flowchart}
    The flowchart of next-neighbor averaging in the transverse plane.
    Lattice points with unphysical eigenvalues from the Landau matching procedure are marked in the Mask.
    As long as the mask is not empty, hierarchical next-neighbor averaging is performed.
    At each iteration, the mask is queried for cells with NN number of valid next neighbors in descending order.
    Next-neighbor averaging is performed for the cells selected in the query.
    If cells with one or fewer valid neighbors remain, an error is returned.}
\end{figure}

The formula in \cref{eq:landau-nn-avg} is applied to all marked cells using a recursive, hierarchical algorithm.
The flowchart of this procedure is given in \cref{fig:landau-matching-flowchart}.
At each iteration, the list of marked values (= Mask) is queried for all cells with a certain number of known-good next neighbors in descending order, starting from 4 and stopping at 2.
This allows to first repair the cells that have the most amount of usable information in their next neighbors and is capable of dissolving clusters of marked cells.
Under normal circumstances, this algorithm should never reach a point where the mask is not empty and the remaining marked cells have less than two known-good neighbors.
Cells with only two valid next neighbors are typically part of a larger cluster and will eventually be replaced in an averaging process with a higher number of neighbors, where some of the previously masked neighbors were replaced in previous iterations.


\chapter{Momentum space picture of the (3+1)D dilute Glasma}\label{ch:momentum-space-picture}

In this chapter, the goal is to derive solutions for the dilute Glasma in momentum space.
These solutions are then used in \cref{sec:gluon-numbers-dilute} to calculate the gluon number distribution.
The ideal starting point is the wave equation for the Glasma field $a$ in \cref{eq:dilute-a-wave} with the formal solution in \cref{eq:dilute-a(x)-formal-solution}.
We transform \cref{eq:dilute-a(x)-formal-solution} to momentum space and get
\begin{align}\label{eq:dilute-tilde-a-Gret}
    \tilde a^\mu(k) = \tGret(k) \tilde{S}^\mu(k),
\end{align}
and equivalently for the solution of $\tilde f$ (which obeys an analogous wave equation in \cref{eq:dilute-fmunu-wave})
\begin{align}\label{eq:dilute-tfmunu-def-propagator}
    \tilde{f}^{\mu\nu}(k) = \tGret(k) \tilde{S}^{\mu\nu}(k),
\end{align}
where we use the Fourier transformation of the retarded propagator with the correct $\epsilon$-prescription that preserves causality
\begin{align}\label{eq:ret-propagator-k}
    \tGret(k) = \frac{1}{k_\mu k^\mu+\ii\epsilon k^t}.
\end{align}

The momentum space solutions are given in terms of the Fourier transformations of the source terms $S^\mu$ in \cref{eq:dilute-source-S+,eq:dilute-source-S-,eq:dilute-source-Si}
\begin{align}
    \tilde{S}^+(x) &= \intop \dd^4 x\, \ee^{\ii k_\mu x^\mu} \\
    & \times \Big( \ii g \left[ \mathcal{A}_\A^-(x^+,\xperp), \partial_- \mathcal{A}_\B^+(x^-,\xperp) \right] + \ii g \intop_{-\infty}^{x^+} \dd y^+ \left[ \mathcal{A}_\A^-(y^+,\xperp) , \mathcal{J}_\B^+(x^-,\xperp) \right] \Big), \label{eq:dilute-source-S+-FT}\\
    \tilde{S}^-(x) &= \intop \dd^4 x\, \ee^{\ii k_\mu x^\mu} \nn \\
    & \times \Big( \ii g \left[ \mathcal{A}_\B^+(x^-,\xperp), \partial_+ \mathcal{A}_\A^-(x^+,\xperp) \right] + \ii g \intop_{-\infty}^{x^-} \dd y^- \left[ \mathcal{A}_\B^+(y^-,\xperp) , \mathcal{J}_\A^-(x^+,\xperp) \right] \Big), \label{eq:dilute-source-S--FT}\\
    \tilde{S}^\bi(x) &= \intop \dd^4 x\, \ee^{\ii k_\mu x^\mu} \nn \\
    & \times \left( \ii g \left[ \mathcal{A}_\A^-(x^+,\xperp), \partial_\bi \mathcal{A}_\B^+(x^-,\xperp) \right] + \ii g \left[ \mathcal{A}_\B^+(x^-,\xperp) , \partial_\bi \mathcal{A}_\A^-(x^+,\xperp) \right] \right), \label{eq:dilute-source-Si-FT}
\end{align}
where we re-expressed the single-nucleus field strengths in the spatial components via derivatives of the gauge fields (in covariant gauge).
We split the calculation into two parts.
First, we Fourier transform the commutator of a single-nucleus gauge field and its derivative, which is common to all components of $S^\mu$,
\begin{align}
    & \ii g \intop \dd^4 x\, \ee^{\ii k_\nu x^\nu} \left[ \mathcal{A}_{\A/\B}^\mp(\xpm,\xperp), \partial_\mu \mathcal{A}_{\B/\A}^\pm(\xmp,\xperp) \right] \nn \\
    &= \frac{\ii g}{(2\pi)^6} \intop \dd^4x\, \dd^4p\, \dd^4q\, \ee^{\ii k_\nu x^\nu} \nn\\
    & \times\left[ \delta(\ppm)\tilde{\mathcal{A}}_{\A/\B}^\mp(\pmp,\pperp) \ee^{-\ii p_\nu x^\nu} , \delta(\qmp) \tilde{\mathcal{A}}_{\B/\A}^\pm(\qpm,\qperp) \partial_\mu \ee^{-\ii q_\nu x^\nu} \right] \nn\\
    &=  \frac{g}{(2\pi)^6} \intop \dd^4x\, \dd^4p\, \dd^4q\, \ee^{\ii k_\nu x^\nu -\ii p_\nu x^\nu -\ii q_\nu x^\nu}  \delta(\ppm) \delta(\qmp) q_\mu \nn \\
    & \times \left[\tilde{\mathcal{A}}_{\A/\B}^\mp(\pmp,\pperp) , \tilde{\mathcal{A}}_{\B/\A}^\pm(\qpm,\qperp) \right] \nn\\
    &=  \frac{g}{(2\pi)^2} \intop \dd^4p\, \dd^4q\, \delta^{(4)}(k-p-q) \delta(\ppm) \delta(\qmp) q_\mu \left[\tilde{\mathcal{A}}_{\A/\B}^\mp(\pmp,\pperp) , \tilde{\mathcal{A}}_{\B/\A}^\pm(\qpm,\qperp) \right] \nn\\
    &= \frac{g}{(2\pi)^2} \intop \dd^2\pperp\, \dd^2\qperp\, \deltaperp(\kperp-\pperp-\qperp) \left( \kpm \delta^\mp_\mu - q^\bi \delta^\bi_\mu \right) \left[\tilde{\mathcal{A}}_{\A/\B}^\mp(\kmp,\pperp) , \tilde{\mathcal{A}}_{\B/\A}^\pm(\kpm,\qperp) \right].
\end{align}
Here, we inserted the Fourier transformations of $\mathcal{A}_{\A/\B}$ and evaluated the derivative acting on the $x$ dependence of the exponential in the second and third lines.
Integrating out $x$ in the fourth and fifth lines generates delta functions, which simplify the momentum integrations to a pair of transverse integrals in the last line.

Next, we Fourier transform the integral of the commutator of a gauge field and current for the $S^\pm$ components
\begin{align}
    & \ii g \intop \dd^4x\, \ee^{\ii k_\nu x^\nu} \intop_{-\infty}^{\xpm} \dd \ypm \left[ \mathcal{A}_{\A/\B}^\mp(\ypm,\xperp) , \mathcal{J}_{\B/\A}^\pm(\xmp,\xperp) \right] \nn \\
    &= \ii g \intop \dd^4x\,\dd \ypm \, \ee^{\ii k_\nu x^\nu} \theta(\xpm-\ypm) \left[ \mathcal{A}_{\A/\B}^\mp(\ypm,\xperp) , \mathcal{J}_{\B/\A}^\pm(\xmp,\xperp) \right] \nn\\
    &= \frac{\ii g}{(2\pi)^7} \intop \dd^4x\,\dd \ypm\,\dd^4q\,\dd^2\pperp\,\dd \pmp\,\dd \lmp\, \ee^{\ii k_\nu x^\nu - \ii q_\nu x^\nu + \ii \pperp\cdot\xperp - \ii \pmp \ypm - \ii \lmp (\xpm-\ypm)} \nn \\
    &\times \delta(\qmp) \frac{\ii}{\lmp + \ii\epsilon} \left[ \tilde{\mathcal{A}}_{\A/\B}^\mp(\pmp,\pperp) , \qperp^2 \tilde{\mathcal{A}}_{\B/\A}^\pm(\qpm,\qperp) \right].\label{eq:tilde-Spm-int-part}
\end{align}
Here, we introduced the step function $\theta$ and changed the upper limit of the $\ypm$ integral to $+\infty$ in the second line.
In the last line, we inserted the Fourier transformations of $\mathcal{A}_{\A/\B}$ and $\mathcal{J}_{\A/\B}$, where the latter can be expressed via the single-nucleus gauge field as a solution of the single-nucleus YM equations (cf.\ \cref{eq:nucl-A-poisson-FT}).
We also use the Fourier representation of the step function
\begin{align}
    \theta(x-y) = \lim_{\epsilon\rightarrow 0^+}\intop \frac{\dd k}{2\pi}\,\ee^{-\ii k (x-y)} \frac{\ii}{k+\ii\epsilon},
\end{align}
where we implicitly keep track of the limit for $\epsilon$ and drop the limit in the following.
Evaluating the $x$ and $\ypm$ integrals generates delta functions
\begin{align}
    \text{\cref{eq:tilde-Spm-int-part}} &= \frac{-g}{(2\pi)^2} \intop\dd^4q\,\dd^2\pperp\,\dd \pmp\,\dd \lmp\, \frac{\qperp^2}{\lmp+\ii \epsilon} \left[ \tilde{\mathcal{A}}_{\A/\B}^\mp(\pmp,\pperp) ,\tilde{\mathcal{A}}_{\B/\A}^\pm(\qpm,\qperp) \right] \nn \\
    &\times \deltaperp(\kperp-\pperp-\qperp)\delta(\kmp-\qmp-\lmp)\delta(\kpm-\qpm)\delta(\pmp-\lmp)\delta(\qmp) \nn\\
    &= \frac{-g}{(2\pi)^2} \intop \dd^2\pperp\,\dd^2\qperp\,\deltaperp(\kperp-\pperp-\qperp) \frac{\qperp^2}{\kmp+\ii \epsilon} \left[ \tilde{\mathcal{A}}_{\A/\B}^\mp(\kmp,\pperp) ,\tilde{\mathcal{A}}_{\B/\A}^\pm(\kpm,\qperp) \right],
\end{align}
and we are left with a pair of transverse momentum integrals in the last line.
We can now assemble $\tilde{S}^\mu$ from these results
\begin{align}
    \tilde{S}^+(k) &= g\intop \frac{\dd^2\pperp\,\dd^2\qperp}{(2\pi)^2}\,\deltaperp(\kperp-\pperp-\qperp) \left[\tilde{\mathcal{A}}_\A^-(k^-,\pperp),\tilde{\mathcal{A}}_\B^+(k^+,\qperp) \right]\left(k^+-\frac{\qperp^2}{k^-+\ii\epsilon}\right), \label{eq:dilute-tilde-S+}\\
    \tilde{S}^-(k) &= g\intop \frac{\dd^2\pperp\,\dd^2\qperp}{(2\pi)^2}\,\deltaperp(\kperp-\pperp-\qperp) \left[\tilde{\mathcal{A}}_\A^-(k^-,\pperp),\tilde{\mathcal{A}}_\B^+(k^+,\qperp) \right]\left(-k^-+\frac{\pperp^2}{k^++\ii\epsilon}\right), \label{eq:dilute-tilde-S-}\\
    \tilde{S}^\bi(k) &= g\intop \frac{\dd^2\pperp\,\dd^2\qperp}{(2\pi)^2}\,\deltaperp(\kperp-\pperp-\qperp) \left[\tilde{\mathcal{A}}_\A^-(k^-,\pperp),\tilde{\mathcal{A}}_\B^+(k^+,\qperp) \right] (p^\bi -q^\bi). \label{eq:dilute-tilde-Si}
\end{align}
All components of $\tilde{S}^\mu$ share the same transverse integral and gauge field structure and only differ in the momentum contributions in parentheses at the ends of each of \cref{eq:dilute-tilde-S+,eq:dilute-tilde-S-,eq:dilute-tilde-Si}.
In fact, the terms in parentheses are identified as the Lipatov effective vertex (see e.g.,~\cite{Dumitru:2008wn,Blaizot:2008yb,Blaizot:2004wu}%
\footnote{%
While~\cite{Balitsky:1978ic,Lipatov:1976zz,Kuraev:1977fs} are the conventionally cited primary sources for the Lipatov vertex, here, we rely on the established modern sources mentioned above.
}%
) which reads
\begin{align}\label{eq:Lipatov-vertex}
    C^+_L(k,\pperp) = k^+ \! -\frac{(\kperp-\pperp)^2}{k^-+\ii\epsilon}, \quad 
    C^-_L(k,\pperp) = -k^- \! +\frac{\pperp^2}{k^++\ii\epsilon}, \quad
    C^\bi_L(k,\pperp) = 2p^\bi -k^\bi,
\end{align}
where we eliminated all occurences of $\qperp$ using $\qperp=\kperp-\pperp$ as enforced by the delta function in $\tilde{S}^\mu$.
Note that the components given in \cref{eq:Lipatov-vertex} satisfy the covariant gauge condition for $\tilde{a}^\mu$ in momentum space
\begin{align}
    k_\mu \tilde{a}^\mu(k) = 0 \quad \Longleftrightarrow \quad  k_\mu C^\mu_L(k,p) = 0 .
\end{align}
Using the Lipatov vertex, we can write all components of $\tilde{S}^\mu$ in \cref{eq:dilute-tilde-S+,eq:dilute-tilde-S-,eq:dilute-tilde-Si} in the compact form
\begin{align}\label{eq:dilute-tilde-Smu}
    \tilde{S}^\mu(k) = g\intop \frac{\dd^2\pperp\,\dd^2\qperp}{(2\pi)^2}\,\deltaperp(\kperp-\pperp-\qperp) \left[\tilde{\mathcal{A}}_\A^-(k^-,\pperp),\tilde{\mathcal{A}}_\B^+(k^+,\qperp) \right] C^\mu_L(k,\pperp),
\end{align}
which provides the Glasma field $\tilde a$ when inserted into \cref{eq:dilute-tilde-a-Gret}.

The Fourier transformation of the source terms $S^{\mu\nu}$ can be obtained from \cref{eq:dilute-tilde-Smu} via the Fourier transformation of \cref{eq:dilute-Smunu-def} as
\begin{align}
    \tilde{S}^{\mu\nu}(k) &= -\ii k^\mu \tilde{S}^\nu(k) + \ii k^\nu \tilde{S}^\mu(k) \label{eq:dilute-tilde-Smunu-def}\\
    &= \ii g\intop \frac{\dd^2\pperp\,\dd^2\qperp}{(2\pi)^2}\,\deltaperp(\kperp-\pperp-\qperp) \left[\tilde{\mathcal{A}}_\A^-(k^-,\pperp),\tilde{\mathcal{A}}_\B^+(k^+,\qperp) \right] \nn \\
    &\times \left( - k^\mu C^\nu_L(k,\pperp) + k^\nu C^\mu_L(k,\pperp) \right). \label{eq:dilute-Smunu-start}
\end{align}
We continue by evaluating the contribution of the last line of \cref{eq:dilute-Smunu-start} to the independent components of $\tilde{S}^{\mu\nu}$ separately.
We will use $\qperp$ instead of $\kperp-\pperp$ to reduce clutter.
The first contribution is
\begin{align}
    - k^+ C^-_L(k,\pperp) + k^- C^+_L(k,\pperp) &= - k^+ \left( -k^- + \frac{\pperp^2}{k^+ +\ii\epsilon} \right) + k^- \left( k^+ - \frac{\qperp^2}{k^-+\ii\epsilon} \right) \nn\\
    &= \left( 2 k^+ k^- - p^\bj p^\bj - q^\bj q^\bj \right) \nn\\
    &= \left( k_\nu k^\nu + k^\bj k^\bj - p^\bj p^\bj - q^\bj q^\bj \right) \nn\\
    &= \left( k_\nu k^\nu + 2 p^\bj q^\bj \right).
\end{align}
Here, in the second line, we took $\epsilon\rightarrow 0^+$ safely, as the poles for $\kmp$ are cancelled.
Then, we introduced the square of the $k$ four-vector in the third line and finally used the identity
\begin{align}
    k^\bj k^\bj = p^\bj p^\bj + q^\bj q^\bj + 2 p^\bj q^\bj,
\end{align}
which follows from the delta function $\deltaperp(\kperp-\pperp-\qperp)$ that is part of $\tilde{S}^\mu$.
Next, we calculate the transverse components
\begin{align}
   - k^\bi C^\bj_L(k,\pperp) + k^\bj C^\bi_L(k,\pperp) &= - k^\bi (p^\bj -q^\bj) + k^\bj (p^\bi -q^\bi) \nn\\
    &= -(p^\bi+q^\bi)(p^\bj-q^\bj) + (p^\bj+q^\bj)(p^\bi-q^\bi) \nn\\
    &= 2 p^\bk q^\bl \left( \delta^{\bi\bk}\delta^{\bj\bl} - \delta^{\bj\bk}\delta^{\bi\bl} \right) \nn\\
    &= 2 p^\bk q^\bl \epsilon^{\bi\bj}\epsilon^{\bk\bl},
\end{align}
where we used $k^\bi = p^\bi + q^\bi$ in the second line, as given again by the delta function.
In the last line, we used the identity
\begin{align}\label{eq:eps-eps-transv-id}
    \epsilon^{\bi\bj}\epsilon^{\bk\bl} = \delta^{\bi\bk}\delta^{\bj\bl} - \delta^{\bj\bk}\delta^{\bi\bl},
\end{align}
which involves the transverse-only Levi-Civita and Kronecker-delta symbols.%
\footnote{%
This identity is easily derived from the well-known identity
\begin{equation*}
    \epsilon_{nij}\epsilon_{nkl} = \delta_{ik}\delta_{jl} - \delta_{il}\delta_{jk},
\end{equation*}
involving the standard Levi-Civita symbol and recognising that $i,j,k$ and $l$ only take transverse index values.
Then, $n=z$ is the only nonzero index, and coincides with the transverse-only case.}
The calculation for the $\tilde{S}^{\pm\bi}$ components is more involved and we treat them separately, starting with
\begin{align}
   - k^+ C^\bi_L(k,\pperp) + k^\bi C^+_L(k,\pperp) &= - k^+ (p^\bi - q^\bi) + k^\bi \left( k^+ - \frac{\qperp^2}{k^-+\ii\epsilon}\right) \nn\\
    &= \frac{1}{k^- +\ii\epsilon}\left( 2k^+ k^- q^\bi - k^\bi q^\bj q^\bj \right) \nn\\
    &= \frac{1}{k^- +\ii\epsilon}\left( k_\nu k^\nu q^\bi + k^\bj k^\bj q^\bi - k^\bi q^\bj q^\bj \right) \nn\\
    &= \frac{1}{k^- +\ii\epsilon}\left( k_\nu k^\nu q^\bi + k^\bj p^\bj q^\bi - k^\bj p^\bi q^\bj + k^\bi q^\bj p^\bj \right) \nn\\
    &= \frac{1}{k^- +\ii\epsilon}\left( k_\nu k^\nu q^\bi + k^\bj p^\bk q^\bl (\delta^{\bi\bj}\delta^{\bk\bl} - \epsilon^{\bi\bj}\epsilon^{\bk\bl}) \right).
\end{align}
Here, in the second line, we used $k^\bi = p^\bi + q^\bi$ and introduced the square of $k$ in the third line.
Then, we replaced $q^\bj = k^\bj-p^\bj$ in the last term and used the identity
\begin{align}
    k^\bj q^\bi = p^\bj q^\bi -p^\bi q^\bj + q^\bj k^\bi,
\end{align}
which holds because of the delta function, to get to the fourth line.
Finally, we used \cref{eq:eps-eps-transv-id} to get to the last line.
Analogously, but with a different overall sign and $p \leftrightarrow q$ and $k^- \leftrightarrow k^+$ swapped, the last component is
\begin{align}
   - k^- C^\bi_L(k,\pperp) + k^\bi C^-_L(k,\pperp) &= - k^- (p^\bi -q^\bi) + k^\bi \left( -k^- + \frac{\pperp^2}{k^+ + \ii\epsilon} \right) \nn\\
    &= \frac{-1}{k^+ + \ii\epsilon} \left( k_\nu k^\nu p^\bi + k^\bj k^\bj p^\bi - k^\bi p^\bj p^\bj \right) \nn\\
    &= \frac{-1}{k^+ + \ii\epsilon} \left( k_\nu k^\nu p^\bi + k^\bj q^\bj p^\bi - k^\bj q^\bi p^\bj + k^\bi p^\bj q^\bj \right) \nn\\
    &= \frac{-1}{k^+ + \ii\epsilon} \left( k_\nu k^\nu p^\bi + k^\bj p^\bk q^\bl (\delta^{\bi\bj}\delta^{\bk\bl} + \epsilon^{\bi\bj}\epsilon^{\bk\bl}) \right),
\end{align}
where, in the third line, we used the corresponding identity
\begin{align}
    k^\bj p^\bi = q^\bj p^\bi - q^\bi p^\bj + p^\bj k^\bi.
\end{align}

With these results, we can finally assemble the independent components of $\tilde{S}^{\mu\nu}$
\begin{align}
    \tilde{S}^{+-}(k) &= \ii g\intop \frac{\dd^2\pperp\,\dd^2\qperp}{(2\pi)^2}\,\deltaperp(\kperp-\pperp-\qperp) \left[\tilde{\mathcal{A}}_\A^-(k^-,\pperp),\tilde{\mathcal{A}}_\B^+(k^+,\qperp) \right] \left( k_\nu k^\nu + 2 p^\bj q^\bj \right), \\
    \tilde{S}^{+\bi}(k) &= \ii g\intop \frac{\dd^2\pperp\,\dd^2\qperp}{(2\pi)^2}\,\deltaperp(\kperp-\pperp-\qperp) \left[\tilde{\mathcal{A}}_\A^-(k^-,\pperp),\tilde{\mathcal{A}}_\B^+(k^+,\qperp) \right] \nn\\
    &\times \frac{1}{k^- +\ii\epsilon}\left( k_\nu k^\nu q^\bi + k^\bj p^\bk q^\bl (\delta^{\bi\bj}\delta^{\bk\bl} - \epsilon^{\bi\bj}\epsilon^{\bk\bl}) \right), \\
    \tilde{S}^{-\bi}(k) &= \ii g\intop \frac{\dd^2\pperp\,\dd^2\qperp}{(2\pi)^2}\,\deltaperp(\kperp-\pperp-\qperp) \left[\tilde{\mathcal{A}}_\A^-(k^-,\pperp),\tilde{\mathcal{A}}_\B^+(k^+,\qperp) \right] \nn\\
    &\times \frac{-1}{k^+ + \ii\epsilon} \left( k_\nu k^\nu p^\bi + k^\bj p^\bk q^\bl (\delta^{\bi\bj}\delta^{\bk\bl} + \epsilon^{\bi\bj}\epsilon^{\bk\bl}) \right),\\
    \tilde{S}^{\bi\bj}(k) &= 2 \ii g\intop \frac{\dd^2\pperp\,\dd^2\qperp}{(2\pi)^2}\,\deltaperp(\kperp-\pperp-\qperp) \left[\tilde{\mathcal{A}}_\A^-(k^-,\pperp),\tilde{\mathcal{A}}_\B^+(k^+,\qperp) \right] p^\bk q^\bl \epsilon^{\bi\bj}\epsilon^{\bk\bl}.
\end{align}
We can rearrange these solutions for $\tilde{S}^{\mu\nu}$ into a familiar structure.
We may replace the factors of Fourier-transformed single-nucleus gauge fields that come with a matching transverse momentum factor by the Fourier transformation of the single-nucleus field strengths (cf.\ \cref{eq:single-nucleus-field-strengths}),
\begin{align}\label{eq:FT-single-nucl-F}
    \tilde{\mathcal{F}}^{\mp\bi}_{\A/\B}(\kmp,\kperp) = \ii k^\bi \tilde{\mathcal{A}}^{\mp}_{\A/\B}(\kmp,\kperp).
\end{align}
Then, we can replace the commutator terms involving only $\tilde{\mathcal{F}}_{\A/\B}$ terms with the Fourier transformation of $V^{\bi\bj}$ defined in \cref{eq:dilute-Vij-integrand} given as
\begin{align}
    \tilde{V}^{\bi\bj}(k) &= -\ii \intop \dd^4 x\, \ee^{\ii k_\nu x^\nu}  \left[ \mathcal{F}_\A^{-\bi}(x^+,\xperp), \mathcal{F}_\B^{+\bj}(x^-,\xperp) \right] \label{eq:tildeV-F(x)F(x)}\\
    &= \frac{-\ii}{(2\pi)^2} \intop \dd^2\pperp\,\dd^2\qperp\,\deltaperp(\kperp-\pperp-\qperp) \left[ \tilde{\mathcal{F}}_\A^{-\bi}(k^-,\pperp), \tilde{\mathcal{F}}_\B^{+\bj}(k^+,\qperp) \right].\label{eq:tildeV-tF-tF}
\end{align}
After this and further replacements, the components of $\tilde{S}^{\mu\nu}$ are
\begin{align}
    \tilde{S}^{+-}(k) &= 2 g \tilde{V}^{\bk\bl}(k) \delta^{\bk\bl} -g \tilde{V}_{\mathcal{A}\mathcal{A}}(k) k_\nu k^\nu, \label{eq:tilde-S+-}\\
    \tilde{S}^{+\bi}(k) &= g \left( \tilde{V}^{\bk\bl}(k) \frac{k^\bj}{k^- +\ii\epsilon} (\delta^{\bi\bj}\delta^{\bk\bl} - \epsilon^{\bi\bj}\epsilon^{\bk\bl}) +\frac{\ii}{k^- +\ii\epsilon} \tilde{V}^\bi_{\mathcal{A}\mathcal{F}}(k) k_\nu k^\nu \right), \label{eq:tilde-S+i}\\
    \tilde{S}^{-\bi}(k) &= -g \left( \tilde{V}^{\bk\bl}(k) \frac{k^\bj}{k^+ + \ii\epsilon} (\delta^{\bi\bj}\delta^{\bk\bl} + \epsilon^{\bi\bj}\epsilon^{\bk\bl}) + \frac{\ii}{k^+ + \ii\epsilon} \tilde{V}^\bi_{\mathcal{F}\mathcal{A}}(k) k_\nu k^\nu \right), \label{eq:tilde-S-i}\\
    \tilde{S}^{\bi\bj}(k) &= 2 g \tilde{V}^{\bk\bl}(k) \epsilon^{\bi\bj}\epsilon^{\bk\bl}, \label{eq:tilde-Sij}
\end{align}
where we used new shorthand fields in analogy to $\tilde{V}^{\bi\bj}$ defined as
\begin{align}
    \tilde{V}_{\mathcal{A}\mathcal{A}}(k) &= - \ii \intop \dd^4 x\, \ee^{\ii k_\nu x^\nu} \left[ \mathcal{A}^-_\A(x^+,\xperp), \mathcal{A}^+_\B(x^-,\xperp) \right], \label{eq:tildeV-AA} \\
    \tilde{V}^\bi_{\mathcal{A}\mathcal{F}}(k) &= - \ii \intop \dd^4 x\, \ee^{\ii k_\nu x^\nu} \left[ \mathcal{A}^-_\A(x^+,\xperp), \mathcal{F}^{+\bi}_\B(x^-,\xperp) \right], \label{eq:tildeV-AF} \\
    \tilde{V}^\bi_{\mathcal{F}\mathcal{A}}(k) &= - \ii \intop \dd^4 x\, \ee^{\ii k_\nu x^\nu} \left[ \mathcal{F}^{-\bi}_\A(x^+,\xperp), \mathcal{A}^+_\B(x^-,\xperp) \right]. \label{eq:tildeV-FA}
\end{align}
The momentum space expressions for $\tilde{S}^{\mu\nu}$ have contributions in the components with at least one temporal index $\pm$ that are only nonzero for off-shell momenta where $k_\nu k^\mu \neq 0$.
These contributions can be expressed as a total derivative acting on an expression that is localized to the nuclear tracks and therefore evaluates to zero.
This calculation is done in \cref{appx:tildeS-knuknu-terms}.

Recall that the Fourier-transformed field strength tensor of the Glasma, $\tilde f$, is given in terms of the product of $\tilde{S}^{\mu\nu}$ and the retarted propagator $\tGret$ as defined in \cref{eq:dilute-tfmunu-def-propagator}.
If we evaluate the field strength tensor for on-shell momenta, we get
\begin{align}
    \lim_{k_\nu k^\nu \rightarrow 0} & k_\nu k^\nu \tilde{f}^{+-}(k) &=&\ \lim_{k_\nu k^\nu \rightarrow 0} \tilde{S}^{+-}(k) &=&\ 2 g \tilde{V}^{\bk\bl}(k) \delta^{\bk\bl},  \label{eq:tilde-S+--onshell}\\
    \lim_{k_\nu k^\nu \rightarrow 0} & k_\nu k^\nu \tilde{f}^{+\bi}(k) &=&\ \lim_{k_\nu k^\nu \rightarrow 0} \tilde{S}^{+\bi}(k) &=&\ 2g \tilde{V}^{\bk\bl}(k) \frac{k^\bj}{2k^-} (\delta^{\bi\bj}\delta^{\bk\bl} - \epsilon^{\bi\bj}\epsilon^{\bk\bl}), \label{eq:tilde-S+i-onshell} \\
    \lim_{k_\nu k^\nu \rightarrow 0} & k_\nu k^\nu \tilde{f}^{-\bi}(k) &=&\ \lim_{k_\nu k^\nu \rightarrow 0} \tilde{S}^{-\bi}(k) &=&\  2g \tilde{V}^{\bk\bl}(k) \frac{k^\bj}{2k^+} (-\delta^{\bi\bj}\delta^{\bk\bl} - \epsilon^{\bi\bj}\epsilon^{\bk\bl}), \label{eq:tilde-S-i-onshell}\\
    \lim_{k_\nu k^\nu \rightarrow 0} & k_\nu k^\nu \tilde{f}^{\bi\bj}(k) &=&\ \lim_{k_\nu k^\nu \rightarrow 0} \tilde{S}^{\bi\bj}(k) &=&\ 2 g \tilde{V}^{\bk\bl}(k) \epsilon^{\bi\bj}\epsilon^{\bk\bl}. \label{eq:tilde-Sij-onshell}
\end{align}
We can safely take $\epsilon \rightarrow 0^+$ for the poles in light cone momenta because, for on-shell gluons, one light cone momentum can only be zero if the gluon is moving along the beam axis (i.e., $k^+=(k^t+k^z)/\sqrt{2}=0$ when moving along negative $z$-direction).
These gluons are localized within the nuclear tracks where we cannot separate off the Glasma field $\tilde a$ or field strength $\tilde f$ in the dilute approximation.
Therefore, we cannot faithfully make use of these solutions in these specific kinematic regions.

We can now compare $\tilde f$ to $f$ from \cref{eq:dilute-f+--sol,eq:dilute-f+i-sol,eq:dilute-f-i-sol,eq:dilute-fij-sol}.
Apart from the off-shell contributions, we have recovered the same structure as $f$.
We have identified the same sources in Fourier space $\tilde{V}^{\bi\bj}$ and factors analogous to the displacement vector for the $\tilde{S}^{\pm\bi}$ components.
We will come back to \cref{eq:tilde-S+--onshell,eq:tilde-S+i-onshell,eq:tilde-S-i-onshell,eq:tilde-Sij-onshell} in \cref{sec:gluon-numbers-dilute}, when we calculate the gluon number distribution of the Glasma.

\section{Gluon numbers in the dilute Glasma}\label{sec:gluon-numbers-dilute}

In the physical interpretation of the dilute approximation that we developed in \cref{sec:dilute-glasma-approx}, we identified that the dilute Glasma field is made up of gluons produced in $2\rightarrow1$ scattering processes involving only the gluons from the single-nucleus fields for the incoming states.
The scatterings only take place in the spacetime region where the single-nucleus fields overlap, i.e., the interaction region.
After production, there are no further interactions between the gluons.
We therefore consider free propagating gluons after the sources have been "switched off", which happens for sufficiently late times after the collision.
In this free-streaming limit, we can calculate the time-independent number of gluons carried by the Glasma field.
The formula for the classical production of gluons reads%
\footnote{%
Compared to other widely used definitions of the "occupation number", especially in the context of EKT (e.g.~\cite{Schlichting:2019abc,Berges:2020fwq,Kurkela:2018vqr}), we do not normalize by degeneracy factors and use the Lorentz-invariant definition of the gluon numbers $n(\vec k)$ common in field theory contexts.}
(see e.g.,~\cite{Peskin:1995ev,Srednicki:2007qs})
\begin{align}\label{eq:gluon-n-def}
    n(\vec k) &= \sum_\lambda b^{* a}_\lambda(\vec k) b^a_\lambda(\vec k) = 2 \tr \sum_\lambda b_\lambda(\vec k) b^\dagger_\lambda(\vec k),
\end{align}
where the sum runs over all polarizations $\lambda$ and the sum over the color index $a$ is implied.
For the second equality, we changed to algebra-valued fields, $b \in \mathfrak{su}(N_c)$ and used the normalisation $2\tr t^a t^b = \delta^{ab}$ for the generators $t^a$ of $\mathfrak{su}(N_c)$.
The fields $b$ are the time-independent expansion coefficients of the time-dependent mode expansion of the Glasma field in Coulomb gauge
\begin{align}
    a^\mu_C(t,\vec x) &= \intop \frac{\dd^3 \vec k}{(2\pi)^3 2|\vec k|} \sum_\lambda \left( b_\lambda(\vec k) \varepsilon^{*\mu}_\lambda(\vec k) \ee^{-\ii |\vec k| t + \ii \vec k \cdot \vec x} + b^\dagger_\lambda(\vec k) \varepsilon^\mu_\lambda(\vec k) \ee^{\ii |\vec k| t - \ii \vec k \cdot \vec x} \right),\! \\
    \tilde{a}_C^\mu(t, \vec k) &= \frac{1}{2 |\vec k|} \sum_\lambda \left( b_\lambda(\vec k) \varepsilon^{*\mu}_\lambda(\vec k) \ee^{-\ii |\vec k| t} + b^\dagger_\lambda(-\vec k) \varepsilon^\mu_\lambda(-\vec k) \ee^{\ii |\vec k| t} \right), \label{eq:coulomb-expansion-full}
\end{align}
with the polarisation vectors $\varepsilon^\mu_\lambda = (0,\vec \varepsilon_\lambda)$, which are transverse $\vec k \cdot \vec \varepsilon_\lambda=0$ such that Coulomb gauge is satisfied via $\partial_i a^i_C = 0$.
The gluons are taken on-shell, i.e., $(k^t)^2 - \vec{k}^2 = 0$, and we eliminated $k^t$ from the expressions.
Setting the temporal component in Coulomb gauge to zero reflects that we do not have any sources.
From \cref{eq:coulomb-expansion-full}, it is straightforward to obtain the expansion coefficients
\begin{align}\label{eq:mode-exp-coeff}
    b_\lambda(\vec k) = \ii \ee^{\ii |\vec k| t} \left( \partial_t -\ii |\vec k| \right) \varepsilon^i_\lambda(\vec k) \tilde{a}^i_C(t, \vec k),
\end{align}
using the orthogonality of the polarisation vectors
\begin{align}\label{eq:polarization-orthogonality}
    \vec{\varepsilon}_\lambda(\vec k) \cdot \vec{\varepsilon}^{\,\,*}_{\lambda'}(\vec k) = \delta_{\lambda\lambda'}.
\end{align}
The polarisation vectors also satisfy the completeness relation
\begin{align}\label{eq:polarisation-completeness}
    \sum_\lambda \varepsilon^{*i}_\lambda(\vec k) \varepsilon^j_\lambda(\vec k) = \delta^{ij} - \frac{k^i k^j}{\vec{k}^2} = P^{ij}_{\vec k}.
\end{align}

\subsubsection{Gauge transformation}

We proceed with the gauge transformation of the Glasma gauge field $a$ to Coulomb gauge and subsequent calculation of the mode expansion coefficients.
The solution in covariant gauge, which we obtained in the previous sections, has the properties
\begin{align}
    \partial_\mu a^\mu(x) = 0, \qquad a^t(x) \neq 0.
\end{align}
We want to perform the infinitesimal non-Abelian gauge transformation
\begin{align}\label{eq:gauge-trafo}
    a^\mu(x) \rightarrow a^\mu(x) + \partial^\mu \omega(x) + ig \left[ \omega(x), a^\mu(x) \right] \equiv a^\mu_C(x),
\end{align}
with the gauge function $\omega \in \mathfrak{su}(N_c)$ such that $ \partial_i a^i_C(x) = 0$.
We realize Coulomb gauge through
\begin{align}
    \partial_\mu a^\mu_C(x) = 0, \qquad a^t_C(x) = 0. \label{eq:coulomb-gauge-conditions}
\end{align}
Note that the gauge function $\omega$ has to be of the same order in the dilute approximation as the gauge field itself.
Therefore, we can use the infinitesimal gauge transformation to the lowest order in $\omega$.
Furthermore, any terms linear in the gauge function in \cref{eq:gauge-trafo} have to contribute to the same order as the Glasma field $a$ itself, i.e., at the lowest order of the dilute expansion $\mathcal{J}_\A\mathcal{J}_\B$.
The commutator term, which is bilinear in $a$ and $\omega$, is of higher order, and we can drop it.
In the dilute approximation, gauge transformations become effectively Abelian, although the fields take values in a non-commuting algebra.
This greatly simplifies \cref{eq:gauge-trafo} and allows us to solve for $a_C$.

To this end, we insert \cref{eq:gauge-trafo} into \cref{eq:coulomb-gauge-conditions} and get two defining equations,
\begin{align}
    a^t_C(x) &= a^t(x) + \partial^t \omega(x) = 0, \\
    \partial_\mu a^\mu_C(x) &= \partial_\mu a^\mu(x) + \partial_\mu \partial^\mu \omega(x) = 0.
\end{align}
From the first equation, we extract
\begin{align}
    \partial^t \omega(x) = - a^t(x), \label{eq:del0-alpha}
\end{align}
and in the second equation, the divergence of the Glasma field drops due to our choice of covariant gauge, and we are left with
\begin{align}
    \nabla^2 \omega(x) &= (\partial^t)^2 \omega(x) \nn\\
    &= - \partial^t a^t(x), \label{eq:laplace-alpha}
\end{align}
where $\nabla^2 = \partial^i\partial^i$ is the Laplace operator and we used \cref{eq:del0-alpha} to replace the gauge function on the right.
We see that the gauge function is given by a Poisson equation where the inhomogeneity is the time derivative of the temporal component of the Glasma field in covariant gauge.

When we act with the Laplace operator on \cref{eq:del0-alpha},
\begin{align}
    \nabla^2 \partial^t \omega(x) =\partial^t \nabla^2 \omega(x) =\partial^t(-\partial^t a^t(x)) &= - \nabla^2 a^t(x),
\end{align}
where we used \cref{eq:laplace-alpha} in the second step, we obtain the wave equation for $a^t$,
\begin{align}
    (\partial^t)^2 a^t(x) - \nabla^2 a^t(x) = S^t(x) = 0. \label{eq:at-wave-eqs}
\end{align}
This seeming contradiction makes it evident that we cannot fix Coulomb gauge in the presence of nonzero sources.
Still, in the free-streaming limit, where we evaluate for $x$ at sufficiently late times and far away from the localized nuclear tracks, the source terms $S^\mu(x) \rightarrow 0$.

The formal solution for the Poisson equation \eqref{eq:laplace-alpha} reads
\begin{align}\label{eq:alpha-solution}
    \omega(x) = -\nabla^{-2} \partial^t a^t(x) + \omega_0(x),
\end{align}
where we added a homogeneous solution $\omega_0$.
Inserting \cref{eq:alpha-solution} into \cref{eq:del0-alpha} yields
\begin{align}
    \partial^t \omega_0(x) = -a^t(x) + \partial^t(\nabla^{-2}\partial^t a^t(x)) = -a^t(x) + \nabla^{-2}(\partial^t)^2 a^t(x),
\end{align}
where we exchanged the inverse Laplace operator with the partial time derivative in the second equation.
This implies that we do not allow for time-dependent homogeneous solutions to $\nabla^{-2}\partial^t a^t$ (cf.~\cref{eq:at-wave-eqs}).
As a result, the right-hand side is zero.
It follows that the function $\omega_0(x) = \omega_0(\vec x)$ and is, therefore, time independent.%
\footnote{%
The contribution of $\omega_0$ drops out in the final expression for the expansion coefficients $b_\lambda$ in \cref{eq:expansion-coeff-solution}.
Still, this discussion illustrates that the gauge transformation does not introduce additional time dependence when solved consistently.}

The final expression for the Glasma field in Coulomb gauge reads
\begin{align}
    a_C^i(x) = a^i(x) - \partial^i\partial^t\nabla^{-2} a^t(x) + \partial^i \omega_0(\vec x).
\end{align}
Next, we perform the Fourier transformation of the spatial $\vec x$-dependence of $a^i_C$ to $\vec k$,
\begin{align}
    \tilde{a}^i_C(t,\vec k) &= \intop \dd^3\vec x\, \ee^{-\ii\vec x \cdot \vec k} a^i_C(t,\vec x) = \nn\\
    &= \tilde{a}^i(t,\vec k) - \intop \frac{\dd^3\vec x\, \dd^3\vec p}{(2\pi)^3}\, \ee^{-\ii \vec x \cdot \vec k} \left( \partial^t \tilde{a}^t(t,\vec p)\, \partial^i\nabla^{-2}\ee^{\ii\vec x \cdot \vec p} - \tilde{\omega}_0(\vec p)\partial^i \ee^{\ii\vec x \cdot \vec p}\right) \nn\\
    &= \tilde{a}^i(t,\vec k) - \frac{-\ii k^i}{-\vec{k}^2} \partial^t \tilde{a}^t(t,\vec k) - \ii k^i \tilde{\omega}_0(\vec k) \nn\\
    &= \tilde{a}^i(t,\vec k) - \frac{\ii k^i}{\vec{k}^2} (-\ii k^j )\tilde{a}^j(t,\vec k) - \ii k^i \tilde{\omega}_0(\vec k),
\end{align}
where we keep track of the explicit arguments of the fields that are not Fourier transformed.
In the third line, we used the covariant gauge condition in $(t,\vec k)$-space
\begin{align}
    \intop \dd^3\vec x\, \ee^{-\ii\vec x \cdot \vec k} \partial_\mu a^\mu(x) = \partial^t \tilde{a}^t(t,\vec k) + \ii k^i \tilde{a}^i(t,\vec k) = 0
\end{align}
to express $\tilde{a}^t(t,\vec k)$ in terms of $\tilde{a}^j(t,\vec k)$.
This allows us to pull out a common prefactor
\begin{align}
    \tilde{a}^i_C(t,\vec k) &= \left(\delta^{ij} - \frac{k^i k^j}{\vec{k}^2}\right) \tilde{a}^j(t,\vec k) - \ii k^i \tilde{\omega}_0(\vec k) \nn\\
    &= P^{ij}_{\vec k} \tilde{a}^j(t,\vec k) - \ii k^i \tilde{\omega}_0(\vec k), \label{eq:tildeaC-tk-res}
\end{align}
which is the projector $P^{ij}_{\vec k}$ into the plane transverse to $\vec k$.

\subsubsection{Mode decomposition}

The Glasma field in covariant gauge and with $(t,\vec k)$-dependence $\tilde{a}(t,\vec k)$ can be obtained in analogy to \cref{eq:dilute-tilde-a-Gret,eq:dilute-a(x)-formal-solution} from the convolution
\begin{align}\label{eq:tildeamu-tk-def}
    \tilde{a}^\mu(t,\vec k) = \intop \dd t'\, \tGret(t-t',\vec k) \tilde{S}^\mu(t',\vec k)
\end{align}
of the source terms and the retarded propagator
\begin{align}
\tGret(t-t',\vec k) = \frac{-\ii}{2|\vec k|} \left( \ee^{-\ii |\vec k|(t-t')} - \ee^{\ii |\vec k|(t-t')} \right) \theta(t-t'),
\end{align}
as the solution of a wave equation.
Inserting into \cref{eq:tildeamu-tk-def} yields
\begin{align}
    \tilde{a}^\mu(t,\vec k) &= \frac{-\ii}{2|\vec k|} \intop \dd t'\, \left( \ee^{-\ii |\vec k|(t-t')} - \ee^{\ii |\vec k|(t-t')} \right) \theta(t-t') \tilde{S}^\mu(t',\vec k).
\end{align}
We can now perform the $t'$-integration to obtain the mode expansion for $\tilde{a}_C(t,\vec k)$,
\begin{align}
    \tilde{a}^i_C(t,\vec k) &= \frac{-\ii P^{ij}_{\vec k}}{2|\vec k|} \intop \dd t'\, \left( \ee^{-\ii |\vec k|(t-t')} - \ee^{\ii |\vec k|(t-t')} \right) \theta(t-t') \tilde{S}^i(t',\vec k) -\ii k^i\tilde{\omega}_0(\vec k) \nn\\
    &= \frac{-\ii P^{ij}_{\vec k}}{2|\vec k|} \intop \dd t'\, \left( \ee^{-\ii |\vec k|(t-t')} - \ee^{\ii |\vec k|(t-t')} \right) \tilde{S}^i(t',\vec k) -\ii k^i\tilde{\omega}_0(\vec k) \nn\\
    &= \frac{-\ii P^{ij}_{\vec k}}{2|\vec k|}\left( \tilde{S}^j(|\vec k|,\vec k) \ee^{-\ii|\vec k|t} - \tilde{S}^j(-|\vec k|,\vec k)\ee^{\ii|\vec k|t} \right)-\ii k^i\tilde{\omega}_0(\vec k).
\end{align}
In the second line, we dropped the step function from the propagator because we are evaluating at sufficiently large $t$ such that the source terms will cut off the integral sooner.
In the third line, we carried out the Fourier transformation in $t'$ and obtained the source terms $\tilde{S}^j(k)$ evaluated at on-shell $k_\mu k^\mu = 0$. We can now use \cref{eq:mode-exp-coeff} to obtain the mode expansion coefficients
\begin{align}
     b_\lambda(\vec k) &= \ii \ee^{\ii |\vec k| t} \left( \partial_0 -\ii |\vec k| \right) \varepsilon^i_\lambda(\vec k) \nn\\
     &\times \left( \frac{-\ii P^{ij}_{\vec k}}{2|\vec k|}\left( \tilde{S}^j(|\vec k|,\vec k) \ee^{-\ii|\vec k|t} - \tilde{S}^j(-|\vec k|,\vec k)\ee^{\ii|\vec k|t} \right)-\ii k^i\tilde{\omega}_0(\vec k) \right) \nn\\
    &= -\ii \varepsilon^i_\lambda(\vec k) P^{ij}_{\vec k} \tilde{S}^j(|\vec k|,\vec k) \nn\\
    &= -\ii \varepsilon^j_\lambda(\vec k) \tilde{S}^j(|\vec k|,\vec k).\label{eq:expansion-coeff-solution} 
\end{align}
In the last line, we used $\vec{k} \cdot \vec{\varepsilon}_\lambda = 0$.
Note that the contribution of $\tilde{\omega}_0(\vec k)$ drops out when projecting $\tilde{a}_C(t,\vec k)$ onto the polarisation vectors.
This removes any ambiguity introduced by $\omega_0$ in the gauge transformation in \cref{eq:alpha-solution}.
The physical content of the mode expansion is purely determined by the projection of the source terms $\tilde{S}^j$ onto the polarisation vectors.

\subsubsection{Gluon number distribution}\label{subsec:gluon-number-distribution}

We now insert the expansion coefficients from \cref{eq:expansion-coeff-solution} into the occupation number in \cref{eq:gluon-n-def}
\begin{align}
    n(\vec k) &= 2 \tr \sum_\lambda \left( -\ii \varepsilon^i_\lambda(\vec k) \tilde{S}^i(|\vec k|, \vec k) \right) \left( -\ii \varepsilon^j_\lambda(\vec k) \tilde{S}^j(|\vec k|, \vec k) \right)^\dagger \nn\\
    &= 2 \tr \sum_\lambda \varepsilon^i_\lambda(\vec k) \varepsilon^{*j}_\lambda(\vec k) \tilde{S}^i(|\vec k|, \vec k) \tilde{S}^j(|\vec k|,\vec k)^\dagger \nn\\
    &= 2 \tr P^{ij}_{\vec k} \tilde{S}^i(|\vec k|, \vec k) \tilde{S}^j(|\vec k|,\vec k)^\dagger \nn\\
    &= 2 \tr P^{ij}_{\vec k} \tilde{S}^i(|\vec k|, \vec k) P^{jk}_{\vec k}\tilde{S}^k(|\vec k|,\vec k)^\dagger. \label{eq:n(k)-PSS}
\end{align}
Here, we used \cref{eq:polarisation-completeness} in the third line and the idempotency of the transverse projector in the last line.
The occupation number is given by the squared magnitude of $P^{ij}_{\vec k} \tilde{S}^i(|\vec k|, \vec k)$, which is closely related to the field strength tensor.
From the covariant gauge condition $\partial_\mu a^\mu = 0$ and from \cref{eq:dilute-tilde-a-Gret} it follows that
\begin{align}
    \ii k^i \tilde{S}^i = \ii k^t \tilde{S}^t.
\end{align}
We can expand
\begin{align}
    P^{ij}_{\vec k} \tilde{S}^i(|\vec k|, \vec k) &= \tilde{S}^j(|\vec k|, \vec k) - \frac{k^i k^j}{|\vec k|^2} \tilde{S}^i(|\vec k|, \vec k) \nn\\
    &= \tilde{S}^j(|\vec k|, \vec k) - \frac{k^t k^j}{|\vec k|^2} \tilde{S}^t(|\vec k|, \vec k) \nn\\
    &= \frac{1}{|\vec k|} \left( k^t \tilde{S}^j(|\vec k|, \vec k) - k^j \tilde{S}^t(|\vec k|, \vec k) \right) \nn\\
    &= \frac{-1}{\ii |\vec k|} \tilde{S}^{tj}(|\vec k|, \vec k),
\end{align}
where we used that $k^t = |\vec k|$ for on-shell momenta in the third line and identified the source terms of the field strength tensor $\tilde{S}^{\mu\nu}$, as defined in \cref{eq:dilute-tilde-Smunu-def} in the last line.
The final expression for the occupation number then reads
\begin{align}\label{eq:n(k)-final}
    n(\vec k) = \frac{2}{|\vec k|^2} \tr \tilde{S}^{tj}(|\vec k|,\vec k) \tilde{S}^{tj}(|\vec k|,\vec k)^\dagger,
\end{align}
and is given by the mixed temporal-spatial components of the source terms $\tilde{S}^{\mu\nu}$.
These components determine the chromo-electric part of the field-strength tensor $\tilde{f}^{\mu\nu}(|\vec k|,\vec k)$ for on-shell gluons as given in \cref{eq:tilde-S+--onshell,eq:tilde-S+i-onshell,eq:tilde-S-i-onshell,eq:tilde-Sij-onshell}.

We continue to evaluate the index contraction in \cref{eq:n(k)-final} in terms of the known expressions in light cone coordinates,
\begin{align}
    n(\vec k) &= \frac{2}{|\vec k|^2} \tr \left( \tilde{S}^{t\bi}\tilde{S}^{t\bi,\dagger} + \tilde{S}^{tz}\tilde{S}^{tz,\dagger} \right) \nn\\
    &= \frac{2}{|\vec k|^2}\tr \left( \frac{1}{\sqrt{2}}(\tilde{S}^{+\bi}+\tilde{S}^{-\bi})\frac{1}{\sqrt{2}}(\tilde{S}^{+\bi,\dagger}+\tilde{S}^{-\bi,\dagger}) + \tilde{S}^{+-}\tilde{S}^{+-,\dagger} \right) \nn\\
    &= \frac{1}{|\vec k|^2}\tr \left( \tilde{S}^{+\bi}\tilde{S}^{+\bi,\dagger} + \tilde{S}^{+\bi}\tilde{S}^{-\bi,\dagger} + \tilde{S}^{-\bi}\tilde{S}^{+\bi,\dagger} + \tilde{S}^{-\bi}\tilde{S}^{-\bi,\dagger} + 2 \tilde{S}^{+-}\tilde{S}^{+-,\dagger} \right),
\end{align}
where we suppress the arguments of $\tilde{S}^{\mu\nu}$ to reduce clutter.
We now insert the expressions from \cref{eq:tilde-S+--onshell,eq:tilde-S+i-onshell,eq:tilde-S-i-onshell,eq:tilde-Sij-onshell} for each of the summands in the trace, starting with
\begin{align}
    \tilde{S}^{+\bi}\tilde{S}^{+\bi,\dagger} &= g^2 \tilde{V}^{\bk\bl}\tilde{V}^{\bi'\bj',\dagger} \frac{k^\bj k^{\bl'}}{\left(k^-\right)^2}\left( \delta^{\bi\bj}\delta^{\bk\bl} - \varepsilon^{\bi\bj}\varepsilon^{\bk\bl} \right)\left( \delta^{\bi\bl'}\delta^{\bi'\bj'} - \varepsilon^{\bi\bl'}\varepsilon^{\bi'\bj'} \right) \nn\\
    &= g^2 \tilde{V}^{\bk\bl}\tilde{V}^{\bi'\bj',\dagger} \frac{k^\bj k^{\bl'}}{\left(k^-\right)^2}\left( \delta^{\bj\bl'}\delta^{\bk\bl}\delta^{\bi'\bj'} - \delta^{\bk\bl}\varepsilon^{\bj\bl'}\varepsilon^{\bi'\bj'} - \delta^{\bi'\bj'}\varepsilon^{\bl'\bj}\varepsilon^{\bk\bl} + \varepsilon^{\bi\bj}\varepsilon^{\bi\bl'}\varepsilon^{\bk\bl}\varepsilon^{\bi'\bj'} \right) \nn\\
    &=  g^2 \tilde{V}^{\bk\bl}\tilde{V}^{\bi'\bj',\dagger} \frac{\kperp^2}{\left(k^-\right)^2}\left( \delta^{\bk\bl}\delta^{\bi'\bj'} + \varepsilon^{\bk\bl}\varepsilon^{\bi'\bj'} \right).
\end{align}
In the second line we used that $k^\bj k^{\bl'} \varepsilon^{\bl'\bj} = 0$ and that the contraction of two transverse Levi-Civita symbols is
\begin{align}
    \varepsilon^{\bi\bj}\varepsilon^{\bi\bl'} = \delta^{\bj\bl'}.
\end{align}
Note that the cross-terms in the product of Kronecker deltas and Levi-Civita symbols vanish for all summands.
In full analogy,
\begin{align}
    \tilde{S}^{-\bi}\tilde{S}^{-\bi,\dagger} &= g^2 \tilde{V}^{\bk\bl}\tilde{V}^{\bi'\bj',\dagger} \frac{\kperp^2}{\left(k^+\right)^2}\left( \delta^{\bk\bl}\delta^{\bi'\bj'} + \varepsilon^{\bk\bl}\varepsilon^{\bi'\bj'} \right),
\end{align}
with the $k^-$ component in the denominator swapped for $k^+$.
Next,
\begin{align}
    \tilde{S}^{+\bi}\tilde{S}^{-\bi,\dagger} &= - g^2 \tilde{V}^{\bk\bl}\tilde{V}^{\bi'\bj',\dagger} \frac{k^\bj k^{\bl'}}{k^+ k^-}\left( \delta^{\bi\bj}\delta^{\bk\bl} - \varepsilon^{\bi\bj}\varepsilon^{\bk\bl} \right)\left( \delta^{\bi\bl'}\delta^{\bi'\bj'} + \varepsilon^{\bi\bl'}\varepsilon^{\bi'\bj'} \right) \nn\\
    &= - 2 g^2 \tilde{V}^{\bk\bl}\tilde{V}^{\bi'\bj',\dagger}\left( \delta^{\bk\bl}\delta^{\bi'\bj'} - \varepsilon^{\bk\bl}\varepsilon^{\bi'\bj'} \right),
\end{align}
where we used that $2k^+ k^- = \kperp^2$ for on-shell momenta.
In full analogy,
\begin{align}
    \tilde{S}^{-\bi}\tilde{S}^{+\bi,\dagger} &= - 2 g^2 \tilde{V}^{\bk\bl}\tilde{V}^{\bi'\bj',\dagger}\left( \delta^{\bk\bl}\delta^{\bi'\bj'} - \varepsilon^{\bk\bl}\varepsilon^{\bi'\bj'} \right),
\end{align}
and the final summand is
\begin{align}
    2 \tilde{S}^{+-}\tilde{S}^{+-,\dagger} = 8g^2 \tilde{V}^{\bk\bl}\tilde{V}^{\bi'\bj',\dagger}\, \delta^{\bk\bl} \delta^{\bi'\bj'}.
\end{align}
Putting it all together, we relabel $\bi'\rightarrow\bi$ and $\bj'\rightarrow\bj$ and obtain for the occupation number
\begin{align}
    n(\vec k) &= \frac{g^2}{|\vec k|^2}\tr \tilde{V}^{\bk\bl}\tilde{V}^{\bi\bj,\dagger} \nn\\
    &\times \left( \delta^{\bk\bl}\delta^{\bi\bj} \left( \frac{\kperp^2}{\left(k^-\right)^2} - 4 + \frac{\kperp^2}{\left(k^+\right)^2} \right) + \varepsilon^{\bk\bl}\varepsilon^{\bi\bj}\left( \frac{\kperp^2}{\left(k^-\right)^2} + 4 + \frac{\kperp^2}{\left(k^+\right)^2} \right) + 8 \delta^{\bk\bl}\delta^{\bi\bj} \right) \nn\\
    &= \frac{g^2}{|\vec k|^2}\tr \tilde{V}^{\bk\bl}\tilde{V}^{\bi\bj,\dagger}\left( \delta^{\bk\bl}\delta^{\bi\bj} + \varepsilon^{\bk\bl}\varepsilon^{\bi\bj} \right)\frac{\kperp^2(k^+)^2 + 4 (k^+k^-)^2 + \kperp^2(k^-)^2}{\left(k^+ k^-\right)^2} \nn\\
    &= \frac{g^2}{|\vec k|^2}\tr \tilde{V}^{\bk\bl}\tilde{V}^{\bi\bj,\dagger}\left( \delta^{\bk\bl}\delta^{\bi\bj} + \varepsilon^{\bk\bl}\varepsilon^{\bi\bj} \right) \frac{\kperp^2(k^+)^2 + 2 (k^+k^-) \kperp^2 + \kperp^2(k^-)^2}{\kperp^4/4} \nn\\
    &= \frac{g^2}{|\vec k|^2}\tr \tilde{V}^{\bk\bl}\tilde{V}^{\bi\bj,\dagger}\left( \delta^{\bk\bl}\delta^{\bi\bj} + \varepsilon^{\bk\bl}\varepsilon^{\bi\bj} \right)\frac{4 \kperp^2}{\kperp^4} \left( k^+ + k^- \right)^2 \nn\\
    &= \frac{g^2}{|\vec k|^2}\tr \tilde{V}^{\bk\bl}\tilde{V}^{\bi\bj,\dagger}\left( \delta^{\bk\bl}\delta^{\bi\bj} + \varepsilon^{\bk\bl}\varepsilon^{\bi\bj} \right) \frac{8(k^t)^2}{\kperp^2} \nn\\
    &= \frac{8g^2}{\kperp^2}\left( \delta^{\bk\bl}\delta^{\bi\bj} + \varepsilon^{\bk\bl}\varepsilon^{\bi\bj} \right) \tr \tilde{V}^{\bk\bl}\tilde{V}^{\bi\bj,\dagger}, \label{eq:n(k)-tVtV}
\end{align}
where $\tilde{V}^{\bi\bj} = \tilde{V}^{\bi\bj}(|\vec k|, \vec k)$ and is evaluated for on-shell momenta $k$.
Next, we insert the expression from \cref{eq:tildeV-F(x)F(x)} for both $\tilde{V}$ and get
\begin{align}
    n(\vec k) &= \frac{8g^2}{\kperp^2}\left( \delta^{\bk\bl}\delta^{\bi\bj} + \varepsilon^{\bk\bl}\varepsilon^{\bi\bj} \right)\intop \dd^4 x\,\dd^4 y\,\ee^{\ii|\vec k|(x^t-y^t) - \ii \vec k \cdot (\vec x - \vec y)} \nn\\
    &\times \tr \left[ \mathcal{F}_\A^{-\bk}(x^+,\xperp), \mathcal{F}_\B^{+\bl}(x^-,\xperp) \right]\left[ \mathcal{F}_\A^{-\bi}(y^+,\yperp), \mathcal{F}_\B^{+\bj}(y^-,\yperp) \right]^\dagger. \label{eq:n(k)-F(x)F(x)}
\end{align}
This result cleanly depicts the structure of the occupation number in terms of the single-nucleus field strengths known prior to the collision.

\subsubsection{Longitudinal structure from nuclear charge distributions}

As the next step, we reintroduce the single-nucleus charge distributions to obtain a direct interpretation of the occupation number in terms of the nuclear structure.
To this end, we insert the other expression for $\tilde{V}^{\bi\bj}$ from \cref{eq:tildeV-tF-tF} into \cref{eq:n(k)-tVtV},
\begin{align}
     n(\vec k) &= \frac{8g^2}{\kperp^2}\left( \delta^{\bk\bl}\delta^{\bi\bj} + \varepsilon^{\bk\bl}\varepsilon^{\bi\bj} \right) \nn\\
     &\times \intop \frac{\dd^2\pperp_\A\, \dd^2\pperp_\B\, \dd^2\qperp_\A\, \dd^2\qperp_\B}{(2\pi)^4} \deltaperp(\kperp-\pperp_\A-\pperp_\B)\deltaperp(\kperp-\qperp_\A-\qperp_\B) \nn\\
     &\times \tr \left[ \tilde{\mathcal{F}}_\A^{-\bk}(k^-,\pperp_\A), \tilde{\mathcal{F}}_\B^{+\bl}(k^+,\pperp_\B) \right]\left[ \tilde{\mathcal{F}}_\A^{-\bi}(k^-,\qperp_\A), \tilde{\mathcal{F}}_\B^{+\bj}(k^+,\qperp_\B) \right]^\dagger \nn\\
     &= \frac{8g^2}{\kperp^2}\left( \delta^{\bk\bl}\delta^{\bi\bj} + \varepsilon^{\bk\bl}\varepsilon^{\bi\bj} \right) \intop \frac{\dd^2\pperp\,\dd^2\qperp}{(2\pi)^4} p^\bk(k^\bl-p^\bl)q^\bi(k^\bj-q^\bj) \nn\\
     &\times \tr \left[ \ii\tilde{\mathcal{A}}_\A^-(k^-,\pperp), \ii\tilde{\mathcal{A}}_\B^+(k^+,\kperp-\pperp) \right]\left[ \ii\tilde{\mathcal{A}}_\A^-(k^-,\qperp), \ii\tilde{\mathcal{A}}_\B^+(k^+,\kperp-\qperp) \right]^\dagger, \label{eq:n(k)-L2-ref}
\end{align}
where we integrated out the delta distributions and dropped the $\A/\B$ labels on the integration variables.
Then, we used \cref{eq:FT-single-nucl-F} to replace the field strengths $\tilde{\mathcal{F}}_{\A/\B}$ with the gauge fields $\tilde{\mathcal{A}}_{\A/\B}$ in the last line, which generated the transverse momentum factors.
Next, we use the solutions of the gauge fields $\tilde{\mathcal{A}}_{\A/\B}$ in terms of the charge distributions $\tilde\rho_{\A/\B}$ given in \cref{eq:single-nucleus-gauge-fields}, where we introduce explicit IR regulation via
\begin{align}
    \tilde{\mathcal{A}}^\mp_{\A/\B}(\kmp,\kperp) = \frac{\tilde{\rho}_{\A/\B}(\kmp,\kperp)}{\kperp^2+m^2},
\end{align}
with the regulator $m$ taking the same value for nucleus $\A$ and $\B$.
This leads to
\begin{align}
     n(\vec k) &= \frac{8g^2}{\kperp^2}\left( \delta^{\bk\bl}\delta^{\bi\bj} + \varepsilon^{\bk\bl}\varepsilon^{\bi\bj} \right) \intop \frac{\dd^2\pperp\,\dd^2\qperp}{(2\pi)^4} \nn\\
     &\times \frac{p^\bk(k^\bl-p^\bl)q^\bi(k^\bj-q^\bj)}{\left(\pperp^2+m^2\right)\left((\kperp-\pperp)^2+m^2\right)\left(\qperp^2+m^2\right)\left((\kperp-\qperp)^2+m^2\right)} \nn\\
     &\times \tr \left[ \tilde{\rho}_\A(k^-,\pperp), \tilde{\rho}_\B(k^+,\kperp-\pperp) \right]\left[ \tilde{\rho}_\A(k^-,\qperp), \tilde{\rho}_\B(k^+,\kperp-\qperp) \right]^\dagger.
\end{align}
We can further simplify the trace in color space by expanding the $\mathfrak{su}(N_c)$-algebra elements in the basis of the generators $t^a$,
\begin{align}
         n(\vec k) &= \frac{8g^2}{\kperp^2}\left( \delta^{\bk\bl}\delta^{\bi\bj} + \varepsilon^{\bk\bl}\varepsilon^{\bi\bj} \right) \intop \frac{\dd^2\pperp\,\dd^2\qperp}{(2\pi)^4} \nn\\
     &\times \frac{
     p^\bk(k^\bl-p^\bl)q^\bi(k^\bj-q^\bj)
     \tilde{\rho}^a_\A(k^-,\pperp) \tilde{\rho}^b_\B(k^+,\kperp-\pperp) \tilde{\rho}^{*c}_\A(k^-,\qperp) \tilde{\rho}^{*d}_\B(k^+,\kperp-\qperp)
     }{\left(\pperp^2+m^2\right)\left((\kperp-\pperp)^2+m^2\right)\left(\qperp^2+m^2\right)\left((\kperp-\qperp)^2+m^2\right)} \nn\\
     &\times \tr \left[ t^a,t^b \right]\left[ t^c,t^d \right]^\dagger, \label{eq:n(k)-tr-tttt}
\end{align}
such that we were able to pull the charge distributions out of the trace.
Note that the Hermitian conjugate acting on $\tilde{\rho}^a_{\A/\B}$ reduces to plain complex conjugation for each color component $a$.
Using the fundamental properties of the generators (cf.~\cref{eq:generators-herm-trace,eq:generators-norm,eq:generators-comm}),
\begin{align}
    \left[ t^a, t^b \right] = \ii f^{abc} t^c, \qquad \tr t^a t^b = \frac{1}{2}\delta^{ab}, \qquad (t^a)^\dagger = t^a,
\end{align}
the last line of \cref{eq:n(k)-tr-tttt} reduces to a contraction of structure constants.
We obtain
\begin{align}
     n(\vec k) &= \frac{4g^2}{\kperp^2} \intop \frac{\dd^2\pperp\,\dd^2\qperp}{(2\pi)^4} f^{abe}f^{cde}\left( \delta^{\bk\bl}\delta^{\bi\bj} + \varepsilon^{\bk\bl}\varepsilon^{\bi\bj} \right) \nn\\
     &\times \frac{p^\bk(k^\bl-p^\bl)q^\bi(k^\bj-q^\bj)}{\left(\pperp^2+m^2\right)\left((\kperp-\pperp)^2+m^2\right)\left(\qperp^2+m^2\right)\left((\kperp-\qperp)^2+m^2\right)} \nn\\
     &\times \tilde{\rho}^a_\A(k^-,\pperp) \tilde{\rho}^b_\B(k^+,\kperp-\pperp) \tilde{\rho}^{*c}_\A(k^-,\qperp) \tilde{\rho}^{*d}_\B(k^+,\kperp-\qperp). \label{eq:n(k)-rrrr}
\end{align}
The structure of this expression is remarkable.
The dependence on the external transverse momentum $\kperp$ is convolved in integrals over the charge distributions.
However, the charge distributions are directly evaluated at the on-shell light cone momenta $k^\mp$.
The longitudinal structure of the nuclei is crucial to describe this dependence.

Furthermore, \cref{eq:n(k)-rrrr} does not contain a CGC expectation value.
The charge distributions $\tilde{\rho}^a_{\A/\B}$ are realisations of the nuclear structure for a single event.
In principle, $n(\vec k)$ can be evaluated on an event-by-event basis, with independent realizations of the nuclear structure in each event.
By assuming different distributions for the ensemble of charge distributions, one can then study event-by-event fluctuations.
In the following, however, we will consider event-averaged gluon numbers, which allows us to exploit the MV model for the ensemble of charge distributions.

\subsubsection{CGC averaged distribution}

The averaged total number of gluons $N$ can be obtained by integrating the occupation number over the entire phase space via
\begin{align}
    N &= \intop \dd^3 \vec k\, \frac{\dd N}{\dd^3 \vec k} = \intop \frac{\dd^3 \vec k}{(2\pi)^3 2 |\vec k|} \langle n(\vec k) \rangle \nn\\
    &= \intop \dd^2 \kperp\,\dd\Y \frac{|\vec k|}{(2\pi)^3 2 |\vec k|} \langle n(\kperp,\Y) \rangle = \intop \dd^2 \kperp\,\dd\Y\, \frac{\dd N}{\dd^2\kperp\, \dd\Y}, \label{eq:dNd2kdY-def}
\end{align}
where we used the Lorentz-invariant phase space volume for on-shell gluons and parametrized the components of the lightlike momentum four-vector (cf.\ \cref{sec:description-initial-state})
\begin{align}
    k^t = |\vec k| = \kperp \cosh(\Y),& \qquad k^z = \kperp \sinh(\Y), \label{eq:ktz-Y-param} \\
    k^+ = \frac{1}{\sqrt{2}}|\kperp|\ee^{\Y},& \qquad k^- = \frac{1}{\sqrt{2}}|\kperp|\ee^{-\Y}, \label{eq:kpm-Y-param}
\end{align}
by the momentum rapidity $\Y$ and transverse momentum $\kperp$.
We also take the CGC expectation value of the occupation number as a means to perform the event average in the MV model (cf.\ \cref{sec:MV-nucl-model}).
\Cref{eq:dNd2kdY-def} defines the gluon number distribution $\dd N/\dd^2\kperp\dd\Y$ that is differential in transverse momentum $\kperp$ and momentum rapidity $\Y$.
We continue by evaluating the CGC average of \cref{eq:n(k)-rrrr} for $\dd N/\dd^2\kperp\dd\Y$,
\begin{align}
    \kperp^2\frac{\dd N}{\dd^2\,\kperp\dd\Y} &= \kperp^2 \frac{\langle n(\kperp,\Y)\rangle}{2(2\pi)^3} \nn\\
    &= \frac{2g^2}{(2\pi)^3} \intop \frac{\dd^2\pperp\,\dd^2\qperp}{(2\pi)^4} f^{abe}f^{cde}\left( \delta^{\bk\bl}\delta^{\bi\bj} + \varepsilon^{\bk\bl}\varepsilon^{\bi\bj} \right) \nn\\
    &\times \frac{p^\bk(k^\bl-p^\bl)q^\bi(k^\bj-q^\bj)}{\left(\pperp^2+m^2\right)\left((\kperp-\pperp)^2+m^2\right)\left(\qperp^2+m^2\right)\left((\kperp-\qperp)^2+m^2\right)} \nn\\
    &\times \langle\tilde{\rho}^a_\A(k^-,\pperp) \tilde{\rho}^{*c}_\A(k^-,\qperp)\rangle\langle\tilde{\rho}^b_\B(k^+,\kperp-\pperp) \tilde{\rho}^{*d}_\B(k^+,\kperp-\qperp)\rangle,
\end{align}
where we multiplied the IR divergent $1/\kperp^2$ to the left and exploited that the expectation value factorizes into two-point correlators of charge distributions of the same nucleus.
To make progress, recall that a basic property of the two-point correlators in the MV model is that the different colors are uncorrelated, i.e., each two-point function is proportional to the Kronecker delta in color space.
Then, the color structure reduces to a prefactor
\begin{align}
    \kperp^2\frac{\dd N}{\dd^2\kperp\,\dd\Y} &= \frac{2g^2}{(2\pi)^3} N_c(N_c^2-1) \intop \frac{\dd^2\pperp\,\dd^2\qperp}{(2\pi)^4} \nn\\
    &\times \frac{\left( \delta^{\bk\bl}\delta^{\bi\bj} + \varepsilon^{\bk\bl}\varepsilon^{\bi\bj} \right)\, p^\bk(k^\bl-p^\bl)q^\bi(k^\bj-q^\bj)}{\left(\pperp^2+m^2\right)\left((\kperp-\pperp)^2+m^2\right)\left(\qperp^2+m^2\right)\left((\kperp-\qperp)^2+m^2\right)} \nn\\
    &\times \frac{\langle\tilde{\rho}^a_\A(k^-,\pperp) \tilde{\rho}^{*a}_\A(k^-,\qperp)\rangle}{N_c^2-1}\frac{\langle\tilde{\rho}^b_\B(k^+,\kperp-\pperp) \tilde{\rho}^{*b}_\B(k^+,\kperp-\qperp)\rangle}{N_c^2-1} \nn\\
    &= \frac{2g^2N_c}{(2\pi)^3(N_c^2-1)} \intop \frac{\dd^2\pperp\,\dd^2\qperp}{(2\pi)^4} \nn\\
    &\times \frac{\left( \delta^{\bk\bl}\delta^{\bi\bj} + \varepsilon^{\bk\bl}\varepsilon^{\bi\bj} \right)\, p^\bk(k^\bl-p^\bl)q^\bi(k^\bj-q^\bj)}{\left(\pperp^2+m^2\right)\left((\kperp-\pperp)^2+m^2\right)\left(\qperp^2+m^2\right)\left((\kperp-\qperp)^2+m^2\right)} \nn\\
    &\times \langle\tilde{\rho}^a_\A(k^-,\pperp) \tilde{\rho}^{a}_\A(-k^-,-\qperp)\rangle \langle\tilde{\rho}^b_\B(k^+,\kperp-\pperp) \tilde{\rho}^{b}_\B(-k^+,-\kperp+\qperp)\rangle, \label{eq:dNd2kdY-result}
\end{align}
where we used that the full contraction of two structure constants $f^{abe}f^{abe} = N_c(N_c^2-1)$ and $\delta^{aa} = (N_c^2 -1)$.
In the last line, we evaluated the complex conjugation on the charge distributions.
For real-valued functions in position space, $\rho^a_{\A/\B}(x^\pm,\xperp) \in \mathbb{R}$, and the complex conjugation of the Fourier transformation reduces to a sign change of the arguments.

Note that \cref{eq:dNd2kdY-result} clearly shows that the longitudinal structure of the nuclei is essential to describe the gluon distribution, especially as a function of momentum rapidity $\Y$.
The dependence of the two-point functions on the longitudinal momenta $k^\mp$ directly translates to the $\Y$-depenence via \cref{eq:kpm-Y-param}.
Hence, the traditional MV model with infinitesimal longitudinal support for the nuclei cannot provide the necessary dependence.
In later chapters, we will study different nuclear models that generalize this longitudinal support and investigate their effects on the gluon distribution $\dd N/\dd^2\kperp\dd\Y$.

\pagebreak
\subsubsection{Discussion}\label{subsec:discussion-gluon-numbers}

The calculation of the gluon distribution discussed until now reproduces well-established results~\cite{Gribov:1983ivg,Kovner:1995ts,Kovner:1995ja,Gyulassy:1997vt,Kovchegov:1997ke,Kovchegov:1998bi,Guo:1998pe,Kharzeev:2001gp,Kovchegov:2001sc,Blaizot:2008yb,Blaizot:2010kh,Dumitru:2001ux,Dumitru:2008wn,Blaizot:2004wu,Gelis:2008rw} and we were able to reach a clear interpretation of the origin of the momentum rapidity dependence of the gluon distribution.
Compared to the literature, we focused on the field strengths of the single-nucleus fields and the Glasma field, rather than on the gauge fields themselves.
This is in contrast to the typical formulation of the classical production of gluons in terms of the squared amplitude $\langle | \mathcal{M}_\lambda|^2\rangle$.
The gluon distribution can then be calculated via
\begin{align}
    \frac{dN}{d^3 \vec k} = \frac{1}{(2\pi)^3 2 |\vec k|} \sum_\lambda \langle | \mathcal{M}_\lambda|^2\rangle,
\end{align}
where the color structure is traced out.
The amplitude is obtained from the on-shell contribution of the Glasma gauge field in Coulomb gauge
\begin{align}
    \mathcal{M}_\lambda(\vec k) = \lim_{k_\nu k^\nu \rightarrow 0} k_\nu k^\nu \tilde a^i_C(k) \epsilon^i_\lambda(\vec k) = P^{ij}_{\vec k} \tilde{S}^j(|\vec k|,\vec k) \epsilon^i_\lambda(\vec k).
\end{align}
For the second equality, we used our results for the gauge transformation in \cref{eq:tildeaC-tk-res}, which introduces the transverse projector.
The on-shell contributions reduce to the source terms given in \cref{eq:dilute-tilde-Smu}, which contain the product of the Lipatov vertex $C^\mu_L(k,\pperp)$ from \cref{eq:Lipatov-vertex} and the fields from the incoming nuclei.
The squared amplitude then translates to taking the square of the Lipatov vertex.
This calculation is performed in \cref{appx:square-of-lipatov} and yields
\begin{align}
    P^{ij}_{\vec k} C^i_L(\vec k,\pperp) C^j_L(\vec k,\qperp) = \frac{4}{\kperp^2}\left( \delta^{\bk\bl}\delta^{\bi\bj} + \varepsilon^{\bk\bl}\varepsilon^{\bi\bj} \right)\, p^\bk(k^\bl-p^\bl)q^\bi(k^\bj-q^\bj),
\end{align}
in full agreement with our results from \cref{eq:n(k)-L2-ref}.

Furthermore, the resulting gluon distribution in \cref{eq:dNd2kdY-result} has a special structure known as $k_T$-factorization (the $T$ stands for \emph{transverse}).
The squared vertex, which encodes the interactions of the incoming nuclei, only enters via a convolution over transverse-only momenta and is factorized from the longitudinal dependence.
The high-energy kinematics encoded in the Lipatov vertex modify the leading-order $1/\kperp^2$ factor from the elementary partonic $2\rightarrow1$ gluon production.
The longitudinal dependence of $\dd N/\dd^2\kperp\dd\Y$ allows for a straightforward generalization to go beyond the boost-invariant approximation for the incoming nuclei.
In the boost-invariant limit (cf.~\cref{sec:boost-inv-glasma,sec:MV-nucl-model}), the Fourier-transformed color charge correlators do not depend on the longitudinal momenta.

In the parton picture, $k_T$-factorization is a powerful tool used in calculations involving high-energy kinematics.
There, the contribution of the single-nucleus gauge fields is interpreted as unintegrated parton distribution functions.
We will explore the parton model interpretation of the results for the gluon distribution in \cref{sec:TMDs-QCD}, where we will identify the parton distribution functions in the dilute limit.
Then, in \cref{ch:nuclear-models}, we will interpret the parton distributions in terms of charge correlators following a generalized MV model.
\pagebreak

\section{Limit of large nuclei in the MV model}\label{sec:limit-large-nucl}

In this section, we derive a simplified expression for the gluon number distribution from \cref{eq:dNd2kdY-result} in a specific limit.
The underlying assumption is that the characteristic size of the nucleus, determined by the radius parameter of the nuclear envelope, is the largest scale in the system.
In particular, the radius is well separated from the scale of the correlations inside the nuclei and from the scale of the inverse IR regulator.
Phenomenologically, this situation corresponds to the configuration of large nuclei, where the fluctuation scales set by the nucleons are much smaller than the overall size of the nucleus.
Additionally, we assume head-on collisions with zero impact parameter to simplify the calculations.

Recall the gluon number distribution from \cref{eq:dNd2kdY-result},
\begin{align}
    \kperp^2\frac{\dd N}{\dd^2\kperp\,\dd\Y} &= \frac{2g^2N_c}{(2\pi)^3(N_c^2-1)} \intop \frac{\dd^2\pperp\,\dd^2\qperp}{(2\pi)^4}\, \omega_m(\kperp,\pperp,\qperp) \nn\\
    &\times \langle\tilde{\rho}^a_\A(k^-,\pperp) \tilde{\rho}^{a}_\A(-k^-,-\qperp)\rangle \langle\tilde{\rho}^b_\B(k^+,\kperp-\pperp) \tilde{\rho}^{b}_\B(-k^+,-\kperp+\qperp)\rangle, \label{eq:dNd2kdY-large-nuclei-start}
\end{align}
where we now introduced the shorthand $\omega_m$ for the IR-regulated effective vertex
\begin{align}
    \omega_m(\kperp,\pperp,\qperp) = \frac{\left( \delta^{\bk\bl}\delta^{\bi\bj} + \varepsilon^{\bk\bl}\varepsilon^{\bi\bj} \right)\, p^\bk(k^\bl-p^\bl)q^\bi(k^\bj-q^\bj)}{\left(\pperp^2+m^2\right)\left((\kperp-\pperp)^2+m^2\right)\left(\qperp^2+m^2\right)\left((\kperp-\qperp)^2+m^2\right)}.\label{eq:effective-omega-m-def}
\end{align}
In the generalized MV nuclear model defined by \cref{eq:general-correlator-t(x)-t(y)} (cf.~\cref{sec:generalized-correl}), the color charge correlators in \cref{eq:dNd2kdY-large-nuclei-start} are given in \cref{eq:FT-general-correlator-t(x)-t(y)} as the convolution of the Fourier-transformed correlation functions $\tilde{\Gamma}_{\A/\B}$ and square-roots of single-nucleus envelopes $\tilde{t}_{\A/\B}$.
The expression for nucleus $\B$ reads
\begin{align}
    \frac{\langle\tilde{\rho}^b_\B(k^+,\kperp-\pperp) \tilde{\rho}^{b}_\B(-k^+,-\kperp+\qperp)\rangle}{N_c^2-1} & \nn\\
    &\hspace{-119.5pt}= \intop \frac{\dd\kappa^+\,\dd^2\kappaperp}{(2\pi)^3}\, \tilde{\Gamma}_\B(\kappa^+,\kappaperp) \tilde{t}_\B(k^+-\kappa^+,\kperp-\pperp-\kappaperp)\tilde{t}_\B(-k^++\kappa^+,-\kperp+\qperp+\kappaperp) \nn\\
    &\hspace{-119.5pt}= \intop \frac{\dd\kappa^+\,\dd^2\kappaperp}{(2\pi)^3}\, \tilde{\Gamma}_\B(k^+-\kappa^+,\kperp-\pperp-\kappaperp)\tilde{t}_\B(\kappa^+,\kappaperp)\tilde{t}_\B(-\kappa^+,-\kappaperp-\pperp+\qperp).\label{eq:rhorhoB-large-nuclei}
\end{align}
For the second equality, we performed the substitutions
\begin{align}
    k^+-\kappa^+ \rightarrow \kappa^+, \qquad \kperp-\pperp-\kappaperp \rightarrow \kappaperp,
\end{align}
such that we shifted the dependencies on the external momenta $k^+$ and $\kperp$ into the correlation function and moved $\pperp$ and $\qperp$ to the same envelope.
The analogous expression for nucleus $\A$ is obtained by swapping the signs of $\pperp$ and $\qperp$, dropping $\kperp$, and changing all longitudinal momenta to the minus components.
Expressing the correlators for both nuclei in this way, we write \cref{eq:dNd2kdY-large-nuclei-start} as
\begin{align}
    \kperp^2\frac{\dd N}{\dd^2\kperp\,\dd\Y} &= \frac{2g^2N_c(N_c^2-1)}{(2\pi)^3} \intop \frac{\dd^2\pperp\,\dd^2\qperp}{(2\pi)^4}\, \omega_m(\kperp,\pperp,\qperp) \nn\\
    &\times \intop \frac{\dd l^-\,\dd^2\lperp}{(2\pi)^3}\, \tilde{\Gamma}_\A(k^- -l^-,\pperp-\lperp)\tilde{t}_\A(l^-,\lperp)\tilde{t}_\A(-l^-,-\lperp+\pperp-\qperp) \nn\\
    &\times \intop \frac{\dd\kappa^+\,\dd^2\kappaperp}{(2\pi)^3}\, \tilde{\Gamma}_\B(k^+-\kappa^+,\kperp-\pperp-\kappaperp)\tilde{t}_\B(\kappa^+,\kappaperp)\tilde{t}_\B(-\kappa^+,-\kappaperp-\pperp+\qperp).
\end{align}
Next, we use the substitutions
\begin{align}
    \uperp = \frac{\pperp+\qperp}{2}, \qquad \vperp = \pperp-\qperp,
\end{align}
and shift the integration over $\vperp$ as
\begin{align}
    \wperp = \vperp-\lperp
\end{align}
to get
\begin{align}
    \kperp^2\frac{\dd N}{\dd^2\kperp\,\dd\Y} &= \frac{2g^2N_c(N_c^2-1)}{(2\pi)^3} \intop \frac{\dd^2\uperp\,\dd^2\vperp}{(2\pi)^4}\frac{\dd l^-\,\dd^2\lperp}{(2\pi)^3} \frac{\dd\kappa^+\,\dd^2\kappaperp}{(2\pi)^3} \nn\\
    &\times \omega_m(\kperp,\uperp+\tfrac{\vperp}{2},\uperp-\tfrac{\vperp}{2}) \nn\\
    &\times \tilde{\Gamma}_\A(k^- -l^-,\uperp+\tfrac{\vperp}{2}-\lperp)\tilde{t}_\A(l^-,\lperp)\tilde{t}_\A(-l^-,-\lperp+\vperp) \nn\\
    &\times \tilde{\Gamma}_\B(k^+-\kappa^+,\kperp-\uperp-\tfrac{\vperp}{2}-\kappaperp)\tilde{t}_\B(\kappa^+,\kappaperp)\tilde{t}_\B(-\kappa^+,-\kappaperp-\vperp) \nn\\
    &= \frac{2g^2N_c(N_c^2-1)}{(2\pi)^3} \intop \frac{\dd^2\uperp\,\dd^2\wperp}{(2\pi)^4} \frac{\dd l^-\,\dd^2\lperp}{(2\pi)^3} \frac{\dd\kappa^+\,\dd^2\kappaperp}{(2\pi)^3} \nn\\
    &\times \omega_m(\kperp,\uperp+\tfrac{\wperp+\lperp}{2},\uperp-\tfrac{\wperp+\lperp}{2}) \nn\\
    &\times \tilde{\Gamma}_\A(k^- -l^-,\uperp+\tfrac{\wperp-\lperp}{2})\tilde{t}_\A(l^-,\lperp)\tilde{t}_\A(-l^-,\wperp) \nn\\
    &\times \tilde{\Gamma}_\B(k^+-\kappa^+,\kperp-\uperp-\tfrac{\wperp+\lperp}{2}-\kappaperp) \tilde{t}_\B(\kappa^+,\kappaperp)\tilde{t}_\B(-\kappa^+,-\kappaperp-\wperp-\lperp).
    \label{eq:dNd2kdY-large-nuclei-scales}
\end{align}
We now turn our attention to the different scales in \cref{eq:dNd2kdY-large-nuclei-scales}.
As we initially assumed, the envelope functions in position space set the largest scale.
In momentum space, this largest position space scale translates to the smallest momentum space scale.
Hence, the Fourier-transformed square roots of the envelopes $\tilde{t}_{\A/\B}$ are highly peaked functions of their arguments.
When carrying out convolution integrals over $\tilde{t}_{\A/\B}$, the envelopes act as Dirac-delta functions and lead to a factorization of the integrals.
We will use these observations to derive an approximate result next.
A rigorous derivation of the following calculation is given in \cref{appx:factorizable-large-R-scales}.

First, we examine the longitudinal scales that enter only via the correlation functions and envelopes.
The integral over $l^-$ only involves the quantities with the label of nucleus $\A$ and the integral over $\kappa^+$ involves only $\B$.
Since the envelopes are highly peaked at $l^-\approx 0$ and $\kappa^+\approx 0$, compared to the change of the correlation functions with said variables, we can effectively evaluate the correlation functions at 0 for the momenta being integrated out.
This leads to a factorization of the integrations over the longitudinal momenta,
\begin{align}
    \kperp^2\frac{\dd N}{\dd^2\kperp\,\dd\Y} &= \frac{2g^2N_c(N_c^2-1)}{(2\pi)^3} \intop \frac{\dd^2\uperp\,\dd^2\wperp}{(2\pi)^4} \frac{\dd^2\lperp}{(2\pi)^2} \frac{\dd^2\kappaperp}{(2\pi)^2} \nn\\
    &\times \omega_m(\kperp,\uperp+\tfrac{\wperp+\lperp}{2},\uperp-\tfrac{\wperp+\lperp}{2}) \tilde{\Gamma}_\A(k^-,\uperp+\tfrac{\wperp-\lperp}{2}) \tilde{\Gamma}_\B(k^+,\kperp-\uperp-\tfrac{\wperp+\lperp}{2}-\kappaperp)\nn\\
    &\times \intop\frac{\dd l^-}{2\pi} \tilde{t}_\A(l^-,\lperp)\tilde{t}_\A(-l^-,\wperp) \nn\\
    &\times \intop\frac{\dd \kappa^+}{2\pi} \tilde{t}_\B(\kappa^+,\kappaperp)\tilde{t}_\B(-\kappa^+,-\kappaperp-\wperp-\lperp),
\end{align}
where we pulled both $\tilde{\Gamma}_{\A/\B}$ in front of the these integrals.

Next, we examine the transverse scales.
Note that there are four transverse momenta being integrated over.
The earlier variable substitutions introducing $\uperp$, $\vperp$ and $\wperp$ allowed us to isolate the transverse momentum dependencies of three of the four envelopes to a single momentum.
Additionally, the momentum $\uperp$ only appears in the effective vertex and the correlation functions.
Using that the $\tilde{t}_{\A/\B}$ are highly peaked in their transverse momentum arguments as well, we assume that the transverse correlation scales and the scale of the effective vertex (given by $1/m$ in position space) are well separated from the scale of the envelopes.
Then, we can factorize each of the $\wperp$, $\lperp$ and $\kappaperp$ integrals, where we evaluate the effective vertex and the correlation functions at 0 for these momenta.
We can now pull the effective vertex and the correlation functions in front of these integrals,
\begin{align}
    \kperp^2\frac{\dd N}{\dd^2\kperp\,\dd\Y} &= \frac{2g^2N_c(N_c^2-1)}{(2\pi)^3} \intop \frac{\dd^2\uperp}{(2\pi)^2} \omega_m(\kperp,\uperp,\uperp) \tilde{\Gamma}_\A(k^-,\uperp) \tilde{\Gamma}_\B(k^+,\kperp-\uperp)\nn\\
    &\times \intop\frac{\dd^2\wperp}{(2\pi)^2} \frac{\dd^2\lperp}{(2\pi)^2} \frac{\dd^2\kappaperp}{(2\pi)^2} \frac{\dd l^-}{2\pi} \tilde{t}_\A(l^-,\lperp)\tilde{t}_\A(-l^-,\wperp) \nn\\
    &\times \intop\frac{\dd \kappa^+}{2\pi} \tilde{t}_\B(\kappa^+,\kappaperp)\tilde{t}_\B(-\kappa^+,-\kappaperp-\wperp-\lperp) \nn\\
    &= \frac{2g^2N_c(N_c^2-1)}{(2\pi)^3} \intop \frac{\dd^2\uperp}{(2\pi)^2} \omega_m(\kperp,\uperp,\uperp) \tilde{\Gamma}_\A(k^-,\uperp) \tilde{\Gamma}_\B(k^+,\kperp-\uperp)\nn\\
    &\times \intop\frac{\dd^2\vperp}{(2\pi)^2}\intop \frac{\dd l^-\,\dd^2\lperp}{(2\pi)^3} \tilde{t}_\A(l^-,\lperp)\tilde{t}_\A(-l^-,-\lperp+\vperp) \nn\\
    &\times \intop \frac{\dd \kappa^+\,\dd^2\kappaperp}{(2\pi)^3} \tilde{t}_\B(\kappa^+,\kappaperp)\tilde{t}_\B(-\kappa^+,-\kappaperp-\vperp),\label{eq:largeR-factorized-envelopes}
\end{align}
where we reintroduced $\vperp = \wperp + \lperp$ for $\wperp$ and rearranged the integrals in the last equation.
The isolated contribution from the envelopes can be identified with the convolution of two Fourier-transformed projected envelopes.
In particular,
\begin{align}
    \tilde{T}^\perp_{\A/\B}(\kperp) &= \intop\dd^2\xperp\,\ee^{-\ii\kperp\cdot\xperp} T^\perp_{\A/\B}(\xperp) \nn\\
    &= \intop\dd^2\xperp\,\ee^{-\ii\kperp\cdot\xperp}\intop\dd\xpm\,t_{\A/\B}(\xpm,\xperp)t_{\A/\B}(\xpm,\xperp) \nn\\
    &= \intop \frac{\dd \kappa^+\,\dd^2\kappaperp}{(2\pi)^3} \tilde{t}_{\A/\B}(\kappa^+,\kappaperp)\tilde{t}_{\A/\B}(-\kappa^+,-\kappaperp+\kperp). \label{eq:FT-Tperp}
\end{align}
With that, the convolution over $\vperp$ in the last two lines of \cref{eq:largeR-factorized-envelopes} reads
\begin{align}
    \intop \frac{\dd^2\vperp}{(2\pi)^2}\,\tilde{T}^\perp_\A(\vperp)\tilde{T}^\perp_\B(-\vperp) &= \intop\frac{\dd^2\vperp\,\dd^2\xperp\,\dd^2\yperp}{(2\pi)^2}\,\ee^{-\ii\vperp\cdot(\xperp-\yperp)}T^\perp_\A(\xperp)T^\perp_\B(\yperp) \nn\\
    &= \intop \dd^2\xperp\,T^\perp_\A(\xperp)T^\perp_\B(\xperp) \nn\\
    &\eqqcolon \Sop, \label{eq:Sop-def}
\end{align}
where we defined the symbol $\Sop$ for the resulting projected transverse overlap area.
Note that $\Sop$ has the dimension of an area because of the normalization of the single-nucleus envelopes in \cref{eq:envelope-norm}, which renders $T^\perp_{\A/\B}$ dimensionless.

Finally, the gluon number distribution in the limit of large nuclei reads
\begin{align}
    \kperp^2\frac{\dd N}{\dd^2\kperp\,\dd\Y} &\approx \frac{2g^2N_c(N_c^2-1)}{(2\pi)^3}\, \Sop \intop \frac{\dd^2\uperp}{(2\pi)^2}\,
     \omega_m(\kperp,\uperp,\uperp) \tilde{\Gamma}_\A(k^-,\uperp)\tilde{\Gamma}_\B(k^+,\kperp-\uperp). \label{eq:dNd2kdY-large-nuclei-result}
\end{align}
The structure of \cref{eq:dNd2kdY-large-nuclei-result} is remarkable.
This result exhibits a clean factorization between contributions that depend on the external momenta $\kpm$ and $\kperp$ and a pure geometry factor $\Sop$.
In the limit of large nuclei, the exact functional form of the envelopes only sets an overall scale for the gluon distribution.
The shape of $\dd N/\dd^2\kperp\dd\Y$, on the other hand, is determined by the convolution of the effective vertex factor $\omega_m$ and the correlation functions.
Furthermore, the momentum rapidity profile is directly given by the product of correlation functions when parametrizing $\kmp = |\kperp|\ee^{\mp\Y}/\sqrt{2}$ for on-shell $k$.
Gluon production is governed by the interplay of the correlation functions and the vertex only.
The initial dependence of the vertex function on the momenta $\pperp$ and $\qperp$ is diagonalized to $\uperp = (\pperp+\qperp)/2$.
Similarly, the dependence on the difference $\vperp=\pperp-\qperp$ drops out of the correlation functions as well.
This is the direct consequence of the separation of scales from the envelopes.
The spectrum is produced from a domain with a size corresponding to the correlation lengths, and then scaled up to the transverse overlap area $\Sop$.
In this sense, gluons produced from different domains, localized at different positions within the overlapping nuclei, are summed together without interference or interaction.

The transverse overlap area also captures all modifications introduced by a small impact parameter $\bperp$.
Restricting the impact parameter to be small compared to the envelope scale ensures that the approximation discussed in this section stays valid and the envelope scale remains the largest scale in the system.
Then, we assume a universal envelope for both nuclei where the impact parameter only enters as a shift in the transverse coordinates,
\begin{align}
    T^\perp_\A(\xperp+\bperp/2) = T^\perp_\B(\xperp-\bperp/2)\equiv T^\perp(\xperp). \label{eq:Tperp-impact-param-def}
\end{align}
As an immediate consequence, the transverse overlap area becomes impact parameter dependent,
\begin{align}
    \Sop(\bperp) = \intop \dd^2\xperp\,T^\perp_\A(\xperp)T^\perp_\B(\xperp) = \intop \dd^2\xperp\,T^\perp(\xperp-\bperp/2)T^\perp(\xperp+\bperp/2). \label{eq:Sop-impact-param}
\end{align}
The correlation functions are unaffected by any impact parameter, because they measure the correlation of two points within a nucleus.
Formally, they depend on the difference of the coordinates of the two points, which is unaffected by shifting with $\bperp$.

\subsubsection{MV nuclear correlations}

We study a particularly simple case for the gluon distribution in \cref{eq:dNd2kdY-large-nuclei-result}, which assumes the traditional MV model for the correlation function in the color charge correlator (cf.\ \cref{eq:MV-correlator-transverse}).
This special case can be reached via the limit in \cref{eq:Gamma-MV-limit-norm},
\begin{align}
    \lim_{\substack{r_l\,\rightarrow0\\r\,\rightarrow\,0}} \Gamma_{\A/\B}(\xpm,\xperp) = g^2 \mu^2 \delta(\xpm)\deltaperp(\xperp),
\end{align}
where the correlation scales in the longitudinal ($r_l$) and transverse ($r$) directions are taken to zero.
This corresponds to a completely uncorrelated color structure where the prefactor $g^2\mu^2$ sets the strength of color charge per unit area.
In momentum space,
\begin{align}
    \lim_{\substack{r_l\,\rightarrow0\\r\,\rightarrow\,0}} \tilde{\Gamma}_{\A/\B}(\kmp,\kperp) = g^2 \mu^2,
\end{align}
which we insert into \cref{eq:dNd2kdY-large-nuclei-result} and get
\begin{align}
    \kperp^2\frac{\dd N}{\dd^2\kperp\,\dd\Y} &\approx
    \frac{2g^2N_c(N_c^2-1)}{(2\pi)^3} \Sop g^4 \mu^4 
    \intop \frac{\dd^2\uperp}{(2\pi)^2}\,\omega_m(\kperp,\uperp,\uperp) \nn\\
    &= \frac{2g^6 \mu^4 N_c(N_c^2-1)}{(2\pi)^3} \Sop \frac{1}{m^2} \Omega(\kperp/m).\label{eq:dNd2kdY-MV-limit}
\end{align}
Here, we defined the scale-free integrated vertex,
\begin{align}
    \Omega(\kperp/m) &\coloneqq m^2 \intop\frac{\dd^2\uperp}{(2\pi)^2}\,\omega_m(\kperp,\uperp,\uperp) \nn\\
    &= \intop\frac{\dd^2\uperp}{(2\pi)^2} \frac{m^2 \uperp^2(\kperp-\uperp)^2}{\left(\uperp^2+m^2\right)^2\left((\kperp-\uperp)^2+m^2\right)^2} \nn\\
    &= \intop\frac{\dd^2\underline{\uperp}}{(2\pi)^2}\, \frac{\underline{\uperp}^2(\kperp/m - \underline{\uperp})^2}{\left(\underline{\uperp}^2+1\right)^2\left((\kperp/m-\underline{\uperp})^2+1\right)^2} \nn\\
    &= \frac{1}{2\pi}\left(\frac{4+12\underline{\kperp}^2+6\underline{\kperp}^4+\underline{\kperp}^6}{\underline{\kperp}^3 \left(4+\underline{\kperp}^2\right)^{5/2}} \ 2\,\mathrm{artanh}(\tfrac{|\underline{\kperp}|}{\sqrt{4+\underline{\kperp}^2}})
    - \frac{2+3\underline{\kperp}^2+\underline{\kperp}^4}{\underline{\kperp}^2\left(4+\underline{\kperp}^2\right)^2}\right) , \label{eq:Omega(k)-def}
\end{align}
after changing to $\underline{\uperp} = \uperp/m$ in the third line and defining $\underline{\kperp}=\kperp/m$ in the last line.
The calculation to obtain the closed-form solution above is performed in \cref{appx:Integrated-effective-vertex}.
Note that $\Omega(\underline{\kperp}) = \Omega(|\underline{\kperp}|)$, i.e., the integrated vertex is isotropic.

In the traditional MV model, the gluon distribution becomes boost-invariant because there is no more dependence on momentum rapidity $\Y$ or collider energy.
The transverse momentum spectrum is given by $\Omega$ alone.
The UV limit of \cref{eq:Omega(k)-def} reduces to $\sim \ln(\underline{\kperp})/\underline{\kperp}^2$ and reproduces established results from the literature (see~\cite{Kovner:1995ts} for the earliest reports).
In the opposite, IR limit, $\Omega$ is finite for $\underline{\kperp}=0$.

\subsubsection{Longitudinally coherent nuclear correlations}

The property of boost-invariance of the MV model is broken as soon as there is a finite longitudinal correlation scale in the nuclei.
The conceptually simplest model to introduce such a scale is by smearing the color structure coherently along the longitudinal extent of the nuclei.
Effectively, this corresponds to pushing the longitudinal scales of the correlation functions $\Gamma_{\A/\B}$ to infinity so that the nuclear envelope regulates the longitudinal correlation length.
In this case,
\begin{align}
    \Gamma_{\A/\B}(\xpm,\xperp) &= \Lambda_{\A/\B}(\xperp), \nn\\
    \tilde{\Gamma}_{\A/\B}(\kmp,\kperp) &= 2\pi \delta(\kmp)\tilde{\Lambda}_{\A/\B}(\kperp). \label{eq:tLambda-keperp}
\end{align}
Note that we assume that the only remaining correlation scale $r$ in $\Lambda_{\A/\B}$ is transverse and is well separated from the nuclear radius.
Contrarily, the longitudinal scale of $\Gamma_{\A/\B}$ is formally infinite.
Therefore, we cannot use the result of the large-nuclei approximation in \cref{eq:dNd2kdY-large-nuclei-result}, even though the nucleus might be large.
Instead, we insert \cref{eq:tLambda-keperp} into the gluon number distribution in \cref{eq:dNd2kdY-large-nuclei-scales}, where the integrals are not factorized yet.
We get
\begin{align}
    \kperp^2\frac{\dd N}{\dd^2\kperp\,\dd\Y} &= \frac{2g^2N_c(N_c^2-1)}{(2\pi)^3} \intop \frac{\dd^2\uperp\,\dd^2\wperp}{(2\pi)^4} \frac{\dd l^-\,\dd^2\lperp}{(2\pi)^3} \frac{\dd\kappa^+\,\dd^2\kappaperp}{(2\pi)^3} \nn\\
    &\times \omega_m(\kperp,\uperp+\tfrac{\wperp+\lperp}{2},\uperp-\tfrac{\wperp+\lperp}{2}) \nn\\
    &\times 2\pi\delta(k^- -l^-)\tilde{\Lambda}_\A(\uperp+\tfrac{\wperp-\lperp}{2})\tilde{t}_\A(l^-,\lperp)\tilde{t}_\A(-l^-,\wperp) \nn\\
    &\times 2\pi\delta(k^+-\kappa^+)\tilde{\Lambda}_\B(\kperp-\uperp-\tfrac{\wperp+\lperp}{2}-\kappaperp) \tilde{t}_\B(\kappa^+,\kappaperp)\tilde{t}_\B(-\kappa^+,-\kappaperp-\wperp-\lperp) \nn\\
    &= \frac{2g^2N_c(N_c^2-1)}{(2\pi)^3} \intop \frac{\dd^2\uperp\,\dd^2\wperp}{(2\pi)^4} \frac{\dd^2\lperp}{(2\pi)^2} \frac{\dd^2\kappaperp}{(2\pi)^2} \nn\\
    &\times \omega_m(\kperp,\uperp+\tfrac{\wperp+\lperp}{2},\uperp-\tfrac{\wperp+\lperp}{2}) \nn\\
    &\times \tilde{\Lambda}_\A(\uperp+\tfrac{\wperp-\lperp}{2})\tilde{t}_\A(k^-,\lperp)\tilde{t}_\A(-k^-,\wperp) \nn\\
    &\times \tilde{\Lambda}_\B(\kperp-\uperp-\tfrac{\wperp+\lperp}{2}-\kappaperp) \tilde{t}_\B(k^+,\kappaperp)\tilde{t}_\B(-k^+,-\kappaperp-\wperp-\lperp),
\end{align}
where we integrated out the delta functions for the second equality.
Now, in full analogy to the derivation of \cref{eq:dNd2kdY-large-nuclei-result}, the vertex function and the transverse correlation functions $\Lambda_{\A/\B}$ factorize from the envelopes,
\begin{align}
    \kperp^2\frac{\dd N}{\dd^2\kperp\,\dd\Y} &\approx \frac{2g^2N_c(N_c^2-1)}{(2\pi)^3} \intop \frac{\dd^2\uperp}{(2\pi)^2}\, 
    \omega_m(\kperp,\uperp,\uperp)
    \tilde{\Lambda}_\A(\uperp) \tilde{\Lambda}_\B(\kperp-\uperp)\nn\\
    &\times \intop \frac{\dd^2\wperp\,\dd^2\lperp\,\dd^2\kappaperp}{(2\pi)^6}\, \tilde{t}_\A(k^-,\lperp)\tilde{t}_\A(-k^-,\wperp) \tilde{t}_\B(k^+,\kappaperp)\tilde{t}_\B(-k^+,-\kappaperp-\wperp-\lperp) \nn\\
    &= \frac{2g^2N_c(N_c^2-1)}{(2\pi)^3} \intop \frac{\dd^2\uperp}{(2\pi)^2}\, 
   \omega_m(\kperp,\uperp,\uperp)
    \tilde{\Lambda}_\A(\uperp) \tilde{\Lambda}_\B(\kperp-\uperp)\nn\\
    &\times \intop \frac{\dd^2\vperp\,\dd^2\lperp\,\dd^2\kappaperp}{(2\pi)^6}\, \tilde{t}_\A(k^-,\lperp)\tilde{t}_\A(-k^-,\vperp-\lperp) \tilde{t}_\B(k^+,\kappaperp)\tilde{t}_\B(-k^+,-\kappaperp-\vperp),
\end{align}
where we shifted back $\vperp = \wperp+\lperp$ in the last line.
The dependence on the external longitudinal momenta $\kmp$ is now consolidated into the last line.
This convolution of the nuclear envelopes scales with the transverse overlap area of the nuclei, but is not identical to $\Sop$.
Instead,
\begin{align}
    \intop\frac{\dd^2\vperp\,\dd^2\lperp\,\dd^2\kappaperp}{(2\pi)^6}\, \tilde{t}_\A(k^-,\lperp)\tilde{t}_\A(-k^-,\vperp-\lperp) \tilde{t}_\B(k^+,\kappaperp)\tilde{t}_\B(-k^+,-\kappaperp-\vperp)& \nn\\
    &\hspace{-226.5pt}= \intop \frac{\dd^2\vperp\,\dd^2\lperp\,\dd^2\kappaperp\,\dd^2\xperp\,\dd^2\yperp\,\dd^2\xperp'\,\dd^2\yperp'}{(2\pi)^2}\,\ee^{-\ii\lperp\cdot(\xperp-\yperp)-\ii\vperp\cdot\yperp}\ee^{-\ii\kappaperp\cdot(\xperp'-\yperp')+\ii\vperp\cdot{\yperp'}} \nn\\
    &\hspace{-226.5pt}\times \tau_\A(k^-,\xperp)\tau_\A(-k^-,\yperp)\tau_\B(k^+,\xperp')\tau_\B(-k^+,\yperp') \nn\\
    &\hspace{-226.5pt}= \intop \dd^2\xperp\,|\tau_\A(k^-,\xperp)|^2|\tau_\B(k^+,\xperp)|^2, \label{eq:coherent-Sop}
\end{align}
where we introduced the partial Fourier transformation of the envelopes,
\begin{align}
    \tau_{\A/\B}(\kmp,\xperp) = \intop \dd\xpm\,\ee^{\ii \kmp\xpm} t_{\A/\B}(\xpm,\xperp).
\end{align}
The integrations of the exponential functions in the second line of \cref{eq:coherent-Sop} produce delta functions, which set the transverse arguments of all envelopes to the same coordinate.
In the last line, we used that the envelopes are real-valued functions in position space.

The final result for the gluon distribution with coherent nuclear correlations reads
\begin{align}
    \kperp^2\frac{\dd N}{\dd^2\kperp\,\dd\Y} &\approx \frac{2g^2N_c(N_c^2-1)}{(2\pi)^3} \intop \frac{\dd^2\uperp}{(2\pi)^2}\, 
    \omega_m(\kperp,\uperp,\uperp)
    \tilde{\Lambda}_\A(\uperp) \tilde{\Lambda}_\B(\kperp-\uperp) \nn\\
    &\times \intop \dd^2\xperp\,|\tau_\A(k^-,\xperp)|^2|\tau_\B(k^+,\xperp)|^2.
\end{align}
Note that all dependence on momentum rapidity is now contained in the envelope factor in the last line when we parametrize $\kmp=|\kperp|\ee^{\mp\Y}/\sqrt{2}$ for on-shell momenta $k$.
However, the $\kperp$-dependence for the transverse momentum spectrum is not factorized.
Even at mid-rapidity, where $\Y=0$, the envelope factor mixes into the spectrum because of the residual dependence on $\kperp$ via $\kmp$.
Introducing a longitudinal scale, therefore, not only breaks boost-invariance, but also modifies the spectrum at mid-rapidity.

\section{Transverse energy}\label{sec:transverse-energy-mom}

In the limit of free streaming, the gluon number distribution $\dd N/\dd^2\kperp\,\dd\Y$ can be used to calculate the total, time-independent transverse energy $\Ep$.
In this section, we derive a relation between this transverse energy, obtained in the momentum space picture, and the transverse energy defined in \cref{eq:Eperp(x)-def} in the position space picture of the dilute Glasma in \cref{sec:EMT-epsilon(x)}.
We employ the kinetic theory description~\cite{Kurkela:2018vqr,Schlichting:2019abc,Berges:2020fwq,Greif:2017bnr,Ambrus:2021fej}.
Here, the phase space density $f$ describes the distribution of gluons in position and momentum space.
In the Milne frame, the total number of gluons on a hypersurface with fixed proper time $\tau$ is then given by%
\footnote{%
As for the occupation number $n(\vec k)$ in \cref{eq:gluon-n-def}, we do not put explicit degeneracy factors.}
\begin{align}
    N = \intop\dd\tau'\,\dd\eta_s\,\dd^2\xperp\, \frac{\dd k^\eta\,\dd^2\kperp}{(2\pi)^3}\,\tau'\tau'\delta(\tau-\tau')f(\tau',\eta_s,\xperp,\kperp,k^\eta). \label{eq:N-ekt-milne-def}
\end{align}
Note the double Jacobian factors $\tau'$ that enter for each of the three-dimensional volume elements in position and momentum space.
Recall that we evaluate the gluon distribution at asymptotically late times, where the total number of gluons does not change anymore.
Therefore, $N$ does not depend on $\tau$.
The components of the momentum vector in the Milne frame read
\begin{align}
    k^\tau &= \frac{\partial\tau}{\partial t} k^t + \frac{\partial\tau}{\partial z} k^z = |\kperp|\cosh(\Y-\eta_s), \\
    k^\eta &= \frac{\partial\eta_s}{\partial t} k^t + \frac{\partial\eta_s}{\partial z} k^z = |\kperp|\frac{\sinh(\Y-\eta_s)}{\tau},
\end{align}
where we used the parametrization of the momentum components with the momentum rapidity $\Y$ from \cref{eq:ktz-Y-param}.
It is straightforward to check that, still, 
\begin{align}
    k^\mu k_\mu = (k^\tau)^2 - \tau^2(k^\eta)^2 - \kperp^2 = 0,
\end{align}
and the gluons are on-shell.
We change to $\Y$-parametrization in \cref{eq:N-ekt-milne-def}, use $\dd k^\eta = \dd\Y\, k^\tau/\tau$ and get
\begin{align}
    N = \intop\dd^2\kperp\,\dd\Y\,\dd^2\xperp\,\dd\eta_s\,\frac{\tau |\kperp|\cosh(\Y-\eta_s)}{(2\pi)^3}f(\tau,\eta_s,\xperp,\kperp,\Y),
\end{align}
where we can read off
\begin{align}
    \frac{\dd N}{\dd^2\kperp\,\dd\Y} = \intop\dd^2\xperp\,\dd\eta_s\,\frac{\tau|\kperp|\cosh(\Y-\eta_s)}{(2\pi)^3}f(\tau,\eta_s,\xperp,\kperp,\Y). \label{eq:dNd2kdY-general-f}
\end{align}
When the gluons are free-streaming at late times, their momentum and spacetime rapidities become equal such that $f\sim\delta(\Y-\eta_s)$.
More formally,
\begin{align}
    f(\tau,\eta_s,\xperp,\kperp,\Y) = \delta(\Y-\eta_s) \frac{(2\pi)^3}{\tau|\kperp|} \frac{\dd N}{\dd^2\xperp\,\dd^2\kperp\,\dd\Y}, \label{eq:phase-space-density-free-streaming}
\end{align}
which we can insert into \cref{eq:dNd2kdY-general-f} or integrate over $\Y$ to get
\begin{align}
    \frac{\dd N}{\dd^2\kperp\,\dd\Y} &= \frac{\tau|\kperp|}{(2\pi)^3}\intop\dd^2\xperp\,\dd\eta_s\, f(\tau,\eta_s,\xperp,\kperp,\Y), \label{eq:dNd2kdY-f-freestream} \\
    \frac{\dd N}{\dd^2\kperp\,\dd\Y}\Bigg|_{\Y=\eta_s} &= \frac{\tau|\kperp|}{(2\pi)^3}\intop\dd^2\xperp\,\dd\Y'\, f(\tau,\eta_s,\xperp,\kperp,\Y'). \label{eq:dNd2kdeta-f-freestream}
\end{align}
The last equation identifies the gluon number distribution evaluated at $\Y=\eta_s$ with the spacetime rapidity dependent distribution on the right hand side.
Due to the Dirac-delta function in \cref{eq:phase-space-density-free-streaming}, the functional form of  $\dd N/\dd^2\kperp\,\dd\Y$ is identical to the spacetime rapidity distribution.

Next, we use the expression for the energy-momentum tensor from kinetic theory,
\begin{align}
    T^{\mu\nu}(\tau,\eta_s,\xperp) = \tau\intop\frac{\dd^2\kperp\,\dd k^\eta}{(2\pi)^3 k^\tau}\, k^\mu k^\nu f(\tau,\eta_s,\xperp,\kperp,k^\eta),
\end{align}
and evaluate
\begin{align}
    \Ep(\tau) &= \intop \dd\tau'\,\dd\eta_s\,\dd^2\xperp\,\tau' \delta(\tau'-\tau) \left(T^{xx}(\tau',\eta_s,\xperp) + T^{yy}(\tau',\eta_s,\xperp)\right) \tag{\ref{eq:Eperp(x)-def}} \\
    &= \intop \dd\tau'\,\dd\eta_s\,\dd^2\xperp\,\tau' \delta(\tau'-\tau) \intop \frac{\dd^2\kperp\,\dd k^\eta}{(2\pi)^3} \frac{\tau'\kperp^2}{k^\tau} f(\tau',\eta_s,\xperp,\kperp,k^\eta) \nn\\
    &= \intop\dd^2\kperp\,\dd\Y\,\dd^2\xperp\,\dd\eta_s\,\frac{\tau \kperp^2}{(2\pi)^3}f(\tau,\eta_s,\xperp,\kperp,\Y).
\end{align}
We can now use \cref{eq:dNd2kdY-f-freestream,eq:dNd2kdeta-f-freestream} to obtain the transverse energy differential in momentum rapidity
\begin{align}
    \frac{\dd\Ep}{\dd\Y} = \intop\dd^2\kperp\, |\kperp|\frac{\dd N}{\dd^2\kperp\,\dd\Y}, \label{eq:dEpdY-ito-dN}
\end{align}
as well as the transverse energy differential in spacetime rapidity
\begin{align}
    \frac{\dd\Ep}{\dd\eta_s} = \intop\dd^2\kperp\,|\kperp|\frac{\dd N}{\dd^2\kperp\,\dd\Y}\Bigg|_{\Y=\eta_s}. \label{eq:dEpdeta-ito-dN}
\end{align}
Note that these two distributions are functionally the same in the free-streaming limit and we identify \cref{eq:dEpdeta-ito-dN} as the equivalent quantity to the transverse energy obtained in the position space picture in \cref{eq:Eperp(x)-def}.

The last two expressions in \cref{eq:dEpdY-ito-dN,eq:dEpdeta-ito-dN} constitute a straightforward definition of transverse energy.
The number of gluons in a phase space volume element $\dd^2\kperp\,\dd\Y$ is multiplied by the modulus of the transverse momentum $|\kperp|$ to get the total transverse energy in that volume element.
Integrating out $\kperp$ sums over all contributions.
However, as the calculation above demonstrates, this definition of transverse energy only corresponds to the sum of the transverse pressure components of the energy-momentum tensor when the gluons are free-streaming.
Then, the correction factor $\cosh(\Y-\eta_s)$ in the general Milne frame expression for $\dd N/\dd^2\kperp\,\dd\Y$ in \cref{eq:dNd2kdY-general-f} becomes 1 as $\Y\sim\eta_s$.
Furthermore, this also entails that the $k^\eta$ component is zero, i.e., in the Milne frame, the gluons have no longitudinal momentum.

Still, caution is necessary when comparing $\dd\Ep/\dd\eta_s$ obtained from \cref{eq:dEpdeta-ito-dN} to the position space result from \cref{eq:Eperp(x)-def}.
The latter result is not independent of proper time because of the explicit $\tau$ dependence in the Glasma field strength tensor.
We expect this dependence to become weaker for later times and, eventually, reach a steady state that agrees with the free-streaming limit.
This will be investigated in \cref{subsec:EMT-vs-dN}.

Using \cref{eq:dNd2kdY-def} we can express $\dd N/\dd^2\kperp\,\dd\Y$ via the occupation numbers and get
\begin{align}
    \frac{\dd \Ep}{\dd \Y} = \intop \dd^2\kperp\, \frac{|\kperp|}{2(2\pi)^3}\langle n(\kperp,\Y) \rangle = \intop_0^\infty \dd|\kperp|\,\frac{\kperp^2}{2(2\pi)^2}\langle n(|\kperp|,\Y) \rangle.
\end{align}
Here, we exploited the cylindrical symmetry of the entire collision geometry and produced Glasma to integrate out the azimuthal angle in the transverse plane, leading to the last equality.
There is no dependence on the azimuthal angle left in $n$.
We assume that the nuclear envelopes are spherically symmetric in their rest frames and Lorentz-contracted along the beam axis in the lab frame.
The interaction vertex itself is isotropic in the transverse plane as well.
The only anisotropy could come from an impact parameter.
However, we choose the impact parameter to be zero in our momentum space calculation.
As a result, we can identify that the expression in \cref{eq:dNd2kdY-result} enters $\dd\Ep/\dd\Y$ and is automatically IR safe for $\kperp\rightarrow0$.

\addtocontents{toc}{\protect\pagebreak} 

\chapter{Gluon production from unintegrated gluon distributions}\label{sec:TMDs-QCD}

The intrinsic phase-space dynamics of partons, which are modeled as the constituents of nucleons, are commonly encoded in distribution functions.
Depending on which properties one aims to describe in this parton model, various different distribution functions exist (see~\cite{Collins:2011zzd,Avsar:2012hj,Petreska:2018cbf,Boussarie:2023izj,Lorce:2025aqp,Szczurek:2003fu} and references therein).
They are categorized based on their functional dependencies on different dimensions of the full phase space.
For example, the Wigner distribution~\cite{Belitsky:2003nz,Ji:2003ak} retains a fully six-dimensional description.
On the other extreme, the (collinear) parton distribution function (PDF)~\cite{Levy:1997pz,Thomas:2001kw} only depends on the longitudinal momentum fraction $\xB = p^\mp/\mathcal{P}^\mp$ of the parton, where $\mathcal{P^\mp}=\sqrt{2}\gamma m_n$ is the longitudinal momentum of the nucleon, and all the rest of the phase space dependence is integrated out.
Different physical processes are sensitive to different parton distributions, which allows the extraction of nuclear properties by comparing perturbative calculations with experiments.

In \cref{sec:gluon-numbers-dilute}, we determined that the (3+1)D dilute approximation leads to gluon production that can be expressed in $k_T$-factorized form.
In this description, transverse momentum distributions (TMDs) capture the relevant parton dynamics.
They describe the parton distribution in terms of $\xB$ and their transverse momentum $\kperp$.
Since we are working in the scope of the CGC effective theory, the small-$\xB$ regime of TMDs, which describes the gluonic contribution to the parton distribution, is of particular interest.

There are numerous accounts in the literature (e.g.,~\cite{Kharzeev:2003wz,Dominguez:2010xd,Dominguez:2011wm}) where the importance of saturation effects in transverse momentum-dependent processes is worked out in a CGC context.
It was discovered, however, that in the small-$\xB$ regime, TMDs are not universal.
Instead, different definitions emerge from the calculations of different processes.
Among these, the Weizs\"acker-Williams TMD $\xG^{(1)}$ and the dipole TMD $\xG^{(2)}$ can be readily found in the literature.
The Weizs\"acker-Williams TMD is given in the operator definition as
\begin{align}
    \xG^{(1)}_{n_\A/n_\B}(\xB,\kperp) &= \frac{4}{\langle \mathcal{P} | \mathcal{P} \rangle} \intop \frac{\dd\xpm\,\dd^2\xperp\,\dd\ypm\,\dd^2\yperp}{(2\pi)^3}\,\ee^{\ii \xB\mathcal{P}^\mp(\xpm-\ypm) - \ii \kperp\cdot(\xperp-\yperp)} \nn\\
    &\times \tr\langle \mathcal{P}| \hat{\mathcal{F}}^{\mp\bi}_{n_\A/n_\B}(x) U_\rightleftharpoons(x,y) \hat{\mathcal{F}}^{\mp\bi}_{n_\A/n_\B}(y) U^\dagger_\rightleftharpoons(x,y) |\mathcal{P} \rangle.
\end{align}
Here, the single-nucleon field strengths $\hat{\mathcal{F}}_{n_\A/n_\B}$ are operators that act on the momentum state $|\mathcal{P}\rangle$.
As usual, $n_\A$ (nucleon $\A$) moves in negative $z$ direction and $n_\B$ in positive $z$ direction.
The states are normalized as
\begin{align}
    \langle \mathcal{P} | \mathcal{P}' \rangle = (2\pi)^3 2\mathcal{P}^\mp \delta(\mathcal{P}^\mp - \mathcal{P}'^{\mp}) \deltaperp(\mathbfcal{P}-\mathbfcal{P}').
\end{align}
The Wilson lines $U_\rightleftharpoons$ ensure that $\xG^{(1)}$ does not depend on the gauge choice.
They connect the evaluation points of the field strengths along paths that are pointing to the future ($\rightharpoonup$, to $+\infty$) or to the past ($\leftharpoondown$, to $-\infty$) depending on the process.
Note that the two Wilson lines point in the same direction in the Weizs\"acker-Williams TMD.
In contrast, the Wilson lines in the definition of the dipole TMD form a loop by pointing in different directions,
\begin{align}
    \xG^{(2)}_{n_\A/n_\B}(\xB,\kperp) &= \frac{4}{\langle \mathcal{P} | \mathcal{P} \rangle} \intop \frac{\dd\xpm\,\dd^2\xperp\,\dd\ypm\,\dd^2\yperp}{(2\pi)^3}\,\ee^{\ii \xB\mathcal{P}^\mp(\xpm-\ypm) - \ii \kperp\cdot(\xperp-\yperp)} \nn\\
    &\times \tr\langle \mathcal{P}| \hat{\mathcal{F}}^{\mp\bi}_{n_\A/n_\B}(x) U_\leftharpoondown(x,y) \hat{\mathcal{F}}^{\mp\bi}_{n_\A/n_\B}(y) U^\dagger_\rightharpoonup(x,y) |\mathcal{P} \rangle.
\end{align}
The Weizs\"acker-Williams TMD is interpreted as the number density of gluons inside the nucleon.
The dipole TMD has no interpretation in the parton picture and results from a Fourier transformation of the color-dipole cross section.

\section{TMDs in the dilute approximation}\label{sec:TMDs-CGC}

The operator definition of the TMDs can be translated to classical fields in the CGC by replacing the normalized expectation value $\langle \mathcal{P}| \cdot |\mathcal{P}\rangle/\langle \mathcal{P}|\mathcal{P}\rangle$ by the averaging procedure with the CGC weight function over the nuclear color charge distributions.
We further examine the contributions of the objects inside the wedges, order-by-order in the dilute expansion.
The single-nucleon field strengths in covariant gauge contribute at linear order in their respective sources.
The Wilson lines contribute to all orders in the sources because of the exponentiated single-nucleon gauge fields.
When we expand the Wilson lines in the sources, the lowest order term is a unit matrix in color space.
Hence, to the lowest order in the dilute approximation, the Weizs\"acker-Williams and dipole TMDs are identical because their only differentiating factor from the Wilson lines is a higher-order effect.

The (leading order) TMD in the dilute approximation reads
\begin{align}
    \xG_{n_\A/n_\B}(\xB,\kperp) &= \frac{4}{(2\pi)^3} \intop\dd\xpm\,\dd\ypm\,\dd^2\xperp\,\dd^2\yperp\,\ee^{\ii \xB\mathcal{P}^\mp(\xpm-\ypm) - \ii \kperp\cdot(\xperp-\yperp)} \nn\\
    &\times \tr\langle \mathcal{F}^{\mp\bi}_{n_\A/n_\B}(\xpm,\xperp)\mathcal{F}^{\mp\bi}_{n_\A/n_\B}(\ypm,\yperp)\rangle \nn\\
    &= \frac{4}{(2\pi)^3} \tr\langle \tilde{\mathcal{F}}^{\mp\bi}_{n_\A/n_\B}(\xB\mathcal{P}^\mp,\kperp) \tilde{\mathcal{F}}^{\mp\bi}_{n_\A/n_\B}(-\xB\mathcal{P}^\mp,-\kperp) \rangle. \label{eq:tmds-xG-FF}
\end{align}
In the second line, we performed the resulting Fourier transformations of the single-nucleon field strengths.
We can re-express \cref{eq:tmds-xG-FF} in terms of the color charge distributions by inserting the single-nucleon solutions, which are analogous to \cref{eq:single-nucleus-gauge-fields,eq:single-nucleus-field-strengths}.
Although we are now considering fields sourced by the nucleonic charge distribution, mathematically, the CGC calculations work out identically to the fields of full nuclei.
Then, the Fourier transformation of the single-nucleon field is
\begin{align}
    \tilde{\mathcal{F}}^{\mp\bi}_{n_\A/n_\B}(\pmp,\pperp) &= \intop \dd\xpm\,\dd^2\xperp\,\ee^{\ii \xpm\pmp - \ii\xperp\cdot\pperp}\, (-\partial^\bi)\intop \frac{\dd^2\kperp}{(2\pi)^2} \frac{\ee^{\ii\kperp\cdot\xperp}}{\kperp^2+m^2} \tilde{\rho}_{n_\A/n_\B}(\xpm,\kperp) \nn\\
    &= \intop \frac{\dd^2\kperp}{(2\pi)^2}\frac{-\ii k^\bi}{\kperp^2+m^2} \tilde{\rho}_{n_\A/n_\B}(\pmp,\kperp) \intop \dd^2\xperp\,\ee^{\ii \xperp\cdot(\kperp-\pperp)}\nn\\
    &= \frac{-\ii p^\bi}{\pperp^2+m^2} \tilde{\rho}_{n_\A/n_\B}(\pmp,\pperp).
\end{align}
In the second line, we carried out the Fourier transformation of the light cone coordinates and let the partial derivative act on the exponential with the $\xperp$-dependence.
The remaining factors containing $\xperp$ yield a transverse delta function that identifies the $\pperp$ and $\kperp$ momenta under the last integration over $\kperp$.
Note that the overall sign depends on the sign of $\pperp$ so that \cref{eq:tmds-xG-FF} is real and positive, yielding
\begin{align}
    \xG_{n_\A/n_\B}(\xB,\kperp) = \frac{4}{(2\pi)^3}\frac{\kperp^2}{\left(\kperp^2+m^2\right)^2} \tr\langle \tilde{\rho}_{n_\A/n_\B}(\xB\mathcal{P}^\mp,\kperp) \tilde{\rho}_{n_\A/n_\B}(-\xB\mathcal{P}^\mp,-\kperp)\rangle. \label{eq:dilute-TMD}
\end{align}
This result establishes the connection of TMDs with the color charge correlator, where the latter is evaluated along the diagonal in the momentum space spanned by the two three-dimensional momentum arguments of both $\tilde{\rho}_{n_\A/n_\B}$.

In the next section and later \cref{ch:nuclear-models,ch:numerical-results}, we explore how the concept of nucleonic TMDs can be applied to the description of HICs in the (3+1)D dilute Glasma.
The conceptual difference lies in modeling heavy ions, where the fields involved describe full nuclei rather than nucleons.%
\footnote{%
In principle, the nuclear charge distributions can be assembled in a hot-spot model, similar to the distribution of nucleons in a nucleus.
Then the TMD concepts would apply to each hot spot independently.
See~\cite{Leuthner:2025vsd} for ideas on how to implement hot spots in the dilute approximation.}
Apart from that, the mathematical procedure is the same.

\section{Effective TMDs for (3+1)D gluon production}\label{sec:effTMDs-gluon}

As it turns out, there is a direct connection between the TMD in the dilute limit given in \cref{eq:dilute-TMD} and the gluon number distribution $\dd N/\dd^2\kperp\dd\Y$.
In particular, in the limit of large nuclei, it can be shown that the TMD is directly proportional to the Fourier transformation of the correlation functions $\tilde{\Gamma}_{\A/\B}$, introduced in \cref{sec:generalized-correl}, which characterizes the generalized correlations in the nuclear model and enters the gluon distribution in \cref{eq:dNd2kdY-large-nuclei-result}.
We continue with the proof next.

For later convenience, we introduce a shorthand for the color charge correlator from \cref{eq:dilute-TMD}, but using the nuclear charge distributions $\tilde{\rho}_{\A/\B}$ instead, 
\begin{align}
    \tilde{B}_{\A/\B}(\xB\mathcal{P}^\mp,\kperp) &= \frac{1}{N_c^2 -1}\langle \tilde{\rho}^a_{\A/\B}(\xB\mathcal{P}^\mp,\kperp)\tilde{\rho}^a_{\A/\B}(-\xB\mathcal{P}^\mp,-\kperp)\rangle \nn\\
    &= \frac{(2\pi)^3 \left( \kperp^2+m^2\right)^2}{2(N_c^2-1)\kperp^2} \xG_{\A/\B}(\xB,\kperp). \label{eq:tildeB-def}
\end{align}
As before, we used that the color charge correlator is diagonal in color space and $\delta^{aa} = (N_c^2-1)$ to rewrite the trace.
In the last line, we introduced the TMD for the nucleus, $\xG_{\A/\B}$, which we implicitly defined via the nuclear color charge correlator in analogy to \cref{eq:dilute-TMD}.
We can evaluate the Fourier-transformed correlator using \cref{eq:FT-general-correlator-t(x)-t(y)}.
After performing the substitution $p-\kappa = k$, the explicit expression reads
\begin{align}
    \frac{1}{N_c^2-1}\langle \tilde{\rho}^a_{\A/\B}(\pmp,\pperp)\tilde{\rho}^a_{\A/\B}(\qmp,\qperp)\rangle &\nn\\
    &\hspace{-135pt}= \intop\frac{\dd\kmp\,\dd^2\kperp}{(2\pi)^3}\,\tilde{t}_{\A/\B}(\kmp,\kperp)\tilde{t}_{\A/\B}(\qmp+\pmp-\kmp,\qperp+\pperp-\kperp) \tilde{\Gamma}_{\A/\B}(\pmp-\kmp,\pperp-\kperp).
    \label{eq:tmd-FT-correl}
\end{align}
Using this result, $\tilde{B}_{\A/\B}$ from \cref{eq:tildeB-def} can be written as
\begin{align}
    \tilde{B}_{\A/\B}(\pmp,\pperp) &= \intop\frac{\dd\kmp\,\dd^2\kperp}{(2\pi)^3}\,\tilde{t}_{\A/\B}(\kmp,\kperp)\tilde{t}_{\A/\B}(-\kmp,-\kperp) \tilde{\Gamma}_{\A/\B}(\pmp-\kmp,\pperp-\kperp). \label{eq:tildeB-approx-step1}
\end{align}

In the limit of large nuclei, we expect that the Fourier-transformed envelopes $\tilde{t}_{\A/\B}$ are sharply peaked functions.
When the scale of the correlations in the correlation function is well separated from the scale of the envelopes given by the nuclear radius,
we can evaluate the correlation function in \cref{eq:tildeB-approx-step1} at $(\kmp=0,\kperp=\zeroperp)$ and pull it in front of the integral,
\begin{align}
    \tilde{B}_{\A/\B}(\pmp,\pperp) &\approx  \tilde{\Gamma}_{\A/\B}(\pmp,\pperp) \intop\frac{\dd\kmp\,\dd^2\kperp}{(2\pi)^3}\,\tilde{t}_{\A/\B}(\kmp,\kperp)\tilde{t}_{\A/\B}(-\kmp,-\kperp). \label{eq:tildeB-scales}
\end{align}
The convolution integral is now factorized into the Fourier-transformed correlation function, which provides the entire dynamics, and a constant geometry factor that depends only on the envelopes.
The latter can be identified with the Fourier transformation of the projected transverse envelope $T^\perp_{\A/\B}$ from \cref{eq:FT-Tperp}.
Then,
\begin{align}
    \tilde{B}_{\A/\B}(\pmp,\pperp) &\approx \tilde{\Gamma}_{\A/\B}(\pmp,\pperp) \tilde{T}^\perp_{\A/\B}(\zeroperp) \nn\\
    &=  \tilde{\Gamma}_{\A/\B}(\pmp,\pperp) \intop \dd^2\xperp\, T^\perp_{\A/\B}(\xperp) \nn\\
    &=  \tilde{\Gamma}_{\A/\B}(\pmp,\pperp) \Sp^{\A/\B}. \label{eq:tildeB-Gamma-sp}
\end{align}
Here, we introduced the symbol $\Sp^{\A/\B}$ to denote the projected transverse area of nucleus $\A$ and $\B$.
Referring back to \cref{eq:tildeB-def}, we can identify that the correlation function is directly proportional to the nuclear TMD scaled by the inverse transverse area of the nuclei,
\begin{align}
    \tilde{\Gamma}_{\A/\B}(\xB\mathcal{P}^\mp,\kperp) \approx \frac{(2\pi)^3 \left( \kperp^2+m^2\right)^2}{2(N_c^2-1)\kperp^2} \frac{\xG_{\A/\B}(\xB,\kperp)}{\Sp^{\A/\B}}. \label{eq:tildeGamma-TMD}
\end{align}

We can use the result from \cref{eq:tildeGamma-TMD} to derive a particularly simple expression for the gluon distribution $\dd N/\dd^2\kperp\,\dd\Y$ in the limit of large nuclei given by \cref{eq:dNd2kdY-large-nuclei-result},
\begin{align}
    \kperp^2\frac{\dd N}{\dd^2\kperp\,\dd\Y} &\approx \frac{2g^2N_c(N_c^2-1)}{(2\pi)^3}\, \Sop \intop \frac{\dd^2\pperp}{(2\pi)^2}
    \omega_m(\kperp,\pperp,\pperp) \tilde{\Gamma}_\A(k^-,\pperp)\tilde{\Gamma}_\B(k^+,\kperp-\pperp) \nn\\
    &= \frac{2g^2N_c(N_c^2-1)}{(2\pi)^3}\, \Sop \intop \frac{\dd^2\pperp}{(2\pi)^2} \frac{\left( \delta^{\bk\bl}\delta^{\bi\bj} + \varepsilon^{\bk\bl}\varepsilon^{\bi\bj} \right)\, p^\bk(k^\bl-p^\bl)p^\bi(k^\bj-p^\bj)}{\left(\pperp^2+m^2\right)^2\left((\kperp-\pperp)^2+m^2\right)^2} \nn\\
    &\times \frac{(2\pi)^3 \left( \pperp^2+m^2\right)^2}{2(N_c^2-1)\pperp^2} \frac{\xG_\A(\xB_\A,\pperp)}{\Sp^\A} \frac{(2\pi)^3 \left( (\kperp-\pperp)^2+m^2\right)^2}{2(N_c^2-1)(\kperp-\pperp)^2} \frac{\xG_\B(\xB_\B,\kperp-\pperp)}{\Sp^\B} \nn\\
    &= \frac{(2\pi)^3 g^2 N_c}{2(N_c^2 -1)} \frac{\Sop}{\Sp^\A \Sp^\B} \left( \delta^{\bk\bl}\delta^{\bi\bj} + \varepsilon^{\bk\bl}\varepsilon^{\bi\bj} \right) \nn\\
    &\times \intop \frac{\dd^2\pperp}{(2\pi)^2} \frac{p^\bk p^\bi (k^\bl-p^\bl)(k^\bj-p^\bj)}{\pperp^2 (\kperp-\pperp)^2} \xG_\A(\xB_\A,\pperp) \xG_\B(\xB_\B,\kperp-\pperp). \label{eq:dNd2kdY-large-nucl-TMDs-vertex}
\end{align}
In the second equation, we replaced $\tilde{\Gamma}_{\A/\B}$ using \cref{eq:tildeGamma-TMD} and parametrized the rapidity dependence via the momentum fractions of each nucleus,
\begin{align}
    \xB_{\A/\B} = k^\mp/\mathcal{P}^\mp = |\kperp|\ee^{\mp\Y}/(2\gamma m_n). \label{eq:xB-Y-AB}
\end{align}
The remaining transverse vector structure reduces to unity (cf.~\eqref{eq:Omega-transverse-vector-identity} in \cref{appx:Integrated-effective-vertex}) so that the convolution over $\pperp$ only contains the two TMDs,
\begin{align}
    \kperp^2\frac{\dd N}{\dd^2\kperp\,\dd\Y} &\approx \frac{(2\pi)^3 g^2 N_c}{2(N_c^2 -1)} \frac{\Sop}{\Sp^\A \Sp^\B} \intop \frac{\dd^2\pperp}{(2\pi)^2}\, \xG_\A(\xB_\A,\pperp) \xG_\B(\xB_\B,\kperp-\pperp). \label{eq:dNd2kdY-large-nucl-TMDs}
\end{align}
This result is completely analogous to the parton picture in the $k_T$-factorized description.
The cross-section of the partonic interaction $\sim 1/\kperp^2$ and is multiplied by the transverse momentum convolution of the gluonic TMDs for each nucleus.
However, together with \cref{eq:tildeGamma-TMD}, we derived a connection between the correlation function in the MV nuclear model and TMDs in the dilute approximation.
We will use this relation in \cref{sec:CCC-bootstrapping-with-TMDs} to construct correlation functions that are matched to phenomenologically established TMDs.

For now, we will generalize \cref{eq:tildeB-def} and identify effective TMDs that enter the gluon number distribution (for arbitrary envelope scales) obtained in the dilute approximation in \cref{eq:dNd2kdY-result}.
We define the effective $\tBeff_{\A/\B}$,
\begin{align}
    \tBeff_{\A/\B}(\kmp,\pperp,\qperp) &= \frac{1}{N_c^2 -1} \langle \tilde{\rho}^a_{\A/\B}(\kmp,\pperp)\tilde{\rho}^a_{\A/\B}(-\kmp,\qperp)\rangle \nn\\
    &= \intop\frac{\dd\kappa^\mp\,\dd^2\kappaperp}{(2\pi)^3}\, \tilde{\Gamma}_{\A/\B}(\kmp-\kappa^\mp,\pperp-\kappaperp) \nn\\
    &\times \tilde{t}_{\A/\B}(\kappa^\mp,\kappaperp)\tilde{t}_{\A/\B}(-\kappa^\mp,\qperp+\pperp-\kappaperp),\label{eq:tildeB-eff-def}
\end{align}
where we allowed for two independent transverse momenta compared to $\tilde{B}_{\A/\B}$ in \cref{eq:tildeB-def}.
By definition,
\begin{align}
    \tBeff_{\A/\B}(\kmp,\kperp,-\kperp) &= \tilde{B}_{\A/\B}(\kmp,\kperp),
\end{align}
but $\tBeff_{\A/\B}$ is also directly related to $\tilde{B}_{\A/\B}$ in the limit of large nuclei where the support of \cref{eq:tildeB-eff-def} is localized along the diagonal $\pperp = - \qperp$.
To see this, we rewrite \cref{eq:tildeB-eff-def} using the parametrization
\begin{align}
    \uperp = \pperp+\qperp, \qquad \vperp = \frac{1}{2}(\pperp-\qperp)
\end{align}
and get
\begin{align}
    \tBeff_{\A/\B}(\kmp,\uperp,\vperp) &= \intop\frac{\dd\kappa^\mp\,\dd^2\kappaperp}{(2\pi)^3}\, \tilde{\Gamma}_{\A/\B}(\kmp-\kappa^\mp,\tfrac{\uperp+2\vperp}{2}-\kappaperp) \nn\\
    &\times \tilde{t}_{\A/\B}(\kappa^\mp,\kappaperp)\tilde{t}_{\A/\B}(-\kappa^\mp,\uperp-\kappaperp) \nn\\
    &\approx  \tilde{\Gamma}_{\A/\B}(\kmp,\tfrac{\uperp+2\vperp}{2}) \intop\frac{\dd\kappa^\mp\,\dd^2\kappaperp}{(2\pi)^3}\, \tilde{t}_{\A/\B}(\kappa^\mp,\kappaperp)\tilde{t}_{\A/\B}(-\kappa^\mp,\uperp-\kappaperp) \nn\\
    &= \tilde{\Gamma}_{\A/\B}(\kmp,\tfrac{\uperp+2\vperp}{2}) \tilde{T}^\perp_{\A/\B}(\uperp) \nn\\
    &= \frac{\tilde{T}^\perp_{\A/\B}(\uperp)}{\Sp^{\A/\B}} \tilde{B}_{\A/\B}(\kmp, \tfrac{\uperp+2\vperp}{2}). \label{eq:tildeBeff-sigma-tildeB}
\end{align}
As in \cref{eq:tildeB-scales} before, we were able to pull the Fourier-transformed correlation function out of the integral because the scales of the correlations and the nuclear envelopes are well separated in the limit of large nuclei.
In the third equation, we identified the Fourier transformation of the projected envelopes from \cref{eq:FT-Tperp} and in the last equation, we used \cref{eq:tildeB-Gamma-sp}.
Note that $\tilde{T}^\perp_{\A/\B}$ is highly peaked at $\uperp = \zeroperp$.

We can formulate \cref{eq:tildeBeff-sigma-tildeB} as a strict equation in the extreme limit
\begin{align}
    T^\perp_{\A/\B}(\xperp) \rightarrow 1,
\end{align}
i.e., in the limit of transversely infinite nuclei.
Because of the normalization of the envelopes in \cref{eq:envelope-norm}, the constant on the right-hand side is unity.
Then,
\begin{align}
    \lim_{T^\perp_{\A/\B}\rightarrow 1} \tilde{T}^\perp_{\A/\B}(\uperp) = (2\pi)^2 \deltaperp(\uperp).
\end{align}
We can now carefully factor out the divergent pieces in \cref{eq:tildeBeff-sigma-tildeB},
\begin{align}
    \lim_{T^\perp_{\A/\B}\rightarrow 1} \tBeff_{\A/\B}(\kmp,\uperp=\pperp+\qperp,\vperp=\tfrac{\pperp-\qperp}{2}) &= \frac{(2\pi)^2 \deltaperp(\uperp)}{\Sp^{\A/\B}} \tilde{B}_{\A/\B}(\kmp, \tfrac{\uperp+2\vperp}{2}) \nn\\
    &= \frac{(2\pi)^2 \deltaperp(\pperp+\qperp)}{\Sp^{\A/\B}} \tilde{B}_{\A/\B}(\kmp, \tfrac{\pperp-\qperp}{2}), \label{eq:tildeBeff-Tperp-inf}
\end{align}
where $\tilde{B}_{\A/\B} / \Sp^{\A/\B}$ is finite even though the transverse area is formally infinite.
Due to the Dirac-delta function, \cref{eq:tildeBeff-Tperp-inf} has to be treated as a distribution and can only be interpreted under an integral over a transverse momentum argument (e.g., for $\dd N/\dd^2\kperp\dd\Y$).

Having established the properties of $\tBeff_{\A/\B}$, we define the effective TMD as
\begin{align}
    \frac{\xGeff_{\A/\B}(\xB,\pperp,\qperp)}{\Sp^{\A/\B}} \coloneqq \frac{1}{(2\pi)^2} \frac{2(N_c^2-1)}{(2\pi)^3} \frac{\pperp^2\qperp^2}{(\pperp^2+m^2)(\qperp^2+m^2)} \tBeff_{\A/\B}(\xB\mathcal{P}^\mp,\pperp,\qperp). \label{eq:xGeff-xpq-def}
\end{align}
Using this definition in the expression of the gluon number distribution in \cref{eq:dNd2kdY-result} yields a concise result reminiscent of \cref{eq:dNd2kdY-large-nucl-TMDs-vertex},
\begin{align}
    \kperp^2\frac{\dd N}{\dd^2\kperp\,\dd\Y} &= \frac{(2\pi)^3g^2N_c}{2(N_c^2-1)} \intop \dd^2\pperp\,\dd^2\qperp \nn\\
    &\times \frac{\left( \delta^{\bk\bl}\delta^{\bi\bj} + \varepsilon^{\bk\bl}\varepsilon^{\bi\bj} \right)\, p^\bk(k^\bl-p^\bl)q^\bi(k^\bj-q^\bj)}{\pperp^2\qperp^2(\kperp-\pperp)^2(\kperp-\qperp)^2} \nn\\
    &\times \frac{\xGeff_\A(\xB_\A,\pperp,-\qperp)}{\Sp^\A} \frac{\xGeff_\B(\xB_\B,\kperp-\pperp,-\kperp+\qperp)}{\Sp^\B}. \label{eq:dNd2kdY-TMDs}
\end{align}
Here, we used the same parametrization via $\xB_{\A/\B}$ given in \cref{eq:xB-Y-AB}.

Compared to \cref{eq:tildeB-def}, the definition of the effective TMD in \cref{eq:xGeff-xpq-def} involves two transverse momenta and contains an additional factor $\Sp^{\A/\B}$.
This factor comes naturally, even in the case of large nuclei, where it is not $\xG_{\A/\B}$ that enters the gluon distribution in \cref{eq:dNd2kdY-large-nucl-TMDs} but $\tilde{\Gamma}_{\A/\B} \sim \tilde{B}_{\A/\B}/ \Sp^{\A/\B} \sim \xG_{\A/\B} / \Sp^{\A/\B}$.
Also, there is an additional factor of $(2\pi)^2$ as a choice of normalization.
It allows the formulation of a consistent limit $T^\perp_{\A/\B}\rightarrow1$ where the effective TMD reproduces the standard $\xG_{\A/\B}$ under the transverse momentum integrations within $\dd N/\dd^2\kperp\dd\Y$.
It is also helpful to define the reduced effective TMD, where we integrate out one transverse momentum,
\begin{align}
    \kperp^2 \xGeff_{\A/\B}(\xB,\kperp) &= \intop \dd^2\qperp\, \xGeff_{\A/\B}(\xB,\kperp,\qperp)\nn\\
    &= \Sp^{\A/\B} \frac{2(N_c^2-1)}{(2\pi)^3} \intop\frac{\dd^2\qperp}{(2\pi)^2} \frac{\kperp^2\qperp^2\, \tBeff_{\A/\B}(\xB\mathcal{P}^\mp,\kperp,\qperp)}{(\kperp^2+m^2)(\qperp^2+m^2)}. \label{eq:xGeff-reduced-def}
\end{align}
On the one hand, this allows studying the effective TMD in a reduced parameter space, where two dimensions are integrated out.
On the other hand, we can perform the limit $T^\perp_{\A/\B}\rightarrow1$,
\begin{align}
    \lim_{T^\perp_{\A/\B}\rightarrow1} \kperp^2 \xGeff_{\A/\B}(\xB,\kperp) &= \Sp^{\A/\B} \frac{2(N_c^2-1)}{(2\pi)^3} \intop\frac{\dd^2\qperp}{(2\pi)^2} \frac{\kperp^2\qperp^2}{(\kperp^2+m^2)(\qperp^2+m^2)} \nn\\
    &\times \frac{(2\pi)^2 \deltaperp(\kperp+\qperp)}{\Sp^{\A/\B}} \tilde{B}_{\A/\B}(\xB\mathcal{P}^\mp, \tfrac{\kperp-\qperp}{2}) \nn\\
    &= \frac{2(N_c^2-1)}{(2\pi)^3} \frac{\kperp^4}{\left(\kperp^2+m^2\right)^2}\tilde{B}_{\A/\B}(\xB\mathcal{P}^\mp, \kperp) \nn\\
    &= \kperp^2 \xG_{\A/\B}(\xB,\kperp),
\end{align}
where we used \cref{eq:tildeBeff-Tperp-inf} to replace $\tBeff_{\A/\B}$ and recovered the standard TMD.
We will study $\xGeff_{\A/\B}$ for different nuclear models in \cref{sec:eff-TMDs}.

On a final note, we reflect on the insights gained from formulating gluon production in the dilute approximation in terms of TMDs.
In the limit of large nuclei, we can provide additional context to the interpretation given in the discussion of \cref{eq:dNd2kdY-large-nuclei-result}.
Indeed, this limit reduces to the simple picture of partonic scattering, where the TMDs encode the probabilities to find a parton with a certain momentum.
Gluon production in this picture can be calculated with the $k_T$-factorized formula, which yields adequate results when the momenta of the produced gluons are moderate to large.
The result from the partonic scattering is then scaled up by the transverse overlap area $\Sop$ to account for the large number of (independent) scatterings that occur during the collision of two large nuclei.
This corresponds to the incoherent addition of single-scatterings, where the produced gluons do not interact further.

The clean extraction of the transverse overlap area $\Sop$ does not work in the general case, where nuclear correlation and envelope scales are not separated.
We were able to define effective TMDs $\xGeff_{\A/\B}$ and write gluon production in a similar way as $k_T$ factorization in \cref{eq:dNd2kdY-TMDs}, but the convolution involves two transverse momenta and a more complex vertex factor.
Still, \cref{eq:dNd2kdY-TMDs} does scale with the transverse overlap area.
We conclude that the definition of the effective TMD in \cref{eq:xGeff-xpq-def} contains information about the size of the nucleus, even though the transverse area $\Sp^{\A/\B}$ is divided out on the left-hand side.
This should not be surprising, because also in the limit of large nuclei, $\tilde{B}_{\A/\B} \sim \Sp^{\A/\B}$ and the transverse area enters the TMD $\xG_{\A/\B}$, as evident from \cref{eq:tildeB-Gamma-sp}.

\chapter{Limiting fragmentation}\label{ch:limiting-fragmentation}

The "hypothesis of limiting fragmentation" dates back to 1969~\cite{Feynman:1969ej,Benecke:1969sh,Chou:1970bj}, where it was argued that there exists a universal kinematic regime where the outgoing particles in ultra-relativistic collisions approach a limiting distribution.
This distribution is limiting in the sense that higher-energy collisions would be limited to the same distribution as lower-energy ones.
Take, for example, the rapidity distributions of observables (e.g., charged particle distributions).
Then, the kinematic regime of interest will be the very forward and backward rapidities that correspond to the fragmentation regions of the collision partners.

The first experimental reports that confirmed the existence of a scaling region for collisions involving heavy ions were published by the experimental collaborations at RHIC~\cite{BRAHMS:2001llo,BRAHMS:2001gci,PHOBOS:2001zjw,Back:2002wb,PHOBOS:2004zne,STAR:2005ips}.
They measured rapidity distributions of the same observable $\mathcal{O}_{[Y]}$ at different collider energies.
We label the results for a particular energy with the beam rapidity $Y = \arcosh(\gamma)$, which can be expressed by the Lorentz $\gamma$ factor.
When shifting these rapidity distributions by their respective beam rapidities, the data points collapse to a universal curve.
The shift by $Y$ corresponds to a boost into the rest frame of one of the incoming beams (depending on the sign of the shift).
The observation of a limiting distribution in this frame of reference confirmed that the physical process of fragmentation is universal.

Competing experiments at the highest available energies at the LHC, unfortunately, lack the necessary rapidity coverage in the detectors to confirm or rule out limiting fragmentation.
Several efforts have been made to extrapolate experimental data to extreme rapidities (e.g.,~\cite{Sahoo:2018osl,Kellers:2019ezz,Basu:2020jbk}), but no definitive conclusion has been reached yet.

The general formulation of limiting fragmentation for an observable $\mathcal{O}_{[Y]}$ as a function of spacetime rapidity%
\footnote{%
Instead of $\eta_s$, any other rapidity measure, such as pseudo-rapidity or momentum rapidity, is also possible.
We will focus on $\eta_s$ dependent observables in the following.}
reads,
\begin{align}
    \mathcal{O}_{[Y_1]}(\eta_s - Y_1) \Big|_{\eta_s \gg 1} = \mathcal{O}_{[Y_2]}(\eta_s-Y_2)\Big|_{\eta_s\gg 1}, \label{eq:limiting-fragmentation-def}
\end{align}
where $Y_{1/2}$ denote the different beam rapidities that correspond to the different energies.
This equation states that in the fragmentation region, where $\eta_s \gg 1$, the shape of the profile is independent of the collider energy.
In the case of large, negative $-\eta_s \gg 1$, the situation is analogous but with the sign of the shift by $Y_\mathrm{beam}$ reversed.

Limiting fragmentation can also be a powerful factor to discriminate between different models for the initial conditions~\cite{Goncalves:2019uod,Nasim:2011ss}.
Standard parton-shower-based event generators seem to fail to describe the fragmentation region~\cite{ALICE:2014rma,CMS:2018lqt,CMS:2019gzk}.
The importance of saturation effects for the theoretical description was soon established by calculations within the CGC effective theory~\cite{Jalilian-Marian:2002yhb,Gelis:2006tb,Kharzeev:2001gp,Kharzeev:2001yq,Kharzeev:2004if}.
Here, the so-called "black disk" limit emerged as the underlying picture.
In the rest frame of one of the collision partners, the other, incoming projectile appears completely saturated with gluons (i.e., "black") so that every struck nucleon interacts with the dense sheet of colored glass.
The predictions for inclusive multiplicity distributions can then be calculated in the $k_T$-factorized approach.

In this chapter, we build upon the discussion initiated in~\cite{Ipp:2025sbc,Leuthner:2025vsd} and provide a rigorous analysis of limiting fragmentation in the (3+1)D dilute Glasma.
First, we derive the property of (coordinate-)local limiting fragmentation for the position space solution presented in \cref{sec:position-space-picture}.
Then we sketch how limiting fragmentation can arise in the momentum space picture discussed in \cref{ch:momentum-space-picture}.
The analytic studies in this chapter provide the foundation for the detailed analysis of the numerical results in \cref{sec:results-limiting-fragmentation}.

\section{Field strength tensor in position space}\label{sec:lim-fag-pos-space}

In this section, we examine the position-space solutions of the field strength tensor $f$ given in \cref{eq:dilute-f+--sol,eq:dilute-f+i-sol,eq:dilute-f-i-sol,eq:dilute-fij-sol} for the fragmentation region of nucleus $\B$ where $\eta_s \gg 1$.
It is sufficient to focus on the integration over $\eta'$ and we fix the value of $\vperp$ to a nonzero constant in the following.
Dropping constant prefactors, we introduce a shorthand notation for the integrand to write the $\eta'$ integral as
\begin{align}
    \mathcal{I}_{\eta'} = \intop \dd\eta'\, h^{\bk\bl}(v^+,v^-,\vperp) \left[ \mathcal{F}^{-\bk}_\A(u^+,\uperp), \mathcal{F}^{+\bl}_\B(u^-,\uperp) \right].
\end{align}
Here, we used the shifted coordinate
\begin{align}
    u^\mu = x^\mu - v^\mu,
\end{align}
and the function $h$ contains all additional factors of $v$ that differ for the different components,
\begin{align}
    h^{\bk\bl}(v) = \begin{cases}
        \delta^{\bk\bl}, & f^{+-} \\
        (\delta^{\bi\bj}\delta^{\bk\bl} - \epsilon^{\bi\bj}\epsilon^{\bk\bl})\, v^\bj/(2v^-), & f^{+\bi} \\
        (-\delta^{\bi\bj}\delta^{\bk\bl}-\epsilon^{\bi\bj}\epsilon^{\bk\bl})\, v^\bj/(2v^+), & f^{-\bi} \\
        \epsilon^{\bi\bj}\epsilon^{\bk\bl}, & f^{\bi\bj}
    \end{cases}.
\end{align}
Note that we suppressed the unpaired indices of the full vector structure of $h$ that match the respective components of $f$.

As a first step, we introduce the Milne coordinates $(\tau, \eta_s)$ for the evaluation point $(x^+,x^-)$ and use the parametrization from \cref{eq:dilute-v-param} for the lightlike displacement vector $v$.
In changing to the Milne coordinates, we restrict the evaluation points $\xpm$ to be inside the future light cone of the collision, and further offset the origin along the $t$-axis to prevent evaluation in the spacetime region where the dilute (3+1)D results are inapplicable (cf.\ \cref{sec:Milne-frame-shift}).
The light cone components $u^\pm$ can now be written as
\begin{align}
    u^\pm &= \xpm - v^\pm \nn \\
    &= \frac{1}{\sqrt{2}}\tau\ee^{\pm\eta_s} - \frac{1}{\sqrt{2}}|\vperp|\ee^{\pm\eta'} \nn\\
    &= \frac{1}{\sqrt{2}}\ee^{\pm\eta_s}(\tau - |\vperp|\ee^{\pm\eta'\mp\eta_s}),
\end{align}
leading to
\begin{align}
        \mathcal{I}_{\eta'} &= \intop \dd\eta'\, h^{\bk\bl}(\tfrac{1}{\sqrt{2}}|\vperp|\ee^{+\eta'},\tfrac{1}{\sqrt{2}}|\vperp|\ee^{-\eta'},\vperp) \nn \\
        &\times \left[ \mathcal{F}^{-\bk}_\A(\tfrac{1}{\sqrt{2}}\ee^{+\eta_s}(\tau - |\vperp|\ee^{+\eta'-\eta_s}),\uperp), \mathcal{F}^{+\bl}_\B(\tfrac{1}{\sqrt{2}}\ee^{-\eta_s}(\tau - |\vperp|\ee^{-\eta'+\eta_s}),\uperp) \right].
\end{align}
Next, we shift the integration by $\eta_s$ by substituting $\eta'' = \eta'-\eta_s$,
\begin{align}
        \mathcal{I}_{\eta''} &= \intop \dd\eta''\, h^{\bk\bl}(\tfrac{1}{\sqrt{2}}|\vperp|\ee^{+\eta''+\eta_s},\tfrac{1}{\sqrt{2}}|\vperp|\ee^{-\eta''-\eta_s},\vperp) \nn \\
        &\times \left[ \mathcal{F}^{-\bk}_\A(\tfrac{1}{\sqrt{2}}\ee^{+\eta_s}(\tau - |\vperp|\ee^{+\eta''}),\uperp), \mathcal{F}^{+\bl}_\B(\tfrac{1}{\sqrt{2}}\ee^{-\eta_s}(\tau - |\vperp|\ee^{-\eta''}),\uperp) \right]. \label{eq:lf-eta''-shift-integral}
\end{align}
While the bounds on $\eta''$ are, in principle, $(-\infty, \infty)$, in practice, the integration is limited by the support of the integrand.
Along the light cone coordinates, the single-nucleus field strengths $\mathcal{F}_{\A/\B}$ directly inherit their structure from the color charge distributions $\rho_{\A/\B}$ (cf.\ \cref{sec:cgc}).
Since we assume compact support for the nuclear envelopes, the arguments $u^\pm$ of $\mathcal{F}_{\A/\B}$ are limited to
\begin{align}
    u^\pm_\mathrm{min} < u^\pm < u^\pm_\mathrm{max}, \label{eq:lf-upm-lims}
\end{align}
where due to the choice of origin, $u^\pm_\mathrm{min/max} < 0$, i.e., both limits are negative, real numbers.
Rewriting the $u^\pm$ coordinates in the Milne parametrization, the bounds imposed by the nuclei $\A$ and $\B$ are
\begin{align}
    \A: \quad & u^+_\mathrm{min} < \frac{1}{\sqrt{2}}\ee^{+\eta_s}(\tau - |\vperp|\ee^{+\eta''}) < u^+_\mathrm{max}, \\
    \B: \quad & u^-_\mathrm{min} < \frac{1}{\sqrt{2}}\ee^{-\eta_s}(\tau - |\vperp|\ee^{-\eta''}) < u^-_\mathrm{max},
\end{align}
and can be rearranged to solve for $\eta''$,
\begin{align}
    \A: \quad & \ln(\tfrac{\tau - \sqrt{2}u^+_\mathrm{min}\ee^{-\eta_s}}{|\vperp|}) > \eta'' > \ln(\tfrac{\tau - \sqrt{2}u^+_\mathrm{max}\ee^{-\eta_s}}{|\vperp|}), \label{eq:lf-nucl-A-eta''-lims}\\
    \B: \quad & \ln(\tfrac{\tau - \sqrt{2}u^-_\mathrm{min}\ee^{+\eta_s}}{|\vperp|}) > \eta'' > \ln(\tfrac{\tau - \sqrt{2}u^-_\mathrm{max}\ee^{+\eta_s}}{|\vperp|}). \label{eq:lf-nucl-B-eta''-lims}
\end{align}

\begin{figure}
    \centering
    \includegraphics{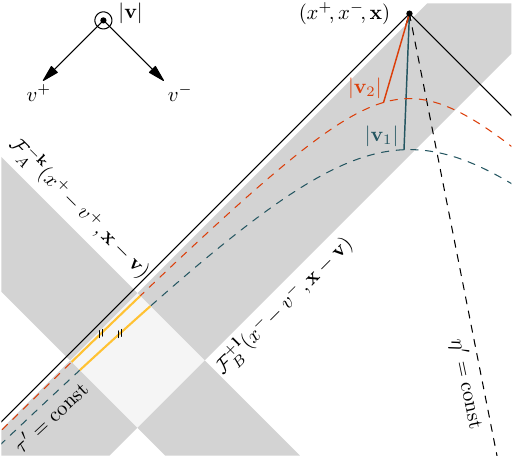}
    \caption[Parallel integration paths in the limiting fragmentation approximation.]{\label{fig:backward-lightcone-lf}
    Analog to \cref{fig:fmunu-backwards-milne-integral}, where now the evaluation point $x = (x^+,x^-,\xperp)$ is placed at large, positive $\eta_s = \ln(x^+/x^-)/2 \gg 1$.
    The segments highlighted in yellow correspond to different values of $|\vperp_{1/2}| = \tau'$ (orange and blue), but are approximately parallel to each other and to the light cone boundary.
    For each possible $\tau' = |\vperp|$, the longitudinal structure of $\mathcal{F}_\B$ is probed with a single value of $v^-=x^-\vperp^2/\tau^2$, whereas for $\mathcal{F}_\A$ the $v^+$-dependence can be integrated out.
    Figure adapted from~\cite{Ipp:2025sbc}.}
\end{figure}

In the fragmentation region of nucleus $\B$, $\eta_s \gg 1$, so that we can exploit that $\ee^{-\eta_s}\ll1$ and approximate the bounds in \cref{eq:lf-nucl-A-eta''-lims} set by nucleus $\A$.
In particular, we demand
\begin{align}
    \tau \gg |\sqrt{2}u^+_\mathrm{min} \ee^{-\eta_s}|, \qquad \tau \gg |\sqrt{2}u^+_\mathrm{max}\ee^{-\eta_s}|, \label{eq:lf-tau-restriction}
\end{align}
which is equivalent to
\begin{align}
    x^+ \gg |u^+_\mathrm{min}| > |u^+_\mathrm{max}|. \label{eq:lf-xpm-restriction}
\end{align}
We emphasize that \cref{eq:lf-tau-restriction} puts constraints on the value of the proper time $\tau$ in addition to $\eta_s$.
Therefore, it is convenient to carry out the calculations in terms of the relation in \cref{eq:lf-xpm-restriction}.
Paraphrasing the equations, we place the evaluation point $(x^+,x^-)$ far away from the track of nucleus $\A$ such that \cref{eq:lf-xpm-restriction} holds.
Simultaneously, the $x^- = \tau \ee^{-\eta_s}/\sqrt{2}$ coordinate is close to the track of nucleus $\B$ since $\eta_s\gg1$.
This geometric arrangement of $x$ and the nuclear tracks is illustrated in \cref{fig:backward-lightcone-lf}.
Note that the required values of $\tau$ are still small, even for moderately large $\eta_s$, because the longitudinal extents of the nuclei are Lorentz contracted $u^+_\mathrm{max} - u^+_\mathrm{min} \sim R/\gamma \approx 10^{-3} \div 10^{-2}$ fm.

Using \cref{eq:lf-tau-restriction}, the bounds on $\eta''$ set by nucleus $\A$ in \cref{eq:lf-nucl-A-eta''-lims} collapse to approximately the same value,
\begin{align}
    \ln(\tau/|\vperp|) \gtrsim \eta'' \gtrsim \ln(\tau/|\vperp|).
\end{align}
Hence, the integration over $\eta''$ is limited to a very narrow interval around
\begin{align}
    \bar{\eta} = \ln(\tau/|\vperp|), \label{eq:lf-eta-bar-def}
\end{align}
which still covers the entire longitudinal support of nucleus $\A$ because of the exponential enhancement of the light cone argument of $\mathcal{F}_\A$ in \cref{eq:lf-eta''-shift-integral}.
On the contrary, the light cone argument of $\mathcal{F}_\B$ is strongly suppressed by $\ee^{-\eta_s}$ and effectively only evaluated at the single value $\eta''=\bar{\eta}$.
We can check if this value is still carried by the longitudinal support of nucleus $\B$ by replacing $\eta''$ in \cref{eq:lf-nucl-B-eta''-lims} by $\bar{\eta}$ from \cref{eq:lf-eta-bar-def}.
This yields
\begin{align}
    \B: \quad -u^-_\mathrm{min} > \frac{\vperp^2 - \tau^2}{\sqrt{2}\tau\ee^{\eta_s}} > -u^-_\mathrm{max}.
\end{align}
Since both $u^-_\mathrm{min}$ and $u^-_\mathrm{max}$ are negative, $|\vperp|$ has to be sufficiently larger than $\tau$ to satisfy the inequalities.
For increasing $\tau$ and/or $\eta_s$, $|\vperp|$ is pushed to larger values as well.
Eventually, the transverse arguments in \cref{eq:lf-eta''-shift-integral} will exceed the transverse support of the nuclei via $\uperp = \xperp-\vperp$, at which point $f$ will fall off to zero.

Having established the weak dependence of $\mathcal{F}_\B$ on $\eta''$, we can expand this single-nucleus field strength in a Taylor series around $\eta''=\bar{\eta}$,
\begin{align}
    \mathcal{F}^{+\bl}_\B(\tfrac{1}{\sqrt{2}}\ee^{-\eta_s}(\tau - |\vperp|\ee^{-\eta''}),\uperp) &= \mathcal{F}^{+\bl}_\B(\tfrac{1}{\sqrt{2}}\ee^{-\eta_s}(\tau - |\vperp|\ee^{-\bar{\eta}}),\uperp) + O(\eta''-\bar{\eta}) \nn\\
    &\approx \mathcal{F}^{+\bl}_\B(x^-(1-\tfrac{\vperp^2}{\tau^2}),\uperp). \label{eq:lf-taylor-FB}
\end{align}
Here, we truncated the series after the term constant in $\eta''$.
As long as the longitudinal correlation scale within nucleus $\B$ is large compared to the $u^-$ interval given by the limits on $\eta''$ in \cref{eq:lf-nucl-A-eta''-lims}, $\mathcal{F}_\B$ will remain constant under the $\eta''$ integral.
In this case, the truncation will introduce minimal error.

We can now factorize the integral in \cref{eq:lf-eta''-shift-integral}, by pulling the contribution of nucleus $\B$ out of the integration,
\begin{align}
    \mathcal{I}_{\eta''} = \left[ \mathcal{H}^{-\bl}_\A(x^+,\vperp,\xperp), \mathcal{F}^{+\bl}_\B(x^-(1-\tfrac{\vperp^2}{\tau^2}),\uperp) \right],
\end{align}
where we defined the shorthand field
\begin{align}
    \mathcal{H}^{-\bl}_\A(x^+,\vperp,\xperp) &= \intop\dd\eta''\, h^{\bk\bl}(\tfrac{|\vperp|}{\sqrt{2}}\ee^{\eta''+\eta_s},\tfrac{|\vperp|}{\sqrt{2}}\ee^{-\eta''-\eta_s},\vperp) \mathcal{F}^{-\bk}_\A(\tfrac{1}{\sqrt{2}}\ee^{\eta_s}(\tau-|\vperp|\ee^{\eta''}),\uperp),
\end{align}
\pagebreak

\vspace*{2\baselineskip}
\noindent
which contains the integral with contributions only from nucleus $\A$.
We further simplify the expression with the substitution
\begin{align}
    u^+ = \frac{1}{\sqrt{2}}\ee^{\eta_s}(\tau-|\vperp|\ee^{\eta''}) = x^+ - v^+, \qquad \dd\eta'' = \frac{1}{u^+-x^-}\dd u^+
\end{align}
and get
\begin{align}
    \mathcal{H}^{-\bl}_\A(x^+,\vperp,\xperp) &= \intop_{x^+}^{-\infty}\dd u^+\, \frac{1}{u^+-x^+} h^{\bk\bl}(x^+-u^+,\tfrac{\vperp^2}{2}\tfrac{1}{x^+-u^+},\vperp)\mathcal{F}^{-\bk}_\A(u^+,\uperp) \nn\\
    &= \intop_{u^+_\mathrm{min}}^{u^+_\mathrm{max}}\dd u^+\, \frac{1}{x^+-u^+} h^{\bk\bl}(x^+-u^+,\tfrac{\vperp^2}{2}\tfrac{1}{x^+-u^+},\vperp)\mathcal{F}^{-\bk}_\A(u^+,\uperp).
\end{align}
In the second line, we flipped the integration bounds and limited the domain to the interval in \cref{eq:lf-upm-lims} that identifies the longitudinal support of $\mathcal{F}_\A$.
Since the coordinate $x^+$ is located outside the track of nucleus $\A$, we were able to replace the bound given by $x^+$ by $u^+_\mathrm{max}$.
Note that the same observation ensures that the denominator stays finite because $x^+ > u^+$.

\pagebreak
Assembling all terms and inserting back into the components of $f$ given in \cref{eq:dilute-f+--sol,eq:dilute-f+i-sol,eq:dilute-f-i-sol,eq:dilute-fij-sol} yields
\begin{align}
    \hspace{-3pt} f^{+-}(x)\Big|_{\eta_s\gg1} &\approx \frac{\ii g}{2\pi} \intop \dd^2\vperp\, \delta^{\bk\bl} \nn\\
    &\times \Big[\intop^{u^+_\mathrm{max}}_{u^+_\mathrm{min}}\dd u^+\, \frac{1}{x^+-u^+}
    \mathcal{F}^{-\bk}_\A(u^+,\xperp-\vperp),
    \mathcal{F}^{+\bl}_\B(x^-(1-\tfrac{\vperp^2}{\tau^2}),\xperp-\vperp) \Big] , \label{eq:lf-f+--large-etas} \\
    \hspace{-3pt} f^{+\bi}(x)\Big|_{\eta_s\gg1} &\approx \frac{\ii g}{2\pi} \intop \dd^2\vperp\, (\delta^{\bi\bj}\delta^{\bk\bl}-\epsilon^{\bi\bj}\epsilon^{\bk\bl})\nn\\
    &\times \Big[\intop^{u^+_\mathrm{max}}_{u^+_\mathrm{min}}\dd u^+\, \frac{v^\bj}{\vperp^2}
    \mathcal{F}^{-\bk}_\A(u^+,\xperp-\vperp),
    \mathcal{F}^{+\bl}_\B(x^-(1-\tfrac{\vperp^2}{\tau^2}),\xperp-\vperp) \Big] , \label{eq:lf-f+i-large-etas} \\
    \hspace{-3pt} f^{-\bi}(x)\Big|_{\eta_s\gg1} &\approx \frac{\ii g}{2\pi} \intop \dd^2\vperp\, (-\delta^{\bi\bj}\delta^{\bk\bl}-\epsilon^{\bi\bj}\epsilon^{\bk\bl})\nn\\
    &\times \Big[\intop^{u^+_\mathrm{max}}_{u^+_\mathrm{min}}\dd u^+\, \frac{v^\bj}{2(x^+-u^+)^2}
    \mathcal{F}^{-\bk}_\A(u^+,\xperp-\vperp),
    \mathcal{F}^{+\bl}_\B(x^-(1-\tfrac{\vperp^2}{\tau^2}),\xperp-\vperp) \Big] , \label{eq:lf-f-i-large-etas} \\
    \hspace{-3pt} f^{\bi\bj}(x)\Big|_{\eta_s\gg1} &\approx \frac{\ii g}{2\pi} \intop \dd^2\vperp\, \epsilon^{\bi\bj}\epsilon^{\bk\bl}\nn\\
    &\times \Big[\intop^{u^+_\mathrm{max}}_{u^+_\mathrm{min}}\dd u^+\, \frac{1}{x^+-u^+}
    \mathcal{F}^{-\bk}_\A(u^+,\xperp-\vperp),
    \mathcal{F}^{+\bl}_\B(x^-(1-\tfrac{\vperp^2}{\tau^2}),\xperp-\vperp) \Big]. \label{eq:lf-fij-large-etas}
\end{align}
Note that the perceived divergence of the factor $v^\bj/\vperp^2$ for $|\vperp| \rightarrow 0$ in \cref{eq:lf-f+i-large-etas} is a coordinate artifact that can be removed when changing to polar coordinates in the transverse $\vperp$-plane.

The following physical picture emerges from this result and is shown in \cref{fig:backward-lightcone-lf}.
For each fixed value $|\vperp| = \tau'$, the integration over $\eta'$ reduces to the yellow-highlighted parts of the hyperbolas, which intersect with the collision region (light gray).
The yellow lines are (almost) parallel to each other and to the $x^-=0$ boundary of the backward light cone that is attached to the evaluation point $x$.
Therefore, these lines cut through the single-nucleus field strength of nucleus $\B$ at a fixed value for the longitudinal argument given by $u^- = x^-(1-\vperp^2/\tau^2)$.
In contrast, the longitudinal structure of nucleus $\A$ is integrated out along the yellow lines, where, depending on the component of $f$, different weighting factors appear.
All of these weights, except for $f^{+\bi}$ in \cref{eq:lf-f+i-large-etas}, scale as a power of $1/x^+ \sim \ee^{-\eta_s}$ and lead to strong suppression of the corresponding components of the dilute Glasma field strength tensor.
The suppression factors are weaker when $u^+$ approaches $x^+$.
This resembles a screening effect, where contributions from parts farther from the evaluation point are shadowed by contributions from parts closer to the evaluation point.
We further study the contribution of $\mathcal{F}_\A$ in a far-field approximation in the next section.

\subsection{Far-field approximation}

Recall \cref{eq:lf-xpm-restriction}, where we placed the evaluation point $x^+$ far away from the track of nucleus $\A$. 
Under this assumption, the weighting factors that appear in \cref{eq:lf-f+--large-etas,eq:lf-f-i-large-etas,eq:lf-fij-large-etas} can be approximated as
\begin{align}
    \frac{1}{(x^+-u^+)^n} \approx \frac{1}{(x^+)^n}, \qquad n \in \{1, 2\},
\end{align}
so that we can pull the remaining powers of $x^+$ out of the $u^+$ integrals.
As a result, all integrations over the longitudinal structure of nucleus $\A$ in \cref{eq:lf-f+--large-etas,eq:lf-f+i-large-etas,eq:lf-f-i-large-etas,eq:lf-fij-large-etas} reduce to the same expression
\begin{align}
    \mathcal{F}^{-\bk}_{\perp,\A}(\uperp) = \intop \dd u^+ \mathcal{F}^{-\bk}_\A(u^+,\uperp), \label{eq:lf-F-proj}
\end{align}
evaluated at $\uperp=\xperp-\vperp$.
Note that we dropped the explicit integration bounds and formally allow $u^+\in(-\infty,\infty)$.
This is possible because the pole at $x^+ = u^+$ does not appear in the far-field.
Still, $u^+$ will be limited by the longitudinal support of the color charge density of the nucleus $\A$.

The single-nucleus field strength in \cref{eq:lf-F-proj} corresponds to the projected, transverse field strength associated with the two-dimensional, transverse color charge distribution $\rho^\perp_\A$ that appears in the boost-invariant MV model (cf.\ \cref{sec:MV-nucl-model}).
To see this, we re-express $\mathcal{F}_\A$ in terms of $\rho_A$ using \cref{eq:single-nucleus-gauge-fields,eq:single-nucleus-field-strengths},
\begin{align}
    \intop \dd u^+ \mathcal{F}^{-\bk}_\A(u^+,\uperp) &= \intop \dd u^+\, (-1) \partial^\bk \mathcal{A}^-_\A(u^+,\uperp) \nn\\
    &= -\partial^\bk \intop\frac{\dd u^+\,\dd^2\kperp}{(2\pi)^2}\frac{\ee^{\ii \kperp\cdot\uperp}}{\kperp^2 + m^2} \tilde{\rho}_\A(u^+,\kperp) \nn\\
    &= -\partial^\bk \intop\frac{\dd^2\kperp}{(2\pi)^2}\frac{\ee^{\ii \kperp\cdot\uperp}}{\kperp^2 + m^2} \tilde{\rho}^\perp_\A(\kperp) \nn\\
    &= - \partial^\bk\mathcal{A}^-_{\perp,\A}(\uperp) = \mathcal{F}^{-\bk}_{\perp,\A}(\uperp).
\end{align}
In the third line, we exchanged the order of the integrals and used the definition of the two-dimensional charge density, $\rho^\perp_\A(\xperp) = \intop \dd x^+ \rho_\A(x^+,\xperp)$.
In the last line, we identified the result with the projected, transverse single-nucleus gauge field $\mathcal{A}_{\perp,\A}$ and field strength.
Note that the partial derivatives are acting only on the transverse arguments of the fields.

The (3+1)D dilute Glasma field strength tensor in the far-field approximation for $\eta_s\gg 1$ reads
\begin{align}
    f^{+-}_\mathrm{LF}(x) &= -\frac{g}{2\pi}\frac{1}{x^+} \intop \dd^2\vperp\, V^{\bk\bl}_\perp(x,\vperp) \delta^{\bk\bl}, \label{eq:lf-f+-} \\
    f^{+\bi}_\mathrm{LF}(x) &= -\frac{g}{2\pi} \intop \dd^2\vperp\, \frac{v^\bj}{\vperp^2} V^{\bk\bl}_\perp(x,\vperp) (\delta^{\bi\bj}\delta^{\bk\bl} - \epsilon^{\bi\bj}\epsilon^{\bk\bl}), \label{eq:lf-f+i} \\
    f^{-\bi}_\mathrm{LF}(x) &= -\frac{g}{2\pi} \frac{1}{2(x^+)^2} \intop \dd^2\vperp\, v^\bj V^{\bk\bl}_\perp(x,\vperp) (-\delta^{\bi\bj}\delta^{\bk\bl} - \epsilon^{\bi\bj}\epsilon^{\bk\bl}), \label{eq:lf-f-i} \\
    f^{\bi\bj}_\mathrm{LF}(x) &= -\frac{g}{2\pi} \frac{1}{x^+} \intop \dd^2\vperp\, V^{\bk\bl}_\perp(x,\vperp) \epsilon^{\bi\bj}\epsilon^{\bk\bl}. \label{eq:lf-fij}
\end{align}
Here, we introduced the shorthand field $V^{\bk\bl}_\perp$ akin to \cref{eq:dilute-Vij-integrand}
\begin{align}
    V^{\bk\bl}_\perp(x,\vperp) = -\ii \left[ \mathcal{F}^{-\bk}_{\perp,A}(\xperp-\vperp) , \mathcal{F}^{+\bl}_\B(x^-(1-\tfrac{\vperp^2}{\tau^2}),\xperp-\vperp) \right]. \label{eq:lf-Vij-def}
\end{align}
The calculation for the fragmentation region of nucleus $\A$, where $-\eta_s \gg 1$, can be done in full analogy and yields the same expressions with the light cone components exchanged, i.e., $x^+\leftrightarrow x^-$ and $f^{+\bi}_\mathrm{LF}\leftrightarrow f^{-\bi}_\mathrm{LF}$.

Compared to the intermediate results for $f$ at $\eta_s\gg 1$ in \cref{eq:lf-f+--large-etas,eq:lf-f-i-large-etas,eq:lf-f+i-large-etas,eq:lf-fij-large-etas}, the weighting factors for nucleus $\A$ disappeared.
Instead, the longitudinal structure of nucleus $\A$ is integrated out and only enters as a boost-invariant charge distribution.
The physical interpretation that comes to mind is that of a one-sided boost-invariant limit.
Indeed, the remaining longitudinal structure in the results (dependence on $\eta_s$) is completely determined by nucleus $\B$.
The integral in the transverse $\vperp$-plane incoherently sums up "collisions" of the collapsed nucleus $\A$ with transverse slices of nucleus $\B$ for each value of $|\vperp|$.

We emphasize the difference between the intermediate results in \cref{eq:lf-f+--large-etas,eq:lf-f-i-large-etas,eq:lf-f+i-large-etas,eq:lf-fij-large-etas} and $f_\mathrm{LF}$ in \cref{eq:lf-f+-,eq:lf-f+i,eq:lf-f-i,eq:lf-fij}.
We started the derivation with the assumption that the coordinates $(x^+,x^-)$ of the evaluation point are far away from the track of nucleus $\A$.
Ultimately, this led to the far-field solutions $f_\mathrm{LF}$ in the fragmentation region of nucleus $\B$.
The discussion of the Taylor series expansion of $\mathcal{F}_\B$ in \cref{eq:lf-taylor-FB}, however, allows for the intermediate results to still be valid in a regime where $\eta_s \gtrsim 1$ as long as the longitudinal structure of nucleus $\B$ is very long ranged.
In this case, the constant term of the Taylor series will still be a good approximation of the full function, because the field will vary only a little within the $\eta''$ interval in \cref{eq:lf-nucl-A-eta''-lims} set by nucleus $\A$.
Pictorially, this would translate to the yellow-highlighted curves in \cref{fig:backward-lightcone-lf} deviating further from being parallel to the light cone boundary.
If the $v^-$ coordinates for the start and end of any given yellow line are still within one coherent region of nucleus $\B$, then the deviation from parallel lines will not matter.

\subsubsection{Proof of limiting fragmentation}

As mentioned at the beginning of this chapter, limiting fragmentation manifests itself as a universal rapidity dependence of observables at different collider energies when shifting the rapidity axis by the respective beam rapidities.
In the Milne frame, a shift in (spacetime) rapidity corresponds to a Lorentz boost.
We will now relate the field strength tensors $f_\mathrm{LF}$ at two different energies via a Lorentz boost and explicitly show that the boost angle corresponds to the difference of the beam rapidities.
This proves limiting fragmentation in the (3+1)D dilute Glasma.

Consider the "reference" setup at a beam rapidity $Y_1$ with the nuclear currents $\mathcal{J}_{\A/\B}$ and the resulting field strength tensor $f_\mathrm{LF}$ in the fragmentation region of nucleus $\B$.
Changing the collider energy amounts to boosting each nucleus along its direction of movement.
Note that this is not equal to boosting the lab frame, because the two boosts for the nuclei are in opposite directions.
Using the light cone components of the boost matrix $\Lambda_\mathrm{LC}$, the boosted currents read
\begin{align}
    \mathcal{J}^\mp_{\A/\B}(\xpm,\xperp) \rightarrow \ee^w \mathcal{J}^\mp(\ee^w \xpm,\xperp),
\end{align}
where we introduced the boost angle $w$.
We apply this boost to the single-nucleus field strengths that enter $V^{\bk\bl}_\perp$ in \cref{eq:lf-Vij-def}.
For nucleus $\A$,
\begin{align}
    \mathcal{F}^{-\bk}_{\perp,\A}(\uperp) \rightarrow & \intop \dd u^+\, \ee^w\mathcal{F}^{-\bk}_\A(\ee^w u^+, \uperp) \nn\\
    &= \intop \dd y^+\, \mathcal{F}^{-\bk}_\A(y^+, \uperp) =  \mathcal{F}^{-\bk}_{\perp,\A}(\uperp).
\end{align}
In the second line, we were able to absorb the boost factors via the substitution $y^+ = \ee^w u^+$.
As expected, the contribution of nucleus $\A$ does not change under boosts (one-sided boost invariant results).
The contribution from nucleus $\B$ acquires boost factors,
\begin{align}
    \mathcal{F}^{+\bl}_\B(u^-,\uperp) \rightarrow \ee^w \mathcal{F}^{+\bl}_\B(\ee^w u^-,\uperp),
\end{align}
which yields
\begin{align}
    V^{\bk\bl}_\perp(x,\vperp) \rightarrow& - \ii \ee^w \left[ \mathcal{F}^{-\bk}_{\perp,A}(\xperp-\vperp) , \mathcal{F}^{+\bl}_\B(\ee^w x^-(1-\tfrac{\vperp^2}{\tau^2}),\xperp-\vperp) \right] \nn\\
    &= \ee^w V^{\bk\bl}_\perp(\ee^w x^-,\tau,\xperp,\vperp).
\end{align}
Next, we apply this result to $f_\mathrm{LF}$ given in \cref{eq:lf-f+-,eq:lf-f+i,eq:lf-f-i,eq:lf-fij}
\begin{align}
    f^{+-}_\mathrm{LF}(x) \rightarrow& -\frac{g}{2\pi}\frac{1}{\ee^{-w}x^+} \intop \dd^2\vperp\, V^{\bk\bl}_\perp(\ee^w x^-,\tau,\xperp,\vperp) \delta^{\bk\bl} \nn\\
    &= f^{+-}_\mathrm{LF}(\ee^{-w}x^+,\ee^w x^-,\xperp), \\
    f^{+\bi}_\mathrm{LF}(x) \rightarrow& -\frac{g}{2\pi} \ee^w \intop \dd^2\vperp\, \frac{v^\bj}{\vperp^2} V^{\bk\bl}_\perp(\ee^w x^-,\tau,\xperp,\vperp) (\delta^{\bi\bj}\delta^{\bk\bl} - \epsilon^{\bi\bj}\epsilon^{\bk\bl}) \nn\\
    &= \ee^w f^{+\bi}_\mathrm{LF}(\ee^{-w}x^+,\ee^w x^-,\xperp), \\
    f^{-\bi}_\mathrm{LF}(x) \rightarrow& -\frac{g}{2\pi} \frac{\ee^{-w}}{2(\ee^{-w}x^+)^2} \intop \dd^2\vperp\, v^\bj V^{\bk\bl}_\perp(\ee^w x^-,\tau,\xperp,\vperp) (-\delta^{\bi\bj}\delta^{\bk\bl} - \epsilon^{\bi\bj}\epsilon^{\bk\bl}) \nn\\
    &= \ee^{-w} f^{-\bi}_\mathrm{LF}(\ee^{-w}x^+,\ee^w x^-,\xperp), \\
    f^{\bi\bj}_\mathrm{LF}(x) \rightarrow& -\frac{g}{2\pi} \frac{1}{\ee^{-w}x^+} \intop \dd^2\vperp\, V^{\bk\bl}_\perp(\ee^w x^-,\tau,\xperp,\vperp) \epsilon^{\bi\bj}\epsilon^{\bk\bl} \nn\\
    &= f^{\bi\bj}_\mathrm{LF}(\ee^{-w}x^+,\ee^w x^-,\xperp),
\end{align}
where we moved the outer $\ee^w$ factors next to the $x^+$ coordinates.
This result is equivalent to boosting $f_\mathrm{LF}$ with the boost matrix
\begin{align}
    \Lambda_{\mathrm{LC}\hphantom{\mu}\nu}^{\hphantom{\mathrm{LC}}\mu} = \mathrm{diag}(\ee^w, \ee^{-w}, 1, 1)
\end{align}
and acting on the coordinates with the inverse boost
\begin{align}
    (\Lambda^{-1})_{\mathrm{LC}\hphantom{\mu}\nu}^{\hphantom{\mathrm{LC}}\mu} x^\nu = (\ee^{-w}x^+, \ee^w x^-, \xperp) = (\tau, \eta_s - w, \xperp),
\end{align}
written compactly as
\begin{align}
    f^{\mu\nu}_\mathrm{LF} \rightarrow \Lambda_{\mathrm{LC}\hphantom{\mu}\rho}^{\hphantom{\mathrm{LC}}\mu}\Lambda_{\mathrm{LC}\hphantom{\nu}\sigma}^{\hphantom{\mathrm{LC}}\nu} f^{\rho\sigma}_\mathrm{LF}(\Lambda^{-1}(x)). \label{eq:lf-boosted-f}
\end{align}

When we assign the beam rapidity $Y_2$ to this second setup in \cref{eq:lf-boosted-f} with boosted currents, we can check the limiting fragmentation condition for observables given in \cref{eq:limiting-fragmentation-def}.
From \cref{eq:lf-boosted-f} it follows that any Lorentz scalars built out of $f_\mathrm{LF}$ are related by the shift $w$ in their $\eta_s$ arguments.
In particular, 
\begin{align}
    \mathcal{O}_{[Y_2]}(\tau, \eta_s - Y_2, \xperp) &= \mathcal{O}_{[Y_1]}(\tau,\eta_s -Y_2-w,\xperp) \nn\\
    &= \mathcal{O}_{[Y_1]}(\tau, \eta_s, -Y_1,\xperp),
\end{align}
where the second equality holds if the beam angle is equal to the difference of the two beam rapidities,
\begin{align}
    w = Y_1 - Y_2.
\end{align}
Hence, we explicitly derived limiting fragmentation in the (3+1)D dilute Glasma.
Furthermore, the limiting fragmentation of the dilute Glasma is local in the transverse plane and valid for event-by-event results.
Any transversely averaged observables, for example, eccentricities or azimuthal (energy) flow coefficients, will inherit the same properties.

\subsection{Numerical implementation}\label{sec:lim-frag-implementation}

The numerical computation of the integrals for $f_\mathrm{LF}$ in \cref{eq:lf-f+-,eq:lf-f+i,eq:lf-f-i,eq:lf-fij} requires less effort than the Monte Carlo procedure presented in \cref{sec:position-space-numerical-implementation}.
Because we were able to reduce the number of integrations to two, it is computationally feasible to pre-compute the integrand on the entire transverse $\vperp$-grid.
Hereby, the same discretization of the single-nucleus fields is applied, which also fixes the extent of the $\vperp$-grid.
From this grid, the discretization of the integrand is deduced.
Given the transverse evaluation point $\xperp$, the transverse argument $\xperp-\vperp$ might not coincide with a lattice point from the single-nucleus grids.
In general, the closest lattice point is used to approximate these values.
Ensuring compatibility with the same single-nucleus initial fields used in \cref{sec:position-space-numerical-implementation} is crucial for comparing the results of the different implementations.
The output for $f_\mathrm{LF}$ is structured the same as well.

The implementation is specifically designed to run on GPU accelerators and is parallelized across all values of the transverse evaluation point $\xperp$.
Additionally, the integrand for all values of $\vperp$ is pre-computed in parallel.
The resulting two-dimensional data array is used to approximate the integral via parallel reduction of the sum of all values scaled by the lattice spacing.
This approach is possible because the data arrays for the setups discussed in \cref{ch:numerical-results} are still small enough to fit into (GPU) memory.

\section{Gluon numbers in momentum space}\label{sec:gluon-numbers-lf}

In this section, we sketch how limiting fragmentation can be interpreted for the gluon number distribution derived in \cref{ch:momentum-space-picture}.
The arguments follow~\cite{Jalilian-Marian:2002yhb,Gelis:2006tb,Kharzeev:2001gp,Kharzeev:2001yq,Kharzeev:2004if} and motivate the term ``black disk''.
We will focus on the limit of large nuclei and the formulation with TMDs where $\dd N/\dd^2\kperp\dd\Y$ is given in \cref{eq:dNd2kdY-large-nucl-TMDs} as
\begin{align}
    \kperp^2\frac{\dd N}{\dd^2\kperp\,\dd\Y} &\approx \frac{(2\pi)^3 g^2 N_c}{2(N_c^2 -1)} \frac{\Sop}{\Sp^\A \Sp^\B} \intop \frac{\dd^2\pperp}{(2\pi)^2}\, \xG_\A(\xB_\A,\pperp) \xG_\B(\xB_\B,\kperp-\pperp). \tag{\ref{eq:dNd2kdY-large-nucl-TMDs}}
\end{align}
The same discussion will apply to the general case given in \cref{eq:dNd2kdY-result} because there is no difference in the longitudinal structure between these two results.
Only the transverse structure changes in the limit of large nuclei.
The formulation with TMDs has the benefit of allowing for cleaner arguments.

In the momentum space picture, limiting fragmentation follows from two assumptions about the behavior of the TMDs:
\begin{enumerate}
    \item TMDs only depend on the collider energy via the explicit arguments $\xB_{\A/\B}$,
    \begin{align}
        \xB_{\A/\B} = \kmp/\mathcal{P}^\mp = \frac{|\kperp|}{2 m_n \gamma}\ee^{\mp\Y} = \frac{|\kperp|}{m_n}\ee^{\mp\Y - \ln(2\gamma)}, \label{eq:xB-limfrag-shape}
    \end{align}
    where $\kmp$ is the longitudinal momentum of the parton and $\mathcal{P}^\mp$ the longitudinal momentum of the nucleon that is fixed by the energy of the collider $\mathcal{P}^\mp = \sqrt{2}m_n \gamma$.
    \item TMDs saturate for small $\xB$, i.e., in the limit $\xB\rightarrow0$, the TMDs are constant w.r.t.\ $\xB$ and highly peaked where the transverse momentum is equal to the saturation momentum $Q_s(\xB)$ (see e.g.~\cite{Iancu:2012xa,Garcia-Montero:2025hys,Andronic:2025ylc,Bartels:2002cj} and references therein).
\end{enumerate}
We can motivate the first assumption based on the generalized MV nuclear model with the correlation function $\Gamma_{\A/\B}$ (cf.~\cref{sec:generalized-correl}).
In the rest frame of the nucleus, we assume a longitudinal correlation scale $r$ such that $\Gamma_{\A/\B}(\xpm,\xperp)$ falls off quickly when $\xpm \gtrsim r$.
When we boost the nucleus to large velocities, the longitudinal correlation scale acquires a Lorentz-$\gamma$ factor and in the laboratory light cone frame, $\xpm \gtrsim r/\gamma$.
Now, recall from \cref{eq:tildeGamma-TMD} that
\begin{align}
    \xG_{\A/\B}(\kmp/\mathcal{P}^\mp,\kperp) \propto \tilde{\Gamma}_{\A/\B}(\kmp,\kperp),
\end{align}
where we dropped transverse momentum factors.
The longitudinal dependence of $\tilde{\Gamma}_{\A/\B}$, and therefore $\xG_{\A/\B}$, can only be a function of $\kmp / \gamma$ to describe the boosted nucleus.
Then,
\begin{align}
    \frac{1}{\gamma}\kmp = \frac{1}{\gamma}\frac{|\kperp|}{\sqrt{2}}\ee^{\mp\Y} \propto |\kperp| \ee^{\mp\Y-\ln(2\gamma)},
\end{align}
which is exactly of the form in \cref{eq:xB-limfrag-shape}.

Next, we investigate how these assumptions modify the longitudinal dynamics of the gluon distribution in \cref{eq:dNd2kdY-large-nucl-TMDs} in the fragmentation region of nucleus $\B$ where $\Y \gg 1$.
The momentum fraction of the gluon from nucleus $\A$ becomes very small,
\begin{align}
    \xB_\A = \frac{|\kperp|}{m_n}\ee^{-\Y-\ln(2\gamma)} \approx 0,
\end{align}
and the TMD for nucleus $\A$ is well within the saturation regime.
Hence, the entire dependence on momentum rapidity $\Y$ is localized to $\xG_\B$ and we write
\begin{align}
    \kperp^2\frac{\dd N}{\dd^2\kperp\,\dd\Y}\Big|_{\Y\gg1} &\approx \frac{(2\pi)^3 g^2 N_c}{2(N_c^2 -1)} \frac{\Sop}{\Sp^\A \Sp^\B} \intop \frac{\dd^2\pperp}{(2\pi)^2}\, \xG_\A(0,\pperp) \xG_\B(\tfrac{|\kperp|}{m_n}\ee^{\Y-\ln(2\gamma)},\kperp-\pperp).
\end{align}
Recall that the beam rapidity is defined as
\begin{align}
    Y_\mathrm{beam} &= \arcosh(\gamma) = \ln(\gamma + \sqrt{\gamma^2-1}) \nn\\
    &\approx \ln(2\gamma),
\end{align}
where in the last line we approximated the $\arcosh$ function for large values of $\gamma$, which are realized at heavy-ion collision experiments.
We immediately see that the functional dependence of $\xG_\B$ contains the difference
\begin{align}
    \Y - \ln(2\gamma) \approx \Y - Y_\mathrm{beam},
\end{align}
which is exactly of the form derived in the previous section with $\eta_s$ exchanged by $\Y$.
Therefore, a change in the energy of the collision can be compensated by an appropriate shift in momentum rapidity, which is limiting fragmentation.

The limiting fragmentation behavior also carries over to the transverse energy $\dd\Ep/\dd\Y$ in \cref{eq:dEpdeta-ito-dN}.
The relevant terms read
\begin{align}
    \frac{\dd \Ep}{\dd\Y}\Big|_{\Y\gg1} &= \intop \dd^2\kperp\, |\kperp| \frac{\dd N}{\dd^2\kperp\,\dd\Y}\Big|_{\Y\gg1} \nn\\
    &\propto \intop \frac{\dd^2\pperp\,\dd^2\kperp}{|\kperp|}\, \xG_\A(\xB_\A,\pperp) \xG_\B(\xB_\B,\kperp-\pperp) \nn\\
    &= \intop \frac{\dd^2\pperp\,\dd^2\kperp}{|\kperp|}\, \xG_\A(\xB_\A,|\kperp-\pperp|) \xG_\B(\xB_\B,|\pperp|),
\end{align}
where we shifted the $\pperp$ integration by $\kperp$ and assumed that the TMDs are isotropic and only depend on the magnitudes of the transverse momentum arguments.
Since $\xB_\A \approx 0$, the TMD of nucleus $\A$ is highly peaked around the saturation momentum $Q_s^\A$.
This sets $|\kperp - \pperp| \approx Q_s^\A$.
However, the saturation momentum for nucleus $\B$ is much smaller, $Q_s^\B \ll Q_s^\A$, and the integration domain for $\pperp$ is restricted to values $|\pperp| \lesssim Q_s^\B$.
Therefore, we can drop the dependence of $\xG_\A$ on $\pperp$ and factorize the integrals,
\begin{align}
    \intop \dd|\kperp|\,\dd\phi\, \xG_\A(\xB_\A\approx0,|\kperp|)\intop \dd^2\pperp\, \xG_\B(\xB_\B,|\pperp|) & \nn\\
    &\hspace{-60pt}\approx 2\pi\, Q_s^\A\,\xG_\A(0,Q_s^\A) \intop \dd^2\pperp\, \xG_\B(\xB_\B,|\pperp|).
\end{align}
Next, we assume that the nuclear TMDs obey geometric scaling,
\begin{align}
    \xG_{\A/\B}(\xB,\kperp) = \xG(\tfrac{\kperp^2}{Q_s(\xB)^2}),
\end{align}
that is, the TMDs are only a function of the ratio $\kperp^2/Q_s(\xB)^2$.
This property is exploited in many phenomenological models (e.g.,~\cite{Golec-Biernat:1998zce,Golec-Biernat:1999qor,Golec-Biernat:2017lfv}).
Then, 
\begin{align}
    \intop \dd|\kperp|\,\dd\phi\, \xG_\A(\xB_\A\approx0,|\kperp|)\intop \dd^2\pperp\, \xG_\B(\xB_\B,|\pperp|) & \nn\\
    &\hspace{-80pt}\approx 2\pi\, Q_s^\A\,\xG_\A(0,Q_s^\A) \intop \dd^2\pperp\, \xG_\B(\tfrac{\pperp^2}{Q_s(\xB_\B)^2}) \nn\\
    &\hspace{-80pt}= 2\pi\, Q_s^\A\,\xG_\A(0,Q_s^\A) Q_s(\xB_\B)^2 \intop \dd^2\kappaperp\, \xG_\B(\kappaperp^2),
\end{align}
where we rescaled the integration variable.
In the end, the dependence on $\xB_\B = \tfrac{|\kperp|}{m_n}\ee^{\Y-Y_\mathrm{beam}}$ is disentangled from the integrations and the particular dependence on $\Y - Y_\mathrm{beam}$ is manifest.

\chapter{Nuclear models}\label{ch:nuclear-models}

In this chapter, we study concrete realizations of the color charge correlators with generalized correlations, which are introduced in \cref{sec:generalized-correl}.
First, we focus on two common shapes for the envelope, the Gaussian and Woods-Saxon, and calculate their geometric properties in \cref{sec:nuclear-envelopes}, which are encoded in the projected transverse (overlap) areas.
Then, we discuss nuclear models with generalized correlations along the longitudinal direction using Gaussian envelopes (\cref{sec:finite-longitudinal-correlations}) and Woods-Saxon (\cref{sec:WS-correlator}).
In \cref{sec:CCC-bootstrapping-with-TMDs}, we introduce a bootstrapping procedure that allows matching the correlation function to phenomenological models of TMDs and examine the example of the Golec-Biernat--W\"usthoff model.
Finally, we compare the effective TMDs, defined in \cref{sec:effTMDs-gluon}, that result from the various nuclear models.

\section{Nuclear envelopes}\label{sec:nuclear-envelopes}

We consider two shapes for the boosted, single-nucleus envelopes $T_{\A/\B}$ introduced in \cref{sec:generalized-correl}.
The Gaussian,
\begin{align}
     H_{\A/\B}(\xpm,\xperp) &\coloneqq \sqrt{\frac{2}{\pi L_l^2}} \exp({-\tfrac{2(\xpm)^2}{L_l^2}} - \tfrac{\xperp^2}{2L_\perp^2}) , \label{eq:envelope-gauss-def}
\end{align}
and the Woods-Saxon (WS),
\begin{align}
    W_{\A/\B}(\xpm,\xperp) &\coloneqq \frac{a}{R_l\ln(1+\ee^a)}\left( 1+ \exp(a\sqrt{\tfrac{(2\xpm)^2}{R_l^2} + \tfrac{\xperp^2}{R_\perp^2}}-a) \right)^{-1}. \label{eq:envelope-ws-def}
\end{align}
Both are parametrized by two scales in the lab frame.
Along the longitudinal (light cone) direction, the parameters $L_l$ and $R_l$ correspond to the Lorentz-contracted diameters.
Along the transverse directions, the parameters $L_\perp$ and $R_\perp$ correspond to the radii.
Assuming that the nuclear envelopes are spherically symmetric in the Cartesian rest frame, the envelopes are parametrized by a single radius parameter, $L$ for Gaussian in \cref{eq:envelope-gauss-def} or $R$ for WS in \cref{eq:envelope-ws-def}.
They are related to $L_l$ and $R_l$ in light cone coordinates via the usual Lorentz contractions,
\begin{align}
    L_l = 2 \frac{L}{\sqrt{2}\gamma} = \frac{\sqrt{2}L}{\gamma}, \\
    R_l = 2 \frac{R}{\sqrt{2}\gamma} = \frac{\sqrt{2}R}{\gamma},
\end{align}
where the Lorentz factor is given by the beam energy $\gamma = \sqrt{s_\mathrm{NN}}/(2m_n)$ and
\begin{align}
    L_\perp = L, \\
    R_\perp = R.
\end{align}
However, the distinction between longitudinal and transverse parameters allows the clean definition of the limit of nuclei that are infinitely extended in the transverse plane.
This case reads
\begin{align}
    \lim_{\substack{L_\perp\,\rightarrow\,\infty \\ R_\perp\,\rightarrow\,\infty}} T_{\A/\B}(\xpm,\xperp) = T_{\A/\B}(\xpm,\zeroperp), \label{eq:envelopes-perp-to-inf}
\end{align}
where the envelope $T_{\A/\B}$ reduces to a one-dimensional function along the longitudinal direction.
The envelopes in \cref{eq:envelope-gauss-def,eq:envelope-ws-def} are normalized according to \cref{eq:envelope-norm},
\begin{align}
    \intop \dd\xpm\,T_{\A/\B}(\xpm,\zeroperp) = 1, \label{eq:envelope-norm-nucl-model}
\end{align}
which also ensures that integrating \cref{eq:envelopes-perp-to-inf} over the longitudinal coordinate $\xpm$ yields unity for transversly infinite nuclei.

The WS envelope in \cref{eq:envelope-ws-def} is defined with the additional parameter $a$ which is related to the skin-depth $d$,
\begin{align}
    a = R/d.
\end{align}
In terms of the skin depth and radius $R$, the spherically symmetric WS function in Cartesian coordinates is commonly written as
\begin{align}
    \mathrm{WS}(\vec x) \sim \left(1+ \exp(\tfrac{|\vec x| -R}{d}) \right)^{-1}.
\end{align}
The shape in \cref{eq:envelope-ws-def} is obtained after changing to light cone coordinates and rewriting the exponential using $a$.
Then, the modulus of the coordinate vector can be written as the square root of the sum of squared components, where each component is divided by a radius parameter.
In \cref{eq:envelope-ws-def}, we allowed different radii for the longitudinal and transverse directions.
Note that in the limit of transversely infinite nuclei given by \cref{eq:envelopes-perp-to-inf}, $a$ is kept fixed.

\subsection{Projected transverse area}\label{sec:proj-transv-area}

We list the explicit expressions for the projected transverse area introduced in \cref{sec:effTMDs-gluon},
\begin{align}
    \Sp^{\A/\B} = \intop\dd^2\xperp\,T^\perp_{\A/\B}(\xperp) = \intop \dd\xpm\,\dd^2\xperp\,T_{\A/\B}(\xpm,\xperp). \label{eq:nucl-models-Sperp}
\end{align}
In the case of Gaussian envelopes in \cref{eq:envelope-gauss-def}, the calculation is straightforward,
\begin{align}
    \Sp^{\A/\B} = \intop \dd\xpm\,\dd^2\xperp\,\sqrt{\frac{2}{\pi L_l^2}} \exp({-\tfrac{2(\xpm)^2}{L_l^2}} - \tfrac{\xperp^2}{2L_\perp^2}) = 2\pi L_\perp^2. \label{eq:Sperp-gauss}
\end{align}
Integrating the WS envelope in \cref{eq:envelope-ws-def} yields,
\begin{align}
    \Sp^{\A/\B} &= \intop\dd\xpm\,\dd^2\xperp\, \frac{a}{R_l\ln(1+\ee^a)}\left( 1+ \exp(a\sqrt{\tfrac{(2\xpm)^2}{R_l^2} + \tfrac{\xperp^2}{R_\perp^2}}-a) \right)^{-1} \nn\\
    &= \frac{R_\perp^2}{2a^2\ln(1+\ee^a)}\intop\dd\zpm\,\dd^2\zperp\,\left( 1 + \exp(\sqrt{(\zpm)^2 + \zperp^2} -a) \right)^{-1},
\end{align}
where we rescaled the integration variables as
\begin{align}
    \zpm = \frac{2a}{R_l}\xpm, \qquad \zperp = \frac{a}{R_\perp}\xperp. \label{eq:rescaling-sop-ws}
\end{align}
After changing to spherical coordinates, $r^2 = (\zpm)^2 + \zperp^2$, and the angular integrations are trivial which yields the result
\begin{align}
     \Sp^{\A/\B} &= \frac{4\pi R_\perp^2}{2a^2\ln(1+\ee^a)} \intop_0^\infty\dd r\, r^2 \left( 1 + \ee^{r-a} \right)^{-1} \nn\\
     &= 2\pi R_\perp^2 \Zp. \label{eq:Sperp-ws}
\end{align}
Here, we extracted the factor $2\pi R_\perp^2$ and defined the modification factor $\Zp$ that contains the remaining coefficient and the solution of the integral.%
\footnote{%
This integral can be solved in terms of elementary functions,
\begin{align}
    \intop_0^\infty\dd r\, r^n \left( 1 + \ee^{r-a} \right)^{-1} = - \Gamma(n+1) \mathrm{Li}_{n+1}(-\ee^a),
\end{align}
where $\Gamma$ is the Euler-gamma function and $\mathrm{Li}_n$ the Poly-Logarithm of degree $n$.}
The dimensionless factor $\Zp$ only depends on the constant $a$.
Compared to the Gaussian transverse area in \cref{eq:Sperp-gauss}, for equal transverse radii $L_\perp = R_\perp$, the WS area in \cref{eq:Sperp-ws} is smaller by the factor $\Zp<1$ for $a = R/d \gtrsim 2.18$ and larger otherwise.

\subsection{Projected transverse overlap area}
We calculate the explicit expressions for the projected transverse overlap area introduced in \cref{eq:Sop-def} in \cref{sec:limit-large-nucl},
\begin{align}
    \Sop = \intop \dd^2\xperp\, T^\perp_\A(\xperp) T^\perp_\B(\xperp) = \intop \dd x^+\,\dd y^-\,\dd^2\xperp\,T_\A(x^+,\xperp)T_\B(y^-,\xperp).
\end{align}
For the Gaussian in \cref{eq:envelope-gauss-def}, the calculation is straightforward,
\begin{align}
    \Sop &= \intop\dd x^+\,\dd y^-\,\dd^2\zperp\,\sqrt{\frac{2}{\pi L_l^2}} \exp({-\tfrac{2(x^+)^2}{L_l^2}} - \tfrac{\zperp^2}{2L_\perp^2})\sqrt{\frac{2}{\pi L_l^2}} \exp({-\tfrac{2(y^-)^2}{L_l^2}} - \tfrac{\zperp^2}{2L_\perp^2}) \nn\\
    &= \intop\dd^2\zperp\,\exp(-\zperp^2/L_\perp^2) = \pi L_\perp^2.
\end{align}
The integration of the WS in \cref{eq:envelope-ws-def} can be evaluated using the same rescaling as in \cref{eq:rescaling-sop-ws},
\begin{align}
    \Sop &= \frac{R_\perp^2}{4a^2\ln(1+\ee^a)^2} \intop \dd x^+\,\dd y^-\,\dd^2\zperp \nn\\
    &\times \left( 1+\exp(\sqrt{(x^+)^2+\zperp^2} -a) \right)^{-1} \left( 1+\exp(\sqrt{(y^-)^2+\zperp^2} -a)\right)^{-1} \nn\\
    &= \frac{\pi R_\perp^2}{2a^2\ln(1+\ee^a)^2)} \intop\dd x^+\,\dd y^-\intop_0^\infty \dd r \nn\\
    & \times r \left( 1+\exp(\sqrt{(x^+)^2+r^2} -a) \right)^{-1} \left( 1+\exp(\sqrt{(y^-)^2+r^2} -a)\right)^{-1} \nn\\
    &= \pi R_\perp^2 \Zop,
\end{align}
where we changed to polar coordinates, $r^2=\zperp^2$, and introduced the overlap modification factor $\Zop$ in the last equation.
In analogy to $\Zp$ in \cref{eq:Sperp-ws}, the dimensionless factor $\Zop$ contains the remaining coefficient and the solution of the integrals.
It only depends on the constant $a$ and leads to a smaller $\Sop$ when the transverse radius $R_\perp$ is matched to the Gaussian $L_\perp$ in \cref{eq:envelope-gauss-def} and $a\gtrsim 2.05$.

\section{Parametrized longitudinal correlations}\label{sec:finite-longitudinal-correlations}

We introduce the normalized Gaussian
\begin{align}
    U^\xi_{\A/\B}(\xpm) = \frac{1}{\sqrt{2\pi \xi^2}}\, \ee^{- \frac{(\xpm)^2}{2\xi^2}}, \label{eq:Uxi-def}
\end{align}
as a model to regulate correlations within the nuclei along the longitudinal direction.
The parameter $\xi$ controls the width of the Gaussian and, therefore, parametrizes the longitudinal correlation length.
We use \cref{eq:Uxi-def} in place of the Dirac-delta shaped longitudinal correlations to generalize the MV color charge correlator in \cref{eq:MV-correlator-T(x+y)} and obtain
\begin{align}
    \langle \rho^a_{\A/\B}(\xpm,\xperp) \rho^b_{\A/\B}(\ypm,\yperp) \rangle = \delta^{ab} g^2 \mu^2 T_{\A/\B}(\tfrac{\xpm+\ypm}{2},\tfrac{\xperp+\yperp}{2})U^\xi_{\A/\B}(\xpm-\ypm)\deltaperp(\xperp-\yperp). \label{eq:Uxi-T(x+y)-correlator}
\end{align}
This correlator still retains the uncorrelated structure in the transverse plane given by the transverse delta function.
The coordinate dependencies on the right-hand side are separated into average, $(x+y)/2$, and difference, $x-y$, coordinates.
In these coordinates, the six-dimensional configuration space of the correlator is parametrized along directions that are orthogonal to each other.
Hence, by moving along those orthogonal directions, one can conceptually decouple the contributions from the longitudinal correlation function $U^\xi_{\A/\B}$ from the modulation with the single-nucleus envelope $T_{\A/\B}$.
For now, we stick to the Gaussian envelope $H_{\A/\B}$ from \cref{eq:envelope-gauss-def}.

To prove the positive semi-definiteness of this model, we refer back to the factorization of the Gaussian envelope in \cref{eq:T(x+y)-factorized}.
We identify the square roots of the single-nucleus envelopes,
\begin{align}
    t_{\A/\B}(\xpm,\xperp) = \sqrt{H_{\A/\B}(\xpm,\xperp)}
\end{align}
and rewrite the correlator in \cref{eq:Uxi-T(x+y)-correlator} in the generalized form given in \cref{eq:general-correlator-t(x)-t(y)} using the correlation function
\begin{align}
    \Gamma^\xi_{\A/\B}(\xpm,\xperp) &= g^2\mu^2 U^\xi_{\A/\B}(\xpm) \ee^{\frac{(\xpm)^2}{2L_l^2} +\frac{\xperp^2}{8L_\perp^2}} \deltaperp(\xperp) \nn\\
    &= \frac{g^2\mu^2}{\sqrt{2\pi \xi^2}}\, \ee^{-(\xpm)^2(\tfrac{1}{2\xi^2} - \tfrac{1}{2L_l^2})} \deltaperp(\xperp), \label{eq:Gamma-xi-def}
\end{align}
where we dropped the exponential containing $\xperp$ due to the delta function.
The resulting correlator defines the "$\Gxi$" model and reads,
\begin{align}
    \Gxi: \qquad \langle \rho^a_{\A/\B}(\xpm,\xperp) \rho^b_{\A/\B}(\ypm,\yperp) \rangle & \nn\\
    &\hspace{-67.3pt}= \delta^{ab}\sqrt{H_{\A/\B}(\xpm,\xperp)}\sqrt{H_{\A/\B}(\ypm,\yperp)} \Gamma^\xi_{\A/\B}(\xpm-\ypm,\xperp-\yperp). \label{eq:gauss-xi-correlator}
\end{align}

Positive semi-definiteness is achieved when the Fourier transformation of $\Gamma^\xi_{\A/\B}$ exists and is non-negative (cf.~\cref{eq:fourier-pos-semi-def-cond}).
This restricts the $\xi$ parameter such that the exponent in \cref{eq:Gamma-xi-def} is negative,
\begin{align}
    0\leq \xi \leq L_l. \label{eq:xi-limits}
\end{align}
It follows that
\begin{align}
    \tilde{\Gamma}^\xi_{\A/\B}(\kmp,\kperp) = \frac{g^2\mu^2 L_l}{\sqrt{L_l^2 - \xi^2}}\,\ee^{-\frac{(\pmp)^2}{2}\frac{L_l^2\xi^2}{L_l^2 - \xi^2}} \geq 0,\label{eq:Gamma-xi-FT}
\end{align}
which proves the positive semi-definiteness of the correlator in \cref{eq:gauss-xi-correlator}.

The bounds for $\xi$ in \cref{eq:xi-limits} can both be saturated.
The limit $\xi \rightarrow 0$ restores the original MV model in the center of the nuclei (i.e., evaluating the correlator at $\xperp = \yperp = \zeroperp$) in accordance with \cref{eq:Gamma-MV-limit-norm},
\begin{align}
    \lim_{\xi\rightarrow\, 0} \Gamma_{\A/\B}(\xpm,\xperp) = g^2\mu^2\delta(\xpm)\deltaperp(\xperp). \label{eq:Gamma-xi-delta-lim}
\end{align}
The opposite limit, $\xi\rightarrow L_l$, yields a correlation function that is constant along the longitudinal direction,
\begin{align}
    \lim_{\xi\rightarrow L_l} \Gamma_{\A/\B}(\xpm,\xperp) = \frac{g^2\mu^2}{\sqrt{2\pi L_l^2}}\deltaperp(\xperp). \label{eq:Gamma-xi-coherent-lim}
\end{align}
This corresponds to coherent smearing of the color charge along the entire longitudinal extent.
Still, there is longitudinal structure in this coherent limit because the envelopes provide a maximal correlation length restricted by their support (the extent of the nuclei).

Next, we calculate the Fourier transformation of the $\Gxi$ correlator using \cref{eq:FT-general-correlator-t(x)-t(y)} and evaluate it to obtain $\xG_{\A/\B}$ from \cref{eq:tildeB-def} and $\xGeff_{\A/\B}$ from \cref{eq:xGeff-xpq-def}.
To this end, we need the Fourier transformation of the square root of $H_{\A/\B}$,
\begin{align}
    \widetilde{\sqrt{H_{\A/\B}}}[\kmp,\kperp] &\coloneqq \intop \dd\xpm\,\dd^2\xperp\, \ee^{\ii\xpm\kmp - \ii\xperp\cdot\kperp} \sqrt{H_{\A/\B}(\xpm,\xperp)} \nn\\
    &= (2\pi)^{1/4}4\pi L_\perp^2 \sqrt{L_l} \ee^{-(\kmp)^2\frac{L_l^2}{4} - \kperp^2 L_\perp^2} \label{eq:widetildeH-AB-def}
\end{align}
The TMD reads
\begin{align}
    \xG_{\A/\B}(\xB_{\A/\B}=\kmp/\mathcal{P}^\mp,\kperp) &= \frac{2(N_c^2-1)}{(2\pi)^3}\frac{\kperp^2}{\left(\kperp^2+m^2\right)^2} \nn\\
    &\times \intop \frac{\dd\kappa^\mp\,\dd^2\kappaperp}{(2\pi)^3} \left| \widetilde{\sqrt{H_{\A/\B}}}[\kmp-\kappa^\mp,\kperp-\kappaperp] \right|^2 \tilde{\Gamma}^\xi_{\A/\B}(\kappa^\mp,\kappaperp) \nn\\
    &= \Sp^{\A/\B}\frac{2(N_c^2-1)g^2\mu^2}{(2\pi)^3}\frac{\kperp^2}{\left(\kperp^2+m^2\right)^2} \ee^{-(\kmp)^2\frac{\xi^2}{2}}, \label{eq:Gauss-xi-TMD}
\end{align}
where the beam momentum $\mathcal{P}^\mp = \sqrt{2}\gamma m_n$ and we used that $H_{\A/\B}$ is a symmetric function in all of its arguments.
Note that the TMD is proportional to the projected transverse area $\Sp^{\A/\B}$.
Remarkably, \cref{eq:Gauss-xi-TMD} is free of the longitudinal envelope scale $L_l$.
The longitudinal dynamics are completely determined by the Fourier transformation of the $U^\xi_{\A/\B}$ function from \cref{eq:Uxi-def}.
In fact, this directly follows from the structure of \cref{eq:Uxi-T(x+y)-correlator}, which, in the $\Gxi$ model, is equivalent to the defining correlator in \cref{eq:gauss-xi-correlator}.
Because of the clean factorization in average and difference coordinates, the envelope will only contribute as a function of $p+q$ for the Fourier transformation of the correlator in \cref{eq:FT-general-correlator-t(x)-t(y)}.
For $\xG_{\A/\B}$, \cref{eq:FT-general-correlator-t(x)-t(y)} is evaluated at $p=-q$ and the envelope factor degenerates to a constant.
This constant is only proportional to $L_\perp^2$ because of the normalization of the envelopes in \cref{eq:envelope-norm-nucl-model}.

The effective TMD reads
\begin{align}
    \frac{\xGeff_{\A/\B}(\xB_{\A/\B}=\kmp/\mathcal{P}^\mp,\pperp,\qperp)}{\Sp^{\A/\B}} &= \frac{2(N_c^2-1)}{(2\pi)^5}\frac{\pperp^2\qperp^2}{(\pperp^2+m^2)(\qperp^2+m^2)} \nn\\
    &\times \intop \frac{\dd\kappa^\mp\,\dd^2\kappaperp}{(2\pi)^3}\,\tilde{\Gamma}^\xi_{\A/\B}(\kappa^\mp,\kappaperp) \nn\\
    &\times \widetilde{\sqrt{H_{\A/\B}}}[\kmp-\kappa^\mp,\pperp-\kappaperp]\widetilde{\sqrt{H_{\A/\B}}}[-\kmp+\kappa^\mp,\qperp+\kappaperp] \nn\\
    &= \Sp^{\A/\B}\frac{2(N_c^2-1)g^2\mu^2}{(2\pi)^5}\frac{\pperp^2\qperp^2}{(\pperp^2+m^2)(\qperp^2+m^2)} \nn\\
    &\times \exp(-(\kmp)^2\tfrac{\xi^2}{2} - (\pperp+\qperp)^2\tfrac{\Sp^{\A/\B}}{4\pi}). \label{eq:Gauss-xi-effTMD(k-p-q)}
\end{align}
This result is again free of the longitudinal envelope scale, but the transverse envelope scale $L_\perp$ enters in the prefactor and in the exponential via $\Sp^{\A/\B}$.
Integrating out one transverse momentum yields the reduced effective TMD,
\begin{align}
    \frac{\xGeff_{\A/\B}(\xB_{\A/\B}=\kmp/\mathcal{P}^\mp,\kperp)}{\Sp^{\A/\B}} &= \Sp^{\A/\B}\frac{2(N_c^2-2)g^2\mu^2}{(2\pi)^5(\kperp^2+m^2)} \ee^{-(\kmp)^2\frac{\xi^2}{2}} \nn\\
    &\times \intop \dd^2\qperp\,\frac{\qperp^2}{\qperp^2+m^2}\ee^{-(\kperp+\qperp)^2\frac{\Sp^{\A/\B}}{4\pi}} \nn\\
    &= \Sp^{\A/\B}\frac{2(N_c^2-2)g^2\mu^2}{(2\pi)^5(\kperp^2+m^2)} \ee^{-(\kmp)^2\frac{\xi^2}{2}} \nn\\
    &\times \intop \dd^2\pperp\,\frac{(\pperp-\kperp)^2}{(\pperp-\kperp)^2+m^2}\ee^{-\pperp^2\frac{\Sp^{\A/\B}}{4\pi}} \nn\\
    &= \Sp^{\A/\B}\frac{2(N_c^2-2)g^2\mu^2}{(2\pi)^5(\kperp^2+m^2)} \ee^{-(\kmp)^2\frac{\xi^2}{2}} \nn\\
    &\times \intop_0^{2\pi}\dd\varphi\intop_0^\infty \dd p\,\frac{p\left(\kperp^2 + p^2 - 2|\kperp|p\cos(\varphi)\right)}{\kperp^2 + p^2 -2|\kperp|p\cos(\varphi)+m^2}\ee^{-p^2\frac{\Sp^{\A/\B}}{4\pi}}.
\end{align}
Here, we first shifted the integration variable $\kperp+\qperp = \pperp$ and then changed to polar coordinates, $\dd^2\pperp = \dd p\,\dd\varphi\, p$.
Since \cref{eq:Gauss-xi-effTMD(k-p-q)} only depends on the relative angle between the two transverse momenta, integrating out $\qperp$ leads to an isotropic result w.r.t.\ the other momentum.
We can further carry out the angular integration,
\begin{align}
    \frac{\xGeff_{\A/\B}(\xB_{\A/\B}=\kmp/\mathcal{P}^\mp,\kperp)}{\Sp^{\A/\B}} &= \Sp^{\A/\B}\frac{2(N_c^2-2)g^2\mu^2}{(2\pi)^5(\kperp^2+m^2)} \ee^{-(\kmp)^2\frac{\xi^2}{2}} \nn\\
    &\times \intop_0^\infty \dd p\, 2\pi p\, \ee^{-p^2\frac{\Sp^{\A/\B}}{4\pi}} \nn\\
    &\times \left( 1 - \frac{m^2}{\sqrt{\left(\kperp^2-\pperp^2\right)^2 + 2m^2(\kperp^2+\pperp^2) + m^4}} \right) \nn\\
    &= \frac{2(N_c^2-2)g^2\mu^2}{(2\pi)^3(\kperp^2+m^2)} \ee^{-(\kmp)^2\frac{\xi^2}{2}} \nn\\
    &\times \left( 1- \intop_0^\infty\frac{\dd p}{2\pi}\frac{m^2\Sp^{\A/\B}\, p\,\ee^{-p^2\frac{\Sp^{\A/\B}}{4\pi}}}{\sqrt{\left(\kperp^2-\pperp^2\right)^2 + 2m^2(\kperp^2+\pperp^2) + m^4}}\right), \label{eq:Gauss-xi-effTMD}
\end{align}
and evaluate the radial integration for the constant term in the big parentheses.
In the final result, only the second term in the parentheses still contains the transverse envelope scale $L_\perp$ via $\Sp^{\A/\B}$.
These factors of $\Sp^{\A/\B}$ cannot be cleanly factorized into a prefactor.
The remaining integral does not scale $\sim\Sp^{\A/\B}$ but shows a more complicated behavior.

\subsection{Reparametrized correlation scale}

The definition of the correlation function $\Gamma^\xi_{\A/\B}$ in \cref{eq:Gamma-xi-def} and its Fourier transformation in \cref{eq:Gamma-xi-FT} highlight that the parametrization with $\xi$ is cumbersome.
On the one hand, the TMD of the $\Gxi$ model is only a function of $\xi$.
But, this is a special case for the Gaussian envelopes in \cref{eq:envelope-gauss-def}.
On the other hand, when generalizing to arbitrary envelopes, the mixing of $\xi$ and $L_l$ in $\Gamma^\xi_{\A/\B}$ seems unnatural.
The longitudinal envelope scale $L_l$ appears in a Gaussian weight factor and regulates the correlation function.
To help with the conceptualization of this model, we introduce a different parametrization of the correlation function,
\begin{align}
    \Gamma^\zeta_{\A/\B}(\xpm,\xperp) &=  \frac{g^2\mu^2}{\sqrt{2\pi}}\sqrt{\frac{1}{\zeta^2}+\frac{1}{L_l^2}}\, \ee^{-(\xpm)^2\frac{1}{2\zeta^2}} \deltaperp(\xperp), \label{eq:Gamma-zeta-gauss-def} \\
    \tilde{\Gamma}^\zeta_{\A/\B}(\kmp,\kperp) &= \frac{g^2\mu^2}{L_l}\sqrt{L_l^2+\zeta^2}\,\ee^{-(\kmp)^2\frac{\zeta^2}{2}}. \label{eq:Gamma-zeta-gauss-FT}
\end{align}
Now, the Gaussian correlations are controlled by a single parameter.
The new scale $\zeta$ can take the value of any non-negative real number,
\begin{align}
    0 \leq \zeta < \infty,
\end{align}
still ensuring that $\tilde{\Gamma}^\zeta_{\A/\B} \geq 0$ and the model to be positive semi-definite.
The normalization is fixed such that \cref{eq:Gamma-zeta-gauss-def,eq:Gamma-zeta-gauss-FT} are equivalent to the $\Gxi$ model correlations in \cref{eq:Gamma-xi-def,eq:Gamma-xi-FT} under the identifications
\begin{align}
    \xi^2 = \frac{\zeta^2 L_l^2}{L_l^2 + \zeta^2}, \qquad \zeta^2 = \frac{\xi^2 L_l^2}{L_l^2 - \xi^2}. \label{eq:xi-zeta-relation}
\end{align}
This also leads to the same limits for $\Gamma^\zeta_{\A/\B}$ given in \cref{eq:Gamma-xi-delta-lim,eq:Gamma-xi-coherent-lim} when $\zeta$ satisfies the bounds,
\begin{align}
    \zeta = 0 \quad& \Longleftrightarrow \quad \xi =0, \\
    \zeta\rightarrow \infty \quad& \Longleftrightarrow \quad \xi \rightarrow L_l.
\end{align}
A comparable model was used in~\cite{Matsuda:2023gle,Matsuda:2024mmr,Matsuda:2024moa} for the color charge correlator of nucleonic hot spots in simulations of the Glasma.

\subsection{Woods-Saxon envelopes}\label{sec:WS-correlator}

We define the "$\WSxi$" model as the generalized correlator from \cref{eq:general-correlator-t(x)-t(y)} where the envelopes are replaced by the WS functions in \cref{eq:envelope-ws-def} and the correlation function from \cref{eq:Gamma-zeta-gauss-def} is used,
\begin{align}
    \WSxi: \qquad\langle \rho^a_{\A/\B}(\xpm,\xperp) \rho^b_{\A/\B}(\ypm,\yperp) \rangle & \nn\\
    &\hspace{-58.7pt}= \delta^{ab}\sqrt{W_{\A/\B}(\xpm,\xperp)}\sqrt{W_{\A/\B}(\ypm,\yperp)} \Gamma^\zeta_{\A/\B}(\xpm-\ypm,\xperp-\yperp). \label{eq:WS-xi-correlator}
\end{align}
This model corresponds to the $\Gxi$ model in \cref{eq:gauss-xi-correlator} but with the envelopes exchanged.
We clarify that the longitudinal Woods-Saxon diameter $R_l$ takes the role of $L_l$ in the normalization factor of $\Gamma^\zeta_{\A/\B}$ in \cref{eq:Gamma-zeta-gauss-def}.
The $\WSxi$ correlator in \cref{eq:WS-xi-correlator} is used in \cite{Ipp:2024ykh}, but there it is formulated in terms of the $\xi$ parameter (cf.\ \cref{eq:xi-zeta-relation}).

The TMD reads
\begin{align}
    \xG_{\A/\B}(\xB_{\A/\B}=\kmp/\mathcal{P}^\mp,\kperp) &= \frac{2(N_c^2-1)}{(2\pi)^3}\frac{\kperp^2}{\left(\kperp^2+m^2\right)^2} \nn\\
    &\times \intop \frac{\dd\kappa^\mp\,\dd^2\kappaperp}{(2\pi)^3} \left| \widetilde{\sqrt{W_{\A/\B}}}[\kmp-\kappa^\mp,\kperp-\kappaperp] \right|^2 \tilde{\Gamma}^\zeta_{\A/\B}(\kappa^\mp,\kappaperp) \nn\\
    &= \frac{2(N_c^2-1)g^2\mu^2}{(2\pi)^3}\frac{\sqrt{R_l^2+\zeta^2}}{R_l}\frac{\kperp^2}{\left(\kperp^2+m^2\right)^2} \nn\\
    &\times \intop \frac{\dd\pmp\,\dd\xpm\,\dd\ypm}{2\pi}\,\ee^{-(\kmp-\pmp)^2\frac{\zeta^2}{2}} \ee^{\ii\pmp(\xmp-\ymp)}\nn\\
    &\times \intop \dd^2\xperp\,\sqrt{W_{\A/\B}(\xpm,\xperp)}\sqrt{W_{\A/\B}(\ypm,\xperp)}.
    \label{eq:WS-zeta-TMD}
\end{align}
Here, we shifted $\kmp-\kappa^\mp=\pmp$ and restored the square roots of the WS functions in position space using the analogous definition of $\widetilde{\sqrt{W_{\A/\B}}}$ as given for Gaussian envelopes in \cref{eq:widetildeH-AB-def}.
Note that the last line contains an integral over the entire transverse plane that scales with the projected transverse area $\Sp^{\A/\B}$.
However, $\Sp^{\A/\B}$ does not factor out cleanly, because the WS envelopes cannot be separated into a product of transverse and longitudinal factors.

Similarly, the effective TMD is
\begin{align}
    \frac{\xGeff_{\A/\B}(\xB_{\A/\B}=\kmp/\mathcal{P}^\mp,\pperp,\qperp)}{\Sp^{\A/\B}} &= \frac{2(N_c^2-1)}{(2\pi)^5}\frac{\pperp^2\qperp^2}{(\pperp^2+m^2)(\qperp^2+m^2)} \nn\\
    &\times \intop \frac{\dd\kappa^\mp\,\dd^2\kappaperp}{(2\pi)^3}\,\tilde{\Gamma}^\zeta_{\A/\B}(\kappa^\mp,\kappaperp) \nn\\
    &\times \widetilde{\sqrt{W_{\A/\B}}}[\kmp-\kappa^\mp,\pperp-\kappaperp]\widetilde{\sqrt{W_{\A/\B}}}[-\kmp+\kappa^\mp,\qperp+\kappaperp] \nn\\
    &= \frac{2(N_c^2-1)g^2\mu^2}{(2\pi)^5}\frac{\sqrt{R_l^2+\zeta^2}}{R_l}\frac{\pperp^2\qperp^2}{(\pperp^2+m^2)(\qperp^2+m^2)} \nn\\
    &\times \intop \frac{\dd\pmp\,\dd\xpm\,\dd\ypm}{2\pi}\,\ee^{-(\kmp-\pmp)^2\frac{\zeta^2}{2}} \ee^{\ii\pmp(\xmp-\ymp)}\nn\\
    &\times \intop \dd^2\xperp\,\sqrt{W_{\A/\B}(\xpm,\xperp)}\sqrt{W_{\A/\B}(\ypm,\xperp)}\ee^{-\ii \xperp\cdot(\pperp+\qperp)} ,
    \label{eq:WS-zeta-effTMD(k-p-q)}
\end{align}
and the reduced effective TMD is
\begin{align}
    \frac{\xGeff_{\A/\B}(\xB_{\A/\B}=\kmp/\mathcal{P}^\mp,\kperp)}{\Sp^{\A/\B}} &= \frac{2(N_c^2-1)g^2\mu^2}{(2\pi)^3}\frac{\sqrt{R_l^2+\zeta^2}}{R_l}\frac{1}{\kperp^2+m^2} \nn\\
    &\times \intop \frac{\dd\pmp\,\dd\xpm\,\dd\ypm}{2\pi}\,\ee^{-(\kmp-\pmp)^2\frac{\zeta^2}{2}} \ee^{\ii\pmp(\xmp-\ymp)}\nn\\
    &\times \intop \frac{\dd^2\xperp\,\dd^2\pperp}{(2\pi)^2} \frac{(\pperp-\kperp)^2}{(\pperp-\kperp)^2+m^2} \nn\\
    &\times \sqrt{W_{\A/\B}(\xpm,\xperp)}\sqrt{W_{\A/\B}(\ypm,\xperp)}\ee^{-\ii \xperp\cdot\pperp} ,
    \label{eq:WS-zeta-effTMD}
\end{align}
where we shifted $\kperp+\qperp = \pperp$.

\section{Color charge correlator bootstrapping with TMDs}\label{sec:CCC-bootstrapping-with-TMDs}

In this section, we present a novel approach to modeling the color charge correlators used in the generalized MV nuclear model (cf.\ \cref{sec:MV-nucl-model}).
The basic idea is to combine information about the shape of a nucleus with information about its momentum content.
While the shape is most naturally described via a single-nucleus envelope, it turns out that the momentum space properties, which are relevant for gluon production in the dilute limit, are encoded in the transverse momentum distributions (TMDs) (cf.\ \cref{sec:effTMDs-gluon}).
Hence, it is desirable to use phenomenologically or experimentally motivated TMDs as additional input for ``bootstrapping'' color charge correlators.
In the context of the dilute Glasma, this approach was first described in~\cite{Schlichting:2020wrv,Singh:2021hct} where a particular ansatz for the color charge correlator was used to simplify the calculations.
Here, the goal is to generalize this procedure to arbitrary envelopes and use the correlation function in correlators of the type given in \cref{eq:general-correlator-t(x)-t(y)}.
From the determination of the $\Gxi$ and $\WSxi$ models in the previous section, it is clear that fixing a correlation function $\Gamma_{\A/\B}$ introduces a dependence of the TMDs on the shape of the envelopes.
The three objects $\Gamma_{\A/\B}$, $T_{\A/\B}$ and $\xG_{\A/\B}$ are intertwined.
Therefore, when fixing the envelope and TMD, the resulting correlation function will be a particular combination of the latter two, which we aim to derive in the following.

\subsection{Bootstrapping with factorizable envelopes}

First, we review the calculations in~\cite{Schlichting:2020wrv,Singh:2021hct}.
We stress that for these calculations, it is essential to assume that the single-nucleus envelopes $T_{\A/\B}$ are factorizable (cf.\ \cref{sec:generalized-correl}) according to
\begin{align}
    T_{\A/\B}(\tfrac{\xpm+\ypm}{2},\tfrac{\xperp+\yperp}{2}) = t_{\A/\B}(\xpm,\xperp)t_{\A/\B}(\ypm,\yperp) f^T_{\A/\B}(\xpm-\ypm,\xperp-\yperp), \label{eq:T(x+y)-factor-f}
\end{align}
where the function $f^T_{\A/\B}$ only depends on the difference coordinates and captures any terms for \cref{eq:T(x+y)-factor-f} to hold.
The generalized color charge correlator in \cref{eq:general-correlator-t(x)-t(y)} can then be written in the form of \cref{eq:general-correlator-T(x+y)},
\begin{align}
        \langle \rho^a_{\A/\B}(\xpm,\xperp) \rho^b_{\A/\B}(\ypm,\yperp) \rangle &= \delta^{ab}\, T_{\A/\B}(\tfrac{\xpm+\ypm}{2},\tfrac{\xperp+\yperp}{2}) \Gamma'_{\A/\B}(\xpm-\ypm,\xperp-\yperp). \label{eq:bootstrap-correl-T(x+y)}
\end{align}
This correlator is the starting point used in~\cite{Schlichting:2020wrv,Singh:2021hct}.
Here, we differentiate between the primed correlation function $\Gamma'_{\A/\B}$ and $\Gamma_{\A/\B}$, where the latter enters the generalized correlator formulated with the square roots $t_{\A/\B}$ in \cref{eq:general-correlator-t(x)-t(y)}.
They are related by the function $f^T_{\A/\B}$,
\begin{align}
    \Gamma_{\A/\B}(\xpm,\xperp) = \Gamma'_{\A/\B}(\xpm,\xperp) f^T_{\A/\B}(\xpm,\xperp). \label{eq:bootstrap-T(x+y)-Gamma'}
\end{align}
To evaluate the TMD in the model given by \cref{eq:bootstrap-correl-T(x+y)}, we need the Fourier transformation of the color charge correlator,
\begin{align}
    \langle \tilde{\rho}^a_{\A/\B}(\pmp,\pperp) \tilde{\rho}^b_{\A/\B}(\qmp,\qperp) \rangle &= \delta^{ab}\, \tilde{T}_{\A/\B}(\pmp+\qmp,\pperp+\qperp) \tilde{\Gamma}'_{\A/\B}(\tfrac{\pmp-\qmp}{2},\tfrac{\pperp-\qperp}{2}).
\end{align}
Note that from the clean separation of the dependencies of $T_{\A/\B}$ and $\Gamma'_{\A/\B}$ on the average and difference coordinates in \cref{eq:bootstrap-correl-T(x+y)} it follows that also the Fourier transformation is separated into average and difference momentum dependencies.
Then, using \cref{eq:tildeB-def}, we evaluate the TMD,
\begin{align}
    \xG_{\A/\B}(\xB=\kmp/\mathcal{P}^\mp,\kperp) &= \frac{2(N_c^2-1)}{(2\pi)^3}\frac{\kperp^2}{\left(\kperp^2 + m^2\right)^2} \frac{1}{N_c^2-1}\langle \tilde{\rho}^a_{\A/\B}(\kmp,\kperp) \tilde{\rho}^a_{\A/\B}(-\kmp,-\kperp) \rangle  \nn\\
    &= \frac{2(N_c^2-1)}{(2\pi)^3}\frac{\kperp^2}{\left(\kperp^2 + m^2\right)^2} \tilde{T}_{\A/\B}(0,\zeroperp) \tilde{\Gamma}'_{\A/\B}(\kmp,\kperp). \label{eq:bootstrap-TMD-T(x+y)}
\end{align}
Note that all IR regulation factors $m$ are explicit.
The Fourier transformation of the single-nucleus envelopes is evaluated at the origin, which yields the projected transverse area of the nuclei given in \cref{eq:nucl-models-Sperp}.
We can now invert \cref{eq:bootstrap-TMD-T(x+y)} and solve for $\tilde{\Gamma}'_{\A/\B}$ in terms of the TMD,
\begin{align}
    \tilde{\Gamma}'_{\A/\B}(\kmp,\kperp) = \frac{(2\pi)^3}{2(N_c^2-2)}\frac{\left( \kperp^2 + m^2 \right)^2}{\kperp^2} \frac{\xG_{\A/\B}(\xB=\kmp/\mathrm{P}^\mp,\kperp)}{\Sp^{\A/\B}}. \label{eq:bootstrap-tildeGamma-TMD-T(x+y)}
\end{align}
The structure of this result is identical to \cref{eq:tildeGamma-TMD}, where we derived the approximate relation of the correlation function and the TMD in the limit of large nuclei.
However, we stress that $\tilde{\Gamma}'_{\A/\B}$ in \cref{eq:bootstrap-tildeGamma-TMD-T(x+y)} is not equivalent to the correlation function that appears in \cref{eq:tildeGamma-TMD}.
They differ by the function $f^T_{\A/\B}$ as given in \cref{eq:bootstrap-T(x+y)-Gamma'}.
Comparing \cref{eq:bootstrap-tildeGamma-TMD-T(x+y)} to \cref{eq:tildeGamma-TMD} lets us anticipate that in the limit of large nuclei, the contribution of $f^T_{\A/\B}$ becomes negligible.
We will revisit this discussion in \cref{sec:pos-semidef-GBW}.

Finally, we study the positive semi-definiteness of the correlator in \cref{eq:bootstrap-correl-T(x+y)} using the correlation function from \cref{eq:bootstrap-tildeGamma-TMD-T(x+y)}.
To this end, we rewrite the correlator using the square roots of the envelopes $t_{\A/\B}$,
\begin{align}
     \langle \rho^a_{\A/\B}(\xpm,\xperp) \rho^b_{\A/\B}(\ypm,\yperp) \rangle &= \delta^{ab} t_{\A/\B}(\xpm,\xperp)t_{\A/\B}(\ypm,\yperp) \nn\\
     &\times f^T_{\A/\B}(\xpm-\ypm,\xperp-\yperp) \Gamma'_{\A/\B}(\xpm-\ypm,\xperp-\yperp).
\end{align}
The condition in \cref{eq:tildeGamma-gt-0} then amounts to
\begin{align}
    &\intop \dd\xpm\,\dd^2\xperp\,f^T_{\A/\B}(\xpm,\xperp) \Gamma'_{\A/\B}(\xpm,\xperp) \ee^{\ii\xpm\kmp - \ii \xperp\cdot\kperp} \nn\\
    &= \intop \frac{\dd\pmp\,\dd^2\pperp}{(2\pi)^3}\, \tilde{f}^T_{\A/\B}(\pmp,\pperp) \tilde{\Gamma}'_{\A/\B}(\kmp-\pmp,\kperp-\pperp)
    \geq 0. \label{eq:bootstrap-T(x+y)-pos-semdef}
\end{align}
We emphasize that it is not enough for the correlation function $\tilde{\Gamma}_{\A/\B}\geq 0$, or equivalently $\xG_{\A/\B}\geq0$.
Instead, the specific combination with the function $f^T_{\A/\B}$ given in \cref{eq:bootstrap-T(x+y)-pos-semdef} needs to be non-negative.
To obtain the second line in \cref{eq:bootstrap-T(x+y)-pos-semdef}, we assumed that the Fourier transformation of $f^T_{\A/\B}$ exists.
In general, this may not be the case and the detour with the position space representation of the correlation function will be necessary to check the condition.%
\footnote{%
In fact, for the example of the factorizable Gaussian envelope in \cref{eq:T(x+y)-factorized}, the Fourier transformation of $f^T_{\A/\B}$ does not exist.}

\subsection{Bootstrapping with general envelopes}

We now generalize the calculation presented in the previous section to single-nucleus envelopes that do not factorize.
The starting point is again the connection between the TMD and the color charge correlator given in \cref{eq:tildeB-def}.
Using the shorthand field $\tilde{B}_{\A/\B}$, the goal is to solve \cref{eq:tildeB-approx-step1}, given as
\begin{align}
    \tilde{B}_{\A/\B}(\pmp,\pperp) &= \intop\frac{\dd\kmp\,\dd^2\kperp}{(2\pi)^3}\,\tilde{t}_{\A/\B}(\kmp,\kperp)\tilde{t}_{\A/\B}(-\kmp,-\kperp) \tilde{\Gamma}_{\A/\B}(\pmp-\kmp,\pperp-\kperp), \tag{\ref{eq:tildeB-approx-step1}}
\end{align}
for the correlation function.
We can disentangle $\tilde{\Gamma}_{\A/\B}$ and the envelopes in position space,
\begin{align}
    B_{\A/\B}(\xpm,\xperp) &= \intop\frac{\dd\pmp\,\dd^2\pperp}{(2\pi)^3}\, \tilde{B}_{\A/\B}(\pmp,\pperp)\nn\\
    &= \intop\frac{\dd\kmp\,\dd^2\kperp\,\dd\qmp\,\dd^2\qperp}{(2\pi)^6}\, \ee^{-\ii \xpm(\kmp+\qmp) + \ii\xperp\cdot(\kperp+\qperp)} \nn\\
    &\times \tilde{t}_{\A/\B}(\kmp,\kperp)\tilde{t}_{\A/\B}(-\kmp,-\kperp) \tilde{\Gamma}_{\A/\B}(\qmp,\qperp) \nn\\
    &= \Gamma_{\A/\B}(\xpm,\xperp) \intop \frac{\dd\kmp\,\dd^2\kperp}{(2\pi)^3}\,\ee^{-\ii\xpm\kmp+\ii\xperp\cdot\kperp}\,\tilde{t}_{\A/\B}(\kmp,\kperp)\tilde{t}_{\A/\B}(-\kmp,-\kperp) \nn\\
    &= \Gamma_{\A/\B}(\xpm,\xperp) \intop \dd\ypm\,\dd^2\yperp\, t_{\A/\B}(\ypm,\yperp)t_{\A/\B}(\ypm-\xpm,\yperp-\xperp).
\end{align}
In the second equation, we shifted the integrations over $\pmp$ as $\qmp = \pmp -\kmp$ and $\pperp$ as $\qperp = \pperp-\kperp$ to separate the momentum variables for the envelopes and the correlation function.
In the third equation, we identified the correlation function in position space and in the last equation, we used that the Fourier transformation of the product of two $\tilde{t}_{\A/\B}$ is equal to a convolution in position space.

We may rearrange for $\Gamma_{\A/\B}$,
\begin{align}
    \Gamma_{\A/\B}(\xpm,\xperp) &= \frac{B_{\A/\B}(\xpm,\xperp)}{\intop \dd\ypm\,\dd^2\yperp\, t_{\A/\B}(\ypm,\yperp)t_{\A/\B}(\ypm-\xpm,\yperp-\xperp)} \nn\\
    &= B_{\A/\B}(\xpm,\xperp) C_{\A/\B}(\xpm,\xperp), \label{eq:Gamma-B-C}
\end{align}
where we introduced the shorthand $C_{\A/\B}$ for the reciprocal of the denominator.
However, this last step is potentially ill-defined and warrants detailed discussion.
The convolution of two $t_{\A/\B}$ as a function of $(\xpm,\xperp)$ is supposed to be well-behaved.
Assuming that the envelopes fall off on a scale set by the nuclear radius, their convolution will also fall off on a scale proportional to that radius.
Contrarily, the behavior of $C_{\A/\B}$ as a function of $(\xpm,\xperp)$ is the opposite.
It is minimal for $(\xpm,\xperp) = (0,\zeroperp)$ and steeply rises when its arguments become comparable to the radius's scale.
Clearly, $\Gamma_{\A/\B}$ can only exist if $B_{\A/\B}$ (related to the TMD in momentum space via \cref{eq:tildeB-def}) is able to regulate this seeming divergence of $C_{\A/\B}$.

Considering the role of $\Gamma_{\A/\B}$ as a measure for the correlation between two points in position space, $\Gamma_{\A/\B}$ is evaluated at the separation of these points in the color charge correlator.
Intuitively, one would expect that the correlations decrease for sufficiently large separations.
This would correspond to a strong regulation of the divergence of $C_{\A/\B}$ by $B_{\A/\B}$.
If that were not the case, correlations would increase for larger separations and eventually diverge.
The latter case, although not excluded by \cref{eq:Gamma-B-C}, is unphysical and imposes a strict constraint on the compatibility of any given TMD with the envelope of choice for model building of the correlation function.
Practically, this constraint can be formulated as a lower bound for the envelope scale (via the radius) given a scale in $B_{\A/\B}$ set by the TMD.
An illustrative example is discussed in \cref{sec:pos-semidef-GBW}.

Note that the possible divergence of \cref{eq:Gamma-B-C} for large $(\xpm,\xperp)$ is not real for the full color charge correlator, even in the case of envelopes with strict compact support that evaluate to 0 exactly.
To see this, consider the simple case of cubic nuclei with transverse edge length $2R$ and longitudinal edge length $2R_l$ that are placed at the coordinate origin,
\begin{align}
    t_{\A/\B}(\xpm,\xperp) \propto \theta(|\xpm| - R_l) \theta(|x^x|-R) \theta(|x^y| - R).
\end{align}
The convolution in $C_{\A/\B}$ will evaluate to zero when $|\xpm| = 2R_l$ or $|x^\bi| = 2R$ and $\Gamma_{\A/\B}$ will diverge.
However, the correlation function is part of the color charge correlator, where it is multiplied by the same envelopes,
\begin{align}
    \langle \rho^a_{\A/\B}(\xpm,\xperp) \rho^b_{\A/\B}(\ypm,\yperp) \rangle &= \delta^{ab}\, t_{\A/\B}(\xpm,\xperp)t_{\A/\B}(\ypm,\yperp) \Gamma_{\A/\B}(\xpm-\ypm,\xperp-\yperp).
\end{align}
If we evaluate $\Gamma_{\A/\B}$ at a separation $|x^\bi-y^\bi| = 2R$, this means that $|x^\bi|\geq R$ or $|y^\bi| \geq R$, and one (or both) of the envelope factors will evaluate to zero and cut off the divergence.
The situation for the longitudinal argument is analogous.
We conclude this discussion with the remark that \cref{eq:Gamma-B-C} will not lead to a divergent correlator in the general case.
Still, it is not a fail-safe recipe for creating physical correlation functions from arbitrary combinations of TMDs and envelopes.

Acknowledging the caveats of \cref{eq:Gamma-B-C}, we continue with the determination of the model.
We insert the Fourier transformation of \cref{eq:Gamma-B-C} into the color charge correlator in momentum space given in \cref{eq:FT-general-correlator-t(x)-t(y)},
\begin{align}
    \langle \tilde{\rho}^a_{\A/\B}(\pmp,\pperp)\tilde{\rho}^b_{\A/\B}(\qmp,\qperp)\rangle &= \delta^{ab}\intop \frac{\dd\kappa^\mp\,\dd^2\kappaperp\,\dd\xpm\,\dd^2\xperp}{(2\pi)^3} \nn\\
    &\times \tilde{t}_{\A/\B}(\pmp-\kappa^\mp,\pperp-\kappaperp)\tilde{t}_{\A/\B}(\qmp+\kappa^\mp,\qperp+\kappaperp) \nn\\
    &\times B_{\A/\B}(\xpm,\xperp)C_{\A/\B}(\xpm,\xperp) \ee^{\ii\xpm\kappa^\mp- \ii\xperp\cdot\kappaperp} \nn\\
    &= \delta^{ab} \intop\frac{\dd\kmp\,\dd^2\kperp}{(2\pi)^3}\, \tilde{B}_{\A/\B}(\kmp,\kperp) \nn\\
    &\times \intop\dd\upm\,\dd^2\uperp\,\dd\vpm\,\dd^2\vperp\, \nn\\
    &\times \ee^{\ii\upm\pmp-\ii\uperp\cdot\pperp + \ii\vpm\qmp - \ii\vperp\cdot\qperp -\ii\kmp(\upm-\vpm) + \ii\kperp(\uperp-\vperp)} \nn\\
    &\times t_{\A/\B}(\upm,\uperp)t_{\A/\B}(\vpm,\vperp) C_{\A/\B}(\upm-\vpm,\uperp-\vperp). \label{eq:FT-rhorho-C}
\end{align}
To obtain the second equation, we first inserted the Fourier transformation of $B_{\A/\B}$ and the position space representations of the envelopes $t_{\A/\B}$.
Then, we evaluated the integrals over $\kappa^\mp$, $\kappaperp$, $\xpm$ and $\xperp$.
We introduce the shorthand
\begin{align}
    K_{\A/\B}(\xpm,\ypm,\xperp,\yperp) = t_{\A/\B}(\xpm,\xperp)t_{\A/\B}(\ypm,\yperp)C_{\A/\B}(\xpm-\ypm,\xperp-\yperp),
\end{align}
whose Fourier transformation replaces the last three lines of \cref{eq:FT-rhorho-C},
\begin{align}
    \langle \tilde{\rho}^a_{\A/\B}(\pmp,\pperp)\tilde{\rho}^b_{\A/\B}(\qmp,\qperp)\rangle &= \delta^{ab} \intop\frac{\dd\kmp\,\dd^2\kperp}{(2\pi)^3}\, \tilde{B}_{\A/\B}(\kmp,\kperp) \nn\\
    &\times \tilde{K}_{\A/\B}(\pmp-\kmp,\qmp+\kmp,\pperp-\kperp,\qperp+\kperp).
\end{align}
Then, we perform a shift of the integration variables,
\begin{align}
    \kappa^\mp = \kmp - \frac{1}{2}(\pmp-\qmp), \qquad \kappaperp = \kperp - \frac{1}{2}(\pperp-\qperp),
\end{align}
to finally obtain
\begin{align}
    \langle \tilde{\rho}^a_{\A/\B}(\pmp,\pperp)\tilde{\rho}^b_{\A/\B}(\qmp,\qperp)\rangle &= \delta^{ab} \intop\frac{\dd\kappa^\mp\,\dd^2\kappaperp}{(2\pi)^3}\, \tilde{B}_{\A/\B}(\tfrac{\pmp-\qmp}{2}+\kappa^\mp,\tfrac{\pperp-\qperp}{2}+\kappaperp) \nn\\
    &\times \tilde{K}_{\A/\B}(\tfrac{\pmp+\qmp}{2}-\kappa^\mp,\tfrac{\pmp+\qmp}{2}+\kappa^\mp,\tfrac{\pperp+\qperp}{2}-\kappaperp,\tfrac{\pperp+\qperp}{2}+\kappaperp). \label{eq:FT-correlator-tB-tK}
\end{align}

According to this result, the color charge correlator is given as the convolution in momentum space of the two objects, $\tilde{B}_{\A/\B}$ and $\tilde{K}_{\A/\B}$.
These two factors cleanly separate into the contributions from the TMD (via $\tilde{B}_{\A/\B}$) and a pure geometry factor $\tilde{K}_{\A/\B}$ that is fixed by the envelopes.
The specific combination of how the momenta of the correlator enter the arguments of $\tilde{K}_{\A/\B}$ shows that its contribution to the charge correlator is isolated along the diagonal in phase space where $\pmp=\qmp$ and $\pperp=\qperp$.
Along the orthogonal diagonal, where $\pmp = -\qmp$ and $\pperp=-\qperp$, the contribution of $\tilde{B}_{\A/\B}$ is isolated.
This contribution of $\tilde B_{\A/\B}$ is ``smeared'' into the full $(p^\pm,q^\mp,\pperp,\qperp)$ phase space by the geometry factor $\tilde K_{\A/\B}$.


The explicit expression for the Fourier transformation of the geometry factor $K_{\A/\B}$ reads,
\begin{align}
    \tilde{K}_{\A/\B}(\pmp,\qmp,\pperp,\qperp) &= \intop \dd\upm\,\dd^2\uperp\, \ee^{-\ii \qmp\upm + \ii\qperp\cdot\uperp}\intop \dd\xpm\,\dd^2\xperp\,\ee^{\ii \xpm(\pmp+\qmp) - \ii\xperp\cdot(\pperp+\qperp)} \nn\\
    &\times \frac{t_{\A/\B}(\xpm,\xperp)t_{\A/\B}(\xpm-\upm,\xperp-\uperp)}{\intop\dd\zpm\,\dd^2\zperp\,t_{\A/\B}(\zpm,\zperp)t_{\A/\B}(\zpm-\upm,\zperp-\uperp)}, \label{eq:FT-K-def}
\end{align}
from which it follows in full generality that
\begin{align}
    \tilde K(-\kappa^\mp,\kappa^\pm,-\kappaperp,\kappaperp) = (2\pi)^3 \delta(\kappa^\mp)\deltaperp(\kappaperp),
\end{align}
and \cref{eq:FT-correlator-tB-tK} recoveres by construction the definiton of $\tilde{B}_{\A/\B}$ from \cref{eq:tildeB-def} (and the TMD).

As noted earlier, only the combined choice of a TMD and single-nucleus envelope uniquely determines $\tilde \Gamma(k^\mp,\kperp)$ from the Fourier transformation of \cref{eq:Gamma-B-C}.
A priori, it might not be clear that the combination of any given TMD and envelope leads to a valid color charge correlator, even if $\Gamma_{\A/\B}$ is well regulated and its Fourier transformation exists.
The problem of the model building presented in this section is that the positive semi-definiteness of the resulting color charge correlator is not enforced.
For a positive semi-definite correlator, it is sufficient that $\tilde{\Gamma}_{\A/\B}\geq 0$ (cf.\ \cref{eq:tildeGamma-gt-0}).
However, this condition can only be checked after the model has been assembled and $\tilde{\Gamma}_{\A/\B}$ can be evaluated.

\section{Phenomenological GBW saturation model}\label{sec:GBW-saturation-model}

The TMD named after Golec-Biernat and W\"usthoff (GBW)~\cite{Golec-Biernat:1998zce,Golec-Biernat:1999qor,Golec-Biernat:2017lfv} reads
\begin{align}
    \xG(\xB,\kperp) &= \frac{2N_c\, \Sp}{\pi^2 g^2} \frac{\kperp^2}{Q_s(\xB)^2}\ee^{-\frac{\kperp^2}{Q_s(\xB)^2}} \label{eq:GBW-TMD}.
\end{align}
Here,
\begin{align}
 \label{eq:Qs}   
 Q_s(\xB) = Q_0\xB^{-\lambda}(1-\xB),
\end{align}
is the saturation momentum parametrized by the momentum fraction $\xB\in (0,1]$.
The constant $\lambda = 0.144$ was originally obtained by fitting the exponent of the \mbox{$\sim\xB^{-\lambda}$} behavior at small $\xB$ to HERA deep inelastic scattering data.
The extension of the simple power law with the factor $(1-\xB)$ is commonly used to regulate $Q_s$ for large $\xB$~\cite{Garcia-Montero:2023gex,Schlichting:2020wrv,Andronic:2025ylc} and extends the GBW model past the small-$\xB$ regime relevant for highly saturated systems.
The prefactor $Q_0$ is given in units of GeV and treated as a model parameter.
\Cref{eq:GBW-TMD} is only a function of the ratio $\kperp^2/Q_s(\xB)^2$, which connects the transverse and longitudinal energy dependencies in a specific way known as geometric scaling~\cite{Stasto:2000er}.
The transverse area $\Sp$ is defined in \cref{eq:nucl-models-Sperp} and contains the residual dependence of the TMD on the specific envelopes%
\footnote{%
It can also be seen as a broad constraint on the envelopes to reproduce a certain transverse area, while their shape is completely unconstrained.}
and does not scale with the longitudinal size of the nucleus.
This is a consequence of the normalization of the envelopes discussed with \cref{eq:envelope-norm}.
In contrast to the original GBW model, here, the TMD has to be considered as a density w.r.t.\ the longitudinal dimension.

The GBW TMD in \cref{eq:GBW-TMD} is IR-safe without explicit regulation and usually does not contain regulation factors.
However, it turns out that the $\sim \kperp^2$ IR behavior is unable to regulate the IR divergence of the integrand of the three-dimensional gluon production formula in \cref{eq:dNd2kdY-TMDs}.%
\footnote{%
Still, these poles are integrable in \cref{eq:dNd2kdY-TMDs} as long as $|\kperp|\neq 0$.}
For practical reasons, we include IR regulation also for the GBW model%
\footnote{%
An alternative way to understand this is that $\tilde{B}_{\A/\B}$ is defined in \cref{eq:tildeB-def} via the bare color charge distributions.
It is also $\tilde{B}_{\A/\B}$ that is used to define the effective TMD $\xGeff_{\A/\B}$ in \cref{eq:xGeff-xpq-def}.
The IR regulation factors of the color charges are pushed into the TMD.
\Cref{eq:GBWIR-TMD} reduces to \cref{eq:GBW-TMD} when $m=0$ but for $\tilde B_{\A/\B}$ in \cref{eq:tildeB-def} we need to obtain the same expression, regardless whether $m\neq0$ or $m=0$ to be consistent.}
and define the IR-regulated GBW TMD
\begin{align}
    \xG^\mathrm{GBW}(\xB,\kperp) &\coloneqq \frac{2N_c\, \Sp}{\pi^2 g^2} \frac{\kperp^6}{Q_s(\xB)^2(\kperp^2+m^2)^2}\ee^{-\frac{\kperp^2}{Q_s(\xB)^2}} \label{eq:GBWIR-TMD}.
\end{align}

Now, we use this TMD from \cref{eq:GBWIR-TMD} for building a nuclear model according to \cref{eq:FT-correlator-tB-tK} discussed in the previous section.
Here, we only use the Gaussian envelope from \cref{eq:envelope-gauss-def}.
The Fourier transformation of the geometry factor defined in \cref{eq:FT-K-def} then yields
\begin{align}
    \tilde K_{\A/\B}(\tfrac{p^\mp+q^\mp}{2}-\kappa^\mp, \tfrac{p^\mp+q^\mp}{2}+\kappa^\mp, \tfrac{\pperp+\qperp}{2}-\kappaperp, \tfrac{\pperp+\qperp}{2}+\kappaperp) &\nn\\
    &\hspace{-200.7pt}= \intop \dd\upm\,\dd^2\uperp\,\ee^{-\ii\upm(\frac{\pmp+\qmp}{2}+\kappa^\mp) +\ii\uperp\cdot(\frac{\pperp+\qperp}{2}+\kappaperp)}\intop\dd\xpm\,\dd^2\xperp\,\ee^{\ii\xpm(\pmp+\qmp)-\ii\xperp\cdot(\pperp+\qperp)} \nn\\
    &\hspace{-200.7pt}\times \frac{\sqrt{H_{\A/\B}(\xpm,\xperp)}\sqrt{H_{\A/\B}(\xpm-\upm,\xperp-\uperp)}}{2\pi L_\perp^2 \exp(-\tfrac{(\upm)^2}{2L_l^2}-\tfrac{\uperp^2}{8L_\perp^2})} \nn\\
    &\hspace{-200.7pt}= \intop \dd\upm\,\dd^2\uperp\,\ee^{-\ii\upm(\frac{\pmp+\qmp}{2}+\kappa^\mp) +\ii\uperp\cdot(\frac{\pperp+\qperp}{2}+\kappaperp)}\nn\\
    &\hspace{-200.7pt}\times \ee^{+\ii\upm\left(\tfrac{\pmp+\qmp}{2}\right)-\ii\uperp\cdot\left(\tfrac{\pperp+\qperp}{2}\right)}\ee^{-\tfrac{L_l^2}{2}\left(\tfrac{\pmp+\qmp}{2}\right)^2-2L_\perp^2\left(\tfrac{\pperp+\qperp}{2}\right)^2} \nn\\
    &\hspace{-200.7pt}= (2\pi)^3 \delta(\kappa^\mp)\deltaperp(\kappaperp)\, \ee^{-\frac{L_l^2}{2}\left(\frac{p^\mp+q^\mp}{2}\right)^2}\ee^{-2L_\perp^2\left(\frac{\pperp+\qperp}{2}\right)^2}.
    \label{eq:tildeK-gauss}
\end{align}
When inserting $\tilde K_{\A/\B}$ from \cref{eq:tildeK-gauss} into the general form of the Fourier-transformed color charge correlator in \cref{eq:FT-correlator-tB-tK}, the convolutions over $\kappa^\mp$ and $\kappaperp$ factorize into a product of $\tilde B_{\A/\B}$ and the exponentials from \cref{eq:tildeK-gauss} because of the Dirac-delta functions.
The result is the color charge correlator of the GBW model in momentum space
\begin{align}
    \langle\tilde \rho^a_{\A/\B}(p^\mp,\pperp) \tilde \rho^b_{\A/\B}(q^\mp,\qperp) \rangle &= \delta^{ab}\frac{8\pi N_c\, \Sp^{\A/\B}}{(N_c^2-1)g^2} \ee^{-\frac{L_l}{2}\left(\frac{\pmp+\qmp}{2}\right)^2}\,\ee^{-2L_\perp^2\left(\frac{\pperp+\qperp}{2}\right)^2}\nn \\
    &\times \left(\frac{\pperp-\qperp}{2}\right)^4 \frac{1}{Q_s(\bar{\xB}_{\A/\B})^2}\exp(-\left(\tfrac{\pperp-\qperp}{2}\right)^2\tfrac{1}{Q_s(\bar{\xB}_{\A/\B})^2}), \label{eq:FT-correlator-GBW}
\end{align}
where the momentum fraction $\bar{\xB}_{\A/\B}$ is evaluated w.r.t\ the difference of the longitudinal momenta in the correlator,
\begin{align}
    \bar{\xB}_{\A/\B} = \frac{|\pmp-\qmp|}{2\mathcal{P}^\mp}.
\end{align}
Even though $Q_s$ only enters quadratically, taking the absolute value of the momentum difference is necessary to ensure that the parametrization of $Q_s$ in \cref{eq:Qs} evaluates to positive numbers.
Additionally, the large-$\xB$ regulation is formulated for positive $\xB$.
Because of the kinematic constraint $\bar{\xB}_{\A/\B}\in(0,1]$, the longitudinal separation in \cref{eq:FT-correlator-GBW}
\begin{align}
    |\pmp-\qmp| \leq 2\mathcal{P}^\mp
\end{align}
and is limited by the total nucleon momentum $\mathcal{P}^\mp$.
This is a hard cutoff and in this limit, the color charge correlator evaluates to zero as $Q_s\rightarrow0$.
However, the bounds on $\bar{\xB}_{\A/\B}$ are not enough to guarantee positive semi-definiteness in accordance with \cref{eq:tildeGamma-gt-0}.
We will study the positive semi-definiteness condition for a simplified GBW model, where the transverse nuclear scale is taken to infinity, in \cref{sec:pos-semidef-GBW}.
For practical reasons, we only enforce the kinematic limit on $|\pmp-\qmp|$ for the results presented in \cref{sec:eff-TMDs,sec:results-gluon-numbers}.

The effective TMD that results from the correlator in \cref{eq:FT-correlator-GBW} reads
\begin{align}
     \frac{\xGeff_{\A/\B}(\xB_{\A/\B},\pperp,\qperp)}{\Sp^{\A/\B}} &= \frac{N_c\,\Sp^{\A/\B}}{2(2\pi)^4g^2} \frac{\pperp^2\qperp^2\left(\pperp-\qperp\right)^4}{Q_s(\xB_{\A/\B})^2(\pperp^2+m^2)(\qperp^2+m^2)} \nn\\
     &\times \exp(-2L_\perp^2\left(\tfrac{\pperp+\qperp}{2}\right)^2- \tfrac{1}{Q_s(\xB_{\A/\B})^2}\left(\tfrac{\pperp-\qperp}{2}\right)^2),
\end{align}
and integrating out the momentum $\qperp$ yields the reduced effective TMD,
\begin{align}
    \frac{\xGeff_{\A/\B}(\xB_{\A/\B},\kperp)}{\Sp^{\A/\B}} &= \frac{N_c\,\Sp^{\A/\B}}{2(2\pi)^2g^2} \frac{1}{Q_s(\xB_{\A/\B})^2}\frac{1}{\kperp^2+m^2} \nn\\
    &\times \intop\frac{\dd^2\pperp}{(2\pi)^2}\frac{\pperp^4(\kperp-\pperp)^2}{(\kperp-\pperp)^2+m^2}\exp(-\tfrac{L_\perp^2}{2}(2\kperp-\pperp)^2 - \tfrac{\pperp^2}{4Q_s(\xB_{\A/\B})^2}), \label{eq:GBW-effTMD}
\end{align}
where we shifted $\pperp = \kperp-\qperp$.
Note that the dependence on the longitudinal momentum is encoded in the dependence of $Q_s$ on the momentum fractions $\xB_{\A/\B}$, given in \cref{eq:xB-Y-AB}.
Additionally, only the transverse nuclear radius $L_\perp$ enters in the effective TMD.
The longitudinal radius drops out when evaluating the correlator in \cref{eq:FT-correlator-GBW} at $\pmp=-\qmp$ for the effective TMD, which is special to the choice of Gaussian envelopes.

\subsection{Positive semi-definiteness for transversely infinite nuclei}\label{sec:pos-semidef-GBW}

We investigate the positive semi-definiteness condition (cf.~\cref{sec:pos-semi-definite-correlators}) for the GBW saturation model.
The goal is to calculate the Fourier transformation of the correlation function $\Gamma_{\A/\B}$.
To obtain straightforward results and keep analytic control of the relevant parameters, we will assume a simplified GBW model where the saturation scale $Q_s$ follows a simpler parametrization.
Still, the important feature of geometric scaling is conserved.
The obtained results will, therefore, only serve as an illustrative example of how the different scales combine in the result.
The equivalent analysis with the phenomenologically tuned GBW model will require numerical methods and be less transparent.
Additionally, to reduce the number of scales relevant for the calculation, we consider the case of transversely infinite nuclei.
We formally split the Gaussian envelope in \cref{eq:envelope-gauss-def} into two parts,
\begin{align}
    T_{\A/\B}(\xpm,\xperp) = H_{\A/\B}(\xpm,\zeroperp) H^\perp_{\A/\B}(\xperp), \label{eq:H-transv-inf}
\end{align}
where the first factor corresponds to the limit $L_\perp\rightarrow\infty$ of $H_{\A/\B}$ given in \cref{eq:envelopes-perp-to-inf}.
The second factor is equal to the projected transverse envelope,
\begin{align}
    \intop \dd\xpm\,T_{\A/\B}(\xpm,\xperp) = T^\perp_{\A/\B}(\xperp) = H^\perp_{\A/\B}(\xperp)
\end{align}
and reduces to unity when the limit is enforced.
For now, we keep $H^\perp_{\A/\B}$ as a means of bookkeeping.

Next, we re-derive \cref{eq:Gamma-B-C} with the envelopes from \cref{eq:H-transv-inf} in the transversely infinite limit.
Recall \cref{eq:tildeB-scales} where we derived the relation between $\tilde{B}_{\A/\B}$ and the Fourier-transformed color charge correlator in the limit of large nuclei.
For the current discussion, we have not specified the relation between the longitudinal scales from the correlation function and the envelopes.
It is important that we keep those scales general.
However, we have specified that the transverse envelope scale is infinite, which allows the partial factorization
\begin{align}
    \tilde{B}_{\A/\B}(\pmp,\pperp) &= \intop\frac{\dd\kmp\,\dd^2\kperp}{(2\pi)^3}\,\tilde{t}_{\A/\B}(\kmp,\kperp)\tilde{t}_{\A/\B}(-\kmp,-\kperp) \tilde{\Gamma}_{\A/\B}(\pmp-\kmp,\pperp-\kperp) \nn\\
    &= \intop\frac{\dd\kmp}{2\pi}\,\tilde{\Gamma}_{\A/\B}(\pmp-\kmp,\pperp)\intop\frac{\dd^2\kperp}{(2\pi)^2}\,\tilde{t}_{\A/\B}(\kmp,\kperp)\tilde{t}_{\A/\B}(-\kmp,\kperp).
\end{align}
After inserting the envelopes from \cref{eq:H-transv-inf}, we can isolate the projected transverse area $\Sp^{\A/\B}$ as follows,
\begin{align}
    \tilde{B}_{\A/\B}(\pmp,\pperp) &= \intop\frac{\dd\kmp}{2\pi}\,\tilde{\Gamma}_{\A/\B}(\pmp-\kmp,\pperp) \nn\\
    &\times \intop\frac{\dd^2\kperp\,\dd\xpm\,\dd\ypm\,\dd^2\xperp\,\dd^2\yperp}{(2\pi)^2}\, \ee^{\ii\kmp(\xpm-\ypm)-\ii\kperp\cdot(\xperp-\yperp)} \nn\\
    &\times \sqrt{H_{\A/\B}(\xpm,\zeroperp)H^\perp_{\A/\B}(\xperp)}\sqrt{H_{\A/\B}(\ypm,\zeroperp)H^\perp_{\A/\B}(\yperp)} \nn\\
    &= \intop\frac{\dd\kmp}{2\pi}\,\tilde{\Gamma}_{\A/\B}(\pmp-\kmp,\pperp) \intop \dd^2\xperp\, \sqrt{H^\perp_{\A/\B}(\xperp)}\sqrt{H^\perp_{\A/\B}(\xperp)}\nn\\
    &\times \intop\dd\xpm\,\dd\ypm\,\dd^2\xperp\, \ee^{\ii\kmp(\xpm-\ypm)} \sqrt{H_{\A/\B}(\xpm,\zeroperp)}\sqrt{H_{\A/\B}(\ypm,\zeroperp)} \nn\\
     &= \Sp^{\A/\B} \intop\frac{\dd\kmp}{2\pi}\,\tilde{\Gamma}_{\A/\B}(\pmp-\kmp,\pperp) \nn\\
    &\times \intop\dd\xpm\,\dd\ypm\,\dd^2\xperp\, \ee^{\ii\kmp(\xpm-\ypm)} \sqrt{H_{\A/\B}(\xpm,\zeroperp)}\sqrt{H_{\A/\B}(\ypm,\zeroperp)}.
\end{align}
Here, the integration over $\kperp$ generated two Dirac-delta functions, which set the transverse arguments of $H^\perp_{\A/\B}$ to the same coordinates in the second equation.
We then identified the remaining integration over the transverse coordinate $\xperp$ with the projected transverse area $\Sp^{\A/\B}$ in \cref{eq:nucl-models-Sperp}.
When transforming this result to position space, \cref{eq:Gamma-B-C} changes to
\begin{align}
    \Gamma_{\A/\B}(\xpm,\xperp) = \frac{B_{\A/\B}(\xpm,\xperp)}{\Sp^{\A/\B}\intop\dd\ypm\,\sqrt{H_{\A/\B}(\ypm,\zeroperp)}\sqrt{H_{\A/\B}(\ypm-\xpm,\zeroperp)}}. \label{eq:Gamma-B-C-transv-inf}
\end{align}
In comparison to \cref{eq:Gamma-B-C}, the limit of transversely infinite nuclei leads to the factorization of the integral in the denominator (the reciprocal of the $C_{\A/\B}$ function) into $\Sp^{\A/\B}$ and the convolution of the longitudinal parts of the envelopes.
As discussed previously, \cref{eq:Gamma-B-C-transv-inf} is only well-defined if the scales of $B_{\A/\B}$ and the envelopes in the denominator are well separated.
In the extreme limit, where the transverse scale of the envelopes is infinite, the transverse scale of $B_{\A/\B}$ becomes irrelevant because of the factorization into $\Sp^{\A/\B}$.
Still, the longitudinal scale of $B_{\A/\B}$ will be crucial to identify the physical validity of the model.

To this end, we need to Fourier transform \cref{eq:Gamma-B-C-transv-inf}.
First, we evaluate the denominator,
\begin{align}
    \intop\dd\ypm\,\sqrt{H_{\A/\B}(\ypm,\zeroperp)}\sqrt{H_{\A/\B}(\ypm-\xpm,\zeroperp)} = \ee^{-\frac{(\xpm)^2}{2L_l^2}},
\end{align}
and use the original GBW TMD from \cref{eq:GBW-TMD} that does not contain the IR regulation factor,%
\footnote{%
Since gluon production in the transversely infinite case involves the TMDs and not the effective TMDs in \cref{eq:dNd2kdY-large-nucl-TMDs}, IR regulation is not necessary to obtain finite results.}
\begin{align}
    \tilde{B}_{\A/\B}(\kmp=\xB_{\A/\B}\mathcal{P}^\mp,\kperp) = \Sp^{\A/\B}\frac{8\pi N_c}{g^2(N_c^2-1)} \frac{\kperp^4}{Q_s(\xB_{\A/\B})^2}\ee^{-\kperp^2/Q_s(\xB_{\A/\B})^2}.
\end{align}
Putting these parts together, the Fourier transformation of \cref{eq:Gamma-B-C-transv-inf} reads
\begin{align}
    \tilde{\Gamma}_{\A/\B}(\kmp,\kperp) &= \frac{1}{\Sp^{\A/\B}}\intop\dd\xpm\,\ee^{\ii\xpm\kmp} \ee^\frac{(\xpm)^2}{2L_l^2} \nn\\
    &\times \intop \frac{\dd^2\xperp\,\dd\kappa^\mp\,\dd^2\kappaperp}{(2\pi)^3}\,\ee^{-\ii\kappa^\mp\xpm -\ii\xperp\cdot(\kperp - \kappaperp)}\tilde{B}_{\A/\B}(\kappa^\mp,\kappaperp) \nn\\
    &= \frac{8\pi N_c}{g^2(N_c^2-1)} \intop\frac{\dd\xpm\,\dd\kappa^\mp}{2\pi}\, \ee^{\ii\xpm(\kmp-\kappa^\mp)}\ee^\frac{(\xpm)^2}{2L_l^2} \frac{\kperp^4}{Q_s(\kappa^\mp/\mathcal{P}^\mp)^2}\ee^{-\frac{\kperp^2}{Q_s(\kappa^\mp/\mathcal{P}^\mp)^2}}.
\end{align}

Furthermore, we assume a parametrization of $Q_s$ that is simple enough to allow for a fully analytic treatment of the Fourier transformation,
\begin{align}
    Q_s(\xB=\kmp/\mathcal{P}^\mp) = Q_0 \frac{\mathcal{P}^\mp}{|\kmp|}.
\end{align}
This corresponds to the scaling exponent $\lambda = 1$ and does not contain the large $\xB$ regulation factor from \cref{eq:Qs}.
We also extend the domain of $Q_s$ to negative longitudinal momentum arguments and use the absolute value to enforce $Q_s>0$ symmetrically.
Then, we can perform the Fourier transformation w.r.t\ $\kappa^\mp$,
\begin{align}
    \tilde{\Gamma}_{\A/\B}(\kmp,\kperp) &= \frac{8\pi N_c}{g^2(N_c^2-1)} \intop\frac{\dd\xpm\,\dd\kappa^\mp}{2\pi}\, \ee^{\ii\xpm(\kmp-\kappa^\mp)}\ee^\frac{(\xpm)^2}{2L_l^2} \frac{\kperp^4(\kappa^\mp)^2}{Q_0^2(\mathcal{P}^\mp)^2}\ee^{-\frac{\kperp^2(\kappa^\mp)^2}{Q_0^2(\mathcal{P}^\mp)^2}} \nn\\
    &= \frac{\sqrt{\pi}N_c\, Q_0 \mathcal{P}^\mp}{g^2(N_c^2-1)}\intop \dd\xpm\,\ee^{\ii\xpm\kmp}\frac{2\kperp^2 - (Q_0\mathcal{P}^\mp \xpm)^2}{|\kperp|}\, \ee^{-(\xpm)^2\frac{(L_l Q_0 \mathcal{P}^\mp)^2 - 2\kperp^2}{4L_l^2\kperp^2}}.
\end{align}
From this intermediate result, we can already derive a constraint on the model parameters.
For $\tilde{\Gamma}_{\A/\B}$ to exist and the remaining Fourier transformation to be well-defined, the exponent of the exponential with $(\xpm)^2$ has to be negative.
Otherwise, the integrand would blow up for large $|\xpm|$.
This leads to a condition for the longitudinal envelope scale $L_l$,
\begin{align}
    L_l > \sqrt{2}\frac{|\kperp|}{Q_0 \mathcal{P}^\mp} \quad \Longleftrightarrow \quad L>\frac{|\kperp|}{\sqrt{2}Q_0 m_n}. \label{eq:Ll-condtition-GBW}
\end{align}
Using $L_l = \sqrt{2} L / \gamma$ and $\mathcal{P}^\mp = \sqrt{2} \gamma m_n$, we see that the dependence on the collider energy drops out.
The parameter $L$ is the radius parameter of the envelope in the rest frame of the nuclei.
\Cref{eq:Ll-condtition-GBW} can be interpreted as a lower bound for $L$.
Remarkably, this lower bound depends on the transverse momentum $\kperp$ as a consequence of geometric scaling in the GBW TMD.
It demonstrates how, in a simplified model, the scales of the envelope and TMD mix in the calculation of $\tilde{\Gamma}_{\A/\B}$.
Note that both the envelope and GBW TMD contain Gaussian exponential factors, which allow the clean comparison of the involved scales in \cref{eq:Ll-condtition-GBW}.

Assuming that \cref{eq:Ll-condtition-GBW} is satisfied for the relevant values of the transverse momentum $\kperp$, we perform the remaining Fourier transformation and obtain
\begin{align}
    \tilde{\Gamma}_{\A/\B}(\kmp,\kperp) &= \frac{8\pi N_c}{g^2(N_c^2-1)}\,\ee^{-(\kmp)^2\frac{\kperp^2 L_l^2}{(L_l Q_0\mathcal{P}^\mp)^2 - 2\kperp^2}}\frac{\kperp^4 L_l Q_0 \mathrm{P}^\mp}{\left((L_l Q_0 \mathcal{P}^\mp)^2 - 2\kperp^2\right)^{5/2}} \nn\\
    &\times \left( 2\kperp^2 - (L_l Q_0 \mathcal{P}^\mp)^2 + (\kmp)^2 L_l^2(L_l Q_0 \mathcal{P}^\mp)^2 \right)
\end{align}
Note that \cref{eq:Ll-condtition-GBW} ensures that the exponent is negative and that the root in the denominator is real.
Now, we can check the condition for positive semi-definiteness, $\tilde{\Gamma}_{\A/\B} \geq 0$, which holds when the term in parentheses in the second line is non-negative.
Solving the inequality for the longitudinal momentum reads
\begin{align}
    |\kmp| \geq \frac{1}{L_l} \sqrt{1-\frac{2\kperp^2}{(L_l Q_0 \mathcal{P}^\mp)^2}} = \frac{\gamma}{\sqrt{2}L}\sqrt{1-\frac{\kperp^2}{2(Q_0 L m_n)^2}}, \label{eq:kmp-condition-GBW}
\end{align}
where we used the expressions for $L_l$ and $\mathcal{P}^\mp$ in terms of the $\gamma$ factor.
We can interpret \cref{eq:kmp-condition-GBW} as a condition for the IR behavior of $\kmp$ in this model.
If the longitudinal momentum drops below this threshold, the model becomes non-physical.
This suggests an incompatibility with the implicit IR regulation imposed by the nuclear envelopes and the IR behavior of the GBW TMD.
Since the scales in the GBW TMD are momentum dependent, it is inevitable that for some region of the phase space, the scales of the envelope become comparable to the scales of the TMD.
We conclude that without further treatment, the longitudinal IR behavior in this simplified GBW model is unphysical.

Finally, we comment on the case where the scales of the nuclear envelopes are well separated from the scales of the correlation function.
As derived earlier in \cref{eq:tildeB-Gamma-sp}, this clear separation of scales leads to a straightforward relation of $\tilde{\Gamma}_{\A/\B}$ and the TMD $\xG$ (via $\tilde{B}_{\A/\B}$).
Then, it follows that the model is positive semi-definite if $\xG \geq 0$.
While the GBW TMD is non-negative, its property of geometric scaling does not allow for a clean separation from the scales of the envelopes.
If we also push $L_l \rightarrow \infty$, the constraints in \cref{eq:Ll-condtition-GBW,eq:kmp-condition-GBW} hold in the entire phase space and reconcile the current discussion with the assumptions used to derive \cref{eq:tildeB-Gamma-sp}.

\section{Numerical results for effective TMDs}\label{sec:eff-TMDs}

In this section, we discuss numerical results for the TMDs $\xG$ and reduced effective TMDs $\xGeff$ that result from the nuclear models introduced in this chapter.
The evaluation of one-dimensional integrals was done using Mathematica%
\footnote{%
See \cref{appx:software} for details on the tools used.\label{ft:software}}
with default parameters for the \emph{NIntegrate} routine.
The evaluation of multi-dimensional integrals was done using the Mathematica bindings of the CUBA Monte Carlo integration library.\footref{ft:software}
The targeted relative error estimate for these cases was set to $0.1\%$.

\begin{table}
    \vspace{-.6\baselineskip}\caption[Model parameters for effective TMDs.]{\label{tab:nucl-model-params}%
    Values of physical model parameters used for numerical evaluation of TMDs.
    Collider energy and nuclear species correspond to Au+Au collisions at $\sqrt{s_\mathrm{NN}}=200$~GeV performed at RHIC.
    WS $R$ and $d$ are taken from~\cite{Schenke:2012hg}.
    The Gauss radius $L$ is tuned s.t.\ both envelopes defined in \cref{eq:envelope-gauss-def,eq:envelope-ws-def} produce the same value of $\Sp^{\A/\B}$ in \cref{eq:nucl-models-Sperp}.
    The correlation length $\xi$ is given as a ratio to the longitudinal WS diameter $R_l$.
    The corresponding value for $\zeta$ results from \cref{eq:xi-zeta-relation} and is listed as the reciprocal value for convenience.}
    \centering\begin{tabular}{lllclcl}
    \hline\hline
     & \textbf{Name} & \textbf{Value(s)} & & \textbf{Unit}\\
    \hline
         $\gamma$ & Lorentz factor & 100 & & $-$ \\
         $R$ & WS radius & 6.38 & & fm \\
         $R$ & WS radius & 32.33 & & GeV${}^{-1}$ \\
         $d$ & WS skin depth & 0.535 & & fm \\
         $d$ & WS skin depth & 2.711 & & GeV${}^{-1}$ \\
         $L$ & Gauss radius & 3.81 & & fm \\
         $L$ & Gauss radius & 19.30 & & GeV${}^{-1}$ \\
         $\xi$ & Correlation length & $\frac{1}{20}$ & & $R_l=\sqrt{2}\,\frac{R}{\gamma}$ \\
         $\zeta^{-1}$ & Inverse correlation length & $\sqrt{399}$ & & $1/R_l$ \\
         $m$ & IR regulator & \{0.2, 2.0\} & & GeV \\
         $\mu$ & MV scale & 1.0 & & GeV \\
         $g$ & CYM coupling & 1.0 & & $-$ \\
         $Q_0$ & $Q_s$ scale factor & 0.5 & & GeV \\
         $\lambda$ & GBW exponent & 0.144 & & $-$ \\
         $N_c$ & Number of colors & 3 & & $-$ \\
    \hline\hline
    \end{tabular}
\end{table}

In \cref{tab:nucl-model-params}, all parameters required by the different nuclear models are listed.
The collider energy and nuclear species are matched to Au+Au collisions at a center of mass energy $2\gamma m_n = \sqrt{s_\mathrm{NN}}=200$ GeV, which corresponds to experiments performed at RHIC.
The mass of a nucleon is $m_n \approx 1$ GeV.
We are only evaluating at a single energy, because the parametrization of the TMDs with the momentum fraction $\xB$ eliminates the dependence on $\gamma$, which is the only way that energy dependence enters the models.
The results presented here are qualitatively equivalent to the LHC setup that is also used for numerical results in \cref{ch:numerical-results}.
Switching to a different ion at the LHC with slightly modified envelope parameters results in minimal variance for the TMDs.

The envelope parameters for WS in \cref{eq:envelope-ws-def} are taken from~\cite{Schenke:2012hg}.
The radius parameter for Gaussian envelopes is then matched to the WS envelopes such that both produce the same value for the projected transverse area $\Sp^{\A/\B}$ according to \cref{eq:nucl-models-Sperp}.
This ensures that the residual dependence of the TMDs on the envelopes is universal across all models.

For the $\Gxi$ and $\WSxi$ models, we fix the values of the longitudinal correlation lengths $\xi$ and $\zeta$ to a value that is motivated by the size of nucleons inside large nuclei.
The value of $\xi$ is given as a ratio to the longitudinal WS diameter $R_l$ and is the same for both models.
The value of $\zeta$ results from \cref{eq:xi-zeta-relation}.
In the case of the small correlation lengths used here, $R_l/\sqrt{399} \approx R_l/20$, and both parameters evaluate to approximately the same value.

\begin{figure}[p]
    \centering
    \includegraphics[width=0.5\linewidth]{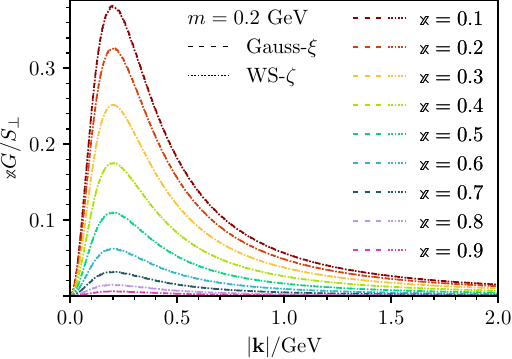}\hfill%
    \includegraphics[width=0.5\linewidth]{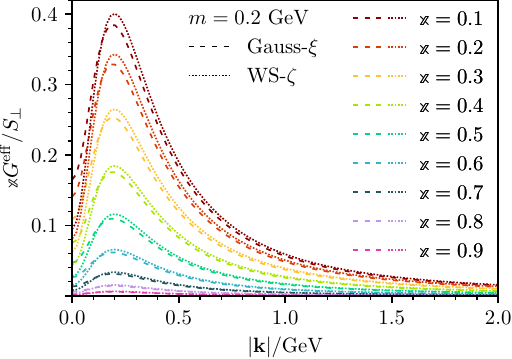}
    \caption[TMD and effective TMD for parametrized longitudinal correlations.]{%
    TMDs (left panel) and reduced effective TMDs (right panel) for $\Gxi$ (dashed) from \cref{eq:Gauss-xi-TMD,eq:Gauss-xi-effTMD} and $\WSxi$ (dotted) from \cref{eq:WS-zeta-TMD,eq:WS-zeta-effTMD}.
    Different values of $\xB$ are color-coded. A single value of $m=0.2$~GeV is used, which sets the $|\kperp|$ value of the peaks.
    Minor differences between the models arise in the infrared.
    }
    \label{fig:TMDs-xi-zeta}
\end{figure}

In \cref{fig:TMDs-xi-zeta}, we compare the TMDs (left panel) and reduced effective TMDs (right panel) of the $\Gxi$ (dashed) model defined in \cref{eq:Gauss-xi-TMD,eq:Gauss-xi-effTMD} and $\WSxi$ (dotted) model defined in \cref{eq:WS-zeta-TMD,eq:WS-zeta-effTMD}.
Both TMDs are normalized to the projected transverse area $\Sp^{\A/\B}$.
The modulus of the transverse momentum varies on the horizontal axis, and a range of momentum fractions $\xB$ is plotted in different colors according to the legend.
The general shape consists of a steep rise and peak when $|\kperp|=m$ with the value of the IR regulator fixed at $m=0.2$ GeV.
The peak transitions into a falloff $\sim1/\kperp^2$, which corresponds to the perturbative tail.
Both $\xG$ and $\xGeff$ show the same large-$|\kperp|$ behavior.
However, they differ in the IR.
Here, the reduced effective TMD starts at a finite value when $|\kperp| =0$ and does not have the same $\sim\kperp^2$ scaling for small $|\kperp|$ as $\xG$.

Similarly, the only qualitative difference between the $\Gxi$ and $\WSxi$ models is visible in the IR behavior of $\xGeff$.
We explain this observation by noting that the IR behavior is expected to be determined by the envelopes because small momenta correspond to large scales in position space.
Also, $\xG$ contains small differences between the envelopes in the IR, which are not visible on the plotted scales.

The variation w.r.t.\ the value of the momentum fraction $\xB$ is a simple exponential for the $\Gxi$ model, which is sensitive to the value of the correlation scale.
This can be read off directly from \cref{eq:Gauss-xi-TMD,eq:Gauss-xi-effTMD}.
In the case of the $\WSxi$ model, the $\xB$-dependence is convolved into integrals.
Due to the similarity of the results between these two models, we conclude that an equivalent exponential can also be factored out in the $\WSxi$ model.
This means that the chosen value for the correlation length is well separated from the scale of the WS envelopes.

\begin{figure}[p]
    \centering
    \includegraphics[width=0.5\linewidth]{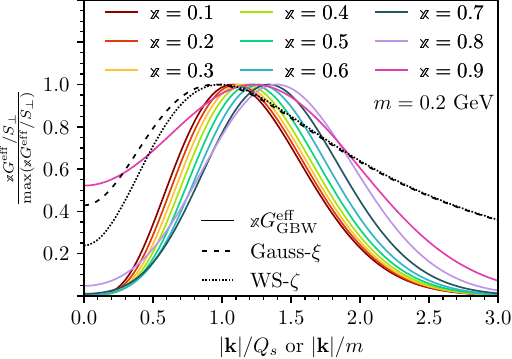}\hfill%
    \includegraphics[width=0.5\linewidth]{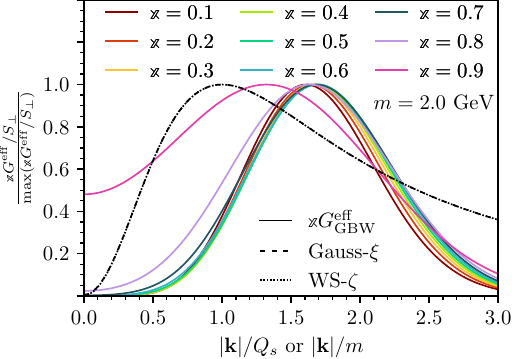}
    \caption[Scaling behavior comparison of effective TMDs.]{%
    Reduced effective TMDs normalized to their maximum values.
    For $\xGeff_\mathrm{GBW}$ (solid lines) from \cref{eq:GBW-effTMD}, the horizontal axis is rescaled by $Q_s$ in \cref{eq:Qs} as a function of $\xB$.
    For $\Gxi$ (dashed, \cref{eq:Gauss-xi-effTMD}) and $\WSxi$ (dotted, \cref{eq:WS-zeta-effTMD}), the horizontal axis is rescaled by the IR regulator $m$.
    Different values of $\xB$ are color-coded.
    The scaling of the original GBW model with $Q_s$ is modified for $\xG^\mathrm{GBW}$.
    $\Gxi$ and $\WSxi$ collapse to a single line per model for all values of $\xB$ and peak at $|\kperp|/m = 1$.
    They only differ in the IR when the regulator $m$ is small.
    }
    \label{fig:effTMDs-rescaled}
\end{figure}

In \cref{fig:effTMDs-rescaled}, we plot the reduced effective TMDs with a particular rescaling.
The black dashed and dotted lines correspond to the $\Gxi$ and $\WSxi$ models.
These curves are normalized to their maximum values and the horizontal axis is rescaled by the IR regulator $m$.
Rescaling $\xGeff$ in this way completely removes the dependence on $\xB$, such that all curves collapse to a single line for each model.
The fact that this also happens for the $\WSxi$ model can be seen as confirmation that the $\xB$-behavior factorizes similarly to the $\Gxi$ case and drops when normalizing.
The value of the IR regulator changes from $m=0.2$~GeV in the left panel to $m=2$~GeV on the right and determines the $|\kperp|$-value of the peak for both models.
In this sense, one can interpret the IR regulator as an effective saturation scale.

Remarkably, the large-$|\kperp|$ behavior of the two models collapses to the same line.
Hence, the small deviations of the models visible in \cref{fig:TMDs-xi-zeta} can only be due to a relative factor that only varies with $\xB$.
In the IR, the difference between the $\Gxi$ and $\WSxi$ models for the smaller value $m=0.2$ GeV is clearly visible.
This difference is eliminated when the IR regulator is increased to $m=2$ GeV.
The latter case indicates that this value is large enough for the IR behavior to be completely dominated by the IR regulation factor, which is shared between the two models.

\begin{figure}[p]
    \centering
    \includegraphics[width=0.5\linewidth]{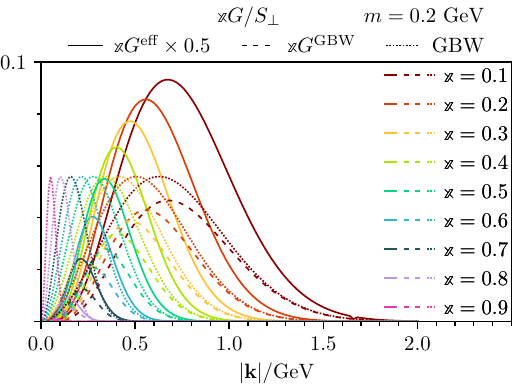}\hfill%
    \includegraphics[width=0.49\linewidth]{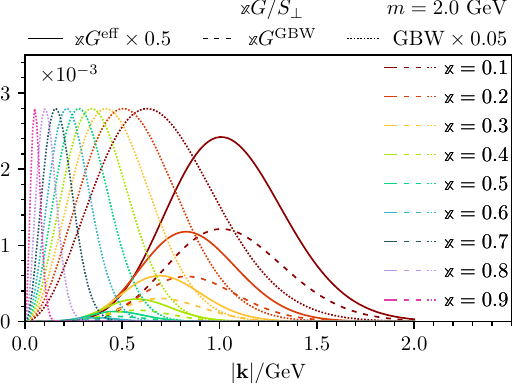}
    \caption[TMD and effective TMD for GBW model.]{%
    Original GBW TMD (dotted, \cref{eq:GBW-TMD}), IR-regulated GBW TMD (dashed, \cref{eq:GBWIR-TMD}) and reduced effective TMD (solid) from \cref{eq:GBW-effTMD}.
    $\xGeff$ scaled by $1/2$.
    GBW scaled by $1/20$ only in the right panel.
    The value of the IR regulator changes from $m=0.2$~GeV on the left to $m=2$~GeV on the right.
    Different values of $\xB$ are color-coded.
    The geometric scaling property of the GBW TMD is modified due to IR regulation and the curves peak at larger values of $|\kperp|$.
    }
    \label{fig:TMDs-GBW}
\end{figure}
\begin{figure}[p]
    \centering
    \includegraphics[width=0.5\linewidth]{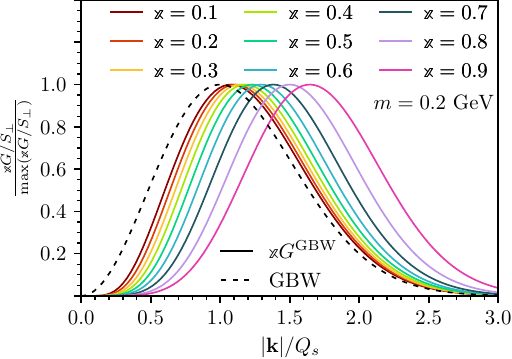}\hfill%
    \includegraphics[width=0.5\linewidth]{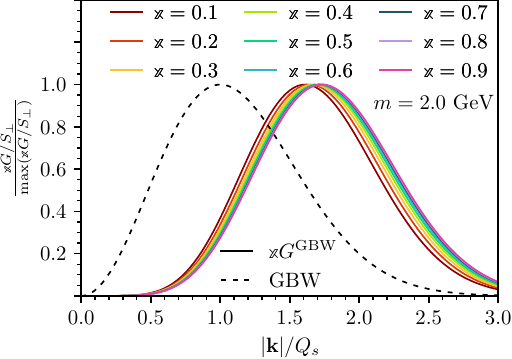}
    \caption[Scaling behavior of GBW TMDs.]{%
    Rescaled original GBW TMD (dashed, \cref{eq:GBW-TMD}) and IR-regulated GBW TMD (solid, \cref{eq:GBWIR-TMD}).
    The curves are normalized to their maximum values and the horizontal axis is rescaled by $Q_s$ from \cref{eq:Qs}.
    Different values of $\xB$ are color-coded.
    The value of the IR regulator changes from $m=0.2$~GeV on the left to $m=2$~GeV on the right.
    IR regulation leads to a different scaling behavior.
    The position of the peaks moves to larger $|\kperp|$ when $m$ increases.
    }
    \label{fig:GBW-TMD-rescaled}
\end{figure}
\afterpage{\FloatBarrier}

\pagebreak
In \cref{fig:TMDs-GBW}, we compare the original GBW TMD (dotted) from \cref{eq:GBW-TMD} to the IR-regulated TMD (dashed) in \cref{eq:GBWIR-TMD} and the reduced effective TMD (solid) in \cref{eq:GBW-effTMD} used in the nuclear model from \cref{sec:GBW-saturation-model}.
All results are normalized to the projected transverse area $\Sp^{\A/\B}$ in \cref{eq:nucl-models-Sperp}.
Additionally, the values for $\xGeff$ are scaled by $\tfrac{1}{2}$ in both panels, and the value of GBW is scaled by $\tfrac{1}{20}$ only in the right panel for better comparability.
In the left panel, the value of $m=0.2$~GeV and in the right panel, $m=2$~GeV.
The IR regulator only affects $\xGeff$ and $\xG^\mathrm{GBW}$.

We highlight the qualitative differences between the original GBW TMD (dotted) and the others.
The IR regulation of the transverse momentum leads to strong deviations at small $|\kperp|$.
Since curves at larger $\xB$ peak at smaller values of $|\kperp|$ and quickly fall off for larger $|\kperp|$, the effect of the IR regulator is most noticeable for large $\xB$.
The peaks of the original GBW TMD are always at the same height, whereas IR regulation leads to suppression for small $|\kperp|$.
The positions of the peaks on the horizontal axis move to smaller $|\kperp|$ when increasing $\xB$.
In the original GBW TMD, the peak is located at $|\kperp| = Q_s$ as a function of $\xB$ given in \cref{eq:Qs} according to geometric scaling.
Introducing an IR regulator changes this scaling.
For $\xGeff$ and $\xG^\mathrm{GBW}$, the positions of the peaks move further to the right when increasing the value of $m$.
Apart from a large difference in the overall scale, the IR-regulated GBW TMD (dashed) and the reduced effective TMD (dotted) behave qualitatively similarly.

\subsection{Effective saturation scale for the IR-regulated GBW model}

We further investigate the different scaling behavior for the IR-regulated GBW model.
In \cref{fig:effTMDs-rescaled}, we demonstrate that rescaling the horizontal axis with $Q_s$ from \cref{eq:Qs} for the reduced effective TMDs $\xGeff_\mathrm{GBW}$ from \cref{eq:GBW-effTMD} does not lock the positions of the peaks.
Furthermore, we observe that for the largest $\xB$ values, the effective TMD does not reach 0 as $|\kperp|=0$.
At that value of the transverse momentum, the integral in \cref{eq:GBW-effTMD} has a closed-form solution and predicts a nonzero value.
However, as $\xB\rightarrow1$ and $Q_s \rightarrow0$, a value of zero is recovered.
Comparing $\xGeff_\mathrm{GBW}$ to the $\Gxi$ and $\WSxi$ models, we see that the perturbative $\sim1/\kperp^2$ tail is missing for the GBW model.
Instead, a much steeper falloff, that is regulated by the exponential in $\kperp^2$, controls the UV.

In \cref{fig:GBW-TMD-rescaled}, we compare the rescaled original GBW TMD (dashed) in \cref{eq:GBW-TMD} to the IR-regulated TMD (solid) in \cref{eq:GBWIR-TMD} for two values of the IR regulator $m$.
The curves are normalized to their maximum values and the horizontal axis is rescaled by $Q_s$ as a function of $\xB$ given in \cref{eq:Qs}.
Here, all curves for the different values of $\xB$ in the original GBW TMD collapse to a single line.
This scaling is changed when introducing IR regulation.
The peaks for $\xG^\mathrm{GBW}$ move to the right and indicate that the value of $Q_s$ is smaller than the actual $|\kperp|$ value of the peaks.

In the case of the IR-regulated GBW TMD from \cref{eq:GBWIR-TMD}, it is straightforward to obtain the parametric form of the $|\kperp|$ value that corresponds to the peaks.
We identify this value with an effective saturation scale,
\begin{align}
    Q_s^\mathrm{eff}(\xB) = \frac{1}{\sqrt{2}}\sqrt{Q_s(\xB)^2 - m^2 + \sqrt{m^4 + 10m^2Q_s(\xB)^2 + Q_s(\xB)^4}}.
    \label{eq:Qseff}
\end{align}
Here, $Q_s$ follows the parametrization given in \cref{eq:Qs} and depends on $\xB$.
This result highlights how the IR regulator $m$ mixes with $Q_s$ and determines the effective saturation scale.

\begin{figure}
    \centering
    \includegraphics[width=0.5\linewidth]{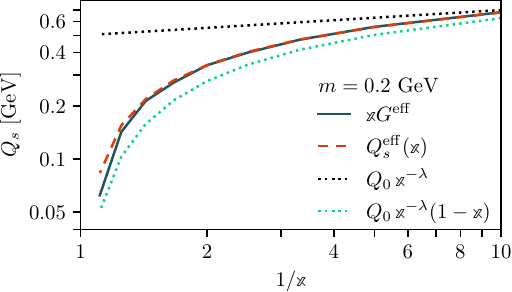}\hfill%
    \includegraphics[width=0.5\linewidth]{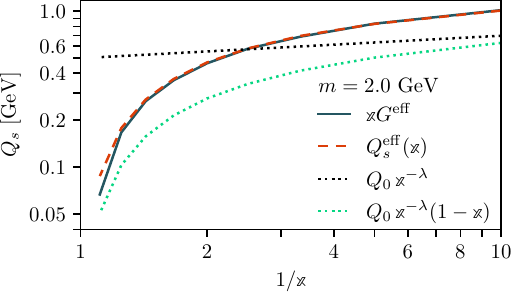}
    \caption[Effective saturation scale and parametrizaions.]{%
    Comparison of the positions of the peaks from the curves in \cref{fig:TMDs-GBW} with different parametrizations of $Q_s$.
    That value of $|\kperp|$ where the TMDs peak is identified with $Q_s$ for each TMD and plotted on the vertical axis on a logarithmic scale.
    On the horizontal axis, the reciprocal value of the momentum fraction $\xB$ is given on a logarithmic scale.
    $\xGeff$ (blue) corresponds to the solid lines in \cref{fig:TMDs-GBW}.
    $Q_s^\mathrm{eff}$ (orange) corresponds to the dashed lines in \cref{fig:TMDs-GBW} and is given in \cref{eq:Qseff}.
    The dotted green line corresponds to the scaling of the original GBW TMD with $Q_s$ given in \cref{eq:Qs}.
    The dotted black line denotes the asymptotic behavior according to the exponent $\lambda$.
    Good agreement between $\xGeff$ and $Q_s^\mathrm{eff}$ highlights the similar behavior of the reduced effective TMD and the IR-regulated GBW TMD.
    }
    \label{fig:Qseff}
\end{figure}

In \cref{fig:Qseff}, we compare different notions of the saturation scale obtained from the TMDs in the GBW model.
On the horizontal axes, the reciprocal value of the momentum fraction $\xB$ is given on a logarithmic scale.
The vertical axis is also scaled logarithmically and corresponds to the different values of $|\kperp|$, identified as the saturation scale.
The solid blue line marks the $|\kperp|$ values of the actual peaks of the solid curves in \cref{fig:TMDs-GBW} for the reduced effective TMD from \cref{eq:GBW-effTMD}.
The dashed orange line is given by $Q_s^\mathrm{eff}$ from \cref{eq:Qseff} and marks the $|\kperp|$ values of the peaks for the IR-regulated GBW TMD (dashed lines in \cref{fig:TMDs-GBW} and \cref{eq:GBWIR-TMD}).
The original parametrization of $Q_s$ from \cref{eq:Qs} is plotted in dotted-green color and the asymptotic behavior according to the $\sim\xB^{-\lambda}$ contribution is shown in dotted-black.

The results show near-perfect agreement between $\xGeff$ and $Q_s^\mathrm{eff}$ with only slight deviations for large $\xB$.
This remains unchanged when the IR regulator is increased by a factor of ten in the right panel.
We conclude that the modification of the scaling behavior of the reduced effective TMD is captured already by the much simpler form of the IR-regulated TMD in \cref{eq:GBWIR-TMD}.

Compared to the original parametrization of $Q_s$ (dotted-green), IR regulation leads to larger values of the identified saturation scale.
The difference is smallest for small $\xB$ and approaches a constant offset for the asymptotic behavior.
This translates to a relative factor between $Q_s$ in \cref{eq:Qs} and $Q_s^\mathrm{eff}$ when $\xB\rightarrow0$ and the identified saturation scale becomes large.

\chapter{Numerical results for the (3+1)D dilute Glasma}\label{ch:numerical-results}

In this chapter, we present numerical results obtained from the position space and momentum space formulations of the (3+1)D dilute Glasma.
In \cref{sec:numerics-param-long-corr}, we focus on the nuclear model with parametrized longitudinal correlations from \cref{sec:finite-longitudinal-correlations}.
First, we review the results for the energy-momentum tensor and local rest frame energy density obtained in the position space framework introduced in \cref{sec:dilute-glasma-approx} that were first published in~\cite{Ipp:2024ykh}.
This discussion largely follows~\cite{Ipp:2024ykh,Leuthner:2025vsd}, where further details are provided.
Then, we present a detailed analysis of the gluon number distribution and transverse energy obtained in the momentum space framework introduced in \cref{ch:momentum-space-picture}.
In \cref{sec:numerics-GBW}, we explore the momentum rapidity dependence of transverse energy in the GBW model from \cref{sec:GBW-saturation-model}.
Finally, in \cref{sec:results-limiting-fragmentation}, we investigate numerical and analytical evidence for the property of limiting fragmentation in the (3+1)D dilute Glasma as derived in \cref{ch:limiting-fragmentation}.

\section{Parametrized longitudinal correlations}\label{sec:numerics-param-long-corr}

In \cref{tab:numerics-zeta-model-params}, the values of all parameters that enter the numerics are listed.
Two different setups are realized that are comparable to heavy-ion collision experiments at RHIC and the LHC.
In the case of RHIC, the center-of-mass energy per nucleon-nucleon pair is set to $\sqrt{s_\mathrm{NN}}=200$~GeV.
This results in a Lorentz factor of $\gamma = \sqrt{s_\mathrm{NN}}/(2m_n) = 100$, where we set the nucleon mass $m_n=1$~GeV.
For the LHC, $\sqrt{s_\mathrm{NN}}=5400$~GeV and $\gamma = 2700$.

Only Woods-Saxon (WS) envelopes and the nuclear model with parametrized longitudinal correlations, as described in \cref{sec:WS-correlator}, are used.
The parameters $R$ and $d$ that characterize the WS envelope function in \cref{eq:envelope-ws-def} are taken from~\cite{Schenke:2012hg} and correspond to gold (Au) nuclei used at RHIC and lead (Pb) nuclei used at the LHC.
The nuclear species matters for parameters derived from the WS radius, leading to slight differences between the setups.
The correlation lengths $\xi$ and $\zeta$ used to parametrize the longitudinal correlations in the $\WSxi$ nuclear model (cf.\ \cref{sec:WS-correlator}) are given as ratios to the Lorentz contracted diameter in light cone coordinates $R_l$.
The value $\xi=R_l$ satisfies the coherent limit and $\xi=R_l/20$ is motivated by the size of nucleons inside the nucleus.
The three largest values are identical to the values used in~\cite{Ipp:2024ykh} (note the different definition of $R_l$).
Additionally, the value $\xi=R_l/40$ at half the nucleonic scale is used.
The corresponding values for $\zeta$ are obtained from \cref{eq:xi-zeta-relation}.

For collisions that incorporate an impact parameter, the nuclei are offset along the $x$-axis and the modulus of the impact parameter $b$ is given in units of the WS radius.
Parameters annotated with an asterisk in \cref{tab:numerics-zeta-model-params} only apply to the position space calculation of the energy-momentum tensor.
Specifically, the shift of the origin of the Milne frame $\delta\tau$ discussed in \cref{sec:Milne-frame-shift} is matched to the nuclear extent via $\delta\tau = (R+d)/(c\gamma)$.
Furthermore, we stress that for the position space calculations, a fixed value of $\Lambda = 10$~GeV is always used to soften the lattice UV cutoff.

A total number of $N_\mathrm{ev}=10$ events is simulated for each set of parameters.
The resulting event statistics for the observables discussed in the following are obtained via Jackknife analysis.
Each simulated event has its own independent realization of the initial conditions.
The nuclear color charge distributions are sampled event-by-event according to the algorithm discussed in \cref{appx:sampling-rho}.
Note that the momentum space calculations for the gluon numbers already evaluate the event-averaged expressions and do not contain event statistics.

\begin{table}
    \vspace{-.6\baselineskip}\caption[Parameters for numerical evaluation of \texorpdfstring{$\WSxi$}{WS-zeta} model results.]{\label{tab:numerics-zeta-model-params}%
    Values of model parameters used for numerical evaluation.
    Collider energies and nuclear species correspond to Au+Au collisions at $\sqrt{s_\mathrm{NN}}=200$~GeV performed at RHIC and Pb+Pb collisions at $\sqrt{s_\mathrm{NN}}=5400$~GeV performed at the LHC.
    WS $R$ and $d$ are taken from~\cite{Schenke:2012hg}.
    The correlation length $\xi$ is given as a ratio to the longitudinal WS diameter $R_l$.
    The corresponding value for $\zeta$ results from \cref{eq:xi-zeta-relation} and is listed as the reciprocal value for convenience.
    Parameters marked with an asterisk (${}^*$) only apply to the position space results of the energy-momentum tensor.}
    \centering\begin{tabular}{lllclcl}
    \hline\hline
     & \textbf{Name} & \textbf{RHIC} & \textbf{LHC} & \textbf{Unit}\\
    \hline
         $\gamma$ & Lorentz factor & 100 & 2700 & $-$ \\
         $R$ & WS radius & 6.38 & 6.62 & fm \\
         $R$ & WS radius & 32.33 & 33.55 & GeV${}^{-1}$ \\
         $d$ & WS skin depth & 0.535 & 0.546 & fm \\
         $d$ & WS skin depth & 2.711 & 2.767 & GeV${}^{-1}$ \\
         $\xi$ & Correlation length & \multicolumn{2}{c}{$\{ \frac{1}{40}, \frac{1}{20}, \frac{1}{4}, 1\}$ } & $R_l=\sqrt{2}\,\frac{R}{\gamma}$ \\
         $\zeta^{-1}$ & Inverse correlation length & \multicolumn{2}{c}{$\{\sqrt{1599}, \sqrt{399}, \sqrt{15}, 0 \}$} & $1/R_l$ \\
         $m$ & IR regulator & \multicolumn{2}{c}{\{0.2, 2.0\}}  & GeV \\
         $\Lambda$ & UV regulator* & \multicolumn{2}{c}{10.0} & GeV \\
         $\mu$ & MV scale & \multicolumn{2}{c}{1.0} & GeV \\
         $g$ & CYM coupling & \multicolumn{2}{c}{1.0} & $-$ \\
         $N_c$ & Number of colors & \multicolumn{2}{c}{3} & $-$ \\
         $N_\mathrm{ev}$ & Number of events* & \multicolumn{2}{c}{10} & $-$ \\
         $b$ & Impact parameter* & \multicolumn{2}{c}{\{ 0., 1.0 \}} & $R$ \\
         $\tau$ & Proper time* & \multicolumn{2}{c}{ \{ 0.2, 0.4, 0.6, 0.8, 1.0 \}} & fm/$c$ \\
         $\delta\tau$ & Milne origin shift* & 0.069 & 0.0027 & fm/$c$ \\
    \hline\hline
    \end{tabular}
\end{table}

\subsection{Three-dimensional distribution of local rest frame energy density}

\begin{figure}[p]
    \centering
    \includegraphics[width=0.6\linewidth]{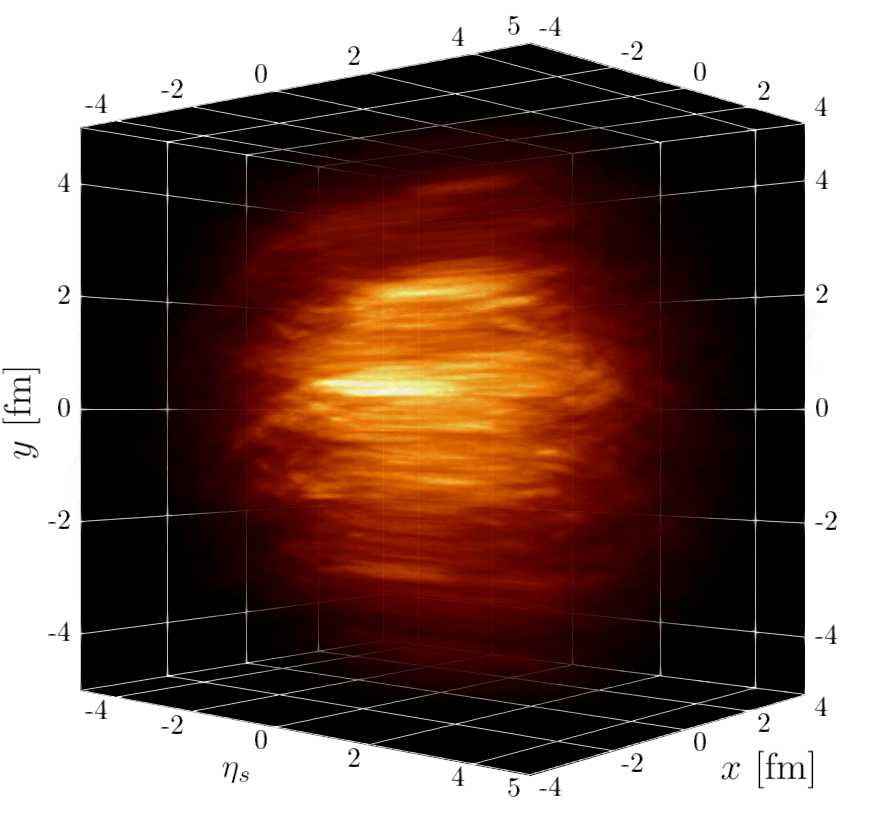}
    \caption[3D perspective plot of local rest frame energy density.]{%
    Three-dimensional plot of the local rest frame energy density $\elrf$ for a single event evaluated at $\tau=0.4$~fm/$c$ using RHIC parameters from \cref{tab:numerics-zeta-model-params}.
    The color palette roughly corresponds to the colorbar of the left panel in \cref{fig:x-etas-slices-elrf}, where brighter colors correlate with larger $\elrf$.
    Transparency has been added for visual enhancement, which skews the colors.
    The other parameters are $b=R$, $m=0.2$~GeV and $\xi=R_l/4$.
    The spacetime rapidity axis protrudes to the right out of the picture plane.
    A movie where the viewing angle is rotated in orbital motion is available in the supplemental material of~\cite{Ipp:2024ykh}.
    The left panels in \cref{fig:x-etas-slices-elrf,fig:x-y-relief-elrf,fig:x-y-slices-elrf-etas=0,fig:x-y-slices-elrf-etas=-Yb} are slices through the depicted $\elrf$ distribution.
    Figure first published in~\cite{Ipp:2024ykh}.
    }
    \label{fig:3D-perspective-elrf}
\end{figure}

\begin{figure}[p]
    \centering
    \includegraphics[width=0.49\linewidth]{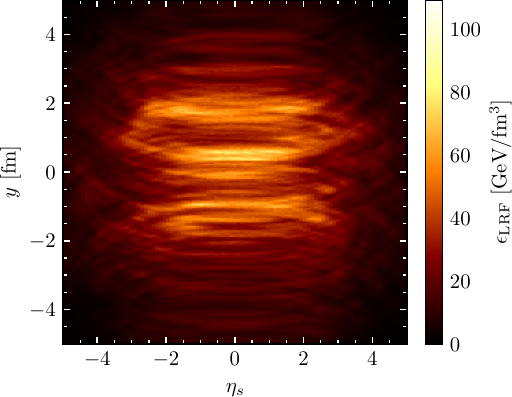}\hfill%
    \includegraphics[width=0.49\linewidth]{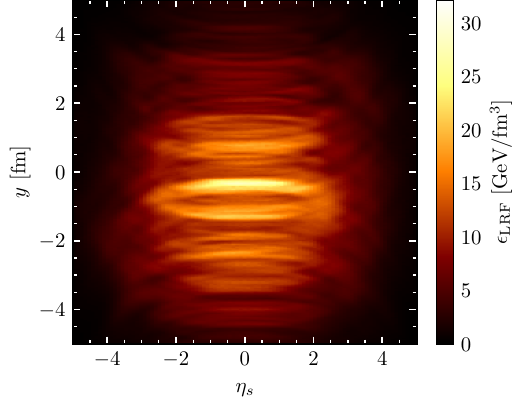}
    \caption[Longitudinal slices of local rest frame energy density.]{%
    Longitudinal slices along $x=0$ through the three-dimensional distribution of local rest frame energy density $\elrf$.
    A single event using RHIC parameters from \cref{tab:numerics-zeta-model-params} evaluated at $b=R$, $m=0.2$~GeV and $\xi=R_l/4$ is shown.
    Left: $\tau=0.4$~fm/$c$ corresponding to \cref{fig:3D-perspective-elrf}. Right: $\tau=1.0$~fm/$c$.
    Extended longitudinal structures reminiscent of ``flux-tubes'' are visible.
    At later times (right panel), the structures become more diffuse.
    Left panel from~\cite{Ipp:2024ykh}.
    }
    \label{fig:x-etas-slices-elrf}
\end{figure}

\begin{figure}[p]
    \centering
    \vspace*{-0.7\baselineskip}%
    \hspace{-1em}\includegraphics[width=0.51\linewidth]{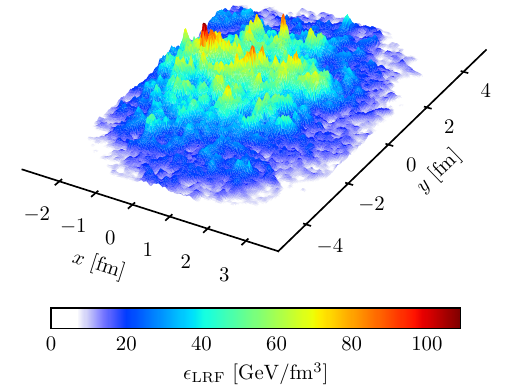}\hspace{-0.5em}%
    \includegraphics[width=0.51\linewidth]{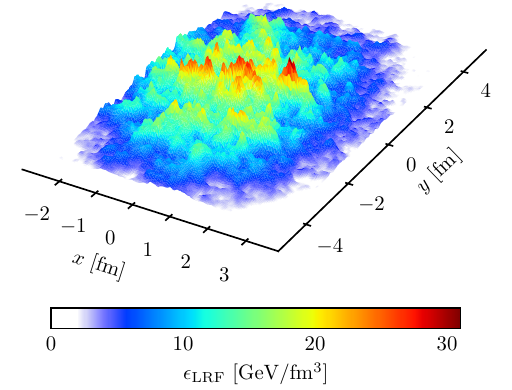}\hspace{-1em}
    \caption[Transverse reliefs of local rest frame energy density at mid-rapidity.]{%
    Transverse reliefs at $\eta_s= 0$ of the three-dimensional distribution of local rest frame energy density $\elrf$.
    A single event using RHIC parameters from \cref{tab:numerics-zeta-model-params} evaluated at $b=R$, $m=0.2$~GeV and $\xi=R_l/4$ is shown.
    Left: $\tau=0.4$~fm/$c$ corresponding to \cref{fig:3D-perspective-elrf}. Right: $\tau=1.0$~fm/$c$.
    The presentation is inspired by figures from~\cite{Schenke:2012hg,Schenke:2012wb} and highlights the ``lumpiness'' of the Glasma.
    Left panel from~\cite{Ipp:2025cdh}.
    }
    \label{fig:x-y-relief-elrf}
\end{figure}

\begin{figure}[p]
    \centering
    \includegraphics[width=0.49\linewidth]{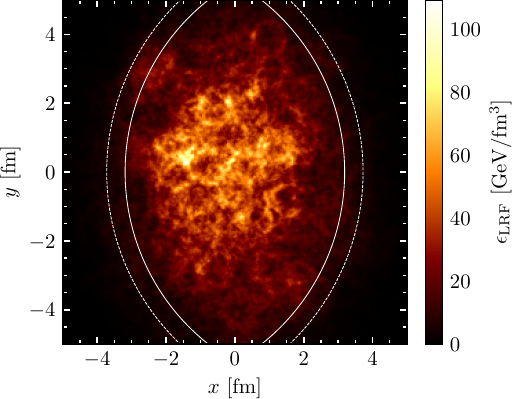}\hfill%
    \includegraphics[width=0.49\linewidth]{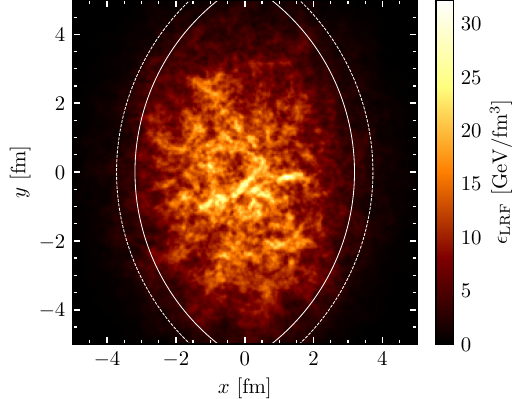}
    \caption[Transverse slices of local rest frame energy density at mid-rapidity.]{%
    Same data as in \cref{fig:x-y-relief-elrf}.
    To guide the eye, white arcs with a radius of $R$ (solid) and $R+d$ (dotted) are drawn from the nuclear centers.
    The almond-shaped overlap region contains most of the deposited $\elrf$.
    Left panel from~\cite{Ipp:2024ykh}.
    }
    \label{fig:x-y-slices-elrf-etas=0}
\end{figure}

\begin{figure}[p]
    \centering
    \includegraphics[width=0.49\linewidth]{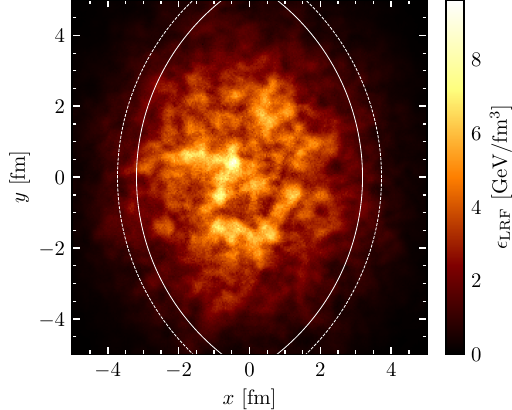}\hfill%
    \includegraphics[width=0.49\linewidth]{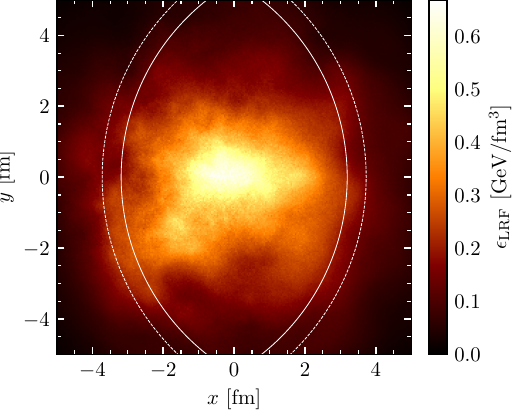}
    \caption[Transverse slices of local rest frame energy density at far-backward rapidity.]{%
    Transverse slices at $\eta_s=-5.36$ (corresponding to the beam rapidity) of the three-dimensional distribution of local rest frame energy density $\elrf$.
    A single event using RHIC parameters from \cref{tab:numerics-zeta-model-params} evaluated at $b=R$, $m=0.2$~GeV and $\xi=R_l/4$ is shown.
    Left: $\tau=0.4$~fm/$c$ corresponding to \cref{fig:3D-perspective-elrf}. Right: $\tau=1.0$~fm/$c$.
    At large $|\eta_s|$ and late times, the structure becomes ``washed out''.
    }
    \label{fig:x-y-slices-elrf-etas=-Yb}
    \vspace*{-1.3\baselineskip}
\end{figure}

Before studying event-averaged results, we visually inspect the distribution of the local rest frame energy density $\elrf$ for a single event in \cref{fig:3D-perspective-elrf,fig:x-etas-slices-elrf,fig:x-y-relief-elrf,fig:x-y-slices-elrf-etas=0,fig:x-y-slices-elrf-etas=-Yb}.
Here, the RHIC parameter set from \cref{tab:numerics-zeta-model-params} is used.
The impact parameter $b=R$, the IR regulator $m=0.2$~GeV and the correlation length $\xi=R_l/4$.
For \cref{fig:x-etas-slices-elrf,fig:x-y-relief-elrf,fig:x-y-slices-elrf-etas=0,fig:x-y-slices-elrf-etas=-Yb}, the left panels contain the same data as \cref{fig:3D-perspective-elrf} obtained at a proper time $\tau=0.4$~fm/$c$.
In the right panels, $\tau=1.0$~fm/$c$.

The rich, fully three-dimensional structure of $\elrf$ is best appreciated in the perspective plot in \cref{fig:3D-perspective-elrf}.
The horizontal plane in the image is spanned by the spacetime rapidity axis and one transverse direction.
The second transverse direction runs upward.
Brighter colors correspond to larger values of $\elrf$.
To improve the visual representation, transparency was added to the color palette, which allows one to peek inside the ``fireball''.

Slicing along the $\eta_s$-axis at a fixed value of $x=0$ reveals extended longitudinal structures with varying transverse extents in \cref{fig:x-etas-slices-elrf}.
They are nearly boost-invariant in an interval $\eta_s\in[-2, 2]$ and reminiscent of the ``flux-tube'' interpretation~\cite{Lappi:2006fp,Dumitru:2008wn,Chen:2013ksa,Lappi:2017skr} of the initial longitudinal fields of the Glasma.
Evaluating the same event at a larger value of $\tau=1.0$~fm/$c$ (right panel) leads to a strong decrease in energy density and the diffusion of the sharp structures.

The effect of the large impact parameter is clearly visible in the transverse slices in \cref{fig:x-y-relief-elrf,fig:x-y-slices-elrf-etas=0,fig:x-y-slices-elrf-etas=-Yb}.
The local rest frame energy density is concentrated in an ``almond-shaped'' region that corresponds to the overlap of the colliding nuclei.
To guide the eye, white arcs are drawn in \cref{fig:x-y-slices-elrf-etas=0,fig:x-y-slices-elrf-etas=-Yb} that correspond to the circumferences of the nuclei at distances $R$ (solid lines) and $R+d$ (dotted lines) from their centers.

\Cref{fig:x-y-relief-elrf,fig:x-y-slices-elrf-etas=0} are evaluated at mid-rapidity, whereas for \cref{fig:x-y-slices-elrf-etas=-Yb} the value of $\eta_s = -5.36$ and corresponds to the beam rapidity.
When increasing $|\eta_s|$, the details in the structure are lost.
In combination with the later evaluation time $\tau=1.0$~fm/$c$, almost none of the transverse structure remains.

\afterpage{\FloatBarrier}

\subsection{Event-averaged energy-momentum tensor in position space}\label{subsec:EbE-WS-xi}

\begin{figure}[p]
    \centering
    \includegraphics[width=0.49\linewidth]{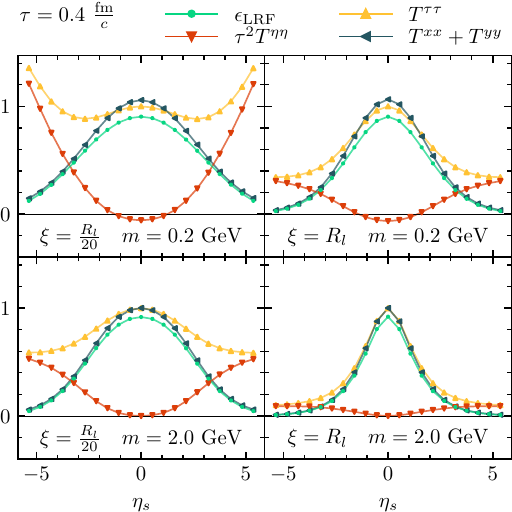}\hfill%
    \includegraphics[width=0.49\linewidth]{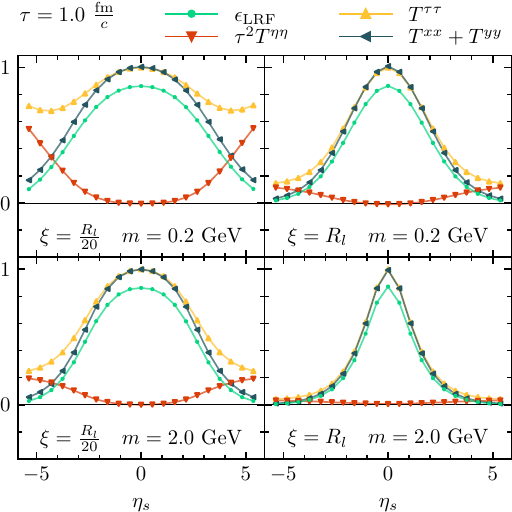}
    \caption[Rapidity profiles of energy-momentum tensor components and local rest frame energy density.]{%
    Event-averaged rapidity profiles of the diagonal components of the energy-momentum tensor $T$ and the local rest frame energy density $\elrf$.
    All quantities are integrated over the entire transverse plane.
    RHIC parameters from \cref{tab:numerics-zeta-model-params} with $b=0$ are used.
    Left: $\tau=0.4$~fm/$c$. Right: $\tau=1.0$~fm/$c$.
    The curves are scaled by a constant factor to normalize $T^{\tau\tau}$ at $\eta_s=0$ and the relative scales are conserved.
    The Milne component $T^{\tau\tau}$ (yellow) deviates from $\elrf$ (green) at large rapidities and renders the Milne frame ill-suited for three-dimensional dynamics.
    Left panel adapted from~\cite{Ipp:2024ykh}.
    }
    \label{fig:Tmunu-elrf-vs-etas}
\end{figure}

Next, we study the rapidity profiles of quantities integrated over the entire transverse plane and averaged over 10 independent events.
In \cref{fig:Tmunu-elrf-vs-etas}, we compare the diagonal components of the energy-momentum tensor and the local rest frame energy density $\elrf$.
The setup corresponds to RHIC parameters from \cref{tab:numerics-zeta-model-params}.
The curves are scaled by the same constant factor for each parameter combination so that the value of $T^{\tau\tau}$ (yellow) at mid-rapidity peaks at 1.
This preserves the relative scales between the different quantities.

The results show significant deviations of $T^{\tau\tau}$ (yellow) from the local rest frame energy density $\elrf$ (green) at large $|\eta_s|$, which come from sizeable off-diagonal components of the energy-momentum tensor.
The three-dimensional dynamics of the Glasma differ from the idealized Bjorken flow, leading to the inability of the Milne frame to capture the symmetries of the system.
Furthermore, the Milne frame tensor components $T^{\tau\tau}$ (yellow) and $T^{\eta\eta}$ (orange) depend strongly on the placement of the origin of the Milne coordinate frame.
Due to the extended collision region in the $t$-$z$ plane, there is no distinguished point to place the origin.
Like $T^{\tau\tau}$, the longitudinal pressure $T^{\eta\eta}$ increases for large $|\eta_s|$, so that the energy-momentum tensor is locally traceless at every evaluation point.
The longitudinal pressure is negative around mid-rapidity at earlier times (left panel) and settles to a plateau with zero value at later times (right panel).
This trend aligns with the time evolution obtained in the boost-invariant limit, albeit at a significantly smaller magnitude.
The behavior of the Milne frame tensor components varies strongly with the correlation length.
Smaller $\xi$ values lead to steeply increasing curves at large $|\eta_s|$, whereas in the coherent limit, this effect is suppressed.

We identify the sum of transverse pressures (blue) as a better-suited notion of energy density that closely matches the frame-independent $\elrf$ across the entire spacetime rapidity range.
The transverse pressures are also unaffected by longitudinal boosts.
\pagebreak

\begin{figure}[p]
    \centering
    \includegraphics[width=0.49\linewidth]{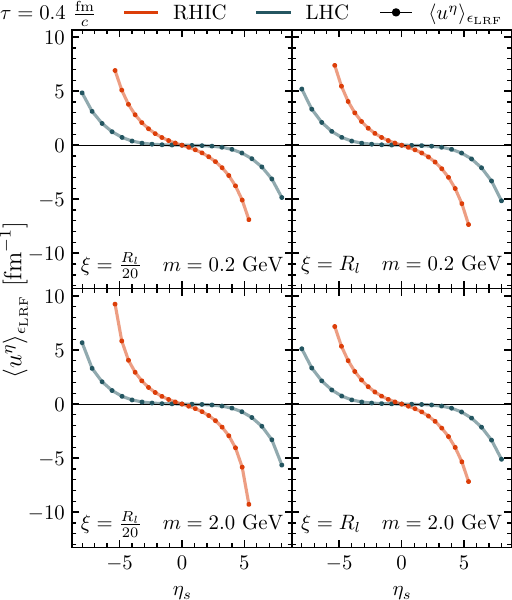}\hfill%
    \includegraphics[width=0.49\linewidth]{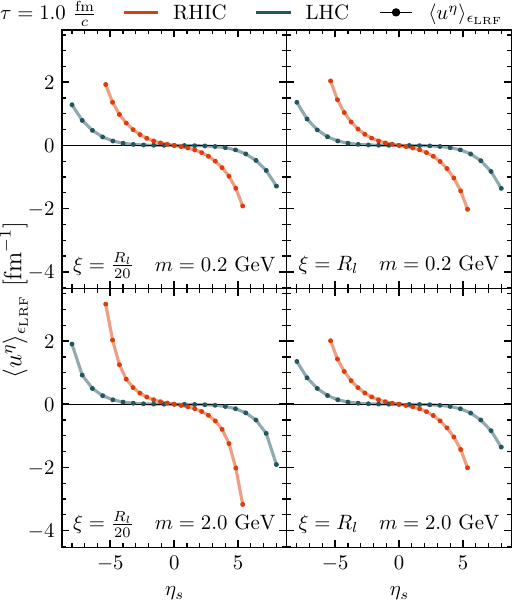}
    \caption[Rapidity profiles of longitudinal flow.]{%
    Event-averaged rapidity profiles of the longitudinal flow $u^\eta$ defined in \cref{eq:ueta-ev-def}.
    RHIC and LHC parameters from \cref{tab:numerics-zeta-model-params} with $b=0$ are used.
    Left: $\tau=0.4$~fm/$c$. Right: $\tau=1.0$~fm/$c$.
    Large values at large rapidities indicate a different flow pattern than Bjorken expansion from a single point.
    Instead, the results match homogeneous Bjorken expansion from an extended collision region.
    Left panel adapted from~\cite{Ipp:2024ykh}.
    }
    \label{fig:ueta-vs-etas}
\end{figure}
\begin{figure}[p]
    \centering
    \includegraphics[width=0.49\linewidth]{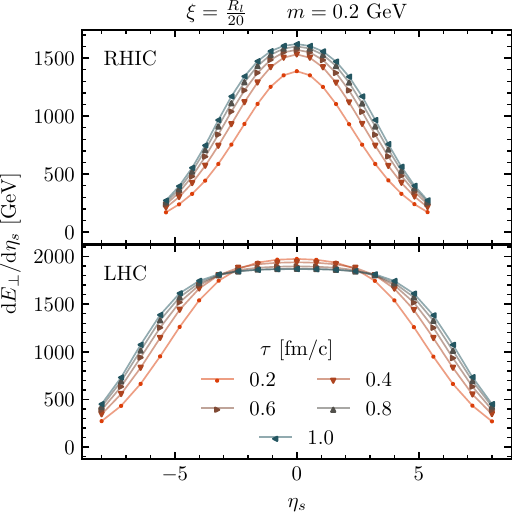}\hfill%
    \includegraphics[width=0.49\linewidth]{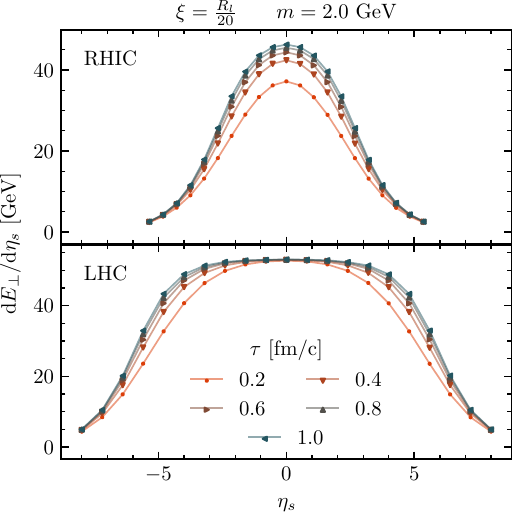}
    \caption[Proper time evolution of transverse energy rapidity profiles.]{%
    Event-averaged rapidity profiles of the transverse energy per unit spacetime rapidity $\dd\Ep/\dd\eta_s$.
    RHIC and LHC parameters from \cref{tab:numerics-zeta-model-params} with $b=0$ and $\xi=R_l/20$ are used.
    Left: $m=0.2$~GeV. Right: $m=2.0$ GeV.
    The profiles stabilize and slightly widen over a duration of $\tau\sim1.0$~fm/$c$.
    Left panel adapted from~\cite{Ipp:2024ykh}.
    }
    \label{fig:dEp-detas-vs-tau}
\end{figure}
\begin{figure}[p]
    \centering
    \includegraphics[width=0.49\linewidth]{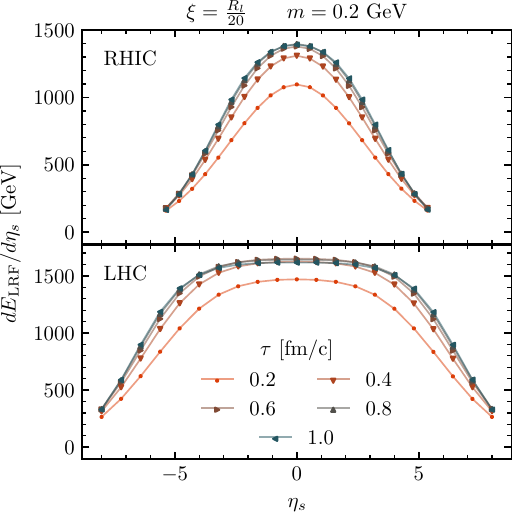}\hfill%
    \includegraphics[width=0.49\linewidth]{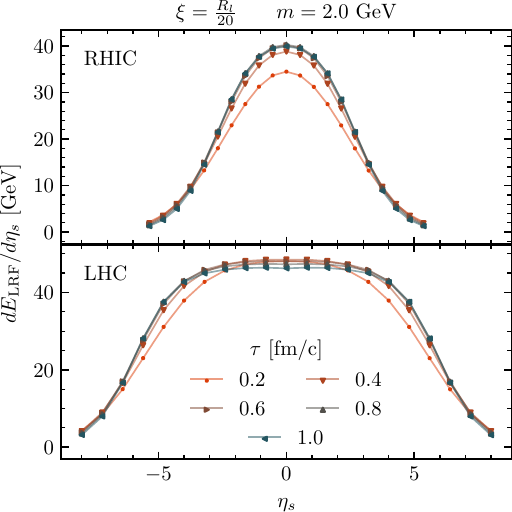}
    \caption[Proper time evolution of local rest frame energy density rapidity profiles.]{%
    Event-averaged rapidity profiles of the local rest frame energy density per unit spacetime rapidity $\dd E_\mathrm{LRF}/\dd\eta_s$ defined in \cref{eq:dElrf-deta-def}.
    RHIC and LHC parameters from \cref{tab:numerics-zeta-model-params} with $b=0$ and $\xi=R_l/20$ are used.
    Left: $m=0.2$~GeV. Right: $m=2.0$ GeV.
    Similarly to \cref{fig:dEp-detas-vs-tau}, the curves stabilize and widen over a duration of $\tau\sim1.0$~fm/$c$.
    }
    \label{fig:dElrf-detas-vs-tau}
\end{figure}
\FloatBarrier

The mismatch of $T^{\tau\tau}$ and $\elrf$ can also be interpreted as the existence of a large longitudinal flow component $u^\eta$.
Unlike in the boost-invariant Glasma, where $u^\eta = 0$ up to local fluctuations, \cref{fig:ueta-vs-etas} demonstrates the existence of sizeable longitudinal flow in the (3+1)D description.
The plotted quantity is the event average of the energy density weighted longitudinal flow, defined as
\begin{align}
    \langle u^\eta \rangle_{\epsilon_\mathrm{LRF}} = \left\langle \frac{\intop\dd^2\xperp\, u^\eta (\tau, \eta_s, \xperp)\, \elrf(\tau, \eta_s, \xperp)}{\intop\dd^2\yperp\, \elrf(\tau, \eta_s, \yperp)} \right\rangle_\mathrm{ev}, \label{eq:ueta-ev-def}
\end{align}
where the integrations cover the entire transverse plane and the result still depends on $\tau$ and $\eta_s$.
The longitudinal flow is negative for positive $\eta_s$ and vice versa.
It is always zero at mid-rapidity, where a flat plateau develops for later times and at the higher LHC energy.
This flow pattern corresponds to longitudinal expansion at velocities slower than those of Bjorken flow.
It can be shown that these particular rapidity profiles are predominantly explained by a simple model that assumes Bjorken flow to originate at every point in the extended interaction region given by the finite longitudinal widths of the colliding nuclei.
The resulting flow pattern is a superposition of flow vectors with different alignments and can be explained by geometric arguments.
Still, a slight deviation from this model remains due to the nonzero transverse flow in the system.
Unlike the Milne frame components of the energy-momentum tensor, the longitudinal flow is largely unaffected by the value of $\xi$, which supports its explanation as a purely geometric effect.

The proper time evolution of the transverse energy per unit spacetime rapidity $\dd\Ep/\dd\eta_s$ defined in \cref{eq:dEperp-detas} is plotted in \cref{fig:dEp-detas-vs-tau}, and the analogous quantity calculated using the local rest frame energy density,
\begin{align}
    \frac{\dd E_\mathrm{LRF}}{\dd\eta_s} = \tau \intop\dd^2\xperp\,\elrf(\tau, \eta_s, \xperp), \label{eq:dElrf-deta-def}
\end{align}
is plotted in \cref{fig:dElrf-detas-vs-tau}.
Note that both quantities are defined with the $\tau$ factor from the Jacobian, which corrects for the increase of the volume element covered by one unit of spacetime rapidity with proper time.
The leading order time dependence of the energy density due to longitudinal expansion is accounted for by this factor of $\tau$.

Over a duration of $\tau\sim 1$~fm/$c$ the rapidity profiles for both RHIC and the LHC stabilize.
For RHIC, the magnitude and width increase.
For the LHC, the magnitude changes considerably less and especially the mid-rapidity plateau is almost independent of proper time.
Hence, the dominant contribution to time evolution is longitudinal expansion, explained by a $1/\tau$ scaling.
Still, the lower energy setup for RHIC deviates from this scaling at early times.

\begin{figure}[p]
    \centering
    \includegraphics[width=0.49\linewidth]{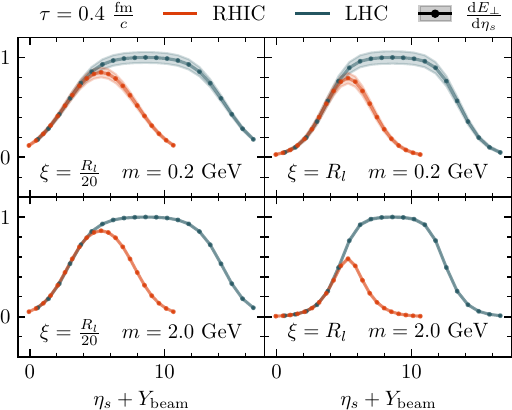}\hfill%
    \includegraphics[width=0.49\linewidth]{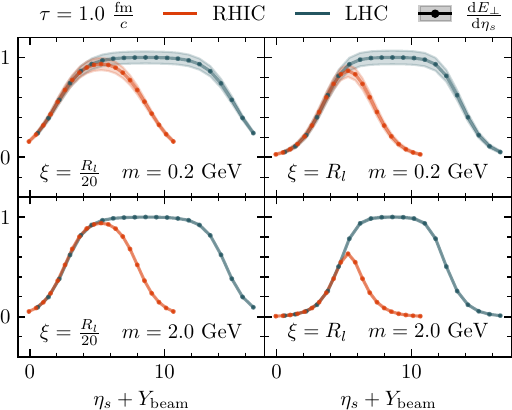}
    \caption[Energy dependence of normalized rapidity profiles of transverse energy.]{%
    Event-averaged rapidity profiles of the transverse energy density $\dd\Ep/\dd\eta_s$.
    The shaded bands denote the standard deviation obtained from the event statistics.
    RHIC and LHC parameters from \cref{tab:numerics-zeta-model-params} with $b=0$ are used.
    Left: $\tau=0.4$~fm/$c$. Right: $\tau=1.0$~fm/$c$.
    The curves are scaled by a constant factor to normalize the LHC results at $\eta_s=0$ and preserve the relative scales.
    The $x$-axis is shifted by the respective, energy-dependent beam rapidities $Y_\mathrm{beam}$.
    LHC energies develop a wide plateau around mid-rapidity and all results demonstrate limiting fragmentation.
    Left panel adapted from~\cite{Ipp:2024ykh}.
    }
    \label{fig:dEp-detas-vs-energy}
\end{figure}
\begin{figure}[p]
    \centering
    \includegraphics[width=0.49\linewidth]{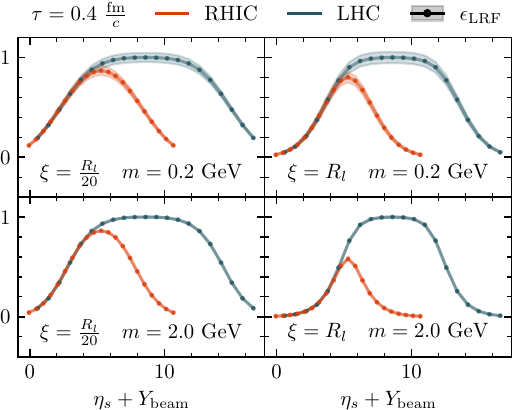}\hfill%
    \includegraphics[width=0.49\linewidth]{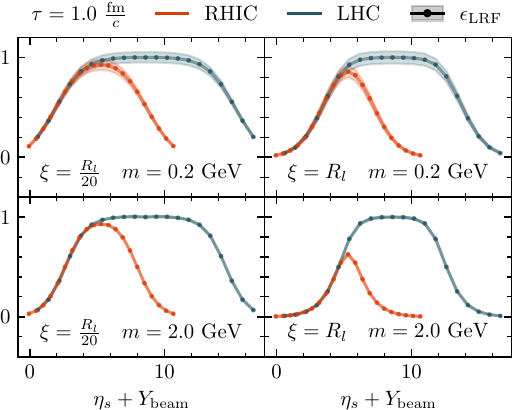}
    \caption[Energy dependence of normalized rapidity profiles of local rest frame energy density.]{%
    Event-averaged rapidity profiles of the local rest frame energy density integrated over the transverse plane.
    The shaded bands denote the standard deviation obtained from the event statistics.
    RHIC and LHC parameters from \cref{tab:numerics-zeta-model-params} with $b=0$ are used.
    Left: $\tau=0.4$~fm/$c$. Right: $\tau=1.0$~fm/$c$.
    The curves are scaled by a constant factor to normalize the LHC results at $\eta_s=0$ and preserve the relative scales.
    The $x$-axis is shifted by the respective, energy-dependent beam rapidities $Y_\mathrm{beam}$.
    Similar to \cref{fig:dEp-detas-vs-energy}, LHC results develop a mid-rapidity plateau and limiting fragmentation is observed.
    }
    \label{fig:dElrf-detas-vs-energy}
\end{figure}
\afterpage{\FloatBarrier}

Next, we discuss the energy dependence of the spacetime rapidity profiles of the transverse energy in \cref{fig:dEp-detas-vs-energy} and the local rest frame energy density in \cref{fig:dElrf-detas-vs-energy}.
Both figures contain shaded bands that mark the width of one standard deviation of the event-by-event fluctuations.
For each parameter set, the data for RHIC and the LHC are overlayed and scaled by the same factor to normalize the LHC results and preserve the relative scales.
Additionally, the spacetime rapidity-axis is shifted by the respective beam rapidity $Y_\mathrm{beam} = \arcosh(\gamma)$.

Across all parameters, the LHC results show a wide central plateau.
The curves for RHIC reach the magnitude of the LHC plateau only for the smaller $\xi$ value and never develop a plateau.
In this sense, larger correlation lengths lead to less energy being deposited during the collision.
Shifting with the beam rapidity highlights the limiting behavior of the results in the fragmentation region at $-\eta_s \gg 1$.
As derived in detail in \cref{ch:limiting-fragmentation}, the (3+1)D dilute Glasma predicts local longitudinal scaling for the field-strength tensor and all derived observables.
These results confirm limiting fragmentation for the transverse energy and $\elrf$ for all used parameters and proper times.
Further numerical results for limiting fragmentation are presented in \cref{sec:results-limiting-fragmentation}.

\subsection{Transverse eccentricity moments}

\begin{figure}[p]
    \centering
    \includegraphics[width=0.49\linewidth]{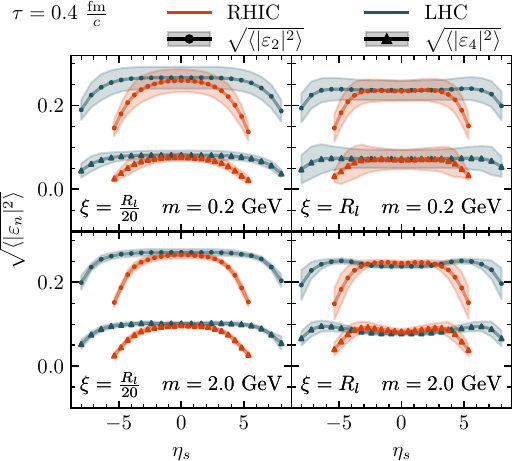}\hfill%
    \includegraphics[width=0.49\linewidth]{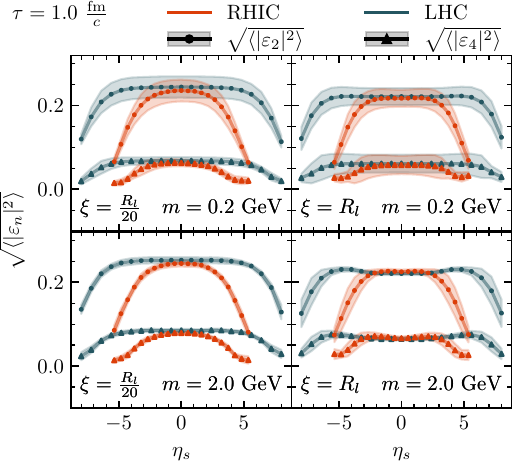}
    \caption[Rapidity profiles of second and fourth order eccentricities.]{%
    Event-averaged rapidity profiles of the second and fourth order eccentricities defined in \cref{eq:eccentricity-def}.
    The absolute square is averaged over the events and then the square root is calculated.
    The shaded bands denote the standard deviation obtained from the event statistics.
    RHIC and LHC parameters from \cref{tab:numerics-zeta-model-params} with a large impact parameter $b=R$ are used.
    Left: $\tau=0.4$~fm/$c$. Right: $\tau=1.0$~fm/$c$.
    The eccentricity is mostly flat along rapidity and falls off at large $|\eta_s|$.
    For later times, the plateau gets narrower.
    Left panel adapted from~\cite{Ipp:2024ykh}.
    }
    \label{fig:ecc-vs-etas}
\end{figure}

One can characterize the global structure in the transverse plane by the eccentricity moments $\varepsilon_n$ defined in \cref{eq:eccentricity-def}.
In \cref{fig:ecc-vs-etas}, the second and fourth order eccentricities are plotted.
The absolute square $|\varepsilon_n|^2$ is averaged over the events and the square root of the result is shown.
Due to the large impact parameter with a value of $b=R$, the produced Glasma exhibits significant eccentricity.
The spacetime rapidity dependence is almost flat and falls off at extreme $\eta_s$ values.
At later times (right panel), the plateaus are narrower for both RHIC and the LHC.
The value of the correlation length does not influence the eccentricity.

\begin{figure}[p]
    \centering
    \includegraphics[width=0.49\linewidth]{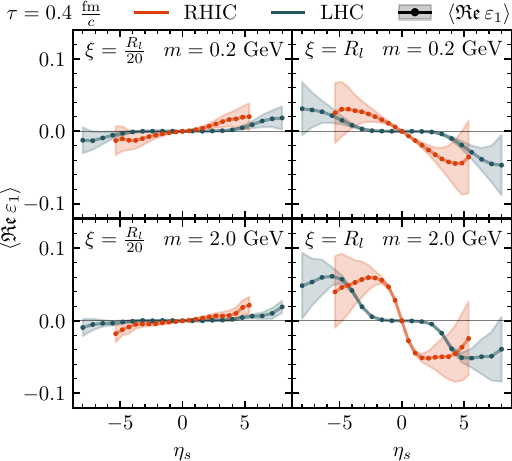}\hfill%
    \includegraphics[width=0.49\linewidth]{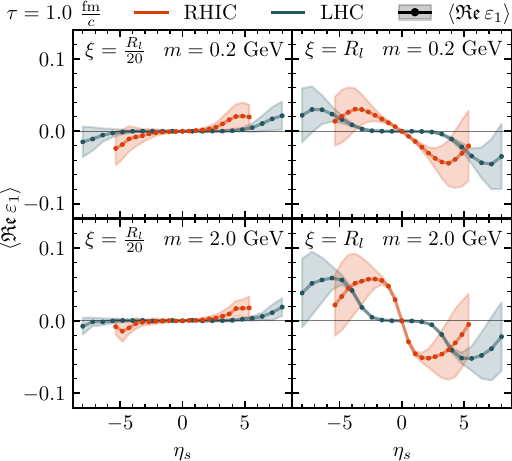}
    \caption[Rapidity profiles of first-order eccentricity.]{%
    Event-averaged rapidity profiles of the real part of the first-order eccentricity defined in \cref{eq:eccentricity-def}.
    The shaded bands denote the standard deviation obtained from the event statistics.
    RHIC and LHC parameters from \cref{tab:numerics-zeta-model-params} with a large impact parameter $b=R$ are used.
    Left: $\tau=0.4$~fm/$c$. Right: $\tau=1.0$~fm/$c$.
    $\varepsilon_1$ shows strong sensitivity to the value of the correlation length $\xi$ and changes signs when going from small $\xi$ to the coherent limit $\xi=R_l$.
    Left panel adapted from~\cite{Ipp:2024ykh}.
    }
    \label{fig:ecc1st-vs-etas}
\end{figure}

We inspect the first-order eccentricity in \cref{fig:ecc1st-vs-etas}.
Only the real part of $\varepsilon_1$ contains structure.
The imaginary part fluctuates around zero because of the alignment of the impact parameter with the $x$-axis and is not shown here.
When focusing on the smaller value of the correlation length $\xi$, we can identify a slightly positive slope for the curves.
Positive values of $\mathfrak{Re}\,\varepsilon_1$ at positive $\eta_s$ denote larger energy densities at positive $x$ coordinates and indicate a shear of the fireball in the direction of the movement of the receding nuclei.
This leads to the picture of the Glasma being dragged behind the nuclei.
We observe little time dependence for this parameter set.

The situation completely changes for the coherent case where $\xi=R_l$.
Now, the slope is negative and much larger values are reached for large $|\eta_s|$.
Both RHIC and the LHC results reach the same magnitudes.
The curves always cross zero at $\eta_s=0$ due to the calculation of the center of mass coordinates at mid-rapidity.
Still, the LHC results exhibit a prominent zero-plateau around mid-rapidity.

The eccentricity moments are connected to the development of flow in the later hydrodynamic stage of the collision and ultimately lead to the azimuthal structure of the detected particle distributions.
For example, the first-order eccentricity leads to directed flow.
By comparing the directed flow resulting from theoretical models with experimental results, various theoretical studies have identified different physical contributions to the slope of the directed flow.
In general, experimentally measured hyperon polarization can be used as a strong constraint on the longitudinal structure and directed flow predicted from theoretical models~\cite{Ryu:2021lnx, Jiang:2023vxp}.
The role of the initial geometry was studied in~\cite{Bozek:2010bi,Jiang:2024ekh}, where a tilted geometry was discussed as a possible explanation for the negative slope in identified particle distributions.
In~\cite{Snellings:1999bt}, the sign of the slope was found to be sensitive to the equation of state used for the hydrodynamic evolution phase.
The details of Baryon stopping also affect the sign of the slope~\cite{Du:2022yok, Ivanov:2014ioa}.
Using the presented results from the (3+1)D dilute Glasma as initial conditions for hydrodynamic simulations could provide a new perspective for these discussions.
The $\WSxi$ model provides parametric control over the longitudinal structure of nuclei and could lead to a deeper understanding of the experimental results.
A thorough phenomenological study of the role of the $\xi$ parameter and longitudinal structure of the $\WSxi$ nuclear model used for these Glasma results is beyond the scope of this thesis.

\pagebreak

\subsection{Gluon numbers and transverse energy}\label{sec:results-gluon-numbers}

In this section, we discuss the results of the gluon number distribution $\dd N/\dd^2\kperp\dd\Y$ and the distribution of transverse energy $\dd\Ep/\dd\Y$ obtained from the momentum space description of the (3+1)D dilute Glasma.
We employ the same $\WSxi$ model from \cref{sec:WS-correlator} that was also used for the position space calculations in the previous section.
The details of the numerical evaluation of the resulting integrals are included in \cref{appx:numerics-dN}.

Before discussing the numerical results, we list the concrete integral expressions and provide further analytical insight that is particular to the $\WSxi$ nuclear model.
This allows us to unravel the individual contributions and provides the motivation for the subsequent numerical study.

The gluon number distribution in the $\WSxi$ nuclear model reads
\begin{align}
    \kperp^2 \frac{\dd N}{\dd^2\kperp\,\dd\Y} &= \frac{2 g^6\mu^4 N_c(N_c^2-1)}{(2\pi)^3}\frac{\sqrt{R_l^2+\zeta_\A^2}\sqrt{R_l^2+\zeta_\B^2}}{R_l^2} \intop\frac{\dd^2\pperp\,\dd^2\qperp}{(2\pi)^4}\,\omega_m(\kperp,\pperp,\qperp) \nn\\
    &\times \intop\frac{\dd\kappa^+\dd\kappa^-}{(2\pi)^2}\,\ee^{-\frac{\zeta_\A^2}{2}(k^- -\kappa^-)^2 - \frac{\zeta_\B^2}{2}(k^+-\kappa^+)^2}\,\widetilde{\mathcal{W}}_\A(\kappa^-,\pperp-\qperp)\widetilde{\mathcal{W}}_\B(\kappa^+,-\pperp+\qperp), \label{eq:dN-WS-zeta-final}
\end{align}
where we used the shorthand for the IR-regulated effective vertex $\omega_m$ defined in \cref{eq:effective-omega-m-def}.
We also used the shorthand
\begin{align}
    \widetilde{\mathcal{W}}_{\A/\B}(\kappa^\mp,\kappaperp) = \intop \dd\xpm\,\dd\ypm\,\dd^2\xperp\, \ee^{\ii\kappa^\mp(\xpm-\ypm)-\ii\kappaperp\cdot\xperp} \sqrt{W_{\A/\B}(\xpm,\xperp)}\sqrt{W_{\A/\B}(\ypm,\xperp)}
\end{align}
to consolidate the contributions of the Woods-Saxon envelopes $W_{\A/\B}$ from \cref{eq:envelope-ws-def} of the nuclei $\A$ and $\B$.
The components of the external momentum associated with the on-shell gluons in \cref{eq:dN-WS-zeta-final} are $\kperp$ and $\kpm = |\kperp|\ee^{\pm\Y}/\sqrt{2}$, which is the only place where the momentum rapidity $\Y$ enters the expression.
The contributions from the correlation function $\tilde{\Gamma}_{\A/\B}^\zeta$ (given in \cref{eq:Gamma-zeta-gauss-FT} with $L_l\leftrightarrow R_l$) introduce the correlation lengths $\zeta_{\A/\B}$ for each nucleus.
Note that the external transverse momentum $\kperp$ only enters the effective vertex $\omega_m$ and $\kpm$.
It drops from the envelope factors $\widetilde{\mathcal{W}}_{\A/\B}$ as a direct consequence of the transverse Dirac-delta function in the correlation function $\Gamma_{\A/\B}$ in \cref{eq:Gamma-zeta-gauss-def}.
This can already be seen on the level of the effective TMD in \cref{eq:WS-zeta-effTMD}, where the contribution of the envelope functions only involves a Fourier transformation evaluated at the sum of the transverse momenta.
For the gluon number distribution, the effective TMD, which contains the contribution of $\kperp$, is evaluated at $(\kperp-\pperp, -\kperp+\qperp)$ (cf.\ \cref{eq:dNd2kdY-TMDs}) for the transverse momenta.
Then, $\kperp$ drops when adding these arguments.
This is reflected in the transverse momentum argument of $\widetilde{\mathcal{W}}_{\A/\B}$ in \cref{eq:dN-WS-zeta-final}.

Furthermore, the approximation of the gluon number distribution in the limit of large nuclei for the $\WSxi$ nuclear model reads
\begin{align}
     \kperp^2 \frac{\dd N}{\dd^2\kperp\,\dd\Y} &\approx \frac{2 g^6\mu^4 N_c(N_c^2-1)}{(2\pi)^3}\frac{\sqrt{R_l^2+\zeta_\A^2}\sqrt{R_l^2+\zeta_\B^2}}{R_l^2} \frac{\Sop}{m^2} \nn\\
     &\times \Omega(\kperp/m) \exp(-\tfrac{1}{4}\kperp^2(\zeta^2_\A \ee^{-2\Y} + \zeta_\B^2 \ee^{+2\Y})). \label{eq:dN-WS-zeta-large-nuclei-final}
\end{align}
Here, we used the scale-free integrated vertex $\Omega$ given in \cref{eq:Omega(k)-def}.
We reiterate that this approximation is valid in the regime where the correlation lengths $\zeta_{\A/\B} \ll R_l$ and $m R_\perp \gg 1$ and the nuclear envelope scales $R_l$ and $R_\perp$ are the largest in the system.
Comparing with the general result in \cref{eq:dN-WS-zeta-final} highlights how the contribution of the envelopes reduces to the prefactor of the transverse overlap area $\Sop$ from \cref{eq:Sop-def}.
Additionally, the transverse momentum convolutions of the effective vertex $\omega_m$ factorize and evaluate to $\Omega$.
The structure of the gluon number distribution in \cref{eq:dN-WS-zeta-large-nuclei-final} exhibits similarities to the boost-invariant MV limit in \cref{eq:dNd2kdY-MV-limit}.
Taking the correlation lengths $\zeta_{\A/\B} \rightarrow0$ in \cref{eq:dN-WS-zeta-large-nuclei-final} restores this limit, where the entire spectrum is fixed by the integrated vertex $\Omega$.
Boost-invariance is broken in the $\WSxi$ nuclear model by the correlation functions $\tilde{\Gamma}_{\A/\B}^\zeta$, which enter as direct multiplicative factors in \cref{eq:dN-WS-zeta-large-nuclei-final}.

As a result, the momentum rapidity profile in \cref{eq:dN-WS-zeta-large-nuclei-final} reduces to the specific double-exponential parametrization
\begin{align}
    \exp(-\tfrac{1}{4}\kperp^2(\zeta^2_\A \ee^{-2\Y} + \zeta_\B^2 \ee^{+2\Y})), \label{eq:WS-zeta-Y-factor}
\end{align}
which is completely fixed by the combination of the external transverse momentum and correlation lengths $|\kperp|\zeta_{\A/\B}$.
One may define a characteristic rapidity $\Y_c$ as the value of $\Y$ where \cref{eq:WS-zeta-Y-factor} reduces to $1/\ee$ of its mid-rapidity value
\begin{align}
    \ee^{-1} = \exp(-\tfrac{1}{4}\kperp^2 ( \zeta_\B^2 (\ee^{+2\Y_c}-1) + \zeta_\A^2 (\ee^{-2\Y_c}-1) )).
\end{align}
The characteristic rapidity $\Y_c$ can be interpreted as a measure of the width of the rapidity profile.
The solution for $\Y_c$ reads
\begin{align}\label{eq:Yc-sol}
    \Y_c &= - \ln(\sqrt{2}) +\frac{1}{2}\ln ( \tfrac{
    4 + \kperp^2(\zeta_\A^2+\zeta_\B^2)}{\kperp^2 \zeta_\B^2}  \pm\tfrac{\sqrt{(4+\kperp^2(\zeta_\A-\zeta_\B)^2)(4+\kperp^2(\zeta_\A+\zeta_\B)^2)}}{\kperp^2 \zeta_\B^2} ),
\end{align}
which allows for positive and negative $\Y_c$.
When choosing symmetric correlation lengths $\zeta_\A = \zeta_\B \equiv \zeta$, \cref{eq:Yc-sol} simplifies to
\begin{align}
    \Y_c = \pm\frac{1}{2}\arcosh\left( 1+2(|\kperp|\zeta)^{-2} \right). \label{eq:Yc-sol-sym}
\end{align}

The behavior of $\Y_c$ is as follows.
When shrinking the correlation scale, $\Y_c$ increases and diverges in the MV limit $\zeta\rightarrow0$.
This describes the development of the boost-invariant plateau, where the rapidity profile is perfectly flat.
Similarly, the IR limit of $|\kperp|$ also approaches the MV limit.
Conversely, large $|\kperp|$ lead to smaller $\Y_c$ and the gluon spectrum is dominated by the mid-rapidity region.

The transverse energy differential in momentum rapidity in the limit of large nuclei for the $\WSxi$ nuclear model reads
\begin{align}
    \frac{\dd\Ep}{\dd\Y} &\approx \frac{2 g^6\mu^4 N_c(N_c^2-1)}{(2\pi)^3}\frac{\sqrt{R_l^2+\zeta_\A^2}\sqrt{R_l^2+\zeta_\B^2}}{R_l^2} \frac{\Sop}{m^2} \nn\\
    &\times \intop\dd^2 \kperp\,\frac{1}{|\kperp|} \Omega(\kperp/m) \exp(-\tfrac{1}{4}\kperp^2(\zeta^2_\A \ee^{-2\Y} + \zeta_\B^2 \ee^{+2\Y})) \nn\\
    &= \frac{2 g^6\mu^4 N_c(N_c^2-1)}{(2\pi)^2}\frac{\sqrt{R_l^2+\zeta_\A^2}\sqrt{R_l^2+\zeta_\B^2}}{R_l^2} \frac{\Sop}{m} \nn\\
    &\times \intop_0^\infty \dd p\, \Omega(p) \exp(-\tfrac{1}{4}p^2 m^2 (\zeta^2_\A \ee^{-2\Y} + \zeta_\B^2 \ee^{+2\Y})). \label{eq:dE-WS-zeta-large-nuclei-final}
\end{align}
In the second equation, we changed to polar coordinates and used the substitution $p=|\kperp|/m$.
Then, we used the fact that the integrated vertex $\Omega$ is isotropic and integrated out the polar angle.
Note that these results assume head-on collisions with zero impact parameter.
The shape of the resulting rapidity profile in \cref{eq:dE-WS-zeta-large-nuclei-final} is fixed by the scaling variable $m\zeta_{\A/\B}$.
This variable combines the only model parameters that affect the shape of the rapidity distribution, in addition to its magnitude in the prefactor.
Compared to the previously identified scaling variable $|\kperp|\zeta$, which fixes the rapidity profile of the gluon number distribution, here, the transverse momentum is replaced by the IR regulator $m$.
This further illustrates that the IR regulator can be associated with the characteristic transverse momentum scale in the dilute Glasma.

\subsubsection{Gluon number spectrum}

\begin{figure}[p]
    \centering
    \includegraphics[width=0.49\linewidth]{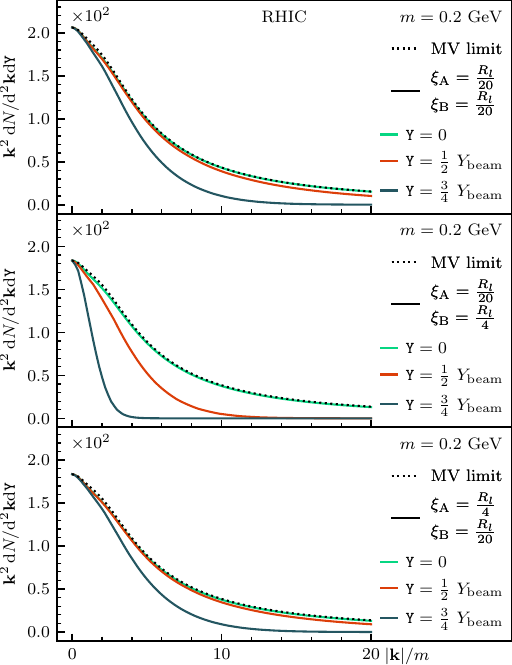}\hfill%
    \includegraphics[width=0.49\linewidth]{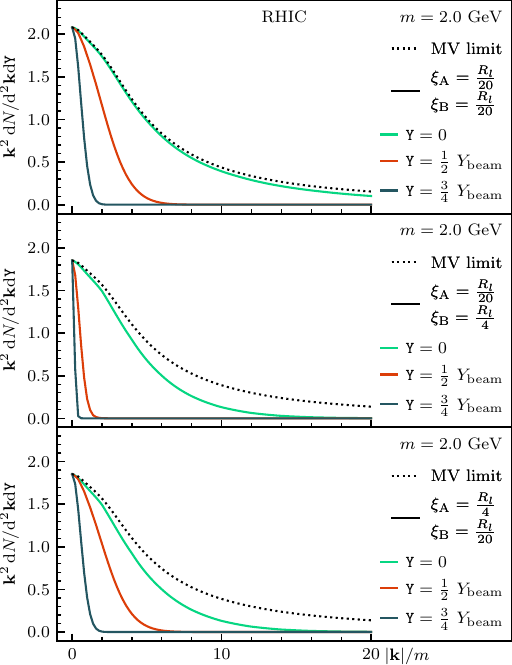}
    \caption[Transverse momentum spectra of gluon numbers for RHIC setup.]{%
    Transverse momentum spectra of rescaled gluon number distributions.
    The $|\kperp|$-values on the horizontal axis are scaled by the IR regulator $m$.
    RHIC parameters from \cref{tab:numerics-zeta-model-params} are used. Left column: $m=0.2$~GeV. Right column: $m=2.0$~GeV.
    Solid colors evaluate \cref{eq:dN-WS-zeta-final} at different momentum rapidities $\Y$.
    The dotted curve is the MV limit $\xi\rightarrow0$.
    The $\WSxi$ nuclear model introduces rapidity dependence.
    }
    \label{fig:dNvk-RHIC}
\end{figure}
\begin{figure}[p]
    \centering
    \includegraphics[width=0.49\linewidth]{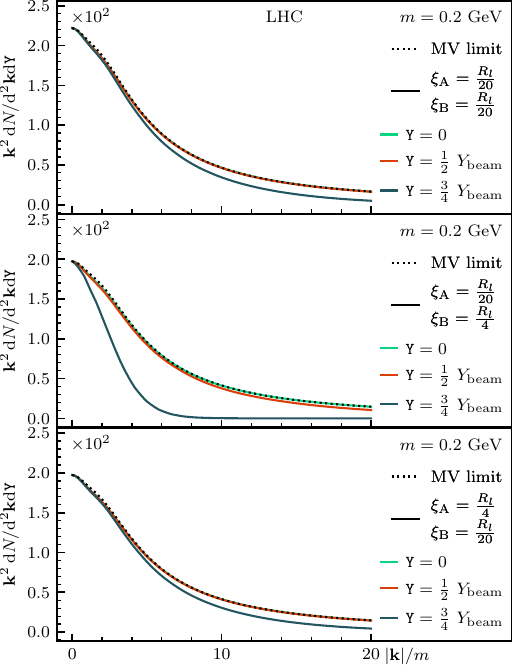}\hfill%
    \includegraphics[width=0.49\linewidth]{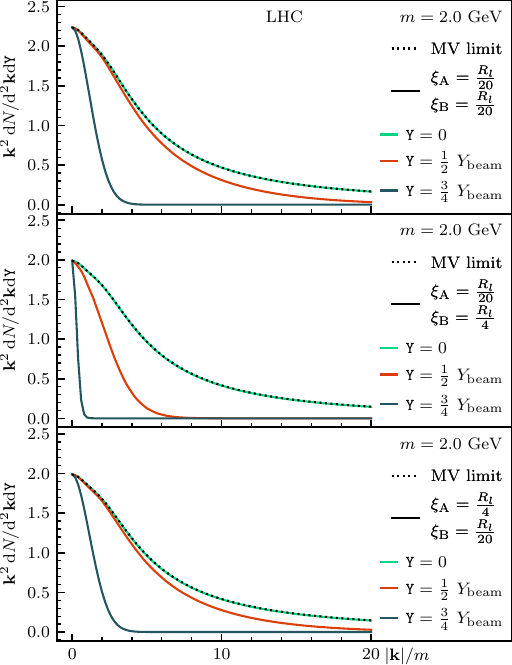}
    \caption[Transverse momentum spectra of gluon numbers for LHC setup.]{%
    Same as \cref{fig:dNvk-RHIC}, but for LHC parameters from \cref{tab:numerics-zeta-model-params}.
    The higher collision energy of the LHC setup leads to results that are closer to the boost-invariant MV limit.
    }
    \label{fig:dNvk-LHC}
\end{figure}
\afterpage{\FloatBarrier}

In \cref{fig:dNvk-RHIC,fig:dNvk-LHC}, the transverse momentum spectra of the gluon number distribution from the $\WSxi$ nuclear model are plotted.
On the vertical axis, the gluon number distribution is scaled by $\kperp^2$ to regulate the IR divergence and obtain a dimensionless quantity.
The horizontal axis for the modulus of the transverse momentum $|\kperp|$ is scaled by the inverse IR regulator $m$.
\Cref{fig:dNvk-RHIC} corresponds to the RHIC setup from \cref{tab:numerics-zeta-model-params} and \cref{fig:dNvk-LHC} to the LHC setup.
The value of the IR regulator changes from $m=0.2$~GeV in the left columns to $m=2.0$~GeV in the right columns for both figures.
The solid lines correspond to the general result of the distribution given in \cref{eq:dN-WS-zeta-final}.
For each color, a different value of the momentum rapidity is used and listed in the legend as a fraction of the beam rapidity.
In each row, the values of the correlation lengths $\xi_{\A/\B}$ are changed.

All curves peak for $|\kperp|=0$ at a value that is the same for all $\Y$ but changes with $\xi_{\A/\B}$.
Increasing one of the correlation lengths slightly reduces the maximum.
Since the inverse of the correlation length is related to a longitudinal momentum scale in the nuclei, larger values of $\xi_{\A/\B}$ suppress the high-momentum modes earlier and result in less energy available for gluon production.
Larger $\xi_{\A/\B}$ also lead to smaller values for the spectra across the entire $|\kperp|$-range.
This is especially noticeable in the middle rows where $\xi_\B$ is increased.
The spectra are symmetric under the simultaneous exchange of $\A\leftrightarrow\B$ and $\Y\leftrightarrow-\Y$.
But, because all values of $\Y\geq0$, the results show an asymmetry w.r.t.\ $\xi_\A$ and $\xi_\B$.
For large, positive rapidities, the fragmentation region of nucleus $\B$ is probed, which leads to the visible asymmetry when varying $\xi_{\A/\B}$.

The dotted curves in \cref{fig:dNvk-RHIC,fig:dNvk-LHC} correspond to the MV limit $\xi_{\A/\B}\rightarrow0$ and are rescaled to match the values of the solid curves at $|\kperp|=0$.
In this limit, the transverse momentum spectrum is given by the integrated vertex $\Omega$ from \cref{eq:Omega(k)-def} and is independent of $\Y$ (i.e., boost-invariant).
At mid-rapidity (solid green), all spectra show similar behavior to the MV case.
However, increasing the IR regulator (right panels) results in a sizable deviation for the RHIC setup.
We conclude that the boost-invariant results are modified in the (3+1)D dilute Glasma even at mid-rapidity.
At smaller values of $\xi_{\A/\B}$ or when increasing the collider energy to the LHC setup in \cref{fig:dNvk-LHC}, the mid-rapidity spectra are indistinguishable from the boost-invariant limit even at the larger value of $m=2.0$~GeV.

The solid curves strongly depend on the momentum rapidity $\Y$.
The spectra are always largest at mid-rapidity.
Larger values of $\Y$ change the asymptotic behavior with $|\kperp|$ from a long tail $\sim\ln(|\kperp|/m)m^2/\kperp^2$ given by the integrated vertex to a steep exponential decay given by the exponentials from the correlation functions.
The variation with $\Y$ demonstrates the effect of breaking boost-invariance with finite nuclei and longitudinal correlation scales.

Finally, we note the scaling dependence on the IR regulator.
Increasing the value of $m$ by a factor of 10 leads to a suppression by a factor of 100, which confirms that the $1/m^2$ scaling extracted in the limit of large nuclei in \cref{eq:dN-WS-zeta-large-nuclei-final} is valid also for the general case.

\begin{figure}[p]
    \centering
    \includegraphics[width=0.49\linewidth]{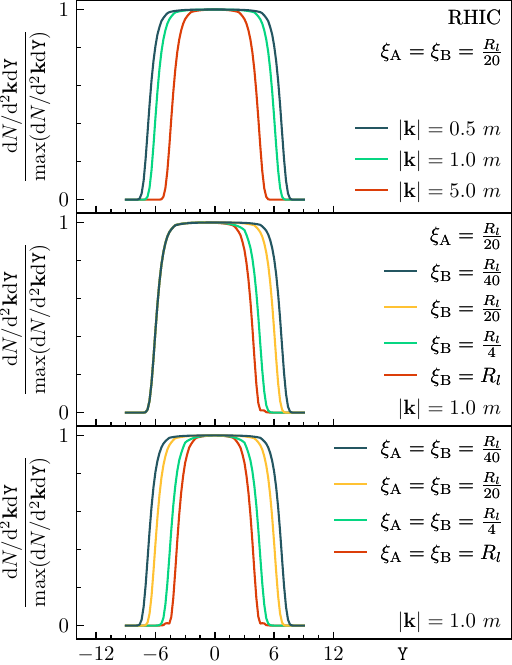}\hfill%
    \includegraphics[width=0.49\linewidth]{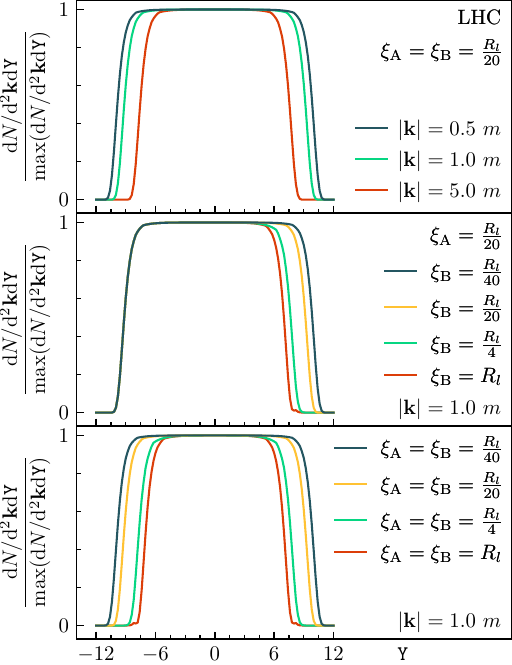}
    \caption[Normalized rapidity profiles of the gluon number distribution.]{%
    Rapidity profiles of the gluon number distribution from \cref{eq:dN-WS-zeta-final}.
    All curves are normalized to 1 at $\Y=0$.
    RHIC (left) and LHC (right) parameters from \cref{tab:numerics-zeta-model-params} are used.
    The value of the IR regulator $m=0.2$~GeV.
    Top row: different colors denote different values of $|\kperp|$ as ratios to $m$.
    Middle row: $\xi_\A$ is fixed and $\xi_\B$ is scanned.
    Bottom row: $\xi_{\A/\B}$ are scanned symmetrically.
    The width of the mid-rapidity plateau increases for smaller $\xi_{\A/\B}$ or smaller $|\kperp|$ and with collider energy.
    Each flank is only sensitive to the $\xi_{\A/\B}$ parameter of the nucleus whose fragmentation region is at that rapidity range.
    }
    \label{fig:dNvY}
\end{figure}
\begin{figure}[p]
    \centering
    \includegraphics{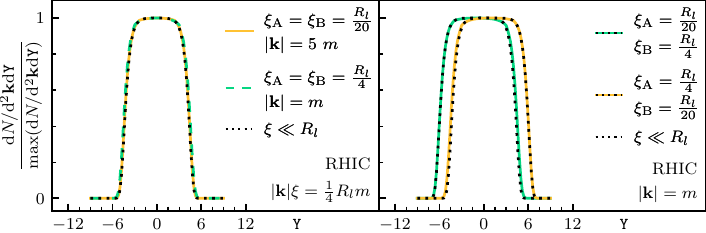}
    \includegraphics{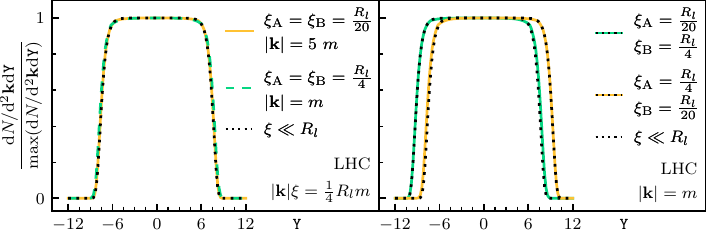}
    \caption[Large nuclei limit and scaling variable of gluon number rapidity distributions.]{%
    Select curves from \cref{fig:dNvY} (colored) are compared with the double exponential profile in \cref{eq:WS-zeta-Y-factor} (dotted), valid for the limit of large nuclei.
    Left panels: the value of the scaling variable $|\kperp|\xi$ is kept fixed for different $|\kperp|$ and symmetric $\xi_{\A/\B}$.
    Righ panels: the value of $|\kperp|$ is fixed and $\xi_{\A/\B}$ varied asymmetrically.
    The scaling variable $|\kperp|\xi$ from the large nuclei limit also holds in the general case.
    }
    \label{fig:dNvY-scaling}
\end{figure}
\afterpage{\FloatBarrier}

Next, we study the momentum rapidity profiles of the gluon number distribution in \cref{eq:dN-WS-zeta-final}.
In \cref{fig:dNvY}, normalized curves for a variety of different parameters are plotted.
The values for the RHIC (left column) and LHC (right column) setups are given in \cref{tab:numerics-zeta-model-params}.
For all panels, the value of the IR regulator is fixed at $m=0.2$~GeV.
Overall, the characteristic shape of the profiles consists of steep flanks and a flat plateau around mid-rapidity ($\Y=0$).

In the top row, the correlation lengths are fixed to $\xi_\A=\xi_\B=R_l/20$ symmetrically and the transverse momentum $|\kperp|$ is varied for the different colors.
Large transverse momenta lead to narrower curves, indicating that the dynamics at large $|\kperp|$ are limited to the mid-rapidity region.
The mirror symmetry of the curves around $\Y=0$ is preserved.

In the middle and bottom rows, the value of $|\kperp| = m = 0.2$~GeV.
Keeping $\xi_\A$ fixed and scanning $\xi_\B$ (middle row) only affects the position of the right flank, which is located at the rapidity values that correspond to the fragmentation region of nucleus $\B$.
This also breaks the $\Y$-symmetry of the curves and the centers of the plateaus shift away from $\Y=0$.
When changing the correlation lengths symmetrically (bottom row), both flanks are affected.
Smaller values of $\xi_{\A/\B}$ lead to wider plateaus.

Similarly, switching to the higher energy for the LHC setup (right column) leads to wider plateaus.
The increase of the Lorentz factor results in smaller physical values for the correlation lengths, since $R_l\sim1/\gamma$.
This reveals again the connection of the available energy in the system with the inverse correlation scale.
Hence, we expect an analogous change in the profiles compared to varying the correlation lengths.
The Lorentz factor also enters the envelopes via $R_l$.
The results appear to be only sensitive to the change in $\xi_{\A/\B}$ and suggest that the modifications due to the envelope are minimal.

We further analyse the contributions of the envelopes by comparing with the result for the limit of large nuclei (dotted) in \cref{fig:dNvY-scaling}.
In this approximation, the entire structure is given by the double exponential in \cref{eq:WS-zeta-Y-factor}.
The contribution of the envelopes reduces to a prefactor that is scaled out in the normalized plots.
We observe perfect agreement of the dotted curves with the colored curves, indicating that for values $\xi_{\A/\B} \lesssim R_l/4$ the shape of the rapidity profiles is given by \cref{eq:WS-zeta-Y-factor}.

In the left column, the value of the previously identified scaling variable $|\kperp|\xi = R_l m /4$ is kept fixed for symmetric $\xi_{\A/\B}$.
The results demonstrate that the shape of the profile is completely determined, also in the general case, by the value of the scaling variable, regardless of the individual values of $|\kperp|$ and $\xi_{\A/\B}$.
In the right column, asymmetric values for $\xi_{\A/\B}$ are used and further confirm that the limit of large nuclei also agrees with the general result for asymmetric profiles.
In this case, there is no unique scaling variable.
Instead, the position of each flank is controlled by one of $|\kperp|\xi_{\A/\B}$ depending on which nucleus' fragmentation region corresponds to those rapidities.
This is supported by the observation that the profile assembled from the orange flank for negative $\Y$ ($\xi_\A=R_l/4$) and green flank for positive $\Y$ ($\xi_\B=R_l/4$) perfectly matches the profile resulting from $\xi_\A=\xi_\B=R_l/4$.

\begin{figure}[p]
    \centering
    \includegraphics{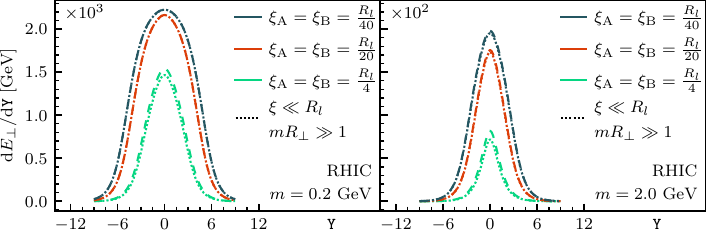}
    \includegraphics{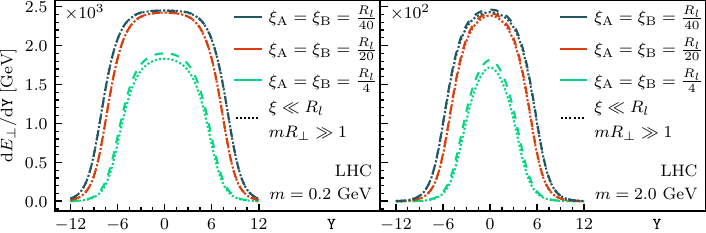}
    \caption[Momentum rapidity distributions of transverse energy.]{%
    Rapidity profiles of the transverse energy obtained from the gluon distribution in the $\WSxi$ nuclear model.
    Solid lines correspond to integrating \cref{eq:dN-WS-zeta-final} over $|\kperp|$, dotted lines to the limit of large nuclei in \cref{eq:dE-WS-zeta-large-nuclei-final}.
    RHIC (top row) and LHC (bottom row) parameters from \cref{tab:numerics-zeta-model-params} with symmetric $\xi_{\A/\B}$ (colors) are used.
    Left column: $m=0.2$~GeV. Right column: $m=2.0$~GeV.
    The overall magnitude scales $\sim 1/m$ and increases for smaller $\xi_{\A/\B}$.
    The large nuclei approximation (dotted) agrees with the general results.
    }
    \label{fig:dEpvY-sym-xi-approx}
\end{figure}
\begin{figure}[p]
    \centering
    \includegraphics{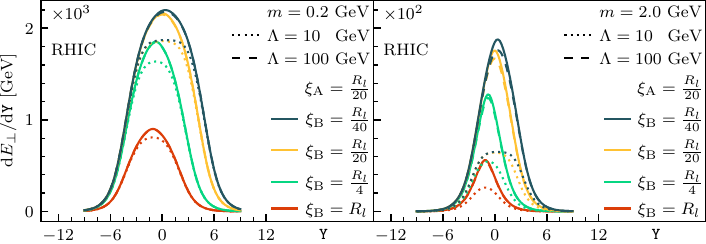}
    \includegraphics{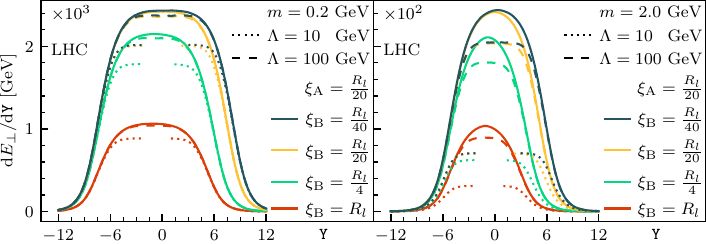}
    \caption[Transverse energy with UV regulation.]{%
    Rapidity profiles of the transverse energy obtained from integrating \cref{eq:dN-WS-zeta-final} over $|\kperp|$.
    RHIC (top row) and LHC (bottom row) parameters from \cref{tab:numerics-zeta-model-params} with fixed $\xi_\A$ and scanning $\xi_\B$ (colors) are used.
    Left column: $m=0.2$~GeV. Right column: $m=2.0$~GeV.
    Changing $\xi_\B$ only affects the overall magnitude and flank in the fragmentation region of nucleus $\B$ ($\Y\gg0$).
    The dotted and dashed curves incorporate UV regulation in the nuclear model (cf.~\cref{sec:position-space-lattice}), which suppresses the peaks and forms flat plateaus.
    Convergence was not achieved for the missing values at mid-rapidity of the dotted curves for LHC.
    }
    \label{fig:dEpvY-UV}
\end{figure}
\FloatBarrier

\subsubsection{Transverse energy distribution}

\begin{figure}[t!]
    \centering
    \includegraphics{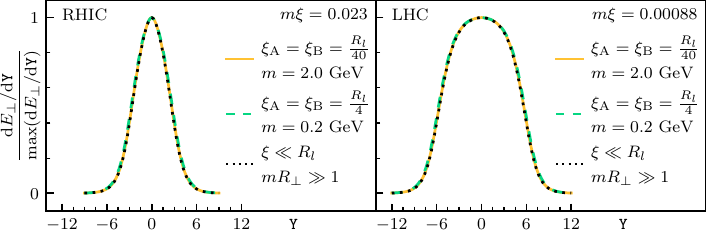}
    \caption[Large nuclei limit and scaling variable for transverse energy.]{%
    Selected curves from \cref{fig:dEpvY-sym-xi-approx}, normalized at $\Y=0$.
    All curves in each panel have the same value of the scaling variable $m\xi$, but different individual values for symmetric $\xi_{\A/\B}$ and $m$ in green and yellow.
    The scaling variable fixes the profile in the limit of large nuclei in \cref{eq:dE-WS-zeta-large-nuclei-final} (dotted lines) and perfectly matches the shape of the general results (solid and dashed lines). 
    }
    \label{fig:dEpvY-mxi-scaling}
\end{figure}

The transverse energy differential in momentum rapidity $\dd\Ep/\dd\Y$ is obtained by integrating the scaled gluon number distribution $|\kperp|\,\dd N/\dd^2\kperp\dd\Y$ over $\kperp$ (cf.~\cref{sec:transverse-energy-mom}).
In \cref{fig:dEpvY-sym-xi-approx}, the transverse energy in the $\WSxi$ nuclear model according to the gluon number distribution in \cref{eq:dN-WS-zeta-final} is plotted (dashed lines).
In the top row, parameters for the RHIC setup, and in the bottom row, those for the LHC setup, listed in \cref{tab:numerics-zeta-model-params}, are used.
Different colors correspond to different symmetric values of $\xi_{\A/\B}$.
As already argued for the gluon numbers themselves, larger correlation lengths suppress the available energy for gluon production and lead to a reduction of the overall magnitudes of $\dd\Ep/\dd\Y$.
The dominant plateaus around mid-rapidity observed for the gluon number distribution are now softened, and only for the LHC setup with small $\xi_{\A/\B}$, a plateau is still visible.
Clearly, a boost-invariant description for the transverse energy seems inappropriate.

We compare the dashed lines with the approximation in the limit of large nuclei given in \cref{eq:dE-WS-zeta-large-nuclei-final} (dotted lines).
Overall, we observe very good agreement with the approximate results across all parameters.
The scales of the curves also agree.
Note that the vertical axes in the right column are scaled by 10 compared to the left column.
The value of the IR regulator is also increased by a factor of 10.
We conclude that the $1/m$ scaling extracted from the approximate results in \cref{eq:dE-WS-zeta-large-nuclei-final} also holds for the general case.
Still, the IR regulator also affects the shape of the curves via the vertex factor $\omega_m$.

We observe a small deviation between the large nuclei approximation and the general result for the largest $\xi_{\A/\B}$ value (green).
In \cref{fig:dEpvY-mxi-scaling}, we further compare these two results by normalizing both at mid-rapidity.
When scaling out any constant prefactors, the large nuclei approximation (dotted lines) agrees with the general results (yellow and green) perfectly.
Even the case with moderate $\xi_{\A/\B}=R_l/4$ is indistinguishable.
Therefore, the approximate results reproduce the shape of the profiles, but not the overall scale.
The modifications due to the finite-size envelopes in the general results do not appear to be significant.
The shape of the profiles is completely fixed by the scaling variable $m\xi$ identified in the approximate expression in \cref{eq:dE-WS-zeta-large-nuclei-final}.
This is demonstrated in \cref{fig:dEpvY-mxi-scaling}, where for each panel, $m\xi$ is fixed to a constant and the values of $m$ and $\xi$ are varied for the different colors.

\pagebreak
We study asymmetric combinations of $\xi_{\A/\B}$ in \cref{fig:dEpvY-UV} with the same panel arrangement as in \cref{fig:dEpvY-sym-xi-approx}.
Keeping $\xi_\A$ fixed and scanning through the values for $\xi_\B$ affects the rapidity position of the flank for $\Y>0$, which corresponds to the fragmentation region of nucleus $\B$.
This is the same behavior already discussed for the gluon number distribution.
Additionally, the overall magnitude decreases for larger $\xi_\B$, but by a smaller factor compared to the symmetric $\xi_{\A/\B}$ values in \cref{fig:dEpvY-sym-xi-approx}.

In \cref{fig:dEpvY-UV}, we also investigate the effect of UV regulation in the nuclear model according to \cref{eq:single-nucleus-gauge-fields-UV-regulated}, which was used for the results obtained in the position space framework.
Here, the exponential UV regulation factor can be absorbed in a redefinition of the effective vertex,
\begin{align}
    \omega_m^\mathrm{UV}(\kperp,\pperp,\qperp) = \omega_m(\kperp,\pperp,\qperp)\,\exp(- \tfrac{1}{2 \Lambda^2}(\pperp^2 + \qperp^2 + (\kperp-\pperp)^2 + (\kperp-\qperp)^2) ),
\end{align}
where $\Lambda$ is the UV regulator.

Two values for the regulator $\Lambda=10$~GeV (dotted) and $\Lambda=100$~GeV (dashed) are shown in \cref{fig:dEpvY-UV}.
For the smaller value of $\Lambda$, the curves are strongly modified.
The peak is cut off by a flat plateau, suggesting that the contribution from the transverse UV modes is most significant in the mid-rapidity region.%
\footnote{%
The numerical results with $\Lambda=10$~GeV for LHC in \cref{fig:dEpvY-UV} (dotted, bottom row) are incomplete at mid-rapidity because the convergence of the numerical integration could not be achieved.}
If UV regulation is combined with a large IR regulator $m=2.0$~GeV (right column), the differences are substantial.
Clearly, limiting the transverse momentum to an interval $2.0\div10.0$~GeV does not capture the important contributions, especially for the setup with larger energy at the LHC (bottom row).
Therefore, the UV regulator is essential for comparing the momentum space and position space results in the next section.

\subsection{Comparison of transverse energy from position space and momentum space calculations}\label{subsec:EMT-vs-dN}

\begin{figure}
    \vspace{-0.2\baselineskip}\centering
    \includegraphics{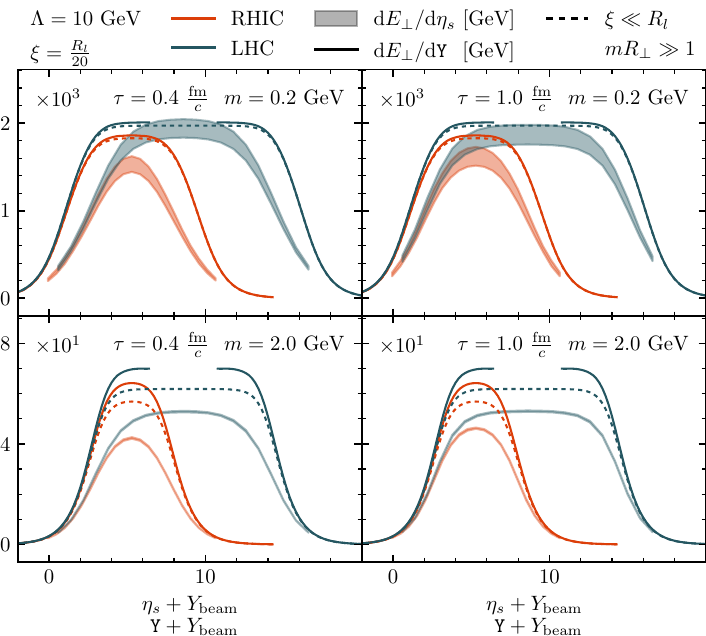}
    \caption[Comparison of transverse energy from position and momentum space calculations.]{%
    Comparison of the transverse energy differential in spacetime rapidity $\dd\Ep/\dd\eta_s$ obtained from the position space calculation (colored bands, same data as in \cref{fig:dEp-detas-vs-energy}) and the transverse energy differential in momentum rapidity $\dd\Ep/\dd\Y$ obtained from the momentum space calculation (colored lines, same data as in \cref{fig:dEpvY-UV}) for the same $\WSxi$ nuclear model.
    RHIC (orange) and LHC (blue) parameters from \cref{tab:numerics-zeta-model-params}.
    The UV regulator is fixed to $\Lambda=10.0$~GeV and the correlation length symmetrically to $\xi=R_l/20$.
    The dashed line is the limit of large nuclei in the momentum space picture with UV regulation.
    Proper time $\tau$ varies per column and only affects $\dd\Ep/\dd\eta_s$.
    The position space results (bands) are narrower and at smaller scales than the momentum space results (lines).
    The differences increase for the larger value of the IR regulator $m=2.0$~GeV (bottom row).
    }
    \label{fig:dEpvY-mom-pos-comp}
\end{figure}

In \cref{fig:dEpvY-mom-pos-comp}, we compare the rapidity profiles for the transverse energy obtained from the position space and momentum space calculations in the (3+1)D dilute Glasma.
The transverse energy differential in spacetime rapidity $\dd\Ep/\dd\eta_s$ (shaded bands) is the same data from \cref{fig:dEp-detas-vs-energy}.
The transverse energy differential in momentum rapidity $\dd\Ep/\dd\Y$ (solid lines) is the same data with UV regulation from \cref{fig:dEpvY-UV}.
Additionally, the dashed lines correspond to the limit of large nuclei in \cref{eq:dE-WS-zeta-large-nuclei-final} where UV regulation was also added.
The same $\WSxi$ nuclear model was used for all calculations.
The parameters for the RHIC (orange) and LHC (blue) setups are listed in \cref{tab:numerics-zeta-model-params}.
The UV regulator is given by $\Lambda = 10$~GeV and the correlation lengths take the value $\xi_\A=\xi_\B=\xi=R_l/20$ for all panels.

Overall, the position and momentum space data are comparable.
For a perfectly free-streaming system, one would expect a match, given the identical parameters and implementation of the nuclear model.
However, the momentum space results are systematically wider and larger.

The value of the proper time changes from $\tau=0.4$~fm/$c$ in the left column to $\tau=1.0$~fm/$c$ in the right column and only affects $\dd\Ep/\dd\eta_s$.
The time evolution of $\dd\Ep/\dd\eta_s$ was studied in \cref{fig:dEp-detas-vs-tau} where it was established that the curves widen as proper time increases.
At later times, the position space results align more closely with the momentum space results.
This is expected, as the momentum space calculation evaluates the Glasma field at asymptotically late times.
In the position space calculation, this asymptotic limit is not reached at the larger value of $\tau=1.0$~fm/$c$.

Furthermore, the (3+1)D dilute Glasma exhibits substantial longitudinal flow (cf.~\cref{fig:ueta-vs-etas}), especially for large rapidities.
Nonzero longitudinal flow breaks the symmetry of the Milne frame and will affect the transverse energy distribution differential in spacetime rapidity.

The IR regulator changes from $m=0.2$~GeV in the top row to $m=2.0$~GeV in the bottom row.
For smaller $m$, the magnitudes of the curves are closer together, whereas increasing $m$ leads to larger differences.
The effect of IR regulation was discussed with \cref{fig:dEpvY-UV}.
In particular, large $m$ and small $\Lambda$ lead to lower plateaus.
The difference of $\dd\Ep/\dd\eta_s$ and $\dd\Ep/\dd\Y$ could come from the additional hard UV cutoff due to the discretization lattice used for the position space calculation.
That cutoff was not included for $\dd\Ep/\dd\Y$.
Still, the shapes of the curves in the bottom row seem to agree well, despite the difference in scale.

Finally, we note that the horizontal axis is shifted by the respective beam rapidity $Y_\mathrm{beam} = \arcosh(\gamma)$.
This demonstrates that the momentum space calculation also exhibits limiting fragmentation.
This topic will be studied in detail in \cref{sec:results-limiting-fragmentation}.

\section{Phenomenological GBW saturation model}\label{sec:numerics-GBW}

\begin{table}
    \caption[Parameters for numerical evaluation of GBW model results.]{\label{tab:numerics-GBW-model-params}%
    Values of model parameters for the GBW nuclear model from \cref{sec:GBW-saturation-model} used for numerical evaluation.
    Collider energies correspond to Au+Au collisions at $\sqrt{s_\mathrm{NN}}=200$~GeV performed at RHIC and Pb+Pb collisions at $\sqrt{s_\mathrm{NN}}=5400$~GeV performed at the LHC.
    The Gauss radius $L$ is tuned s.t.\ the Gaussian envelope in \cref{eq:envelope-gauss-def} produces the same $\Sp^{\A/\B}$ given in \cref{eq:nucl-models-Sperp} as the Woods-Saxon envelope in \cref{eq:envelope-ws-def} used in the previous section.}
    \centering\begin{tabular}{lllclcl}
    \hline\hline
     & \textbf{Name} & \textbf{RHIC} & \textbf{LHC} & \textbf{Unit}\\
    \hline
         $\gamma$ & Lorentz factor & 100 & 2700 & $-$ \\
         $L$ & Gauss radius & 3.81 & 3.95 & fm \\
         $L$ & Gauss radius & 19.30 & 20.01 & GeV${}^{-1}$ \\
         $m$ & IR regulator & \multicolumn{2}{c}{0.2}  & GeV \\
         $g$ & CYM coupling & \multicolumn{2}{c}{1.0} & $-$ \\
         $Q_0$ & $Q_s$ scale factor & \multicolumn{2}{c}{ \{0.25, 0.5, 0.75, 1.0\} } & GeV \\
         $\lambda$ & GBW exponent & \multicolumn{2}{c}{0.144} & $-$ \\
         $m_n$ & Nucleon mass & \multicolumn{2}{c}{1.0} & GeV \\
         $N_c$ & Number of colors & \multicolumn{2}{c}{3} & $-$ \\
    \hline\hline
    \end{tabular}
\end{table}

In this section, we present exploratory results for the transverse energy obtained from the gluon number distribution of the GBW saturation model introduced in \cref{sec:GBW-saturation-model}.
As previously discussed, positive semi-definiteness of the two-point correlator is not guaranteed in this model.
The presented results are, therefore, not unconditionally physical.
Still, they serve as an example for fundamentally different TMDs than were discussed for the gluon number distributions until now.
Furthermore, we compare the GBW nuclear model with three-dimensional, finite envelopes to the limit where the envelope scales are pushed to infinity.
In this limit, the GBW model is positive semi-definite.

We list the concrete integral expressions below, which are numerically evaluated for the plots.
The details of the implementation are discussed in \cref{appx:numerics-dN}.
The gluon number distribution of the GBW model reads
\begin{align}
    \kperp^2 \frac{\dd N}{\dd\kperp^2\,\dd\Y} &= \frac{N_c^3 \Sp^\A \Sp^\B}{16\pi g^2 (N_c^2-1)}\intop\frac{\dd^2\pperp\,\dd^2\qperp}{(2\pi)^4}\frac{\omega_m(\kperp,\pperp,\qperp)}{Q_s^\A(\xB_\A)^2 Q_s^\B(\xB_\B)^2}(\pperp+\qperp)^4(2\kperp-\pperp-\qperp)^4 \nn\\
    &\times \ee^{-L_\perp^2(\pperp-\qperp)^2} \exp(-(\tfrac{\pperp+\qperp}{2})^2\,Q_s^\A(\xB_\A)^{-2} - (\tfrac{2\kperp-\pperp-\qperp}{2})^2\,Q_s^\B(\xB_\B)^{-2}). \label{eq:dN-GBW-numerics}
\end{align}
Here, we introduced the saturation scales $Q_s^{\A/\B}$ for each nucleus defined as
\begin{align}
    Q_s^{\A/\B}(\xB_{\A/\B}) = Q_0^{\A/\B}\xB_{\A/\B}^{-\lambda}(1-\xB_{\A/\B}), \qquad \xB_{\A/\B} = \frac{|\kperp|}{2 m_n \gamma}\ee^{\mp\Y}.
\end{align}
This parametrization of the saturation scale requires that the momentum fractions $\xB_{\A/\B}\leq1$.
In practice, this imposes limits on the maximal momentum rapidity which depend on the value of $|\kperp|$ and $\gamma$.
As $\xB_{\A/\B}\rightarrow1$, the exponential factors in \cref{eq:dN-GBW-numerics} smoothly push the gluon distribution to zero.
The values for all model parameters in this model are listed in \cref{tab:numerics-GBW-model-params}.

Surprisingly, the longitudinal envelope scale drops out of the expression in \cref{eq:dN-GBW-numerics} completely.
Only the transverse Gauss radius $L_\perp$ remains.
This feature is unique for the Gaussian envelopes used in this model.
Note that the projected transverse area $\Sp^{\A/\B}$ does not scale with the longitudinal size of the nuclei due to the normalization of the Gaussian envelope functions (cf.~\cref{sec:proj-transv-area}).

In the limit of large nuclei, the gluon number distribution is given by \cref{eq:dNd2kdY-large-nucl-TMDs} as a convolution of two TMDs.
To be precise, that formula is only valid when the nuclear scales are the largest in the system.
However, in the GBW model, the scales introduced by the TMD are dynamic and can come into conflict with the envelope scales (cf.~\cref{sec:pos-semidef-GBW}).
To circumvent this problem, we formally push the envelope scales to infinity.
Then, the gluon number distribution using the original GBW TMD from \cref{eq:GBW-TMD} is given as
\begin{align}
    \frac{\kperp^2}{\Sop} \frac{\dd N}{\dd\kperp^2\,\dd\Y} &\approx \frac{16 N_c^3}{\pi g^2(N_c^2-1)}\intop\frac{\dd^2\pperp}{(2\pi)^2} \nn\\
    &\times \frac{\pperp^2(\kperp-\pperp)^2}{Q_s^\A(\xB_\A)^2 Q_s^\B(\xB_\B)^2} \exp(-\pperp^2 Q_s^\A(\xB_\A)^{-2} - (\kperp-\pperp)^2 Q_s^\B(\xB_\B)^{-2}) \nn\\
    &= \frac{16 N_c^3}{(2\pi)^2 g^2(N_c^2-1)}\, \ee^{-\frac{\kperp^2}{Q_\Sigma^2}} \left( \kperp^4 \frac{ Q_\Pi^8}{Q_\Sigma^{10}}+\kperp^2 \frac{Q_\Delta^4 Q_\Pi^4}{Q_\Sigma^8}+\frac{2 Q_\Pi^8}{Q_\Sigma^6}\right), \label{eq:dN-GBW-large-nucl-numerics}
\end{align}
where we used the shorthands
\begin{align}
    Q_\Sigma^2 = Q_s^\A(\xB_\A)^2 + Q_s^\B(\xB_\B)^2, \quad Q_\Delta^2 = Q_s^\A(\xB_\A)^2-Q_s^\B(\xB_\B)^2, \quad Q_\Pi^2 = Q_s^\A(\xB_\A) Q_s^\B(\xB_\B). \label{eq:gbw-Q-defs}
\end{align}
We also normalized the gluon number distribution by the formally infinite transverse overlap area $\Sop$ to regularize for the infinite volume.
Hence, this result corresponds to the gluon number density.
The expression in \cref{eq:dN-GBW-large-nucl-numerics} is also consistent with the limit $L_\perp \rightarrow \infty$ of the general result in \cref{eq:dN-GBW-numerics}, if the IR regulator is set to zero.
The GBW TMDs used for the limit do not contain IR regulation factors, in contrast to \cref{eq:dN-GBW-numerics}.
This simplifies the expression, allowing for a closed-form solution of the integral.
The necessary steps are presented in \cref{appx:dN-1DGBW}.

\begin{figure}
    \centering
    \includegraphics[width=0.49\linewidth]{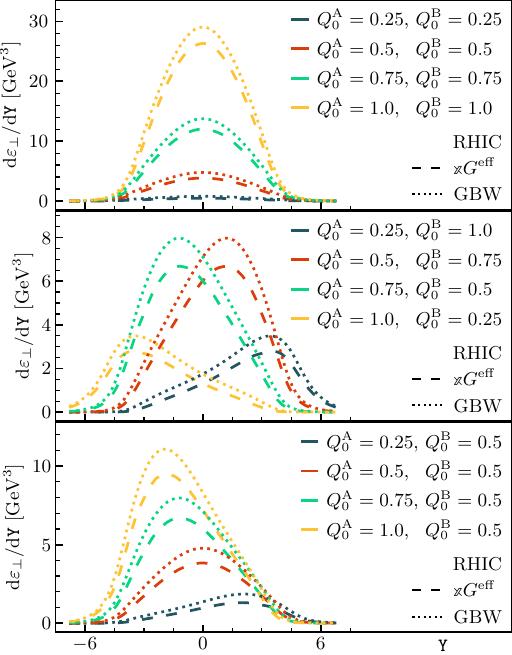}\hfill%
    \includegraphics[width=0.49\linewidth]{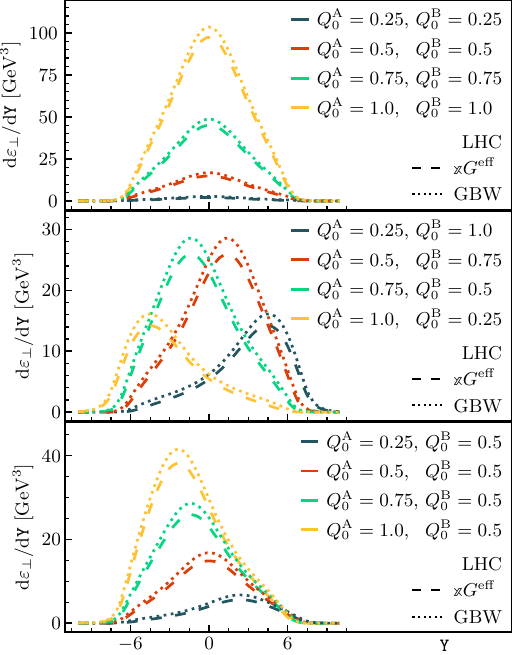}
    \caption[Momentum rapidity distributions of transverse energy.]{%
    Transverse energy differential in momentum rapidity normalized to the transverse overlap area for the GBW model.
    RHIC (left column) and LHC (right column) parameters from \cref{tab:numerics-GBW-model-params} are used.
    The different saturation scale factors used for the nuclei are given in units of GeV and are color-coded in the legend.
    The dashed lines correspond to \cref{eq:dN-GBW-numerics} with three-dimensional Gaussian envelopes and using the IR-regulated GBW TMD in \cref{eq:GBWIR-TMD}.
    The dotted lines correspond to the limit of large nuclei in \cref{eq:dN-GBW-large-nucl-numerics} using the original GBW TMD from \cref{eq:GBW-TMD} and are unaffected by the IR regulator.
    These two different results align closely with each other for all parameters used.
    The modifications due to the envelopes and the IR regulator are moderate.
    }
    \label{fig:dEpvY-GBW}
\end{figure}

In \cref{fig:dEpvY-GBW}, the transverse energy of the GBW model is shown, where the results are normalized to the transverse overlap area and denoted as $\dd\ep/\dd\Y$.
The dashed lines correspond to the model with finite-sized, three-dimensional envelopes and use \cref{eq:dN-GBW-numerics}.
For these results, the GBW TMD contains IR regulation factors, as shown in \cref{eq:GBWIR-TMD}.
The dotted lines are obtained from the limit of large nuclei given in \cref{eq:dN-GBW-large-nucl-numerics} and are unaffected by the IR regulator.
The left column uses RHIC parameters and the right column LHC parameters listed in \cref{tab:numerics-GBW-model-params}.
Each panel shows different combinations of the saturation scale factors $Q_0^{\A/\B}$ in units of GeV for the various colors.

The shapes of the profiles are in stark contrast to the results from the $\WSxi$ model discussed before.
For symmetric $Q_0^{\A/\B}$, there is a peak at mid-rapidity which is more pronounced for the higher energy of the LHC setup (right column).
There is no boost-invariant plateau.
For asymmetric saturation scale factors, the distributions are skewed.
The peaks move closer to the fragmentation region of the nucleus with the larger value of $Q_0^{\A/\B}$.
Generally, larger values of the saturation scale result in more transverse energy.

Comparing the general result (dashed) with the limit of large nuclei (dotted) reveals good agreement overall.
The modifications due to finite, three-dimensional envelopes seem to be moderate.
Still, it is not straightforward to disentangle the effects of the envelopes from the effect of the IR regulator.
When increasing the collider energy, larger values of $Q_s^{\A/\B}$ will allow for larger $|\kperp|$ to contribute to the final transverse energy.
This is because $\xB_{\A/\B} \sim 1/\gamma$ and the saturation scale increases for smaller momentum fractions.
In the case of LHC (right column), the differences between the dashed and dotted curves are smaller.
This can be explained by the fact that the IR regulator $m$ is further separated from the saturation momentum $Q_s^{\A/\B}$ than for the lower energy in the RHIC setup, and $Q_s^{\A/\B}$ becomes the dominant transverse momentum scale.
For the limit of large nuclei (dotted), the only transverse momentum scale is set by $Q_s^{\A/\B}$.
In the case of RHIC, $m$ and $Q_s^{\A/\B}$ are closer together and the results are affected more by the IR regulator.

The dashed curves are always below the dotted curves, indicating that the transverse energy density is slightly lower in the case of finite, three-dimensional envelopes.
This can be explained by the less dense boundary region of the WS envelopes.
The large nuclei approximation corresponds to the homogeneous distribution of transverse energy density in the transverse plane.
However, due to the smooth falloff of the WS envelopes, the transverse energy density in this case is not homogeneous.
It is larger in the center of the transverse plane, where the nuclear model is matched to reproduce the limit of large nuclei.
Hence, integrating the transverse energy over the entire transverse plane and normalizing it with the transverse overlap area yields lower values overall.

\afterpage{\FloatBarrier}
\pagebreak

\section{Limiting fragmentation}\label{sec:results-limiting-fragmentation}

\subsection{Parametrized longitudinal correlations}

\begin{figure}[t!]
    \centering
    \includegraphics[width=0.49\linewidth]{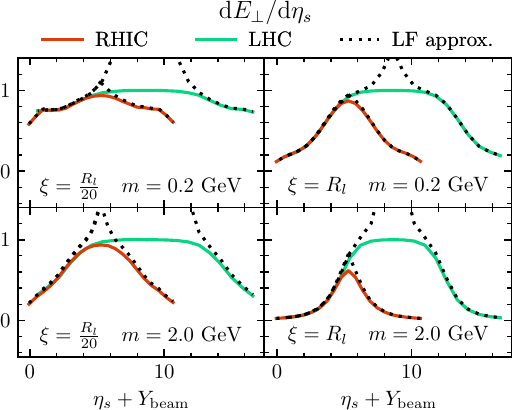}\hfill%
    \includegraphics[width=0.49\linewidth]{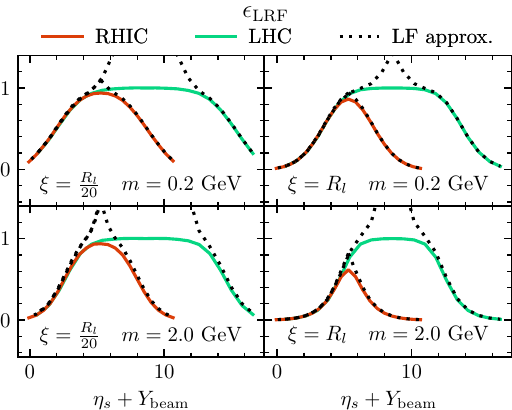}
    \caption[Rapidity profiles for the limiting fragmentation approximation.]{%
    Transverse energy (left panel) and local rest frame energy density (right panel) differential in spacetime rapidity for a single event.
    RHIC (orange) and LHC (green) setups utilize the parameters listed in \cref{tab:numerics-zeta-model-params}.
    The value of $\tau=1.0$~fm/$c$.
    The solid lines correspond to results obtained from the position space formulation.
    The dashed lines are obtained from the limiting fragmentation (LF) approximation discussed in \cref{sec:lim-frag-implementation}.
    The approximation agrees with the full numerical results in the fragmentation regions.
    For the larger value of the correlation length $\xi$, the approximation remains valid for smaller $|\eta_s|$.
    Figure adapted from~\cite{Ipp:2025sbc}.
    }
    \label{fig:lf-numerics}
\end{figure}

In this section, we present numerical results for the limiting fragmentation behavior of the (3+1)D dilute Glasma in the $\WSxi$ nuclear model (cf.~\cref{sec:finite-longitudinal-correlations}).
In \cref{fig:dEp-detas-vs-energy,fig:dElrf-detas-vs-energy,fig:dEpvY-mom-pos-comp}, the property of limiting fragmentation in the $\WSxi$ model was already apparent.
Here, we first compare the full numeric results obtained in the position space framework to the approximation discussed in \cref{sec:lim-fag-pos-space}.
Then, we analyze the expression for the transverse energy distribution obtained from the momentum space formulation and provide a simple proof for limiting fragmentation within our framework.

In \cref{fig:lf-numerics}, the rapidity profiles of the transverse energy $\dd\Ep/\dd\eta_s$ (left panel) and local rest frame energy density $\elrf$ (right panel) are plotted.
Both quantities are integrated over a small patch in the transverse plane.
The data correspond to a single event evaluated at $\tau=1.0$~fm/$c$ with the parameters for the RHIC (orange) and LHC (green) setups listed in \cref{tab:numerics-zeta-model-params}.
The curves are normalized by the same factor for each parameter combination to normalize the LHC results and preserve the relative scale between the two setups.
The horizontal axes show the spacetime rapidity $\eta_s$ shifted by the beam rapidity $Y_\mathrm{beam}=\arcosh(\gamma)$ for each setup.

The solid lines depict the full numerical results, which were discussed in \cref{sec:numerics-param-long-corr}.
The dotted lines are obtained by evaluating the expressions derived for large $|\eta_s|$ in \cref{sec:lim-fag-pos-space} as described in \cref{sec:lim-frag-implementation}.
Note that the initial nuclear fields used in both calculations are identical.
Unsurprisingly, the limiting fragmentation (LF) approximation perfectly describes the flanks of the rapidity distributions.
Compared to \cref{fig:dEp-detas-vs-energy,fig:dElrf-detas-vs-energy}, the shapes of the curves are different because the quantities are only integrated over a small patch in the transverse plane.
This highlights the local nature of limiting fragmentation.

The value of the correlation length $\xi$ critically affects the spacetime rapidity range where the LF approximation agrees with the full result.
For small $\xi$, the dotted curves break away at $\eta_s$-values where the plateau should set in.
For large $\xi$, the onset of the plateau also emerges in the LF approximation.
This behavior was already predicted in \cref{sec:lim-fag-pos-space}, where it was identified that larger correlation lengths allow for the approximation to already be valid at smaller $|\eta_s|$.

\subsubsection{Transverse energy differential in momentum rapidity}

Next, we derive limiting fragmentation in the momentum space formulation.
In \cref{sec:results-gluon-numbers}, it was established that the momentum rapidity distribution of the transverse energy in the $\WSxi$ model is fully described by the simple expression in \cref{eq:dE-WS-zeta-large-nuclei-final}, obtained in the limit of large nuclei.
Now, we assume symmetric $\zeta_\A=\zeta_\B=\zeta$ and write
\begin{align}
    \frac{\dd\Ep}{\dd\Y} &\approx \frac{2 g^6\mu^4 N_c(N_c^2-1)}{(2\pi)^2}\frac{R_l^2+\zeta^2}{R_l^2} \frac{\Sop}{m} \intop_0^\infty \dd p\, \Omega(p) \exp(-\tfrac{1}{2}p^2 m^2 \zeta^2 \cosh(2\Y)).
\end{align}
The integrand consists of the integrated effective vertex $\Omega$, which is only a function of the dimensionless integration variable $p$, and an exponential factor.

In the limit of large $|\Y|$, this exponential suppresses the integrand for large values of $p$, because of the strong dependence of the hyperbolic cosine on the momentum rapidity.
More formally, we assume
\begin{align}
    \frac{1}{2}m^2\zeta^2\cosh(2\Y) \gg 1. \label{eq:WSzeta-lf-largeY}
\end{align}
We will investigate this condition in terms of the scaling variable $m\zeta$ later.
Since the integrand is strongly suppressed at large $p$, the dominant contributions to the integral come from values $p\lesssim1$.
We can expand the integrated vertex in a Taylor series around $p=0$,
\begin{align}
    \Omega(p\ll1) = \frac{1}{2\pi}\left( \frac{1}{6} - \frac{1}{60}p^2 + O(p^4) \right),
\end{align}
which allows us to solve the integral term by term in this expansion.
The result up to next-to-leading order in $1/\cosh(2\Y)$ is
\begin{align}
    \frac{\dd\Ep}{\dd\Y} &\approx \frac{2 g^6\mu^4 N_c(N_c^2-1)}{(2\pi)^{5/2}}\frac{R_l^2+\zeta^2}{R_l^2} \frac{\Sop}{m} \nn\\
    &\times \frac{1}{\sqrt{m^2\zeta^2\cosh(2\Y)}}\left( \frac{1}{12} - \frac{1}{120}\frac{1}{m^2\zeta^2\cosh(2\Y)} + O(\cosh(2\Y)^{-2}) \right).\label{eq:dE-WSzeta-large-Y}
\end{align}
Note that the entire energy dependence of \cref{eq:dE-WSzeta-large-Y} is hidden in the correlation length $\zeta$.
For fixed nuclear parameters, $\zeta \propto R/\gamma$ where the proportionality constant is related to the longitudinal correlation length in the rest frame of the nuclei.
This specific dependence on the collider energy via $\gamma$ can be recast into a shift of the momentum rapidity, in full analogy to the limiting fragmentation behavior discussed in \cref{ch:limiting-fragmentation}.
To this end,
\begin{align}
    \frac{1}{\sqrt{m^2\zeta^2\cosh(2\Y)}} &\propto \frac{1}{\sqrt{m^2 R^2 \gamma^{-2} \cosh(2\Y)}} \nn\\
    &\approx \frac{\sqrt{2}}{mR}\,\gamma \ee^{-|\Y|} = \frac{\sqrt{2}}{mR}\,\ee^{-|\Y|+\ln(\gamma)} = \frac{1}{\sqrt{2}mR}\,\ee^{-|\Y|+\ln(2\gamma)},
\end{align}
where we approximated the hyperpolic cosine for large $|\Y|$ via the exponential function.
In the last term, we can identify the logarithm of the Lorentz factor with the beam rapidity, where $Y_\mathrm{beam} =\arcosh(\gamma) \approx \ln(2\gamma)$ for large collider energies.
Clearly, the functional dependence of the transverse energy in \cref{eq:dE-WSzeta-large-Y} on the momentum rapidity and the beam rapidity is of the form $\dd\Ep(\Y,Y_\mathrm{beam})/\dd\Y \equiv \dd\Ep(\Y-Y_\mathrm{beam})/\dd\Y$ and predicts limiting fragmentation.
This particular dependence on the beam rapidity already arises for the gluon number distribution in \cref{eq:dN-WS-zeta-large-nuclei-final} (for symmetric $\zeta_\A=\zeta_\B$).

We conclude that limiting fragmentation is manifest in the momentum space solutions of the (3+1)D dilute Glasma for the $\WSxi$ nuclear model.
Furthermore, to leading order in the expansion, \cref{eq:dE-WSzeta-large-Y} exhibits a universal shape,
\begin{align}
    \frac{\dd\Ep}{\dd\Y} &\sim \ee^{-|\Y|+Y_\mathrm{beam}},
\end{align}
where all model parameters only enter as a prefactor.
In the fragmentation regions, where $|\Y|\gg 1$, the shape of the rapidity profile of the transverse energy is exponential.

Finally, we discuss the condition in \cref{eq:WSzeta-lf-largeY}.
We solve for the momentum rapidity and get
\begin{align}
    |\Y| \gg \frac{1}{2}\arcosh(2(m\zeta)^{-2}).
\end{align}
Similar to the characteristic rapidity $\Y_c$ in \cref{eq:Yc-sol-sym}, we can identify that the regime of limiting fragmentation is determined by the scaling variable $m\zeta$.
For small values of $m\zeta$, the exponential flanks of the rapidity distribution of $\dd\Ep/\dd\Y$ are pushed to larger $\Y$, where the changeover from a plateau-like flat profile to the steep falloff in the flanks occurs.
For large values of $m\zeta$, the exponential falloff can already begin to set in near mid-rapidity and prevent the formation of a plateau.
Since $\zeta$ is inversely proportional to the collider energy (for fixed nuclear parameters), this effect is enhanced for the RHIC setup at smaller $\gamma$ discussed in \cref{sec:results-gluon-numbers}.

\subsection{Black disk limit in the GBW model}

In this section, we analyze the limiting fragmentation behavior of the phenomenological GBW model introduced in \cref{sec:GBW-saturation-model} (cf.~\cref{sec:numerics-GBW}).
In \cref{sec:numerics-GBW}, it was established that in the limit of large nuclei, the gluon distribution normalized to the transverse overlap area is given by the closed-form expression in \cref{eq:dN-GBW-large-nucl-numerics}.
We continue to work in this limit where the transverse energy density reads
\begin{align}
    \frac{\dd\ep}{\dd\Y} = \frac{8 N_c^3}{\pi g^2 (N_c^2-1)} \intop_0^{k^\mathrm{max}} \dd|\kperp|\, \ee^{-\frac{\kperp^2}{Q_\Sigma^2}} \left( \kperp^4 \frac{ Q_\Pi^8}{Q_\Sigma^{10}}+\kperp^2 \frac{Q_\Delta^4 Q_\Pi^4}{Q_\Sigma^8}+\frac{2 Q_\Pi^8}{Q_\Sigma^6}\right) \label{eq:dep-GBW-large-nuclei}
\end{align}
and where we used the shorthands from \cref{eq:gbw-Q-defs}.
Recall that the transverse momentum $\kperp$ and the momentum rapidity $\Y$ enter the momentum fractions $\xB_{\A/\B}$ associated with each nucleus.
Since $\xB_{\A/\B} \leq 1$, this leads to an upper limit for the integration over $|\kperp|$, given a fixed value of $\Y$.
In particular, the upper limit is
\begin{align}
    k^\mathrm{max} = m_n \ee^{-|\Y|+Y_\mathrm{beam}},
\end{align}
where we used $Y_\mathrm{beam} \approx \ln(2\gamma)$.

We investigate \cref{eq:dep-GBW-large-nuclei} in the fragmentation region of nucleus $\A$ where the momentum rapidity $-\Y\gg 1$.
In this rapidity regime, the momentum fraction $\xB_\B \ll \xB_\A$ and the saturation momentum of nucleus $\B$ becomes the largest transverse momentum scale in the system,
\begin{align}
    Q_s^\B(\xB_\B) \gg Q_s^\A(\xB_\A), \qquad Q_s^\B(\xB_\B) \gg k^\mathrm{max}.
\end{align}
This corresponds to the ``black disk'' limit of nucleus $\B$ (cf.~\cref{sec:gluon-numbers-lf}).
Next, we expand the integrand in \cref{eq:dep-GBW-large-nuclei} to lowest order in $Q_s^\A/Q_s^\B$ and $|\kperp|/Q_s^\B$ and obtain
\begin{align}
    \frac{\dd\ep}{\dd\Y} &\approx \frac{8 N_c^3}{\pi g^2 (N_c^2-1)} \intop_0^{k^\mathrm{max}} \dd|\kperp|\, \ee^{-\kperp^2/Q_s^\B(\xB_\B)^2} \nn\\
    &\times \left( \left(\frac{|\kperp|}{Q_s^\B(\xB_\B)}\right)^4\left(\frac{Q_s^\A(\xB_\A)}{Q_s^\B(\xB_\B)}\right)^2 + \left(\frac{|\kperp|}{Q_s^\B(\xB_\B)}\right)^2 + 2 \left(\frac{Q_s^\A(\xB_\A)}{Q_s^\B(\xB_\B)}\right)^2 \right)Q_s^\A(\xB_\A)^2 \nn\\
    &\approx \frac{8 N_c^3}{\pi g^2 (N_c^2-1)} \intop_0^{k^\mathrm{max}} \dd|\kperp|\, \left(\frac{Q_s^\A(\xB_\A)}{Q_s^\B(\xB_\B)}\right)^2 \left(\kperp^2 + 2 Q_s^\A(\xB_\A)^2 \right). \label{eq:dep-gbw-step1}
\end{align}
The exponential only contributes in this order as the constant factor of unity.
We also dropped the first term in the large parentheses because it is of higher order.

To isolate the dependence of $\dd\ep/\dd\Y$ on the collider energy, recall that the beam rapidity enters the parametrizations of the saturation scales $Q_s^{\A/\B}$ via the momentum fractions $\xB_{\A/\B} = |\kperp| \ee^{\mp\Y-Y_\mathrm{beam}}/m_n$.
In the fragmentation region of nucleus $\A$,
\begin{align}
    \frac{Q_s^\A(\xB_\A)}{Q_s^\B(\xB_\B)} \approx \frac{Q_0^\A}{Q_0^\B}\left(\frac{\xB_A}{\xB_\B}\right)^{-\lambda}(1-\xB_\A)
\end{align}
where we used $1-\xB_\B \approx 1$.
The ratio of the momentum fractions $\xB_\A / \xB_\B = \ee^{-2\Y}$.
We can further simplify \cref{eq:dep-gbw-step1},
\begin{align}
    \frac{\dd\ep}{\dd\Y} &\approx \frac{8 N_c^3}{\pi g^2 (N_c^2-1)} \left( \frac{Q_0^\A}{Q_0^\B} \right)^2 \ee^{4\lambda \Y} \intop_0^{k^\mathrm{max}} \dd|\kperp|\, \left( 1- \frac{|\kperp|}{m_n}\ee^{-\Y-Y_\mathrm{beam}} \right)^2 \nn\\
    &\times \left( \kperp^2 + 2 (Q_0^\A)^2 \left(\frac{|\kperp|}{m_n}\ee^{-\Y-Y_\mathrm{beam}}\right)^{-2\lambda}\left( 1 - \frac{|\kperp|}{m_n}\ee^{-\Y-Y_\mathrm{beam}}\right)^2 \right). \label{eq:dep-gbw-step2}
\end{align}
It should not come as a surprise that the momentum rapidity dependence of the integrand is of the form $\Y - Y_\mathrm{beam}$ that leads to limiting fragmentation.
In fact, as discussed in \cref{sec:gluon-numbers-lf}, this follows from the parametrization of the GBW TMDs via $\xB_{\A/\B}$ and how the beam rapidity is connected to the Lorentz factor.
However, the prefactor in \cref{eq:dep-gbw-step2} contains an isolated contribution of $\Y$.
We may shift the momentum rapidity by the beam rapidity
\begin{align}
    \Y' = \Y+Y_\mathrm{beam},
\end{align}
such that the integral, including its upper bound, no longer depends on the beam rapidity.
Then, 
\begin{align}
    \frac{\dd\ep}{\dd\Y} &\approx \frac{8 N_c^3}{\pi g^2 (N_c^2-1)} \left( \frac{Q_0^\A}{Q_0^\B} \right)^2 \ee^{4\lambda(\Y' - Y_\mathrm{beam})} \intop_0^{k^\mathrm{max}} \dd|\kperp|\, \left( 1- \frac{|\kperp|}{m_n}\ee^{-\Y'} \right)^2 \nn\\
    &\times \left( \kperp^2 + 2 (Q_0^\A)^2 \left(\frac{|\kperp|}{m_n}\ee^{-\Y'}\right)^{-2\lambda}\left( 1 - \frac{|\kperp|}{m_n}\ee^{-\Y'}\right)^2 \right). \label{eq:depsperp-limfrag-res}
\end{align}
This result exhibits limiting fragmentation.
For $-\Y\gg 1$, the transverse energy density scales according to
\begin{align}
    \frac{\dd\ep}{\dd\Y} \sim \ee^{\alpha\Y' - 4\lambda Y_\mathrm{beam}} = \ee^{\alpha\Y'} (2\gamma)^{-4\lambda}, \label{eq:dep-gbw-lf-scaling}
\end{align}
where we expressed the beam rapidity in terms of the Lorentz factor for the second equality.
The constant $\alpha = 4\lambda +1$ and captures the scaling of the exponential prefactor in \cref{eq:depsperp-limfrag-res} ($\sim \ee^{4\lambda\Y'}$) and the leading order $\sim \ee^{\Y'}$ scaling of the integral for $-\Y \gg 1$.
Still, there is residual dependence on the collider energy via the $\gamma$-dependent prefactor.
Previous studies of limiting fragmentation in the CGC have motivated a similar factor (e.g., \cite{Jalilian-Marian:2002yhb,Gelis:2006tb}).

\begin{figure}[p]
    \centering
    \includegraphics[width=0.6\linewidth]{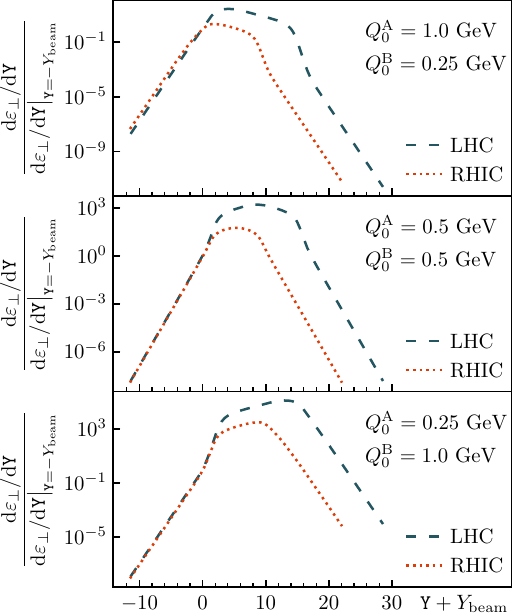}
    \caption[Limiting fragmentation for transverse energy density in the GBW model.]{%
    Transverse energy differential in momentum rapidity normalized to the transverse overlap area for the GBW model.
    The data correspond to selected curves marked $\xGeff$ in \cref{fig:dEpvY-GBW}.
    The values of the used saturation scale factors are given in the legend.
    Dashed, blue curves represent the LHC setup and dotted, orange curves represent the RHIC setup.
    All curves are normalized to their values at $\Y=-Y_\mathrm{beam}$.
    The horizontal axis denotes the momentum rapidity shifted by the beam rapidity.
    The vertical axis is scaled logarithmically.
    The curves exhibit universal exponential behavior in the fragmentation region of nucleus $\A$, where $-\Y \gg 1$.
    }
    \label{fig:GBW-lf}
\end{figure}

In \cref{fig:GBW-lf}, the limiting fragmentation behavior of the transverse energy density is demonstrated.
The plotted data correspond to selected curves from \cref{fig:dEpvY-GBW} where the GBW model with finite, three-dimensional envelopes was used ($\xGeff$). 
The values for the saturation scale factors are listed in the legend.
We compare the energy dependence of the results for the LHC (dashed blue) and RHIC (dotted orange) setups.
The horizontal axis marks the momentum rapidity shifted by the respective beam rapidity.
The vertical axis is scaled logarithmically and clearly shows the exponential behavior of $\dd\ep/\dd\Y$ in the fragmentation regions of both nuclei.
However, note that the rapidity regime where limiting fragmentation sets in is located at very extreme values of $\Y$ where the transverse energy density has already fallen off significantly.

The curves for both setups are each normalized to their values at $\Y = Y_\mathrm{beam}$.
This normalization factor is energy dependent.
In \cref{tab:GBW-black-disk}, we compare the relative error of the normalization factor obtained from the black disk limit in \cref{eq:dep-gbw-lf-scaling} to the numerical results in \cref{fig:GBW-lf}.
The table contains two lines where the error is calculated for the normalization factor at different values of $\Y$.
When normalizing at the beam rapidity, there is a strong dependence on the saturation scale factors.
The more favorable choice for the black disk limit of nucleus $\B$, where $Q_0^\B \gg Q_0^\A$, yields a smaller error than the opposite choice.
The dependence on the saturation scale factors is not predicted by the black disk approximation.
When normalizing at the more extreme momentum rapidity value of $-\Y = Y_\mathrm{beam} + 10$ (second row), the dependence of the relative error on $Q_0^{\A/\B}$ becomes minimal and the values of the error are comparable for all choices.

\begin{table}[p]
    \caption[Relative errors for limiting fragmentation in the GBW black disk limit.]{\label{tab:GBW-black-disk}%
    Relative error of the prediction of the energy-dependent relative prefactor between the RHIC and LHC results in the fragmentation region of nucleus $\A$ ($-\Y\gg 1$) from \cref{fig:GBW-lf}.
    The prediction is obtained in the black disk limit of nucleus $\B$ in \cref{eq:dep-gbw-lf-scaling}.
    For extremely large momentum rapidities ($-\Y = Y_\mathrm{beam} + 10$) the error is comparable across all $Q_0^{\A/\B}$ pairs.
    }
    \centering\begin{tabular}{l|lll}
    \hline\hline
    \rule{0pt}{3ex} &  $Q_0^\A = 1.0$~GeV &  $Q_0^\A=0.5$~GeV & $Q_0^\A = 0.25$~GeV\\
    \textbf{Rapidity} $\Y$ &  $Q_0^\B=0.25$~GeV & $Q_0^\B=0.5$~GeV & $Q_0^\B=1.0$~GeV \\
    \hline
         \rule{0pt}{3ex} $-Y_\mathrm{beam}$ & 171.8\% & 6.8\% & 19.3\% \\
         \rule{0pt}{2ex} $-Y_\mathrm{beam}-10$ & 14.0\% & 16.0\% & 14.7\% \\
    \hline\hline
    \end{tabular}
\end{table}

\addtocontents{toc}{\protect\pagebreak} 

\chapter{Conclusion and Outlook}\label{ch:conclusion}

\section{Summary}

In this thesis, various aspects of the (3+1)D dilute Glasma were studied that pertain to the theoretical treatment of the initial stages of relativistic heavy-ion collisions.
At its core, the description relies on the Color Glass Condensate (CGC) effective theory to model the individual colliding nuclei in the initial state.
An introduction to this framework is provided in \cref{sec:description-initial-state}.
The interaction of the initial, single-nucleus fields obtained from the CGC leads to the time dynamics of the Glasma stage, where strong classical fields evolve according to non-Abelian Yang-Mills theory.
Historically, boost invariance was assumed to reduce the dimensionality of the system from (3+1) to (2+1) in the Milne coordinate frame, such that the description is independent of rapidity.
While this allows for efficient numerical simulations of the time dynamics of the Glasma, it is only applicable at mid-rapidity of the full collision problem due to the lack of rapidity dependence.

Going beyond the boost-invariant limit of the Glasma requires solving the nonlinear Yang-Mills equations, including the dynamics along the longitudinal direction.
Hereby, the longitudinal structure of the nuclei in the initial state is essential.
In \cref{sec:dilute-glasma-approx}, the dilute approximation is introduced as a means to perturbatively expand the Yang-Mills equations in powers of weak sources.
To leading order, the dynamics of the Glasma then become effectively Abelian.
An important simplification in this approach is the assumption of recoilless currents, which move at the speed of light and are formed by the valence charges of the colliding nuclei.
This fully (3+1)D treatment of the dilute Glasma is formulated in terms of concise integral expressions for the components of the field strength tensor of the dilute Glasma.
These analytic expressions lead to the interpretation of the encoded physics in the position space picture.
The gluons of the Glasma field are produced in the interaction region of the single-nucleus fields via $2\rightarrow1$ scattering processes.
Then, they propagate along lightlike paths without further interactions, which corresponds to free streaming.

The integrals in the position space formulation of the dilute Glasma can be efficiently solved on modern hardware.
A suitable implementation is discussed in \cref{sec:position-space-numerical-implementation}.
Compared to traditional Glasma simulations that use time-stepping methods, one benefit of the (3+1)D dilute Glasma is that each calculation for a fixed value of proper time takes the same amount of computation time.
Essentially, the results at later proper times are computationally independent of the results from previous times.

The numerical evaluation is crucial for the event-by-event simulation of heavy-ion collisions.
The event statistics correspond to the CGC expectation values of the calculated observables.
Here, the statistics imposed by the CGC weight function translate to sampling independent initial configurations of the nuclear color charge distributions.
The procedure discussed in this thesis is based on the McLerran-Venugopalan (MV) model (cf.~\cref{sec:MV-nucl-model}) where the CGC weight function is assumed to be Gaussian.
The entire nuclear model is fixed by providing the one-point and two-point correlation functions for the color charge distributions.
However, not all correlators will lead to a positive semi-definite CGC weight function required for physical results.
In \cref{sec:pos-semi-definite-correlators}, a suitable condition on the nuclear model is derived that ensures physical results.
This leads to the interpretation of the two-point correlator in terms of well-defined single-nucleus envelopes and a correlation function that depends only on the difference in coordinates.

In addition to the position space interpretation of the (3+1)D dilute Glasma, the momentum space solutions in the dilute approximation are derived in \cref{ch:momentum-space-picture}.
They are then used to calculate the gluon number distribution in Coulomb gauge.
The necessary gauge transformation from covariant to Coulomb gauge is evaluated at asymptotically late times, where the effects of gluon production from time-dependent sources have stopped and the gluon field is propagating freely.
In the dilute limit, this gauge transformation reduces to the projection of the Glasma field onto the transverse momentum modes.
The final result is analogous to the dynamics captured by the Lipatov vertex.

In the momentum space picture of the (3+1)D dilute Glasma, the importance of the longitudinal structure imposed by the nuclear model becomes manifest.
The momentum rapidity dependence of the gluon number distribution is directly given by the longitudinal structure of the color charge correlators in momentum space.
The dynamics involving the Lipatov vertex are restricted to the transverse momentum structure.
In the limit where there is no longitudinal or transverse structure in the nuclei, the boost-invariant results from the original MV nuclear model are recovered.

Furthermore, in \cref{sec:TMDs-QCD}, the results for the gluon number distribution are reformulated in terms of effective transverse momentum distributions (TMDs) of the gluons in the colliding nuclei.
These effective TMDs generalize the standard TMDs used in perturbative QCD applications to two transverse momentum arguments.
In the dilute limit, (effective) TMDs are directly related to the Fourier-transformed color charge correlators and connect the longitudinal structure of the nuclear model to the dependence of the gluon number distribution on the momentum fraction $\xB$.
In particular, in the limit of large nuclei, where the correlation scales introduced by the correlation function in the nuclear model are well separated from the scales of the nuclear envelopes, gluon production is described in full analogy to the $k_T$-factorization formula.
The effective TMDs reduce to standard TMDs, which are then the only objects that enter the transverse momentum convolution.
The squared Lipatov vertex degenerates to an IR-divergent factor $\sim1/\kperp^2$.
This result corresponds to the leading order partonic cross-section for gluon production.
For heavy-ion collisions, the envelope factors produce an overall prefactor that is interpreted as the transverse overlap area of the nuclei.

The similarities of these results with perturbative QCD schemes allow for further interpretation of the validity of the dilute approximation for heavy-ion collisions.
In the kinematic regime where large transverse momenta dominate gluon production, the dynamics are described well by the (3+1)D dilute Glasma.
This corresponds to the broader mid-rapidity region of the gluon number distribution.
While the (3+1)D dilute Glasma can be used to evaluate the distribution up to extreme rapidities, the infrared limit for transverse kinematics poses conceptual challenges.
The IR regulator introduced for solving the single-nucleus fields in the CGC becomes a relevant phenomenological parameter that significantly influences the results.

In \cref{ch:limiting-fragmentation}, the phenomenon of limiting fragmentation is derived from the position space and momentum space solutions of the (3+1)D dilute Glasma.
In position space, local longitudinal scaling of the field strength tensor predicts that all observables constructed from the field strength tensor will show limiting fragmentation behavior.
In momentum space, limiting fragmentation is tightly connected to the saturation of TMDs for very small momentum fractions.
Numerical evidence for both formulations is given in \cref{sec:results-limiting-fragmentation}.

After having established the framework of the (3+1)D dilute Glasma, concrete realizations for three-dimensional nuclear models with longitudinal structure are discussed in \cref{ch:nuclear-models}.
On the one hand, a model with a straightforward interpretation in position space is introduced.
The correlation function in this model is parametrized by a Gaussian in the longitudinal direction and defines a phenomenological parameter that determines the longitudinal correlation length inside the nuclei.
The transverse direction is uncorrelated and follows the original nuclear model by McLerran and Venugopalan.
The advantage of this model is its conceptual simplicity and clearly defined scales, which yield a positive semi-definite CGC weight function and physical event-by-event statistics.

On the other hand, it is explored how phenomenologically motivated TMDs can be used in conjunction with three-dimensional nuclear envelopes to construct the color charge correlator for the nuclear model.
The example of the TMD named after Golec-Biernat and W\"usthoff (GBW) is studied in detail.
Due to the momentum dependence of the longitudinal and transverse scales in the GBW TMD, the resulting nuclear model is found to violate the positive semi-definiteness condition for certain kinematic regimes.
Still, in the limit of infinitely large nuclei, the envelope scales can be separated from the dynamical scales of the TMD and the nuclear model becomes physically viable.

A large portfolio of numerical results using these nuclear models is presented in \cref{ch:numerical-results}.
Two setups at different collider energies, comparable to Au+Au collisions at RHIC and Pb+Pb collisions at the LHC, are evaluated.
First, the nuclear model with parametrized longitudinal correlations is used for event-by-event simulations.
The focus of the discussion is on the longitudinal structure and spacetime rapidity profiles of observables.
The commonly used Milne frame in position space is found to be ill-suited for capturing the symmetries of the (3+1)D system.
The finite longitudinal size of the nuclei leads to an extended collision region in the $t$-$z$ plane and to the ambiguous choice of the origin for the Milne frame.
In particular, the Milne components of the energy-momentum tensor of the Glasma show unphysical behavior.
In contrast, the local rest frame energy density is a robust observable for the Glasma stage and largely agrees with the sum of transverse pressures.

Two observables are found to be strongly influenced by the longitudinal structure of the nuclei.
The extended geometry of the collision region results in a distinct pattern of longitudinal flow.
The existence of large values of this flow component further indicates the inapplicability of the Milne frame symmetries.
In the transverse plane, the first-order eccentricity is identified as particularly sensitive to the value of the longitudinal correlation parameter.

The gluon number distribution and transverse energy in the nuclear model with longitudinal correlations show remarkable scaling behavior.
The shapes of the momentum rapidity profiles follow universal parametrizations and are each completely fixed by a scaling variable.
For the gluon number distribution, the scaling variable is the product of the longitudinal correlation length and the transverse momentum of the gluons.
For the transverse energy, the transverse momentum is replaced by the IR regulator in the scaling variable.
In the dilute approximation, the IR regulator becomes a relevant phenomenological parameter.
These results are derived analytically in the limit of large nuclei and are verified to agree perfectly with the full numerical results for the employed setups.

Finally, numerical results for the gluon number distribution obtained from the nuclear model with the GBW TMD are explored.
The comparison in this model with finite nuclear envelopes to the limit of infinitely large nuclei reveals overall insignificant differences and the same effects when adjusting the parameters of the TMD.
Still, the validity of the finite GBW model is questioned by the violation of positive semi-definiteness of the color charge correlators.
This prevents the GBW model from being used for event-by-event simulations in the position space framework because the individual realizations of the color charge distributions cannot be sampled from an invalid probability distribution.

The issues with the phenomenological GBW TMD can be interpreted as an incompatibility with finite-sized envelopes.
It remains to be studied whether TMDs in general cannot be coerced into color charge correlators with the peculiar structure required for positive semi-definiteness.
A possible alternative to TMDs could be generalized parton distributions or generalized transverse-momentum-dependent distributions.
Recent theoretical progress~\cite{Lorce:2025aqp,Kovchegov:2025yyl,Bhattacharya:2025fnz,Boer:2025ixc} could lead to the development of phenomenologically motivated models in the future.

\section{Outlook -- Coupling to hydrodynamics}\label{sec:coupling-to-hydro}

The (3+1)D dilute Glasma framework for heavy-ion collisions allows for initial conditions with genuine, three-dimensional nuclear models.
It also provides solutions for the early Glasma stage that exists for a period of $\tau\lesssim 1$~fm/$c$ after the initial collision.
The description of the evolution of the collision for proper times $\tau\gtrsim 1$~fm/$c$ is not properly included in the (3+1)D dilute Glasma.
Hence, the observables that can be extracted from the (3+1)D dilute Glasma are not directly comparable with the experimental results reported by the large collaborations at RHIC and the LHC.

The later evolution of the collision is first described by relativistic hydrodynamics for the quark-gluon plasma stage, before a description of the produced particles as a hadron gas and further evolution via parton cascades becomes applicable.
The community has established robust frameworks for simulating these later stages, enabling a direct comparison with experiments.
One example is the iEBE-MUSIC%
\footnote{%
The source code for iEBE-MUSIC is hosted on \url{https://github.com/chunshen1987/iEBE-MUSIC}.}
framework that provides an event-by-event simulation pipeline for heavy-ion collisions.

The first step to include the (3+1)D dilute Glasma in such frameworks is to use the energy-momentum tensor of the Glasma as initial conditions for the hydrodynamic evolution.
In this outlook, preliminary results for the hydrodynamic evolution of the (3+1)D dilute Glasma using MUSIC%
\footnote{%
The source code for MUSIC is hosted on \url{https://github.com/MUSIC-fluid}.}
are presented.
The sudden changeover from strong classical fields to collective hydrodynamic behavior that occurs at the switching time $\tau_\mathrm{s}$ can be improved by an intermediate stage described by QCD effective kinetic theory~\cite{Kurkela:2018vqr,Schlichting:2019abc,Berges:2020fwq,Greif:2017bnr,Ambrus:2021fej,Du:2025bhb}.
However, this is not considered in the discussion below and left for future work.

We briefly review the canonical treatment of the hydrodynamic stage to the degree relevant for this outlook.
The implementation of the hydrodynamic evolution in MUSIC is described in~\cite{Schenke:2010nt,Schenke:2010rr,Schenke:2011bn,McDonald:2016vlt,Schenke:2019pmk,Schenke:2020mbo}.
The equations of motion are formulated as the Levi-Civita covariant conservation equation for the energy-momentum tensor%
\footnote{%
Note that we use the calligraphic letter $\mathcal{T}$ for the energy-momentum tensor in the hydrodynamic description to distinguish it from the energy-momentum tensor of the (3+1)D dilute Glasma, denoted as $T$.
We also suppress the spacetime arguments of the dynamical fields whenever they take general values.}
\begin{align}
    \partial_\mu \mathcal{T}^{\mu\nu} = 0. \label{eq:hydro-cons}
\end{align}
To solve \cref{eq:hydro-cons}, the energy-momentum tensor is parametrized in terms of various dynamical fields as
\begin{align}
    \mathcal{T}^{\mu\nu} = T^{\mu\nu}_\mathrm{ideal} - \Pi(g^{\mu\nu}-u^\mu u^\nu) + \pi^{\mu\nu}.
\end{align}
Here, the ideal energy-momentum tensor is given as
\begin{align}
    T^{\mu\nu}_\mathrm{ideal} = (\elrf + P)u^\mu u^\nu - P g^{\mu\nu},
\end{align}
where $\elrf$ is the local rest frame energy density, $u^\mu$ the flow four-velocity, and $P$ the local pressure.
The ideal energy-momentum tensor evolves according to ideal hydrodynamics without shear or bulk viscous effects.
The additional contributions to $\mathcal{T}$ come from the bulk pressure $\Pi$ and the viscous stress tensor $\pi^{\mu\nu}$.
These are introduced as dynamical fields whose time evolution is given by the second-order constitutive relations~\cite{Denicol:2014vaa,Denicol:2012cn}
\begin{align}
    \tau_\Pi \Dot{\Pi} + \Pi &= -\zeta \nabla_\mu u^\mu - \delta_{\Pi\Pi} \nabla_\mu u^\mu + \lambda_{\Pi\pi}\pi^{\mu\nu}\sigma_{\mu\nu}, \label{eq:bulk-pressure} \\
    \tau_\pi \Dot{\pi}^{\langle\mu\nu\rangle} + \pi^{\mu\nu} &= 2\eta \sigma^{\mu\nu} - \delta_{\pi\pi}\pi^{\mu\nu}\nabla_\alpha u^\alpha + \varphi_7 \pi_\alpha^{\hphantom{\alpha}\langle\mu}\pi^{\nu\rangle\alpha} - \tau_{\pi\pi}\pi_\alpha^{\hphantom{\alpha}\langle\mu}\sigma^{\nu\rangle\alpha} + \lambda_{\pi\Pi}\Pi\sigma^{\mu\nu}, \label{eq:viscous-stress}
\end{align}
where $\langle \mu\nu \rangle$ indicates the symmetrized and traceless projection w.r.t.\ the indices $\mu$ and $\nu$.
The shear tensor is defined as
\begin{align}
    \sigma^{\mu\nu} = \frac{1}{2}\left( \nabla^\mu u^\nu + \nabla^\nu u^\mu - \frac{2}{3}(g^{\mu\nu} - u^\mu u^\nu)(\nabla_\alpha u^\alpha) \right)
\end{align}
using $\nabla_\mu = (g_{\mu\nu} - u_\mu u_\nu) \partial^\nu$.
The first-order transport coefficients are the shear viscosity $\eta$ and bulk viscosity $\zeta$.
The second-order transport coefficients and the shear and bulk relaxation times $\tau_\pi$ and $\tau_\Pi$ appearing in \cref{eq:bulk-pressure,eq:viscous-stress} can be expressed in terms of the first-order coefficients and other dynamical fields.
Together with the equation of state (EOS), which provides a relation between the local pressure and the local rest frame energy density, the system of equations is closed.

\subsubsection{Matching the (3+1)D dilute Glasma to hydrodynamics}

To switch to the hydrodynamic description at $\tau=\tau_s$, the initial values for all the dynamical fields introduced above have to be determined from the energy-momentum tensor of the (3+1)D dilute Glasma $T$.
The standard matching procedure~\cite{Schenke:2019ruo,Schenke:2020mbo,Mantysaari:2017cni} is as follows.
First, the Landau condition in \cref{eq:Landau-condition} is solved.
The resulting local rest frame energy density and flow four-velocity are used to construct $T^{\mu\nu}_\mathrm{ideal}$, where the ideal EOS for a conformal field theory (CFT) is used to determine the value of the local pressure
\begin{align}
    P_\mathrm{CFT}(\elrf) = \frac{\elrf}{3}.
\end{align}
However, the conformal EOS does, in general, not match the pressure given by the EOS used in the hydrodynamic description.
A reasonable, modern choice for the hydrodynamic EOS is given by matching lattice QCD calculations by the HotQCD collaboration to the hadron resonance gas model at low temperatures~\cite{Moreland:2015dvc,HotQCD:2014kol}.
The difference between these two EOSs at the switching time is used to initialize the bulk pressure
\begin{align}
    \Pi(\tau=\tau_s) = \frac{\elrf(\tau=\tau_s)}{3} - P(\elrf).
\end{align}
At this point, not all the information included in the energy-momentum tensor of the (3+1)D dilute Glasma is used and the stress tensor is not yet determined.
The full energy-momentum tensor $T$ enters the initial value of the stress tensor via
\begin{align}
    \pi^{\mu\nu}(\tau=\tau_s) = T^{\mu\nu}(\tau=\tau_s) - T^{\mu\nu}_\mathrm{ideal}(\tau=\tau_s).
\end{align}

\subsubsection{Preliminary results for a single event}

\begin{figure}[p]
    \centering
    \includegraphics[width=0.49\textwidth]{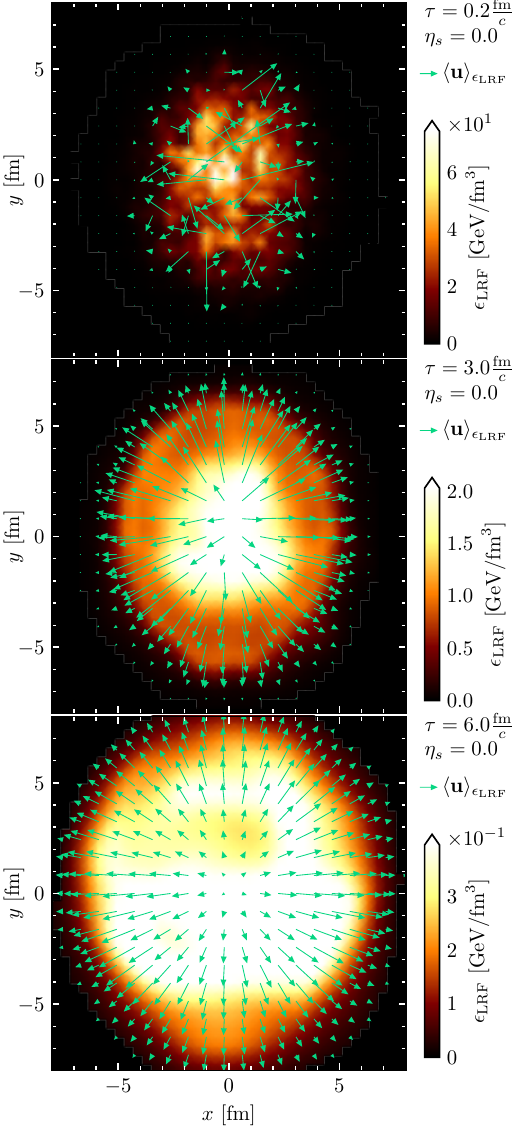}
    \includegraphics[width=0.49\textwidth]{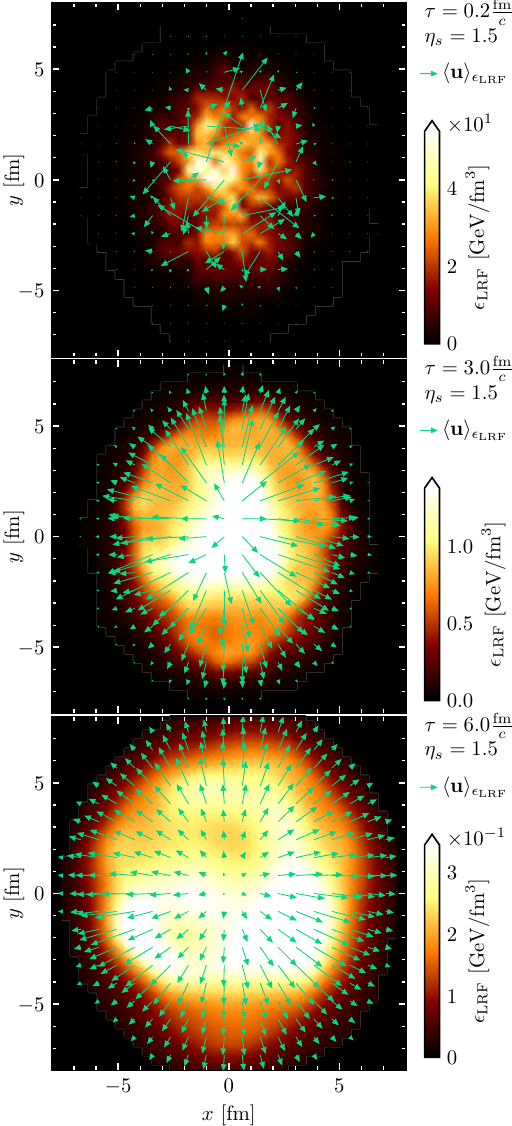}
    \caption[Hydrodynamic evolution of local rest frame energy density and flow in the transverse plane.]{%
    Hydrodynamic evolution of the local rest frame energy density $\elrf$, colored according to its values given by the colorbars, and flow $\uperp$ (green vectors) in the transverse plane.
    The flow vectors are weighted by the local value of $\elrf$ and scaled by an overall factor for illustrative purposes.
    The spacetime rapidity varies from $\eta_s=0$ (left column) to $\eta_s=1.5$ in the right column.
    Proper time advances from the switching time $\tau_s=0.2$~fm/$c$ in the top row to $3.0$~fm/$c$ and $6.0$~fm/$c$ in the middle and bottom rows.
    For the (3+1)D dilute Glasma, the RHIC setup from \cref{tab:numerics-zeta-model-params} was used.
    The value of the IR regulator $m=0.2$~GeV, the UV regulator $\Lambda=10$~GeV.
    The longitudinal correlation length $\xi=0.3 R_l$.
    A large impact parameter $b=R$ was used.
    MUSIC was initialized on a three-dimensional grid with 54 cells spanning $21.45$~fm in $x$-direction, 69 cells spanning $27.55$~fm in $y$-direction, and 13 cells spanning $3$ units in $\eta_s$-direction.
    Second-order coupling terms with bulk and shear viscous effects were enabled. 
    The value of $\eta/s = 0.08$ with temperature dependence according to the default settings.
    The value of $\zeta/s$ and its temperature dependence were set to default.
    No baryon current was included.
    }
    \label{fig:music-uperp}
\end{figure}

\begin{figure}[p]
    \centering
    \includegraphics[width=\textwidth]{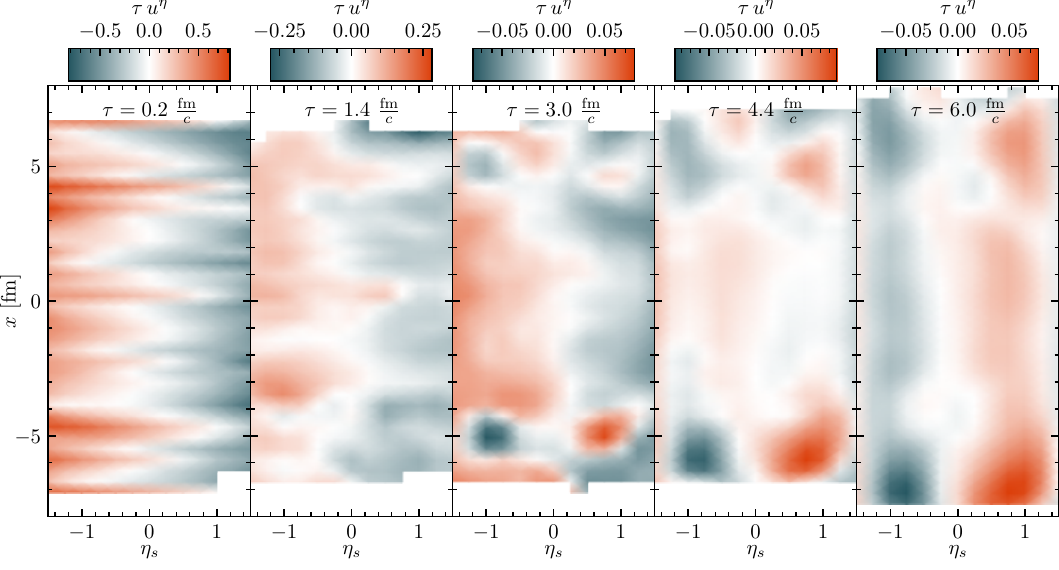}
    \caption[Hydrodynamic evolution of the longitudinal flow.]{%
    Hydrodynamic evolution of the longitudinal flow scaled by proper time in the $x$-$\eta_s$ plane.
    The data correspond to the same event from \cref{fig:music-uperp}.
    Negative values of flow are colored blue, while positive values are colored red.
    The data are cut (white batches) where the local rest frame energy density is smaller than $0.02$~GeV/fm${}^3$.
    }
    \label{fig:music-ueta}
\end{figure}

In \cref{fig:music-uperp,fig:music-ueta}, preliminary results for the hydrodynamic evolution of a single event initialized from the energy-momentum tensor of the (3+1)D dilute Glasma at $\tau_s=0.2$~fm/$c$ are shown.
The $\WSxi$ nuclear model from \cref{sec:finite-longitudinal-correlations} was used and further details of the configured parameters are given in the captions of the figures.
The evolution of the dynamics in the transverse plane in \cref{fig:music-uperp} can be inferred from the three snapshots at proper times $\tau=\tau_s=0.2$~fm/$c$, $3.0$~fm/$c$, and $6.0$~fm/$c$ in the three rows.
The inhomogeneous distribution of local rest frame energy density (black to orange colors) at the switching time is accompanied by an irregular pattern of transverse flow (green arrows).
The collective dynamics in the hydrodynamic phase wash out the structure in the transverse plane and a characteristic flow pattern develops, which corresponds to ordered expansion from the center of the fireball.
Still, at $\tau=3.0$~fm/$c$, there is some irregular flow in the center of the fireball visible.
The value of the spacetime rapidity $\eta_s$ changes from mid-rapidity in the left column to $\eta_s = 1.5$ in the right column.
The dynamics appear similar overall, with lower values of the local rest frame energy density and a more ragged shape for the contours of equal value.

The time evolution of the longitudinal flow in the $x$-$\eta_s$ plane is shown in the panels of \cref{fig:music-ueta}.
As discussed in \cref{subsec:EbE-WS-xi}, the (3+1)D dilute Glasma predicts a negative slope for the rapidity curve of $\tau u^\eta$ (left-most panel).
Along the transverse direction, the value of the flow varies on small scales.
Over the course of the hydrodynamic evolution, the features in the transverse plane are washed out.
Interestingly, for late times, the sign of the gradient along the $\eta_s$ direction flips.
This results in longitudinal flow that is faster than Bjorken expansion.
Still, the values of $\tau u^\eta$ drop by one order of magnitude across the depicted snapshots.
These observations provide the starting point for detailed studies of the hydrodynamic evolution of the (3+1)D dilute Glasma.

On a final note, it is worth mentioning that the hydrodynamic evolution of the (3+1)D dilute Glasma using MUSIC proved to be unstable for many tested configurations.
The strong anisotropies and large gradients of the dynamical fields obtained from the matching procedure could be too far from local thermodynamic equilibrium to allow for a numerically stable simulation.
It remains to be studied if an intermediate QCD effective kinetic theory stage can provide the necessary hydrodynamization of the (3+1)D dilute Glasma.


\appendix
\chapter{Conventions}
\label{appx:conventions}

The conventions for vector notation and the metric are introduced in \cref{sec:description-initial-state} and are repeated here for reference:
\begin{align}
    \hspace{-.3em}\text{Coordinate vector:}& \quad x\equiv x^\mu =(t,x,y,z)^\mu, \quad \mu \in \{t,x,y,z\}, \\
                                           & \quad \vec x = x^i=(x,y,z)^i, \quad i \in \{x,y,z\}, \\
                                           & \quad \xperp = x^\bi = (x,y)^\bi, \quad \bi \in \{x, y\} \\
    \text{General vector:}& \quad p \equiv p^\mu = (p^t,p^x,p^y,p^z)^\mu \\
    \text{Metric:}& \quad g_{\mu\nu} = \mathrm{diag}(+1,-1,-1,-1)_{\mu\nu} \\
    \text{Inner product:}& \quad x^\mu x_\mu = x^\mu x^\nu g_{\mu\nu}, \quad x^i x^i = \vec x \cdot \vec x = \vec x^2,  \quad x^\bi x^\bi = \xperp \cdot \xperp = \xperp^2 \hspace{-.4em} \\
    \text{Line element:}& \quad \dd s^2 = \dd t^2 - \dd x^2 -\dd y^2 -\dd z^2 \\
    \text{Light cone frame:}& \quad \xpm = \frac{1}{\sqrt{2}}(t\pm z), \\
                            & \quad \dd s^2 = \dd x^+ \dd x^- + \dd x^- \dd x^+ - \dd x^2 - \dd y^2 \\
    \text{Milne frame:}& \quad \tau = \sqrt{t^2-z^2} = \sqrt{2x^+x^-}, \\
                       &\quad \eta_s = \artanh(z/t) = \frac{1}{2}\ln(x^+/x^-), \\
                       &\quad t=\tau\cosh(\eta_s), \quad z=\tau\sinh(\eta_s), \\
                       &\quad \xpm = x_{\mp} = \frac{1}{\sqrt{2}}\tau\ee^{\pm\eta_s}, \\
                       &\quad g^\mathrm{Milne}_{\mu\nu} = \mathrm{diag}(+1,-1,-1,-\tau^2)_{\mu\nu}, \\
                       &\quad \dd s^2 = \dd \tau^2 - \dd x^2 - \dd y^2 - \tau^2\dd\eta_s^2 \\
                       &\quad {\renewcommand*{\arraystretch}{1.5} \left( \begin{matrix}
                           \frac{\partial \tau}{\partial t} & \frac{\partial \tau}{\partial z} \\
                           \frac{\partial \eta_s}{\partial t} & \frac{\partial \eta_s}{\partial z} 
                       \end{matrix} \right) = \left( \begin{matrix}
                           \cosh(\eta_s) & - \sinh(\eta_s) \\
                           -\frac{1}{\tau}\sinh(\eta_s) & \frac{1}{\tau}\cosh(\eta_s) 
                       \end{matrix} \right)}
\end{align}

\section{Integrals and Fourier transformations}

Integrals without explicit boundaries run from $-\infty$ to $\infty$.
Multidimensional integrals are abbreviated as
\begin{align}
    \intop \dd^4 x = \intop \dd t\,\dd x\,\dd y\,\dd z, \qquad \intop \dd^3 \vec x = \intop \dd x\, \dd y\, \dd z, \qquad \intop \dd^2\xperp = \intop \dd x\, \dd y.
\end{align}
The prefactor and sign convention for the Fourier transformation are
\begin{align}
    \tilde{F}(k) &= \intop \dd^4 x\, \ee^{+\ii x^\mu k_\mu} F(x), &\qquad F(x) &= \intop \frac{\dd^4 k}{(2\pi)^4}\,\ee^{-\ii x^\mu k_\mu} \tilde{F}(k), \\
    \tilde{F}(\vec k) &= \intop \dd^3\vec x\, \ee^{-\ii \vec x \cdot \vec k} F(\vec x), &\qquad F(\vec x) &= \intop\frac{\dd^3\vec x}{(2\pi)^3}\,\ee^{+\ii \vec x\cdot\vec k} \tilde{F}(\vec k), \\
    \tilde{F}(\kperp) &= \intop \dd^2\xperp\, \ee^{-\ii \xperp\cdot\kperp} F(\xperp), &\qquad F(\xperp) &= \intop \frac{\dd^2\kperp}{(2\pi)^2}\,\ee^{+\ii\xperp\cdot\kperp}\tilde{F}(\kperp),
\end{align}
where $\ii$ is the imaginary unit.
Quantities with the tilde symbol "$\sim$" are understood to be Fourier transformed to momentum space for at least one of their arguments.
If no arguments are explicitly written, all arguments are Fourier transformed.
For example, $\tilde{F} \equiv \tilde{F}(\vec k)$, but
\begin{align}
    \tilde{F}(\xperp,k^z) = \intop \dd z\,\ee^{-\ii z k^z} F(\xperp,z).
\end{align}

\section{Natural units}
Natural units are defined by setting the reduced Planck constant $\hbar$ and the vacuum speed of light $c$ to 1.
This allows us to express quantities in powers of a single base unit, which we choose to be GeV (Gigaelectronvolts).
Using the conversion factor
\begin{align}
    \hbar c \approx 0.197\ 327 \text{ GeV fm} =1,
\end{align}
one can translate between GeV and fm (Femtometers).
It is customary to measure time in fm/c and densities in $\mathrm{fm}^{-3}$, i.e.,
\begin{align}
    1\ \mathrm{fm}/c \approx 5.07 \text{ GeV}^{-1}, \qquad 1\ \text{fm}^{-3} \approx 7.86\times10^{-3} \text{ GeV}^3.
\end{align}

\chapter{Boundary terms for Fourier-transformed sources}\label{appx:tildeS-knuknu-terms}

The Fourier-transformed source terms $\tilde{S}^{\mu\nu}$ given in \cref{eq:tilde-S+-,eq:tilde-S+i,eq:tilde-S-i} for the components with at least one temporal index ($+-,\ +\bi,\ -\bi$) have contributions that are proportional to $k_\nu k^\nu$.
As we will show below, these terms reduce to boundary terms which we can set to zero for nuclei with compact support that are localized to tracks.
We use the placeholder symbol $V$ for the shorthand fields defined in \cref{eq:tildeV-AA,eq:tildeV-AF,eq:tildeV-FA} in the following.
We calculate:
\begin{align}
    \tilde{V}(k) k_\nu k^\nu &= k_\nu k^\nu \intop \dd^4 x\, \ee^{\ii k_\mu x^\mu} V(x) \nn\\
    &= -\intop \dd^4 x\, \left( \partial_\nu \partial^\nu \ee^{\ii k_\mu x^\mu} \right) V(x) \nn\\
    &= -\intop \dd^4 x\, \ee^{\ii k_\mu x^\mu} \partial_\nu \partial^\nu V(x) \nn\\
    &= -\frac{1}{2} \intop \dd^4 x\, \left( \left( \partial_\nu \partial^\nu \ee^{\ii k_\mu x^\mu} \right) V(x) + \ee^{\ii k_\mu x^\mu} \partial_\nu \partial^\nu V(x) \right) \nn\\
    &= -\frac{1}{2} \intop \dd^4 x\, \partial_\nu \partial^\nu \left( \ee^{\ii k_\mu x^\mu} V(x) \right) \nn\\
    &= -\frac{1}{2} \ointop_{\partial} \dd^4 x\, \hat{n}_\nu \partial^\nu \left( \ee^{\ii k_\mu x^\mu} V(x) \right) = 0.
\end{align}
In the third line, we moved the derivatives over to act on $V$ via partially integrating twice.
Recall that $V$ contains the commutator of two single-nucleus fields, each localized to the track of one of the collision partners.
Evaluating $V$ outside of the spacetime region where the two tracks overlap, i.e., the interaction region, yields zero.
Then, the boundary terms from the partial integration in the $x^+$-$x^-$ plane are zero because $V$ is evaluated outside of the interaction region.
The transverse boundary terms are zero because we assume nuclei that have compact support in the transverse plane, enforced by a nuclear envelope.
In the fourth line, we used the identification of the second and third lines to write the integrand as a total derivative in the fifth line.
By virtue of the divergence theorem, we transform the volume integral spanning the entire spacetime over the divergence of a field to an integration along the spacetime boundary of the field itself.
This field is given by the derivative of $V$ multiplied by the Fourier exponential.
Because $V$ itself is constant and zero when approaching the boundary, its derivative is as well.
We conclude that the entire expression is zero.

\chapter{Square of the Lipatov vertex}\label{appx:square-of-lipatov}

In this thesis, we use the representation of the Lipatov effective vertex $C^\mu_L(k,p,q)$ in covariant gauge as given in \cref{eq:Lipatov-vertex} with the light cone components
\begin{align}
    C^+_L(k,\pperp) = k^+ \! -\frac{(\kperp-\pperp)^2}{k^-+\ii\epsilon}, \quad 
    C^-_L(k,\pperp) = -k^- \! +\frac{\pperp^2}{k^++\ii\epsilon}, \quad
    C^\bi_L(k,\pperp) = 2p^\bi -k^\bi,
    \tag{\ref{eq:Lipatov-vertex}}
\end{align}
where the momentum $k$ is external and $\pperp$ denotes an integration label.
In this appendix, we perform the calculation of the squared vertex
\begin{align}\label{eq:appx-Lipatov-square-def}
    P^{ij}_{\vec k} C^i_L(\vec k,\pperp) C^j_L(\vec k,\qperp),
\end{align}
which enters the expression for the occupation number of the gluon field in \cref{eq:gluon-n-def} (cf.\ \textbf{Discussion} on page~\pageref{subsec:discussion-gluon-numbers}).
The transverse projector in \cref{eq:appx-Lipatov-square-def} effectively puts the Lipatov vertex to Coulomb gauge.
The direct calculation of the squared vertex requires more algebra than the method discussed in the main text in Sec.~\textbf{Gluon number distribution} starting on page~\pageref{subsec:gluon-number-distribution}.
We evaluate the vertex for on-shell external momenta $k_\nu k^\nu = 0$, which also allows us to drop the $\ii \epsilon$ prescription for the poles (as argued with \cref{eq:tilde-S+--onshell,eq:tilde-S+i-onshell,eq:tilde-S-i-onshell,eq:tilde-Sij-onshell}).
When squaring the vertex, the external momentum $k$ is shared between both factors and we need two names for the integration labels,  $\pperp$ and $\qperp$.

First, we calculate the $z$-component of the Lipatov vertex from the light cone components in \cref{eq:Lipatov-vertex},
\begin{align}
    C^z_L(k,\pperp) &= \frac{1}{\sqrt{2}}\left( C^+_L(k,\pperp) - C^-_L(k,\pperp) \right) \nn\\
    &= \frac{1}{\sqrt{2}}\left( k^+ - \frac{(\kperp-\pperp)^2}{k^-} + k^- - \frac{\pperp^2}{k^+} \right) \nn\\
    &= \frac{(k^++k^-)}{\sqrt{2}} - \frac{k^+(\kperp^2 + \pperp^2 - 2 \kperp\cdot\pperp)}{\sqrt{2}k^+k^-} - \frac{k^- \pperp^2}{\sqrt{2}k^+k^-} \nn\\
    &= |\vec k| - \frac{1}{\kperp^2}\left( (|\vec k|+k^z)(\kperp^2 + \pperp^2 - 2\kperp\cdot\pperp) + (|\vec k|-k^z)\pperp^2 \right) \nn\\
    &= \frac{1}{\kperp^2}\left( |\vec k|(\kperp^2 - 2\pperp^2) - (|\vec k|+k^z)\kperp\cdot(\kperp-2\pperp)\right),
\end{align}
where we used $(k^++k^-)/\sqrt{2}=k^t=|\vec k|$.
Next, we calculate the contraction of the projector and one vertex, where we distinguish between boldface indices for the transverse components and the longitudinal component with index $z$,
\begin{align}
    P^{ij}_{\vec k} C^j_L(k,\pperp) &= \left( \delta^{i\bi} - \frac{k^i k^\bj}{\vec{k}^2} \right)C^\bj_L(k,\pperp) + \left( \delta^{iz} - \frac{k^i k^z}{\vec{k}^2} \right)C^z_L(k,\pperp) \nn\\
    &= \delta^{i\bj}(2p^\bj-k^\bj) - \frac{k^ik^\bj}{\vec{k}^2}(2p^\bj-k^\bj) \nn\\
    &+ \left( \delta^{i3} - \frac{k^i k^z}{\vec{k}^2} \right)\frac{1}{\kperp^2}\left( |\vec k|(\kperp^2 - 2\pperp^2) - (|\vec k|+k^z)\kperp\cdot(\kperp-2\pperp) \right).
\end{align}
With that, the squared vertex reads
\begin{align}
   &P^{ij}_{\vec k} C^i_L(\vec k,\qperp) C^j_L(\vec k,\pperp) = P^{\bi j}_{\vec k}C^\bi_L(k,\qperp)C^j_L(k,\pperp) + P^{3j}_{\vec k}C^z_L(k,\qperp)C^j(k,\pperp) \nn\\
    &= \Bigg( \delta^{\bi\bj}(2p^\bj-k^\bj) - \frac{k^\bi k^\bj}{\vec{k}^2}(2p^\bj-k^\bj) \nn\\
    & - \frac{k^\bi k^z}{\vec{k}^2} \frac{1}{\kperp^2}\left( |\vec k|(\kperp^2 - 2\pperp^2) - (|\vec k|+k^z)\kperp\cdot(\kperp-2\pperp) \right) \Bigg)(2q^\bi-k^\bi) \nn\\
    &+ \left( -\frac{k^z k^\bj}{\vec{k}^2}(2p^\bj-k^\bj)+\frac{1}{\vec{k}^2\kperp^2}(\vec{k}^2-(k^z)^2)\left( |\vec k|(\kperp^2-2\pperp^2) -(|\vec k|+k^z)\kperp\cdot(\kperp-2\pperp) \right) \right) \nn\\
    &\times \frac{1}{\kperp^2}\left( |\vec k|(\kperp^2-2\qperp^2) - (|\vec k|+k^z)\kperp\cdot(\kperp-2\qperp) \right) \nn\\
    &= (2\pperp-\kperp)\cdot(2\qperp-\kperp) - \frac{1}{\vec{k}^2}\kperp\cdot(2\pperp-\kperp)\kperp\cdot(2\qperp-\kperp) \nn\\
    &-\frac{k^z}{\vec{k}^2\kperp^2}\kperp\cdot(2\qperp-\kperp)\left( |\vec k|(\kperp^2-2\pperp^2) - (|\vec k|+k^z)\kperp\cdot(\kperp-2\pperp) \right) \nn\\
    &-\frac{k^z}{\vec{k}^2\kperp^2}\kperp\cdot(2\pperp-\kperp)\left( |\vec k|(\kperp^2-2\qperp^2)-(|\vec k|+k^z)\kperp\cdot(\kperp-2\qperp) \right) \nn\\
    &+\frac{1}{\vec{k}^2\kperp^2}\left(|\vec k|(\kperp^2-2\pperp^2) - (|\vec k|+k^z)\kperp\cdot(\kperp-2\pperp)\right)\nn\\
    &\hphantom{\frac{1}{\vec{k}^2\kperp^2}\,}\times\left(|\vec k|(\kperp^2-2\qperp^2)-(|\vec k|+k^z)\kperp\cdot(\kperp-2\qperp)\right) \nn\\
    &= (2\pperp-\kperp)\cdot(2\qperp-\kperp) - \frac{1}{\vec{k}^2}\kperp\cdot(2\pperp-\kperp)\kperp\cdot(2\qperp-\kperp) -\frac{1}{\vec{k}^2\kperp^2} \Bigg(\nn\\
    & + k^z|\vec k|(\kperp^2-2\pperp^2)\kperp\cdot(2\qperp-\kperp) + k^z|\vec k|(\kperp^2-2\qperp^2)\kperp\cdot(2\pperp-\kperp) \nn\\
    & + (k^z|\vec k|+\vec{k}^2-\kperp^2)\kperp\cdot(2\qperp-\kperp)\kperp\cdot(2\pperp-\kperp) \nn\\
    & + (k^z|\vec k|+\vec{k}^2-\kperp^2)\kperp\cdot(2\pperp-\kperp)\kperp\cdot(2\qperp-\kperp)\nn\\
    & - (2\vec{k}^2-\kperp^2+2k^z|\vec k|)\kperp\cdot(2\pperp-\kperp)\kperp\cdot(2\qperp-\kperp)\nn\\
    & - \vec{k}^2(\kperp^2-2\pperp^2)(\kperp^2-2\qperp^2) \nn\\
    & - (\vec{k}^2+k^z|\vec k|)\kperp\cdot(2\pperp-\kperp)(\kperp^2 - 2\qperp^2) - (\vec{k}^2+k^z|\vec k|)\kperp\cdot(2\qperp-\kperp)(\kperp^2-2\pperp^2)
    \Bigg) \nn\\
    &= \frac{1}{\kperp^2}\Bigg( \kperp^2(2\pperp-\kperp)\cdot(2\qperp-\kperp) + (\kperp^2-2\pperp^2)(\kperp^2-2\qperp^2) \nn\\
    &\hphantom{\frac{1}{\kperp^2}\Bigg(\,} + (\kperp^2-2\pperp^2)\kperp\cdot(2\qperp-\kperp) + (\kperp^2-2\qperp^2)\kperp\cdot(2\pperp-\kperp) \Bigg) \nn\\
    &= \frac{4}{\kperp^2}\left( \kperp^2 \pperp\cdot\qperp + \pperp^2\qperp^2 - \pperp^2\kperp\cdot\qperp - \qperp^2\kperp\cdot\pperp \right) \nn\\
    &= \frac{4}{\kperp^2}\left( \pperp\cdot(\pperp-\kperp)\qperp\cdot(\qperp-\kperp) + \pperp\cdot\qperp(\pperp-\kperp)\cdot(\qperp-\kperp) - \pperp\cdot(\qperp-\kperp)\qperp\cdot(\pperp-\kperp) \right)\nn\\
    &= \frac{4}{\kperp^2}p^\bi(p^\bj-k^\bj)q^\bk(q^\bl-k^\bl)\left( \delta^{\bi\bj}\delta^{\bk\bl} + \delta^{\bi\bk}\delta^{\bj\bl} - \delta^{\bi\bl}\delta^{\bj\bk}\right) \nn\\
    &= \frac{4}{\kperp^2}p^\bi(p^\bj-k^\bj)q^\bk(q^\bl-k^\bl)\left( \delta^{\bi\bj}\delta^{\bk\bl} + \varepsilon^{\bi\bj}\varepsilon^{\bk\bl}\right).
\end{align}

\chapter{Factorizable gluon distributions for large nuclear scales}\label{appx:factorizable-large-R-scales}

In this appendix, we provide the rigorous derivation of the gluon distribution in the limit of large nuclei discussed in \cref{sec:limit-large-nucl}.
At its core, the argument involves a separation of scales for the quantities involved.
The starting point is \cref{eq:dNd2kdY-large-nuclei-scales},
\begin{align}
    \kperp^2\frac{\dd N}{\dd^2\kperp\,\dd\Y} &= \frac{2g^2N_c(N_c^2-1)}{(2\pi)^3} \intop \frac{\dd^2\uperp\,\dd^2\wperp}{(2\pi)^4} \frac{\dd l^-\,\dd^2\lperp}{(2\pi)^3} \frac{\dd\kappa^+\,\dd^2\kappaperp}{(2\pi)^3} \nn\\
    &\times \omega_m(\kperp,\uperp+\tfrac{\wperp+\lperp}{2},\uperp-\tfrac{\wperp+\lperp}{2}) \nn\\
    &\times \tilde{\Gamma}_\A(k^- -l^-,\uperp+\tfrac{\wperp-\lperp}{2})\tilde{t}_\A(l^-,\lperp)\tilde{t}_\A(-l^-,\wperp) \nn\\
    &\times \tilde{\Gamma}_\B(k^+-\kappa^+,\kperp-\uperp-\tfrac{\wperp+\lperp}{2}-\kappaperp) \tilde{t}_\B(\kappa^+,\kappaperp)\tilde{t}_\B(-\kappa^+,-\kappaperp-\wperp-\lperp). \tag{\ref{eq:dNd2kdY-large-nuclei-scales}}
\end{align}
To formalize this argument, we change to scale-free functions and write out all of the involved scales explicitly.
The scale of the single-nucleus envelopes $T_{\A/\B}$ is parametrized by the radius $R$.
We stress that there are no other scales in the envelopes, which implies that there is no impact parameter.
In the rest frame of the nucleus, its envelope is spherically symmetric.
Hereby, we assume that the nuclear envelopes $T_{\A/\B}$ have the same functional form.
When boosting to the lab frame and changing to light cone coordinates, the scale along the longitudinal direction becomes $R/(\sqrt{2}\gamma)$ and acquires a $\gamma$-factor due to Lorentz contraction.
In contrast, the scale in the transverse directions stays $R$.
We define the scale-free envelopes
\begin{align}
    \underline{T}_{\A/\B}(\underline{x}^\pm,\underline{\xperp}) &\coloneqq \frac{R}{\gamma} T_{\A/\B}(\tfrac{R}{\gamma}\underline{x}^\pm,R \underline{\xperp}), \\
    \underline{t}_{\A/\B}(\underline{x}^\pm,\underline{\xperp}) &\coloneqq \sqrt{\underline{T}_{\A/\B}(\underline{x}^\pm,\underline{\xperp})} = \sqrt{\frac{R}{\gamma}}t_{\A/\B} (\tfrac{R}{\gamma}\underline{x}^\pm,R \underline{\xperp}), \label{eq:scale-free-t}
\end{align}
where underlined quantities are considered dimensionless.
Here, all scales that are hidden in the envelopes are made explicit.
The underlined profile functions $\underline{T}_{\A/\B}$ correspond to spherically symmetric profiles with radius $1$.
The additional prefactor stems from the chosen normalization of the envelopes (cf.~\cref{eq:envelope-norm}) that carries a longitudinal scale.
We also absorbed the factor of $\sqrt{2}$ from the conversion to light cone coordinates into the definition of the scale-free envelopes to reduce clutter.
The Fourier transformation of the envelopes' square roots is\pagebreak
\begin{align}
    \tilde{t}_{\A/\B}(\kappa^\mp,\kappaperp) &= \intop \dd\xpm\,\dd^2\xperp\,\ee^{\ii\xpm\kappa^\mp - \ii\xperp\cdot\kappaperp} t_{\A/\B}(\xpm,\xperp) \nn\\
    &= \intop \dd\underline{x}^\pm\,\dd^2\underline{\xperp}\, \frac{R^3}{\gamma} \exp(\ii\tfrac{R}{\gamma}\underline{x}^\pm\kappa^\mp -\ii R\underline{\xperp}\cdot\kappaperp) t_{\A/\B}(\tfrac{R}{\gamma}\underline{x}^\pm,R\underline{\xperp}) \nn\\
    &= \sqrt{\frac{R^5}{\gamma}}\intop \dd\underline{x}^\pm\,\dd^2\underline{\xperp}\, \exp(\ii\tfrac{R}{\gamma}\underline{x}^\pm\kappa^\mp -\ii R\underline{\xperp}\cdot\kappaperp) \underline{t}_{\A/\B}(\underline{x}^\pm,\underline{\xperp}) \nn\\
    &= \sqrt{\frac{R^5}{\gamma}} \underline{\tilde{t}}_{\A/\B}(\tfrac{R}{\gamma}\kappa^\mp,R\kappaperp), \label{eq:tildet-scalefree}
\end{align}
where we rescaled the integration variables in the second line, used \cref{eq:scale-free-t} in the third line and identified the Fourier transformation of the scale-free envelopes w.r.t.\ their arguments in the last line.

Similarly, we parametrize the scale of the correlations in $\Gamma_{\A/\B}$ in the rest frame of the nuclei by the longitudinal scale $r_l$ and the transverse scale $r$ and define
\begin{align}
    \underline{\Gamma}_{\A/\B}(\underline{x}^\pm,\underline{\xperp}) &\coloneqq \vartheta \frac{r_l r^2}{\gamma}\Gamma_{\A/\B}(\tfrac{r_l}{\gamma}\underline{x}^\pm,r \underline{\xperp}), \\
    \tilde{\Gamma}_{\A/\B}(\kappa^\mp,\kappaperp) &= \frac{1}{\vartheta}\underline{\tilde{\Gamma}}_{\A/\B}(\tfrac{r_l}{\gamma}\kappa^\mp,r\kappaperp), \label{eq:tildeGamma-scalefree}
\end{align}
where the proportionality constant $\vartheta$ has the dimension of energy squared and the factors of $r_l$ and $r$ appear because $\Gamma_{\A/\B}$ is normalized w.r.t.\ the integration over the spacetime coordinates (cf.\ the limit in \cref{eq:Gamma-MV-limit-norm}).

In the case of the IR-regulated effective vertex $\omega_m$, the only scale is introduced by $m$.
We define the scale-free vertex function
\begin{align}
    \underline{\omega}(\underline{\kperp},\underline{\pperp},\underline{\qperp}) \coloneqq m^4 \omega_m(m\underline{\kperp},m\underline{\pperp},m\underline{\qperp}), \label{eq:omega-scale-free}
\end{align}
in full analogy to before, where the functional form of $\underline{\omega}$ is given by $\omega_m$ with $m=1$.

All of these scale-free functions have the common property that they vary on the typical scales of $\sim 1$ for their dimensionless arguments and quickly fall off%
\footnote{%
The falloff rate for the IR-regulated vertex $\omega_m$ is only $1/\pperp^2$ for $|\pperp|\gg m$ (same for $\qperp$).
Still, we consider this rate fast enough for the purpose of this approximation.}
for larger values.
By explicitly keeping track of all scale factors, we can formulate clean arguments.
As will follow next, changing to dimensionless variables introduces ratios of scale parameters that lead to the suppression of dependencies on the integration variables.

We insert \cref{eq:tildet-scalefree,eq:tildeGamma-scalefree,eq:omega-scale-free} back into \cref{eq:dNd2kdY-large-nuclei-scales} and get
\begin{align}
    \kperp^2\frac{\dd N}{\dd^2\kperp\,\dd\Y} &= \frac{2g^2N_c(N_c^2-1)}{(2\pi)^3} \intop \frac{\dd^2\uperp\,\dd^2\wperp}{(2\pi)^4}\frac{\dd l^-\,\dd^2\lperp}{(2\pi)^3} \frac{\dd\kappa^+\,\dd^2\kappaperp}{(2\pi)^3}\frac{R^{10}}{\gamma^2} \nn\\
    &\times \frac{1}{m^4}\underline{\omega}(\tfrac{\kperp}{m},\tfrac{\uperp}{m}+\tfrac{\wperp+\lperp}{2m},\tfrac{\uperp}{m}-\tfrac{\wperp+\lperp}{2m}) \nn\\
    &\times \frac{1}{\vartheta}\underline{\tilde{\Gamma}}_\A(\tfrac{r_l}{\gamma}k^- -\tfrac{r_l}{\gamma}l^-,r\uperp+r\tfrac{\wperp-\lperp}{2}) \nn\\
    &\times \frac{1}{\vartheta}\underline{\tilde{\Gamma}}_\B(\tfrac{r_l}{\gamma}k^+ -\tfrac{r_l}{\gamma}\kappa^+,r(\kperp-\uperp)-r(\tfrac{\wperp+\lperp}{2}+\kappaperp)) \nn\\
    &\times \underline{\tilde{t}}_\A(\tfrac{R}{\gamma}l^-,R\lperp)\underline{\tilde{t}}_\A(-\tfrac{R}{\gamma}l^-,R\wperp) \underline{\tilde{t}}_\B(\tfrac{R}{\gamma}\kappa^+,R\kappaperp)\underline{\tilde{t}}_\B(-\tfrac{R}{\gamma}\kappa^+,-R(\kappaperp+\wperp+\lperp)), \nn\\
    \kperp^2\frac{\dd N}{\dd^2\kperp\,\dd\Y} &= \frac{2g^2N_c(N_c^2-1)}{(2\pi)^3} \intop \frac{\dd^2\uperp\,\dd^2\underline{\wperp}}{(2\pi)^4}\frac{\dd \underline{l}^-\,\dd^2\underline{\lperp}}{(2\pi)^3} \frac{\dd\underline{\kappa}^+\,\dd^2\underline{\kappaperp}}{(2\pi)^3} \nn\\
    &\times \frac{1}{m^4}\underline{\omega}(\tfrac{\kperp}{m},\tfrac{\uperp}{m}+\tfrac{\underline{\wperp}+\underline{\lperp}}{2mR},\tfrac{\uperp}{m}-\tfrac{\underline{\wperp}+\underline{\lperp}}{2mR}) \nn\\
    &\times \frac{1}{\vartheta}\underline{\tilde{\Gamma}}_\A(\tfrac{r_l}{\gamma}k^- -\tfrac{r_l}{R}\underline{l}^-,r\uperp+\tfrac{r}{R}\tfrac{\underline{\wperp}-\underline{\lperp}}{2}) \nn\\
    &\times \frac{1}{\vartheta}\underline{\tilde{\Gamma}}_\B(\tfrac{r_l}{\gamma}k^+ -\tfrac{r_l}{R}\underline{\kappa}^+,r(\kperp-\uperp)-\tfrac{r}{R}(\tfrac{\underline{\wperp}+\underline{\lperp}}{2}+\underline{\kappaperp})) \nn\\
    &\times R^2 \underline{\tilde{t}}_\A(\underline{l}^-,\underline{\lperp})\underline{\tilde{t}}_\A(-\underline{l}^-,\underline{\wperp}) \underline{\tilde{t}}_\B(\underline{\kappa}^+,\underline{\kappaperp})\underline{\tilde{t}}_\B(-\underline{\kappa}^+,-\underline{\kappaperp} -\underline{\wperp} -\underline{\lperp})
    , \label{eq:dNd2kdY-large-nuclei-scale-approx}
\end{align}
where in the second equation we rescaled all integration variables that evaluate the envelopes by the envelope radius,
\begin{align}
    & & \underline{\kappa}^+ &= \frac{R}{\gamma} \kappa^+, & \underline{\kappaperp} &= R \kappaperp, & & \\
    & & \underline{l}^- &= \frac{R}{\gamma} l^-, & \underline{\lperp} &= R \lperp, & & \\
    & & & & \underline{\wperp} &= R \wperp. & &
\end{align}
Now, the limit of large nuclei formulated in terms of the explicit scales reads
\begin{align}
   r_l/R \ll 1, \qquad r/R \ll 1, \qquad 1/(mR) \ll 1. \label{eq:scale-prefactors}
\end{align}
We stress again that there are no other scales than the ones listed in \cref{eq:scale-prefactors}.
These ratios always appear as prefactors to the dimensionless variables in the correlation functions and the vertex function.
The dimensionless variables are limited to values $\lesssim 1$ because they appear as isolated arguments of the scale-free envelopes.
The previous substitutions and shifts lead to single-variable dependencies for three of the envelope factors and allow us to directly read off the support of the integrand in terms of these variables.
Hence, the prefactors in \cref{eq:scale-prefactors} strongly suppress the dependencies on the dimensionless variables.

We investigate each of the terms in \cref{eq:dNd2kdY-large-nuclei-scale-approx} one by one, starting with
\begin{align}
    \underline{\tilde{\Gamma}}_\A(\tfrac{r_l}{\gamma}k^- -\tfrac{r_l}{R}\underline{l}^-,r\uperp+\tfrac{r}{R}\tfrac{\underline{\wperp}-\underline{\lperp}}{2}) \underline{\tilde{t}}_\A(\underline{l}^-,\underline{\lperp})\underline{\tilde{t}}_\A(-\underline{l}^-,\underline{\wperp}). \label{eq:utidleGamma-A-focus}
\end{align}
Along the longitudinal direction, the variable $\underline{l}^-$ is limited to $\underline{l}^-\lesssim 1$ because of $\underline{\tilde{t}}_\A$.
The longitudinal dependence of $\underline{\tilde{\Gamma}}_\A$ involves the difference
\begin{align}
    \frac{r_l}{\gamma}k^- -\frac{r_l}{R}\underline{l}^- \approx \frac{r_l}{\gamma}k^-, \label{eq:utildeGamma-A-args}
\end{align}
where we used \cref{eq:scale-prefactors} and assumed that the longitudinal correlation scale $r_l$ is much smaller than $R$.
For any values of $\underline{l}^- \lesssim 1$ we can, therefore, neglect the contibution of $\underline{l}^-$ to \cref{eq:utildeGamma-A-args}.
As a result, we can drop the dependence of $\underline{\tilde{\Gamma}}_\A$ on $\underline{l}^-$ and pull it out of the integration over $\underline{l}^-$.
Along the transverse direction, the dependence of $\underline{\tilde{\Gamma}}_\A$ is evaluated at
\begin{align}
    r\uperp+\frac{r}{R}\frac{\underline{\wperp}-\underline{\lperp}}{2} \approx r\uperp.
\end{align}
Again, we used \cref{eq:scale-prefactors} and assumed that the transverse correlation scale $r$ is much smaller than $R$.
Each one of $\underline{\wperp}$ and $\underline{\lperp}$ is limited to values $\lesssim 1$ because of one of the $\underline{\tilde{t}}_\A$ factors in \cref{eq:utidleGamma-A-focus}.
Then, we can drop the dependence of $\underline{\tilde{\Gamma}}_\A$ on $\underline{\wperp}$ and $\underline{\lperp}$.
The situation for $\underline{\tilde{\Gamma}}_\B$ is analogous, but involves three transverse variables,
\begin{align}
    \underline{\tilde{\Gamma}}_\B(\tfrac{r_l}{\gamma}k^+ -\tfrac{r_l}{R}\underline{\kappa}^+,r(\kperp-\uperp)-\tfrac{r}{R}(\tfrac{\underline{\wperp}+\underline{\lperp}}{2}+\underline{\kappaperp}))  \underline{\tilde{t}}_\A(\underline{l}^-,\underline{\lperp})\underline{\tilde{t}}_\A(-\underline{l}^-,\underline{\wperp})\underline{\tilde{t}}_\B(\underline{\kappa}^+,\underline{\kappaperp}).
\end{align}
Here, we can drop the longitudinal dependence on $\underline{\kappa}^+$, and the transverse dependencies on $\underline{\wperp}$, $\underline{\lperp}$ and $\underline{\kappaperp}$.

Similarly, the scale-free vertex function involves two transverse momenta,
\begin{align}
    \underline{\omega}(\tfrac{\kperp}{m},\tfrac{\uperp}{m}+\tfrac{\underline{\wperp}+\underline{\lperp}}{2mR},\tfrac{\uperp}{m}-\tfrac{\underline{\wperp}+\underline{\lperp}}{2mR}) \underline{\tilde{t}}_\A(\underline{l}^-,\underline{\lperp})\underline{\tilde{t}}_\A(-\underline{l}^-,\underline{\wperp}).
\end{align}
Using \cref{eq:scale-prefactors}, and assuming that the scale $1/m$ of the vertex is much smaller than $R$,
\begin{align}
    \frac{\uperp}{m}\pm\frac{\wperp+\lperp}{2m} \approx \frac{\uperp}{m},
\end{align}
and we can drop the dependencies on $\underline{\wperp}$ and $\underline{\lperp}$.

Note the prominent appearance of the envelope of nucleus $\A$ in the above discussion.
This is a result of how we chose to shift around and isolate the dependencies.
By symmetry, the same discussion can be brought forward where the envelope of nucleus $\B$ contains only single-momentum dependencies.
Since we initially assumed that there is no impact parameter and that the envelopes of nucleus $\A$ and $\B$ behave equally with the same scale $R$, we have a symmetric system w.r.t.\ the exchange of the labels $\A$ and $\B$.

After applying the results of the above discussion to \cref{eq:dNd2kdY-large-nuclei-scale-approx} the expression for the gluon number distribution reads
\begin{align}
     \kperp^2\frac{\dd N}{\dd^2\kperp\,\dd\Y} &\approx \frac{2g^2N_c(N_c^2-1)}{(2\pi)^3} \intop \frac{\dd^2\uperp}{(2\pi)^2}
     \frac{1}{m^4}\underline{\omega}(\tfrac{\kperp}{m},\tfrac{\uperp}{m},\tfrac{\uperp}{m}) \nn\\
     &\times \frac{r^3}{\vartheta\gamma}\underline{\tilde{\Gamma}}_\A(\tfrac{r}{\gamma}k^-,r\uperp) 
     \frac{r^3}{\vartheta\gamma}\underline{\tilde{\Gamma}}_\B(\tfrac{r}{\gamma}k^+,r(\kperp-\uperp)) \nn\\
     &\times R^2 \intop\frac{\dd^2\underline{\wperp}\,\dd\underline{l}^-\,\dd^2\underline{\lperp}\,\dd\underline{\kappa}^+\,\dd^2\underline{\kappaperp}}{(2\pi)^8} \nn\\
     &\times \underline{\tilde{t}}_\A(\underline{l}^-,\underline{\lperp})\underline{\tilde{t}}_\A(-\underline{l}^-,\underline{\wperp}) \underline{\tilde{t}}_\B(\underline{\kappa}^+,\underline{\kappaperp})\underline{\tilde{t}}_\B(-\underline{\kappa}^+,-\underline{\kappaperp} -\underline{\wperp} -\underline{\lperp}). \label{eq:R2-Sop-scale-free}
\end{align}
Changing back to dimensionful integration variables and absorbing the explicit scales to restore the correlation functions and envelopes yields
\begin{align}
     \kperp^2\frac{\dd N}{\dd^2\kperp\,\dd\Y} &\approx \frac{2g^2N_c(N_c^2-1)}{(2\pi)^3} \intop \frac{\dd^2\uperp}{(2\pi)^2}
     \frac{1}{m^4}\underline{\omega}(\tfrac{\kperp}{m},\tfrac{\uperp}{m},\tfrac{\uperp}{m}) \tilde{\Gamma}_\A(k^-,\uperp)\tilde{\Gamma}_\B(k^+,\kperp-\uperp) \nn\\
     &\times \intop \frac{\dd^2\wperp\,\dd l^-\,\dd^2\lperp\,\dd\kappa^+\,\dd^2\kappaperp}{(2\pi)^8} \nn\\
     &\times \tilde{t}_\A(l^-,\lperp)\tilde{t}_\A(-l^-,\wperp)\tilde{t}_\B(\kappa^+,\kappaperp)\tilde{t}_\B(-\kappa^+,-\kappaperp-\wperp-\lperp) \nn\\
     &= \frac{2g^2N_c(N_c^2-1)}{(2\pi)^3} \intop \frac{\dd^2\uperp}{(2\pi)^2}
     \frac{1}{m^4}\underline{\omega}(\tfrac{\kperp}{m},\tfrac{\uperp}{m},\tfrac{\uperp}{m}) \tilde{\Gamma}_\A(k^-,\uperp)\tilde{\Gamma}_\B(k^+,\kperp-\uperp) \nn\\
     &\times \intop \frac{\dd^2\vperp\,\dd l^-\,\dd^2\lperp\,\dd\kappa^+\,\dd^2\kappaperp}{(2\pi)^8} \nn\\
     &\times \tilde{t}_\A(l^-,\lperp)\tilde{t}_\A(-l^-,\vperp-\lperp)\tilde{t}_\B(\kappa^+,\kappaperp)\tilde{t}_\B(-\kappa^+,-\vperp-\kappaperp), \nn\\
     \kperp^2\frac{\dd N}{\dd^2\kperp\,\dd\Y} &= \frac{2g^2N_c(N_c^2-1)}{(2\pi)^3} \intop \frac{\dd^2\uperp}{(2\pi)^2}
     \frac{1}{m^4}\underline{\omega}(\tfrac{\kperp}{m},\tfrac{\uperp}{m},\tfrac{\uperp}{m}) \tilde{\Gamma}_\A(k^-,\uperp)\tilde{\Gamma}_\B(k^+,\kperp-\uperp) \nn\\
     &\times \intop \frac{\dd^2\vperp}{(2\pi)^2}\, \tilde{T}^\perp_\A(\vperp)\tilde{T}^\perp_\B(-\vperp)
     , \label{eq:dNd2kdY-sigma-sigma-factorized}
\end{align}
where we shifted back $\wperp$ to $\wperp +\lperp = \vperp$ for the second equality and, in the last line, restored the Fourier-transformed projected transverse envelopes $\tilde{T}^\perp_{\A/\B}$ defined in \cref{eq:FT-Tperp}.
This reproduces the result given in \cref{eq:dNd2kdY-large-nuclei-result}.

Lastly, we motivate how a small impact parameter influences the result in \cref{eq:dNd2kdY-sigma-sigma-factorized} (or \cref{eq:dNd2kdY-large-nuclei-result}) and introduces the impact parameter dependence of the transverse overlap area $\Sop$ defined in \cref{eq:Sop-impact-param}.
We assume that the impact parameter dependence can be described by an appropriate shift of the transverse coordinates at which the envelopes $t_{\A/\B}$ are evaluated.
In particular, we replace (as in \cref{eq:Tperp-impact-param-def})
\begin{align}
    t_{\A/\B}(\xpm,\xperp) \rightarrow t_{\A/\B}(\xpm, \xperp \mp \bperp/2), \label{eq:t-impact-param-def}
\end{align}
where each nucleus is shifted by half the impact parameter $\bperp$ in the transverse plane.
However, we maintain the functional form of the envelopes and their scaling with the radius parameter $R$.
This leads to a phase factor modulated by $\bperp$ for the Fourier transformation of the envelopes,
\begin{align}
    \tilde{t}_{\A/\B}(\pmp,\pperp; \bperp) &= \intop \dd\xpm\,\dd^2\xperp\, t_{\A/\B}(\xpm,\xperp \mp\bperp/2) \ee^{\ii \xpm\pmp - \ii\xperp\cdot\pperp} \nn\\
    &= \ee^{\mp\ii\bperp\cdot\pperp} \tilde{t}_{\A/\B}(\pmp,\pperp). \label{eq:tildet-impact-param}
\end{align}
Now, for the factorization of the integral in \cref{eq:dNd2kdY-sigma-sigma-factorized} to still be valid, the impact parameter has to be small compared to the nuclear radius,
\begin{align}
    |\bperp| \ll R.
\end{align}
This ensures that the Fourier-transformed envelope in \cref{eq:tildet-impact-param} still peaks at a single, highly localized value for $\pperp$.
Otherwise, the phase factor would oscillate within the (very narrow) support of $\pperp$ and lead to ill-defined peaks.
After the integral is factorized to the form in \cref{eq:dNd2kdY-sigma-sigma-factorized}, it is straightforward to obtain $\Sop$ in \cref{eq:Sop-impact-param} using \cref{eq:t-impact-param-def} (or equivalently \cref{eq:Tperp-impact-param-def}).

\chapter{Integration of the effective transverse vertex}\label{appx:Integrated-effective-vertex}

\emph{The calculation presented in this appendix was performed by an undergraduate student, \emph{Katharina Gaunersdorfer}, supervised by the author.}\newline

\noindent
In this appendix, we perform the integration of the effective transverse vertex factor that appears in the calculation of the gluon number distribution $\dd N/\dd^2\kperp\dd\Y$ in \cref{eq:dNd2kdY-MV-limit}.
This factor is closely related to the squared Lipatov vertex (cf.\ \cref{appx:square-of-lipatov}), but contains additional factors of inverse momenta with IR regulation.
These momentum factors are generated when assuming the MV model and inserting the color charge correlators of the nuclei.
For detailed discussion see \cref{sec:gluon-numbers-dilute,sec:limit-large-nucl}.

The definition of the integral from \cref{eq:Omega(k)-def} reads
\begin{align}
    \Omega(\kappaperp) = m^2\intop \frac{\dd^2\qperp}{(2\pi)^2}\, \omega_m(\kappaperp,\qperp,\qperp),
\end{align}
where the IR regulated effective vertex $\omega_m$ is given in \cref{eq:effective-omega-m-def}.
First, we rescale the momenta
\begin{align}
    \kappaperp/m = \kperp, \qquad \qperp/m = \pperp
\end{align}
by $m$ to obtain a scale-free expression,
\begin{align}
    \Omega(\kperp) &= \intop \frac{\dd^2\pperp}{(2\pi)^2} \frac{\left( \delta^{\bk\bl}\delta^{\bi\bj} + \varepsilon^{\bk\bl}\varepsilon^{\bi\bj} \right) p^\bk(k^\bl-p^\bl)p^\bi(k^\bj-p^\bj)}{(\pperp^2 + 1)^2\left((\kperp-\pperp)^2 +1 \right)^2}.
\end{align}
We can simplify the vector structure in the numerator,
\begin{align}
   \left( \delta^{\bk\bl}\delta^{\bi\bj} + \varepsilon^{\bk\bl}\varepsilon^{\bi\bj} \right) p^\bk(k^\bl-p^\bl)p^\bi(k^\bj-p^\bj)
    &= \left( \pperp\cdot(\kperp-\pperp)\right)^2 + \left( \pperp\times(\kperp-\pperp)\right)^2 \nn\\
    &= \pperp^2\left(\kperp-\pperp\right)^2 \cos^2(\phi) + \pperp^2\left(\kperp-\pperp\right)^2\sin^2(\phi) \nn\\
    &= \pperp^2\left(\kperp-\pperp\right)^2, \label{eq:Omega-transverse-vector-identity}
\end{align}
where we used the cross product in the transverse plane $\pperp\times\kperp \coloneqq \varepsilon^{\bi\bj}p^\bi k^\bj$ in the first line and used a vector identity in the last line.
Then, the integral simplifies to
\begin{align}
    \Omega(\kperp) &= \intop \frac{\dd^2\pperp}{(2\pi)^2} \frac{\pperp^2\left(\kperp-\pperp\right)^2}{(\pperp^2 + 1)^2\left((\kperp-\pperp)^2 +1 \right)^2}.
\end{align}
To make progress, we resort to Feynman parametrization for the integral, i.e.,
\begin{align}
    \frac{1}{A^2 B^2} = 6 \intop_0^1\dd x\, \frac{x(1-x)}{\left( (1-x)A + xB \right)^4},
\end{align}
with the identifications
\begin{align}
    A = \pperp^2 +1, \qquad B = \left( \kperp - \pperp \right)^2 + 1.
\end{align}
For the case at hand, this yields
\begin{align}
    \Omega(\kperp) = 6 \intop \frac{\dd^2\pperp}{(2\pi)^2}\, \pperp^2\left( \kperp-\pperp \right)^2 \intop_0^1\dd x\, \frac{x(1-x)}{\left( x\left( (\kperp-\pperp)^2 +1\right) + (1-x)(\pperp^2+1) \right)^4}.
\end{align}
The denominator can be simplified as
\begin{align}
    x\left( (\kperp-\pperp)^2 +1\right) + (1-x)(\pperp^2+1) &= \pperp^2 - 2x \kperp\cdot\pperp + x\kperp^2 +1\nn\\
    &= \left( \pperp- x\kperp\right)^2 + x(1-x)\kperp^2 +1.
\end{align}
Using the substitutions
\begin{align}
    \xi(x) = x(1-x)\kperp^2 +1, \qquad \qperp = \pperp-x\kperp,
\end{align}
the integral reads
\begin{align}
    \Omega(\kperp) &= 6\intop_0^1\dd x\, x(1-x) \intop \frac{\dd^2\qperp}{(2\pi)^2}\frac{(\qperp+x\kperp)^2((1-x)\kperp-\qperp)^2}{\left( \qperp^2 + \xi(x) \right)^4} \nn\\
    &= \frac{6}{(2\pi)^2}\intop_0^1\dd x\, x(1-x) \intop_0^\infty\dd |\qperp|\intop_0^{2\pi}\dd\phi\, \frac{|\qperp|}{\left( \qperp^2 + \xi(x) \right)^4} \nn\\
    &\times \left(\qperp^2 + x^2\kperp^2 +2x|\qperp||\kperp|\cos(\phi)\right)\left(\qperp^2+(1-x)^2\kperp^2-2(1-x)|\qperp||\kperp|\cos(\phi)\right).
\end{align}
For the second equality, we changed to polar coordinates for $\qperp$ and introduced the angle $\phi\coloneqq \angle(\qperp,\kperp)$.
In the angular integration, the odd powers of $\cos(\phi)$ drop out.
We are left with
\begin{align}
    \Omega(\kperp) &= \frac{6}{(2\pi)^2}\intop_0^1\dd x\, x(1-x) \intop_0^\infty\dd |\qperp|\intop_0^{2\pi}\dd\phi\, \frac{1}{\left( \qperp^2 + \xi(x) \right)^4} \nn\\
    &\times \left( \qperp^5 + \qperp^3\kperp^2(2x^2-2x+1) +|\qperp|\kperp^4x^2(1-x)^2 - \qperp^3\kperp^2\cos^2(\phi)4x(1-x) \right) \nn\\
    &= \frac{6}{2\pi}\intop_0^1\dd x\, x(1-x) \intop_0^\infty\dd |\qperp|\, \frac{\qperp^5 + \qperp^3\kperp^2(1-2x)^2 + |\qperp|\kperp^4(x(1-x))^2}{\left( \qperp^2 + \xi(x) \right)^4},
\end{align}
after integrating out the angle $\phi$ in the last line.
The $|\qperp|$-integral reduces to a linear combination of the three elementary integrals
\begin{align}
    \intop_0^\infty \dd z\, \frac{z^5}{(z^2+\xi)^4} = \frac{1}{6\xi}, \quad \intop_0^\infty \dd z\, \frac{z^3}{(z^2+\xi)^4} = \frac{1}{12\xi^2}, \quad \intop_0^\infty \dd z\, \frac{1}{(z^2+\xi)^4} = \frac{1}{6\xi^3}.
\end{align}
We obtain
\begin{align}
    \Omega(\kperp) &= \frac{6}{2\pi}\intop_0^1\dd x\, x(1-x) \left( \frac{1}{\xi(x)} + \frac{\kperp^2(1-2x)^2}{12\xi(x)^2} + \frac{\kperp^4(x(1-x))^2}{6 \xi(x)^3} \right) \nn\\
    &= \frac{1}{4\pi} \intop_0^1\dd x\, \frac{x(1-x)(\kperp^4 x(1-x) + \kperp^2 +2)}{(\kperp^2 x(1-x) + 1)^3}.
\end{align}
We continue with the substitution
\begin{align}
    x - \frac{1}{2} = y, \qquad x(1-x) = -y^2 + \frac{1}{4},
\end{align}
leading to
\begin{align}
    \Omega(\kperp) &= \frac{1}{4\pi}\intop_{-1/2}^{1/2} \dd y\, \frac{y^4\kperp^4 - y^2\frac{1}{2}( \kperp^4 + 2\kperp^2 + 4 ) +\frac{1}{16}(\kperp^4 + 4\kperp^2 + 8)}{\left( -y^2\kperp^2 + \frac{\kperp^2}{4} + 1 \right)^3} \nn\\
    &=  \frac{1}{4\pi}\intop_{-1/2}^{1/2} \dd y \Big( - \frac{y^4}{\kperp^2 \left(y^2-\frac{1}{4}-\frac{1}{\kperp^2}\right)^3} \nn\\
    &\hphantom{=\frac{1}{4\pi}\intop_{-1/2}^{1/2} \dd y \Big(\ } + \frac{\kperp^4+2\kperp^2+4}{2\kperp^6}\frac{y^2}{\left(y^2-\frac{1}{4}-\frac{1}{\kperp^2}\right)^3} \nn\\
    &\hphantom{=\frac{1}{4\pi}\intop_{-1/2}^{1/2} \dd y \Big(\ } - \frac{\kperp^4+4\kperp^2+8}{16\kperp^6}\frac{1}{\left(y^2-\frac{1}{4}-\frac{1}{\kperp^2}\right)^3} \Big),
\end{align}
which reduces to the linear combination of the three elementary integrals
\begin{align}
    \intop_{-1/2}^{1/2}\dd y\,\frac{y^4}{\left(y^2-\frac{1}{4}-\frac{1}{\kperp^2}\right)^3} &= -\frac{1}{16}\left( \kperp^4 - 6\kperp^2 + \artanh(\tfrac{|\kperp|}{\sqrt{4+\kperp^2}})\frac{24|\kperp|}{\sqrt{4+\kperp^2}}\right), \\
    \intop_{-1/2}^{1/2}\dd y\,\frac{y^2}{\left(y^2-\frac{1}{4}-\frac{1}{\kperp^2}\right)^3} &= - \frac{\kperp^4(\kperp^2+2)}{4(4+\kperp^2)} + \artanh(\tfrac{|\kperp|}{\sqrt{4+\kperp^2}})\frac{2\kperp^3}{(4+\kperp^2)^{3/2}}, \\
    \intop_{-1/2}^{1/2}\dd y\,\frac{1}{\left(y^2-\frac{1}{4}-\frac{1}{\kperp^2}\right)^3} &= - \frac{\kperp^6(\kperp^2+10)}{(4+\kperp^2)^2} - \artanh(\tfrac{|\kperp|}{\sqrt{4+\kperp^2}})\frac{24\kperp^5}{(4+\kperp^2)^{5/2}}.
\end{align}
Collecting the coefficients in front of the $\artanh$ yields
\begin{align}
    &\frac{1}{2\pi}\frac{1}{\kperp^3(4+\kperp^2)^{5/2}}\frac{1}{8} \left( 6\kperp^2(4+\kperp^2)^2 + 4(\kperp^4+2\kperp^2+4)(4+\kperp^2) \right.\nn\\
    &\hphantom{\frac{1}{2\pi}\frac{1}{\kperp^3(4+\kperp^2)^{5/2}}\frac{1}{8}\Big(}\left. + 6\kperp^2(\kperp^4+4\kperp^2+8) \right) \nn\\
    &= -\frac{2}{2\pi}\frac{4+12\kperp^2+6\kperp^4+\kperp^6}{\kperp^3(4+\kperp^2)^{5/2}},
\end{align}
and the remaining terms sum up to
\begin{align}
    &\frac{1}{2\pi}\frac{1}{\kperp^2(4+\kperp^2)^2}\frac{1}{32} \left( (\kperp^4-6\kperp^2)(4+\kperp^2)^2 - 2(\kperp^4+2\kperp^2+4)(\kperp^2+2)(4+\kperp^2) \right.\nn\\
    &\hphantom{\frac{1}{2\pi}\frac{1}{\kperp^2(4+\kperp^2)^2}\frac{1}{32}\Big(} \left. + \kperp^2(\kperp^2+10)(\kperp^4+4\kperp^2+8) \right) \nn\\
    &= \frac{1}{2\pi}\frac{2+3\kperp^2+\kperp^4}{\kperp^2(4+\kperp^2)^2} .
\end{align}
Putting all together, the solution of the integral is
\begin{align}
    \Omega(\kperp) = \frac{1}{2\pi}\left(\frac{4+12\kperp^2+6\kperp^4+\kperp^6}{\kperp^3 \left(4+\kperp^2\right)^{5/2}} \ 2\,\mathrm{artanh}(\tfrac{|\kperp|}{\sqrt{4+\kperp^2}})
    - \frac{2+3\kperp^2+\kperp^4}{\kperp^2\left(4+\kperp^2\right)^2}\right).
\end{align}

\chapter{Event-by-event sampling of color charge densities}\label{appx:sampling-rho}

To generate a realization of the color charge distribution $\rho^a_{\A/\B}$ following the MV nuclear model (cf.~\cref{sec:MV-nucl-model}), we need a practical implementation for sampling from the Gaussian probability functional fixed by the one- and two-point functions.
A suitable procedure is discussed in~\cite{Ipp:2024ykh,Leuthner:2025vsd} where a slightly simplified model for the two-point function was used.%
\footnote{%
That model parametrized the transverse correlation via a Dirac-delta function.
Here, we keep the transverse correlations completely general as part of the correlation function $\Gamma_{\A/\B}$.}
Given the general form of the color charge correlator in \cref{eq:general-correlator-t(x)-t(y)},
\begin{align}
    \langle \rho^a_{\A/\B}(\xpm,\xperp) \rho^b_{\A/\B}(\ypm,\yperp) \rangle &= \delta^{ab}\, t_{\A/\B}(\xpm,\xperp)t_{\A/\B}(\ypm,\yperp) \Gamma_{\A/\B}(\xpm-\ypm,\xperp-\yperp), \tag{\ref{eq:general-correlator-t(x)-t(y)}}
\end{align}
we make the ansatz
\begin{align}\label{eq:rho-t-zeta}
    \rho^a_{\A/\B}(\xpm,\xperp) = t_{\A/\B}(\xpm,\xperp) \zeta^a_{\A/\B}(\xpm,\xperp),
\end{align}
where we introduce a new stochastic field $\zeta^a_{\A/\B}$.
We take $\zeta^a_{\A/\B}$ to follow a Gaussian probability functional, analogous to $\rho^a_{\A/\B}$ in the MV nuclear model, that is fixed by
\begin{align}
    \langle \zeta^a_{\A/\B}(\xpm,\xperp) \rangle &= 0, \\
    \label{eq:zeta-gamma-correlator}
    \langle \zeta^a_{\A/\B}(\xpm,\xperp) \zeta^b(\ypm,\yperp) \rangle &= \delta^{ab}\, \Gamma_{\A/\B}(\xpm-\ypm,\xperp-\yperp).
\end{align}
Essentially, we are factoring out the square roots of the single-nucleus envelopes $t_{\A/\B}$ and continue only with the correlation function $\Gamma_{\A/\B}$.
This is possible because the $t_{\A/\B}$ that enter the final correlator are already cleanly separated into $(\xpm,\xperp)$ or $(\ypm,\yperp)$ dependence and can be associated with one of the color charge densities on the left-hand side.
The new field $\zeta^a_{\A/\B}$ is related to a random noise field $\chi^a_{\A/\B}$ in momentum space
\begin{align}\label{eq:tzeta-tchi}
    \tilde\zeta^a_{\A/\B}(\pmp,\pperp) = \tilde\chi^a_{\A/\B}(\pmp,\pperp)\sqrt{\tilde{\Gamma}_{\A/\B}(\pmp,\pperp)},
\end{align}
where $\tilde{\Gamma}_{\A/\B}\geq0$ according to the positive semi-definiteness condition discussed in \cref{sec:pos-semi-definite-correlators}.
The random noise is characterized by a Gaussian distribution with
\begin{align}
    \langle \chi^a_{\A/\B}(\xpm,\xperp) \rangle &= 0, \\
    \langle \chi^a_{\A/\B}(\xpm,\xperp) \chi^b_{\A/\B}(\ypm,\yperp) \rangle &= \delta^{ab}\, \delta(\xpm-\ypm)\deltaperp(\xperp-\yperp),\\
    \langle \tilde\chi^a_{\A/\B}(\pmp,\pperp) \tilde\chi^b_{\A/\B}(\qmp,\qperp) \rangle &= (2\pi)^3\delta^{ab}\, \delta(\pmp+\qmp)\deltaperp(\pperp+\qperp).
\end{align}
We can now insert the Fourier transformation of \cref{eq:tzeta-tchi} into the left-hand side of \cref{eq:zeta-gamma-correlator} and check that this construction satisfies the right-hand side,
\begin{align}
    \langle \zeta^a_{\A/\B}(\xpm,\xperp) \zeta^b_{\A/\B}(\ypm,\yperp) \rangle &= \int \frac{\dd\pmp\,\dd^2\pperp\,\dd\qmp\,\dd^2\qperp}{(2\pi)^6}\, \ee^{-\ii(\pmp\xpm + \qmp\ypm)+\ii(\pperp\cdot\xperp+\qperp\cdot\yperp)} \nn\\
    &\times \langle\tilde\zeta^a_{\A/\B}(\pmp,\pperp)\tilde\zeta^b_{\A/\B}(\qmp,\qperp)\rangle \nn\\
    &= \int \frac{\dd\pmp\,\dd^2\pperp\,\dd\qmp\,\dd^2\qperp}{(2\pi)^6}\, \ee^{-\ii(\pmp\xpm + \qmp\ypm)+\ii(\pperp\cdot\xperp+\qperp\cdot\yperp)} \nn\\
    &\times \sqrt{\tilde\Gamma_{\A/\B}(\pmp,\pperp)\tilde\Gamma_{\A/\B}(\qmp,\qperp)}\, \langle\tilde\chi^a_{\A/\B}(\pmp,\pperp)\tilde\chi^b_{\A/\B}(\qmp,\qperp)\rangle \nn\\
    &= \int \frac{\dd\pmp\,\dd^2\pperp\,\dd\qmp\,\dd^2\qperp}{(2\pi)^6}\, \ee^{-\ii(\pmp\xpm + \qmp\ypm)+\ii(\pperp\cdot\xperp+\qperp\cdot\yperp)} \nn\\
    &\times (2\pi)^3 \delta^{ab} \delta(\pmp+\qmp)\deltaperp(\pperp+\qperp) \nn\\
    &\times \sqrt{\tilde\Gamma_{\A/\B}(\pmp,\pperp)\tilde\Gamma_{\A/\B}(\qmp,\qperp)} \nn\\
    &= \delta^{ab} \int \frac{\dd\pmp\,\dd^2\pperp}{(2\pi)^3}\, \ee^{-\ii\pmp(\xpm -\ypm)+\ii\pperp\cdot(\xperp-\yperp)} \nn\\
    &\times \sqrt{\tilde\Gamma_{\A/\B}(\pmp,\pperp)\tilde\Gamma_{\A/\B}(-\pmp,-\pperp)} \nn\\
    &= \delta^{ab} \int \frac{\dd\pmp\,\dd^2\pperp}{(2\pi)^3}\, \ee^{-\ii\pmp(\xpm -\ypm)+\ii\pperp\cdot(\xperp-\yperp)} \tilde\Gamma_{\A/\B}(\pmp,\pperp) \nn\\
    &= \delta^{ab} \Gamma_{\A/\B}(\xpm-\ypm,\xperp-\yperp). \label{eq:zeta-correl-res}
\end{align}
In the second-to-last line we used that $\Gamma_{\A/\B} \in \mathbb{R}$ so that $\tilde\Gamma_{\A/\B}\tilde\Gamma^*_{\A/\B} = |\tilde\Gamma_{\A/\B}|^2$ and evaluated the square root on the positive branch.
Finally, we plug \cref{eq:rho-t-zeta} into the color charge correlator in \cref{eq:general-correlator-t(x)-t(y)} and use \cref{eq:zeta-correl-res} to get
\begin{align}
    \langle \rho^a_{\A/\B}(\xpm,\xperp) \rho^b_{\A/\B}(\ypm,\yperp) \rangle &= t_{\A/\B}(\xpm,\xperp)t_{\A/\B}(\ypm,\yperp)\,\langle \zeta^a_{\A/\B}(\xpm,\xperp) \zeta^b_{\A/\B}(\ypm,\yperp) \rangle \nn\\
    &= \delta^{ab}\, t_{\A/\B}(\xpm,\xperp)t_{\A/\B}(\ypm,\yperp) \Gamma_{\A/\B}(\xpm-\ypm,\xperp-\yperp). 
\end{align}

Using the detour with the field $\zeta^a_{\A/\B}$, the steps to sample a charge distribution $\rho^a_{\A/\B}$ on a discretized spacetime lattice are as follows:
\begin{enumerate}
    \item Generate a random noise field $\chi^a_{\A/\B}$ for each color index $a$.
    \item Calculate the Fourier-transformed noise field $\tilde\chi^a_{\A/\B}$.
    \item Modulate $\tilde\chi^a_{\A/\B}$ with $\sqrt{\tilde\Gamma_{\A/\B}}$ to get $\tilde\zeta^a_{\A/\B}$ in \cref{eq:tzeta-tchi}.
    \item Fourier transform back to get $\zeta^a_{\A/\B}$.
    \item Shape the field $\zeta^a_{\A/\B}$ with the single-nucleus envelope $t_{\A/\B}$ to get $\rho^a_{\A/\B}$ in \cref{eq:rho-t-zeta}.
\end{enumerate}

\chapter[Numerics for gluon numbers and transverse energy]{Implementation details for gluon number and transverse energy integrals} \label{appx:numerics-dN}

This appendix serves as a reference for the numerical implementation of the integrals that were evaluated for the results in \cref{sec:results-gluon-numbers,sec:numerics-GBW}.
These results correspond to the momentum-space formulation of the (3+1)D dilute Glasma, where the gluon number distribution $\dd N/\dd^2\kperp\dd\Y$ and the transverse energy $\dd\Ep/\dd\Y$ are studied.
These quantities involve integrals that were solved numerically.
For one-dimensional integrals, the \emph{NIntegrate} routine from Mathematica%
\footnote{%
See \cref{appx:software} for a list of the references for the tools used.\label{fn:tools2}}
was used.
These cases are not discussed further in this appendix.
For multi-dimensional integrals, the CUBA\footref{fn:tools2} Monte Carlo integration library was used.
Specifically, its Mathematica interface was used to evaluate the concrete integrals listed below.
For all results, the targeted value for the relative error estimate returned by the integration routines was set to $0.1\%$.
We emphasize that all realizations of the parameters, as well as the individual evaluation points for $\Y$ and $\kperp$ for the plots, are obtained from statistically independent integrations.
Therefore, the number of Monte Carlo samples was different for each evaluation and was determined by the adaptive algorithms of the routines to be large enough to reach the targeted error.
Additionally, the integrations were parallelized across the parameter space to improve performance.

The integral for the gluon number distribution in the $\WSxi$ nuclear model is
\begin{align}
    \frac{\kperp^2}{\Sop} \frac{\dd N}{\dd^2\kperp\,\dd\Y} &= \frac{4 \gamma^2 g^6 \mu^4 N_c (N_c^2-1) a^2}{(2 \pi )^{10} R^2 Z_\perp^\between \ln(\ee^{a}+1)^2}\frac{1}{\sqrt{2 R^2-\xi_\A^2 \gamma^2}}\frac{1}{\sqrt{2 R^2- \xi_\B^2 \gamma^2}} \nn\\
    & \times \intop_0^\infty\dd u \intop_0^\infty \dd v \intop_0^{2\pi}\dd \varphi \intop_0^{2\pi}\dd\phi \, u\, v \, \Bigg( -4 \kperp^2 u^2 \sin ^2(\varphi )+ \kperp^2 v^2 \sin ^2(\phi ) \nn\\
    & \qquad + \left(-\kperp^2+u^2+u\,v \cos (\phi -\varphi )+\tfrac{v^2}{4}\right) \nn\\
    & \qquad \times \left(-\kperp^2+u^2-u\,v \cos (\phi -\varphi )+\tfrac{v^2}{4}\right) \Bigg)\nn\\
    &\times 
    \left(4 m^2+\kperp^2+2 |\kperp|\,u \cos (\varphi )+|\kperp|\,v \cos (\phi )+u^2+u\,v \cos (\phi -\varphi )+\tfrac{v^2}{4}\right)^{-1}\nn\\
    &\times
    \left(4 m^2+\kperp^2-2 |\kperp|\,u \cos (\varphi )-|\kperp|\,v \cos (\phi )+u^2+u\,v \cos (\phi -\varphi )+\tfrac{v^2}{4}\right)^{-1}\nn\\
    &\times
    \left(4 m^2+\kperp^2+2 |\kperp|\,u \cos (\varphi )-|\kperp|\,v \cos (\phi )+u^2-u\,v \cos (\phi -\varphi )+\tfrac{v^2}{4}\right)^{-1}\nn\\
    &\times
    \left(4 m^2+\kperp^2-2 |\kperp|\,u \cos (\varphi )+ |\kperp|\,v \cos (\phi )+u^2-u\,v \cos (\phi -\varphi )+\tfrac{v^2}{4}\right)^{-1} \nn\\
    &\times \widetilde{\mathcal{W}}_\A^\xi(|\kperp| \ee^{-\Y} / \sqrt{2}, u)\, \widetilde{\mathcal{W}}_\B^\xi(|\kperp| \ee^{\Y}/\sqrt{2},u). \label{eq:WS-dN-implementation}
\end{align}
Note that this expression uses the parametrization of the longitudinal correlation length with $\xi_{\A/\B}$ as introduced in \cref{sec:finite-longitudinal-correlations}.
Also, the longitudinal diameter $R_l$ and transverse radius $R_\perp$ are rewritten in terms of the Woods-Saxon radius $R$ as defined in \cref{sec:nuclear-envelopes}.

The shorthands in the last line are defined as
\begin{align}
    \widetilde{\mathcal{W}}^\xi_{\A/\B}(k^\mp,\kappaperp) &= \frac{1}{2\pi}\intop \dd\xpm\,\dd\ypm\,\dd^2\xperp\,\dd\kappa^\mp\, \ee^{\ii\kappa^\mp(\xpm-\ypm)-\ii\kappaperp\cdot\xperp} \nn\\
    &\times \exp(-\tfrac{\xi_{\A/\B}^2 R^2}{2R^2 - \xi_{\A/\B}^2\gamma^2}(k^\mp - \kappa^\mp)^2) \nn\\
    &\times \left( 1+ \exp(a\sqrt{\tfrac{2\gamma^2(\xpm)^2}{R^2} + \tfrac{\xperp^2}{R^2}}-a) \right)^{-1/2} \nn\\
    &\times \left( 1+ \exp(a\sqrt{\tfrac{2\gamma^2(\ypm)^2}{R^2} + \tfrac{\xperp^2}{R^2}}-a) \right)^{-1/2}.
\end{align}
The $\widetilde{\mathcal{W}}^\xi_{\A/\B}$ objects involve the Fourier transformation of the square roots of the Woods-Saxon envelopes.
They are performed via FFT routines on a grid.
For the result, the cylindrical symmetry of the envelopes is exploited and the transverse momentum argument is taken to align with the second component of the transverse momentum vector (along one dimension of the discretized grid),
\begin{align}
    \widetilde{\mathcal{W}}^\xi_{\A/\B}(k^\mp,\kappa) \coloneqq \widetilde{\mathcal{W}}^\xi_{\A/\B}(k^\mp,\kappaperp=\begin{psmallmatrix}
        0 \\ \kappa
    \end{psmallmatrix}).
\end{align}
This result is tabulated on a grid for $\kmp$ and $\kappa$.
Then, for the integration, the tabulated values for $\widetilde{\mathcal{W}}^\xi_{\A/\B}$ are linearly interpolated during the Monte Carlo sampling steps of the integrand.

The expression in \cref{eq:WS-dN-implementation} is only valid for correlation lengths that are smaller than the coherent limit $\xi_{\A/\B} < \sqrt{2}R/\gamma$.
When this limit is saturated, the integral over $\kappa^\mp$ for $\widetilde{\mathcal{W}}^\xi_{\A/\B}$ becomes trivial and the expression simplifies.

The integrals in \cref{eq:WS-dN-implementation} are parametrized via $u$, $v$, $\varphi$, and $\phi$.
Compared to the expression given in the main text in \cref{eq:dN-WS-zeta-final}, the original integrations over $\pperp$ and $\qperp$ are changed via variable transformations.
First, the integration variables are shifted,
\begin{align}
    \pperp' = \pperp - \kperp/2, \qquad \qperp' = -\qperp + \kperp /2.
\end{align}
Then, average and difference coordinates are introduced as
\begin{align}
    \uperp = (\pperp'+\qperp')/2, \qquad \vperp = (\pperp'-\qperp').
\end{align}
The two-dimensional integrations of $\uperp$ and $\vperp$ are then parametrized in polar coordinates.
Since the only dependence on the polar angles of these vectors is via inner products with $\kperp$, the polar angle $\varphi$ is defined as the angle between $\uperp$ and $\kperp$ and the polar angle $\phi$ is defined as the angle between $\vperp$ and $\kperp$.
As a result, the final expression in \cref{eq:WS-dN-implementation} is isotropic w.r.t.\ $\kperp$.
Note that these steps are only possible because the envelopes for nucleus $\A$ and $\B$ share the same cylindrical symmetry and there is no impact parameter.

The transverse energy (normalized to the transverse overlap area) for the $\WSxi$ nuclear model is obtained 
by integrating \cref{eq:WS-dN-implementation} over the modulus of $\kperp$ and multiplying by $2\pi$ to account for the trivial angular integration (cf.~\cref{sec:transverse-energy-mom}).
The integrand from \cref{eq:WS-dN-implementation} is used unchanged.

The integral for the transverse energy density in the GBW model is
\begin{align}
    \frac{\dd\ep}{\dd\Y}&=\frac{N_c^3 R^2}{\pi^3 2^{13} g^2 (N_c^2-1)} \intop_0^\infty\dd|\kperp|\intop_0^\infty\dd u \intop_0^\infty \dd v \intop_0^{2\pi}\dd \varphi \intop_0^{2\pi}\dd\phi \, u\, v \, \ee^{-R^2 u^2} \nn\\
    &\times \exp(-\tfrac{1}{2^4} (4 \kperp^2+v^2) \left(\tfrac{1}{Q_s^\A(\xB_\A)^2}+\tfrac{1}{Q_s^\B(\xB_\B)^2}\right)) \nn\\
    &\times \exp(-\tfrac{1}{2^4}4 |\kperp|\, v \cos(\phi) \left(\tfrac{1}{Q_s^\A(\xB_\A)^2}-\tfrac{1}{Q_s^\B(\xB_\B)^2}\right)) \nn\\
    &\times (Q_s^\A(\xB_\A) Q_s^\B(\xB_\B))^{-2} \nn\\
    &\times \left(4 \kperp^2 +4 |\kperp|\, v \cos (\phi )+v^2\right)^2 \left(4 \kperp^2-4 |\kperp|\, v \cos (\phi )+v^2\right)^2 \nn\\
    &\times \Bigg( -4 \kperp^2 u^2 \sin^2(\varphi )+ \kperp^2 v^2 \sin^2(\phi ) \nn\\
    &\qquad + \left(-\kperp^2+u^2+u\, v \cos(\phi -\varphi )+\tfrac{v^2}{4}\right) \left(-\kperp^2+u^2-u\, v \cos (\phi -\varphi )+\tfrac{v^2}{4}\right)\Bigg) \nn\\
    & \times \left(4 m^2+\kperp^2+2 |\kperp|\, u \cos(\varphi )+ |\kperp|\,v \cos(\phi )+u^2+u\, v \cos(\phi -\varphi )+\tfrac{v^2}{4}\right)^{-1} \nn\\
    &\times \left(4 m^2+\kperp^2-2 |\kperp|\, u \cos(\varphi )- |\kperp|\,v \cos(\phi )+u^2+u\, v \cos(\phi -\varphi )+\tfrac{v^2}{4}\right)^{-1} \nn\\
    &\times \left(4 m^2+\kperp^2+2 |\kperp|\, u \cos(\varphi )-|\kperp|\, v \cos(\phi )+u^2-u\, v \cos(\phi -\varphi )+\tfrac{v^2}{4}\right)^{-1} \nn\\
    & \times \left(4 m^2+\kperp^2-2 |\kperp|\,u \cos(\varphi )+|\kperp|\, v \cos(\phi )+u^2-u\, v \cos(\phi -\varphi )+\tfrac{v^2}{4}\right)^{-1}.
\end{align}
The definitions of $Q_s^{\A/\B}$ and $\xB_{\A/\B}$ are given in \cref{sec:numerics-GBW}.
To get from $\pperp$ and $\qperp$ in \cref{eq:dN-GBW-numerics} to the integrals over $|\kperp|$, $u$, $v$, $\varphi$, and $\phi$ the same substitutions as for \cref{eq:WS-dN-implementation} apply.

\chapter{GBW gluon number distribution for large nuclei}\label{appx:dN-1DGBW}

\emph{The calculation presented in this appendix was performed by an undergraduate student, \emph{Katharina Gaunersdorfer}, supervised by the author.}\newline

\noindent
In this appendix, we perform the integration for the gluon number distribution in the GBW nuclear model (cf.~\cref{sec:GBW-saturation-model}).
In particular, we consider the limit of large nuclei, where the gluon number distribution is given by \cref{eq:dNd2kdY-large-nucl-TMDs}, and the original GBW TMD in \cref{eq:GBW-TMD}.
The resulting expression is discussed in \cref{sec:numerics-GBW}.
The integral at hand reads
\begin{align}
   I &= \frac{16 N_c^3}{\pi g^2(N_c^2-1)}\intop\frac{\dd^2\pperp}{(2\pi)^2}\,  \frac{\pperp^2(\kperp-\pperp)^2}{Q_s^\A(\xB_\A)^2 Q_s^\B(\xB_\B)^2} \ee^{-\pperp^2 Q_s^\A(\xB_\A)^{-2} - (\kperp-\pperp)^2 Q_s^\B(\xB_\B)^{-2}}.
\end{align}
In the following, we use the shorthand symbols
\begin{align}
    Q_{\A/\B} \coloneqq Q_s^{\A/\B}(\xB_{\A/\B})
\end{align}
for the parametrized saturation scales of nucleus $\A$ and $\B$.
We start by completing the square for $\pperp$ in the exponent,
\begin{align}
    -\frac{\pperp^2}{Q_\A^2} - \frac{(\kperp-\pperp)^2}{Q_\B^2} = -\frac{\kperp^2}{Q_\A^2+Q_\B^2} - \frac{Q_\A^2+Q_\B^2}{Q_\A^2 Q_\B^2}\left(\pperp - \kperp \frac{Q_\A^2}{Q_\A^2+Q_\B^2} \right)^2,
\end{align}
such that we can pull an exponential factor out of the integral, leading to
\begin{align}
    I =  \frac{16 N_c^3}{\pi g^2(N_c^2-1)}\, \ee^{-\frac{\kperp^2}{Q_\A^2+Q_\B^2}} \intop\frac{\dd^2\pperp}{(2\pi)^2}\, \frac{\pperp^2(\kperp-\pperp)^2}{Q_\A^2 Q_\B^2}\, \ee^{-\frac{Q_\A^2+Q_\B^2}{Q_\A^2 Q_\B^2}\left(\pperp - \kperp \frac{Q_\A^2}{Q_\A^2+Q_\B^2} \right)^2}.
\end{align}
Next, we shift the integration variable by substituting
\begin{align}
    \qperp = \pperp - \kperp \frac{Q_\A^2}{Q_\A^2+Q_\B^2}.
\end{align}
Carrying this substitution through to the numerator yields
\begin{align}
    \pperp^2(\kperp-\pperp)^2 &= \left( \qperp + \kperp \frac{Q_\A^2}{Q_\A^2+Q_\B^2} \right)^2\left( \kperp -\qperp - \kperp \frac{Q_\A^2}{Q_\A^2+Q_\B^2} \right)^2 \nn\\
    &= \left( \qperp\cdot\kperp \frac{2 Q_\A^2 }{Q_\A^2+Q_\B^2}+\kperp^2\frac{Q_\A^4}{\left(Q_\A^2+Q_\B^2\right)^2}+\qperp^2 \right) \nn\\
    &\times\left( \kperp^2-2\qperp\cdot\kperp +\qperp^2 -\kperp^2\frac{2 Q_\A^2}{Q_\A^2+Q_\B^2}+\qperp\cdot\kperp\frac{2 Q_\A^2}{Q_\A^2+Q_\B^2}+\kperp^2\frac{Q_\A^4}{\left(Q_\A^2+Q_\B^2\right)^2} \right) \nn\\
    &= \qperp^4
    + \qperp^2 \kperp^2\frac{Q_\A^4+Q_\B^4}{\left(Q_\A^2+Q_\B^2\right)^2}
    - (\qperp\cdot\kperp)^2 \frac{4 Q_\A^2 Q_\B^2}{\left(Q_\A^2+Q_\B^2\right)^2}
    + \kperp^4\frac{Q_\A^4 Q_\B^4}{\left(Q_\A^2+Q_\B^2\right)^4} \nn\\
    &
    + (\qperp\cdot\kperp) \kperp^2 \frac{2 Q_\A^2 Q_\B^2 \left(Q_\B^2-Q_\A^2\right)}{\left(Q_\A^2+Q_\B^2\right)^3}
    + \qperp^2 (\qperp\cdot\kperp) \left(\frac{4 Q_\A^2}{Q_\A^2+Q_\B^2}-2\right).
\end{align}
Note that the only dependence on the angle between $\qperp$ and $\kperp$ is in the terms of the numerator.
Here, the terms proportional to $\qperp\cdot\kperp$ in the last line drop out of the angular integration.
Changing to polar coordinates,
\begin{align}
    q=|\qperp|, \qquad \phi = \angle(\qperp,\kperp),
\end{align}
the integral reads
\begin{align}
    I &= \frac{16 N_c^3}{\pi g^2(N_c^2-1)} \frac{\ee^{-\frac{\kperp^2}{Q_\A^2+Q_\B^2}}}{Q_\A^2 Q_\B^2} \frac{1}{(2\pi)^2} \intop_0^\infty \dd q\, q\,\ee^{-q^2\frac{Q_\A^2+Q_\B^2}{Q_\A^2 Q_\B^2}} \nn\\
    &\times \intop_0^{2\pi}\dd\phi\, \left(  q^4
    + q^2 \kperp^2\frac{Q_\A^4+Q_\B^4}{\left(Q_\A^2+Q_\B^2\right)^2}
    - q^2\kperp^2\cos^2(\phi) \frac{4 Q_\A^2 Q_\B^2}{\left(Q_\A^2+Q_\B^2\right)^2}
    + \kperp^4\frac{Q_\A^4 Q_\B^4}{\left(Q_\A^2+Q_\B^2\right)^4} \right) \nn\\
    &= \frac{16 N_c^3}{2\pi^2 g^2(N_c^2-1)} \frac{\ee^{-\frac{\kperp^2}{Q_\A^2+Q_\B^2}}}{Q_\A^2 Q_\B^2} \nn\\
    &\times \intop_0^\infty \dd q\, \ee^{-q^2\frac{Q_\A^2+Q_\B^2}{Q_\A^2 Q_\B^2}} \left(  q^5
    + q^3 \kperp^2 \frac{\left(Q_\A^2-Q_\B^2\right)^2}{\left(Q_\A^2+Q_\B^2\right)^2}
    + q\kperp^4\frac{Q_\A^4 Q_\B^4}{\left(Q_\A^2+Q_\B^2\right)^4} \right),
\end{align}
where we carried out the $\phi$-integration.
The terms proportional to $\cos^2(\phi)$ generated a factor $1/2$ compared to the others, such that we pulled out a common factor of $2\pi$.
To solve the remaining integrations, we use the standard formula
\begin{align}
    \intop_0^\infty\dd q\, q^n\,\ee^{-c q^2} = \frac{1}{2}\Gamma(\tfrac{1+n}{2})c^{-\frac{1+n}{2}}, \qquad n\in \mathbb{N}, \qquad c > 0,
\end{align}
where $\Gamma$ is the Euler-Gamma function.
Finally, the result is
\begin{align}
    I &= \frac{16 N_c^3}{2\pi^2 g^2(N_c^2-1)} \frac{\ee^{-\frac{\kperp^2}{Q_\A^2+Q_\B^2}}}{Q_\A^2 Q_\B^2} \nn\\
    &\times \left(
    \kperp^4\frac{Q_\A^6 Q_\B^6}{2 \left(Q_\A^2+Q_\B^2\right)^5}+\kperp^2\frac{Q_\A^4 Q_\B^4 \left(Q_\A^2-Q_\B^2\right)^2}{2 \left(Q_\A^2+Q_\B^2\right)^4}+\frac{Q_\A^6 Q_\B^6}{\left(Q_\A^2+Q_\B^2\right)^3} \right) \nn\\
    &= \frac{16 N_c^3}{(2\pi)^2 g^2(N_c^2-1)}\, \ee^{-\frac{\kperp^2}{Q_\Sigma^2}} \left( \kperp^4 \frac{ Q_\Pi^8}{Q_\Sigma^{10}}+\kperp^2 \frac{Q_\Delta^4 Q_\Pi^4}{Q_\Sigma^8}+\frac{2 Q_\Pi^8}{Q_\Sigma^6}\right),
\end{align}
where we introduced the shorthands
\begin{align}
    Q_\Sigma^2 = Q_\A^2 + Q_\B^2, \qquad Q_\Delta^2 = Q_\A^2-Q_\B^2, \qquad Q_\Pi^2 = Q_\A Q_\B.
\end{align}
\chapter{Software and tools}%
\label{appx:software}

A (non-exhaustive) list of programming languages, software, and tools that were used during the research endeavors presented in this thesis is given below:

\paragraph{Bash:}
\url{https://www.gnu.org/software/bash/}
\paragraph{Python:}
\url{https://www.python.org/}
\paragraph{Numpy:}
\url{https://numpy.org/}
\paragraph{Jupyter:}
\url{https://jupyter.org/}
\paragraph{Matplotlib:}
\url{https://matplotlib.org/}
\paragraph{Nvidia CUDA:}
\url{https://developer.nvidia.com/cuda-toolkit}
\paragraph{Numba:}
\url{https://numba.pydata.org/}
\paragraph{CuPy:}
\url{https://cupy.dev/}
\paragraph{Wolfram Mathematica:}
\url{https://www.wolfram.com/mathematica/}
\paragraph{CUBA MC integrator~\cite{Hahn:2004fe}:}
\url{https://github.com/JohannesBuchner/cuba}
\paragraph{Slurm workload manager:}
\url{https://slurm.schedmd.com/}
\paragraph{Apptainer:}
\url{https://github.com/apptainer/apptainer}

\subsubsection{Note on the use of AI tools:}
The author clarifies that no generative AI tools were used to produce the text or content of this thesis.
Still, AI-assisted checkers for spelling and grammar were used for correcting the English language.

\printbibliography

\end{document}